\documentclass{aa}  

\usepackage[version=3]{mhchem}   
\usepackage[nottoc, notlof, notlot]{tocbibind}   
\usepackage{enumitem}
\usepackage{graphicx}
\usepackage{xcolor}
\usepackage{colortbl}
\usepackage{lscape}
\usepackage{txfonts}
\begin{document} 

\newcommand{\Msol}{\ensuremath{M_{\odot}}}
\newcommand{\Lsol}{\ensuremath{L_{\odot}}}
\def\as {\ifmmode {\rlap.}$\,$''$\,$\! \else ${\rlap.}$\,$''$\,$\!$\fi}
\def\decsec {\ifmmode {\rlap.}$\,$^{\rm s}$\,$\! \else ${\rlap.}$\,$^{\rm s}$\,$\!$\fi}\def\decs {\ifmmode {\rlap.}$\,$^{\rm s}$\,$\! \else ${\rlap.}$\,$^{\rm s}$\,$\!$\fi}

   \title{The complex chemistry of hot cores in Sgr B2(N): \\ Influence of cosmic-ray ionization and thermal history}


  \author{M. Bonfand\inst{1}, A. Belloche\inst{1}, R.~T. Garrod\inst{2}\inst{3}, K. M. Menten\inst{1}, E. Willis\inst{2}, G. St\'{e}phan\inst{2}, H.~S.~P. M\"{u}ller\inst{4}
          }
  \authorrunning{M. Bonfand et al.}
  \titlerunning{Complex chemistry in Sgr B2(N)'s hot cores}
   \institute{\inst{1} Max-Planck-Institut f\"{u}r Radioastronomie, Auf dem H\"{u}gel 69, Bonn, Germany \\
              \email{bonfand@mpifr-bonn.mpg.de}  \\
              \inst{2} Departments of Chemistry, University of Virginia, Charlottesville, VA 22904, USA \\
              \inst{3} Departments of Astronomy, University of Virginia, Charlottesville, VA 22904, USA \\
              \inst{4} I. Physikalisches Institut, Universit\"{a}t zu K\"{o}ln, Z\"{u}lpicher Str. 77, 50937 K\"{o}ln, Germany
             }

   \date{}


  \abstract
  {As the number of complex organic molecules (COMs) detected in the interstellar medium increases, it becomes ever more important to place meaningful constraints on the origins and formation pathways of such chemical species. The molecular cloud Sagittarius~B2(N) is host to several hot molecular cores in the early stage of star formation, where a great variety of COMs are detected in the gas phase. Because of its exposure to the extreme conditions of the the Galactic center (GC) region, Sgr~B2(N) is one of the best targets to study the impact of environmental conditions on the production of COMs.}
 {Our main goal is to characterize the physico-chemical evolution of Sgr~B2(N)'s sources in order to explain their chemical differences and constrain their environmental conditions.}  
 {The chemical composition of Sgr~B2(N)'s hot cores, N2, N3, N4, and N5 is derived by modeling their 3~mm emission spectra extracted from the EMoCA imaging spectral line survey performed with the Atacama Large Millimeter/submillimeter Array (ALMA). We derive the density distribution in the envelope of the sources based on the masses computed from the ALMA dust continuum emission maps. We use the radiative transfer code RADMC-3D to compute temperature profiles and infer the current luminosity of the sources based on the COM rotational temperatures derived from population diagrams. We use published results of 3D radiation-magnetohydrodynamical (RMHD) simulations of high-mass star formation to estimate the time evolution of the sources properties. We employ the astrochemical code MAGICKAL to compute time-dependent chemical abundances in the sources and investigate how physical properties and environmental conditions influence the production of COMs.}
   {The analysis of the abundances of 11 COMs detected toward Sgr~B2(N2-N5) reveals that N3 and N5 share a similar chemical composition while N2 differs significantly from the other sources. We estimate the current luminosities of N2, N3, N4, and N5 to be 2.6$\times$10$^5$~$\Lsol$, 4.5$\times$10$^4$~$\Lsol$, 3.9$\times$10$^5$~$\Lsol$, and 2.8$\times$10$^5$~$\Lsol$, respectively. We find that astrochemical models with a cosmic-ray ionization rate of 7$\times$10$^{-16}$~s$^{-1}$ best reproduce the abundances with respect to methanol of ten COMs  observed toward Sgr~B2(N2-N5). We also show that COMs still form efficiently on dust grains with minimum dust temperatures in the prestellar phase as high as 15~K, but that minimum temperatures higher than 25~K are excluded.}
   {The chemical evolution of Sgr~B2(N2-N5) strongly depends on their physical history. A more realistic description of the hot cores' physical evolution requires a more rigorous treatment with RMHD simulations tailored to each hot core.}


   \maketitle

\section{Introduction}
        \label{section-intro}

To date more than 200 molecules\footnote{see https://cdms.astro.uni-koeln.de/classic/molecules} have been discovered in the interstellar medium (ISM) or in circumstellar envelopes of evolved stars \citep[see, e.g.,][]{mcguire2018}. These molecules are mostly organic (that is they contain at least one atom of carbon) and among them about one third are composed of six or more atoms and are considered as complex in terms of astrochemistry \citep{herbst2009}. As the inventory of complex organic molecules (COMs) expands, they have attracted a lot of attention, in particular recently due to their possible link to amino acids, the chemical building blocks of proteins which constribute to life as we know it on Earth. Numerous extraterrestrial amino acids have been discovered in meteorites \citep{botta2002} and one in comets \citep{altwegg2016}. However, the simplest amino acid, glycine (\ce{NH2CH2COOH}), has not yet been detected in the ISM. Nonetheless, many COMs have been discovered toward the warm and dense regions associated with high-mass star formation, known as hot cores, by investigating their rotational spectrum at (sub)millimeter wavelengths. Despite these numerous detections, the precise origins of these complex species as well as the mechanisms leading to their formation are still subject to debate. The rise of astrochemical models solving coupled kinetic rate equations built from networks of thousands of chemical reactions has allowed significant progress in our understanding of the complex chemistry in the ISM. For instance, it has been shown that dust-grain chemical processes play an important role in the formation of COMs \citep{garrod2006, garrod2007, garrod2008, herbst2009}, the abundances of some of which cannot currently be explained by gas-phase chemistry. Therefore, it is crucial to better understand the chemical processes that take place in the interstellar ices at early time and low temperature to explain the later high gas-phase chemical abundances observed toward regions forming high-mass stars. The numerous COMs detected toward hot molecular cores, right after desorption from the dust grains into the gas phase, may thus be used to trace the physical conditions of their parent molecular cloud since they hold information on the collapse history.

The Galactic center (GC) cloud Sagittarius B2 (Sgr~B2), located at $\sim$~100~pc in projection from the central supermassive black hole Sgr~A$^*$, exhibits one of the richest molecular inventory observed to date and thus appears to be an excellent target to search for COMs and probe interstellar chemistry. Many of the first detections of interstellar molecules at radio and (sub)mm wavelengths were made toward Sgr~B2 \citep{menten2004}, in particular some of the most complex species detected in the ISM, such as acetic acid \ce{CH3COOH} \citep{mehringer1997}, glycolaldehyde \ce{CH2(OH)CHO} \citep{hollis2000}, acetamide \ce{CH3C(O)NH2} \citep{hollis2006}, aminoacetonitrile \ce{NH2CH2CN} \citep{belloche2008}, and N-methylformamide \ce{CH3NHCHO} \citep{belloche2017}. With a mass of $\sim$ 10$^7 \Msol$ in a diameter of $\sim$ 40~pc \citep{lis1990}, Sgr~B2 is also one of the most prominent regions forming high-mass stars in our Galaxy. Its two main centers of star-formation activity, Sgr~B2(N)orth and Sgr~B2(M)ain both host a cluster of (ultra)compact HII regions \citep{mehringer1993, gaume1995, depree1998, depree2015} and several 6.7~GHz class II methanol masers \citep{caswell1996}, which uniquely mark the locations of massive young (proto-)stellar objects.

Because of its exceptional characteristics and is strong, ongoing star-formation activity \citep[e.g.,][]{bonfand2017, sanchezmonge2017, ginsburg2018}, Sgr~B2 stands out from the other clouds located in the central molecular zone (CMZ; inner 200--500 pc) of the Galaxy. Indeed, recent studies find that despite its large reservoir of dense gas, the CMZ appears to be overall deficient in star formation compared to other high-mass star-forming regions in the galactic disk with similar density and age \citep[e.g.,][]{rathborne2014, kruijssen2014, kauffmann2017, walker2018}. This apparent inactivity suggests that the critical density for star formation in the CMZ is $\sim$~10$^7$~cm$^{-3}$ \citep[see, e.g.,][]{rathborne2014, kruijssen2014, kauffmann2017b, ginsburg2018, barnes2019} and thus much higher than in the rest of the galactic disk \citep[$\sim$~10$^4$~cm$^{-3}$, see, e.g.,][]{lada2010, lada2012}. The environmental conditions in the CMZ are also known to be extreme compared to the rest of the galactic disk, in particular with a stronger interstellar radiation field \citep[ISRF,][]{lis2001, clark2013}, and higher cosmic-ray fluxes \citep{oka2005, vandertak2006, yusef2007, clark2013}. The exceptional physical conditions of the CMZ provide an excellent test of current models of high-mass star formation and can advance our understanding of star-formation mechanisms within extreme environments. Due to its proximity to the GC, Sgr~B2 provides us with an interesting case study to investigate how extreme physical and environmental factors influence the high-mass star formation process, whether they tend to enhance or suppress star formation activity.

In a previous analysis of the 3~mm imaging line survey EMoCA (Exploring Molecular Complexity with ALMA) we reported the detection of three new hot cores, Sgr~B2(N3), N4, and N5 \citep{bonfand2017}. These new hot cores are all associated with 6.7 GHz class II methanol masers and Sgr~B2(N5) also with an ultra-compact HII (UCHII) region. From the analysis of the 3~mm emission spectra observed toward the new hot cores, we derived their chemical composition, revealing the presence of up to 24 different species of which about half are complex \citep{bonfand2017}. Because the spectrum observed toward the main hot core Sgr~B2(N1) is close to the confusion limit and severely affected by line blending, here we focus our analysis on the four other hot cores Sgr~B2(N2-N5). That all of these hot cores most probably originate from the same cloud material, with similar chemical and physical properties (e.g., initial chemical abundances, gas-to-dust mass ratio), makes comparing them with each other particularly meaningful. By comparing the predictions of astrochemical models with the observations, we want to identify the most efficient pathways leading to the formation of COMs (see for instance \citet{belloche2017} for the case of N-methylformamide) and explore the impact of the environment on the production of these complex species.

The extreme environmental conditions of the CMZ most likely also affect the chemistry of Sgr~B2(N)'s hot cores. In particular, in dense regions shielded from direct UV irradiation, cosmic rays play an important role in the gas-phase chemistry as they represent the main source of ionization. The direct ionization of atomic and molecular hydrogen results in the formation of secondary electrons (see Eq.~\ref{eq-CR-reaction1}), which can cause additional ionization \citep{goldsmith1978}. In addition, collisions of H and \ce{H2} with secondary electrons inside dense clouds lead to the production of ultra-violet (UV) radiation \citep{prasad1983}. This cosmic ray-induced UV field can dissociate molecules, both in the gas phase and on dust-grain surfaces inside the clouds, where external UV photons cannot penetrate. 

The so-called cosmic-ray ionization rate (CRIR), $\zeta^{\rm H_2}$, describes the total rate of ionization per \ce{H2} molecule per second, including secondary ionizations. The first theoretical determination of the CRIR was made by \citet{hayakawa1961} and revised later by \citet{spitzer1968}. The latter derived a lower limit of $\zeta^H$~=~6.8$\times$10$^{-18}$~s$^{-1}$, which corresponds to $\zeta^{H_2}$~$\sim$1.3$\times$10$^{-17}$~s$^{-1}$ according to the simple approximation $\zeta^{H_2}$~$\sim$~2$\zeta^H$ \citep{glassgold1974} commonly used for dense gas in which most hydrogen is in its molecular form. This value of $\zeta^{\rm H_2}$ is often referred to as the standard CRIR. An accurate measurement of the CRIR is difficult to obtain through direct detection because low-energy particles ($<$ 1~GeV) are prevented from entering the heliosphere by the solar wind, making them not directly measurable from Earth \citep{parker1958}. However, the CRIR can be estimated with astrochemical models, by analyzing specific molecules whose formation and destruction processes are driven by cosmic rays. Because of its relatively simple chemistry, the \ce{H3}$^+$ ion is commonly used as an indirect probe for constraining the CRIR: 
\begin{equation}
\label{eq-CR-reaction1}
\mathrm{H}_2 \: + \mathrm{CR} \rightarrow \mathrm{H}_2^+ \: + e^-
\end{equation}
\begin{equation}
\label{eq-CR-reaction2}
\mathrm{H}_2^+ \: + \mathrm{H}_2 \rightarrow \mathrm{H}_3^+ \: + \mathrm{H} 
\end{equation}
Recently, \citet{lepetit2016} used the Meudon photodissociation region (PDR) code to reproduce the large \ce{H3}$^+$ column densities observed in the diffuse gas component of the line of sight to the CMZ ($n$~=~100~cm$^{-2}$, $T$~=~200--500~K). They derived a CRIR of $\zeta^{H_2}$~$\sim$1 -- 11$\times$10$^{-14}$~s$^{-1}$ that is in line with the values that \citet{indriolo2015} obtain from an analysis of data for OH$^+$, \ce{H2O}$^+$, and \ce{H3O}$^+$ observed by Herschel for various clouds in the CMZ. However, it has been shown that low energy cosmic-ray particles can penetrate diffuse clouds easily but that they lose energy on their path toward the center of denser clouds, as they ionize and excite the matter they travel through \citep{umebayashi1981}. Many previous studies have sought to quantify the attenuation of the cosmic-ray flux in dense clouds \citep[see, e.g.,][]{padovani2009, rimmer2012, neufeld2017}. However, despite the numerous attempts to constrain the CRIR in diverse regions of the ISM, huge uncertainties still remain. Therefore, the CRIR is usually treated as a free parameter in astrochemical models, ranging from $\zeta^{\rm H_2}$~$\sim$~10$^{-17}$~s$^{-1}$ to 10$^{-14}$~s$^{-1}$. The impact of variations in the CRIR value on calculated chemical abundances can then be used to identify functional probes for the CRIR in dense clouds and determine the ionization rate that best reproduces the observations \citep[see, e.g.,][]{albertsson2018, allen2018}.

In this paper we use the chemical kinetics code, MAGICKAL \citep{garrod2013}, coupled with the radiative transfer code RADMC-3D \citep{dullemond2012} and the published results of radiation-magnetohydrodynamical (RMHD) simulations of high-mass star formation \citep{peters2011} to model the physico-chemical evolution of the four dense molecular cores, Sgr~B2(N2-N5). We investigate the impact of physical properties and environmental conditions on the formation of specific COMs observed toward Sgr~B2(N2-N5) in order to constrain the CRIR and the minimum dust temperature reached during the prestellar phase that best characterize Sgr~B2(N). 

The article is structured as follows: the observations and methods of analysis are presented in Sect.~2. In Sect.~3 we determine the physical properties of Sgr~B2(N2-N5) and build up the physical models used as inputs for the chemical modeling discussed in Sect.~4. The results of our standard models are presented in Sect.~5. In Sect.~6 we discuss the impact of varying the minimum dust temperature and the CRIR on the production of COMs. We compare the model results to the observations to constrain the environmental conditions in Sgr~B2(N2-N5). Finally our main results are summarized in Sect.~7.

\section{Observations and methods of analysis}
         \label{section-obs}

         \subsection{The EMoCA survey}
                     \label{section-EMoCA}

The EMoCA imaging spectral line survey was conducted with ALMA in its cycles 0 and 1 at high angular resolution ($\sim$1.6$\arcsec$) with a sensitivity of $\sim$3~mJy/beam per 488~kHz (1.7 to 1.3~km~s$^{-1}$) wide channel. The survey is divided into five spectral setups covering the frequency range between 84.1~GHz and 114.4~GHz (Table~\ref{TAB-appendix-data-redu}). At these frequencies, the size (half power beam width -- HPBW) of the primary beam of the 12~m antennas varies between 69$\arcsec$ at 84~GHz and 51$\arcsec$ at 114~GHz \citep{remijan2015}. The phase center is located half way between the two main hot molecular cores Sgr~B2(N1) and N2 ($\alpha _{\rm J2000}$  = 17$^h$47$^m$19$\decsec$87, $\delta _{\rm J2000}$ = -28$^{\rm o}$22$'$16$\as$0). A detailed description of the observations, the calibration, and the imaging procedures is presented in \citet{belloche2016}.

         \subsection{Data reduction}
                     \label{section-data-reduction}

\cite{belloche2016} used the spectra observed toward the main hot cores Sgr~B2(N1-N2) to split the line and continuum emission into separated datacubes. In comparison, the spectra observed toward the hot cores Sgr~B2(N3-N5) contain significantly less emission lines \citep[that is more emission-free channels,][]{bonfand2017}. We use the spectra extracted toward their peak positions to split the spectral lines from the continuum emission more accurately, using the CLASS\footnote{See http://www.iram.fr/IRAMFR/GILDAS} software. In each spectral window of each setup, a first-order baseline is subtracted to the channels that are free of strong line emission to separate the spectral lines from the continuum. The noise level measured in the new continuum-subtracted spectra is listed in Table~\ref{TAB-appendix-data-redu} along with the values measured by \citet{belloche2016} in the channel maps of the full field of view for comparison.

         \subsection{Radiative transfer modeling of the line survey}
                     \label{section-modelling-obs}

In order to derive the chemical composition of Sgr~B2(N3-N5) we perform a radiative transfer modeling of the EMoCA line survey. Given the high densities observed in the hot cores \citep[$\sim$~10$^7$~cm$^{-3}$,][]{bonfand2017}, collisions in the gas phase are frequent enough for the local thermodynamic equilibrium (LTE) approximation to be valid. We use \textit{Weeds} \citep{maret2011}, which is part of the CLASS software, to produce LTE synthetic spectra that we compare to the observed continuum-subtracted spectra, after correction for the primary beam attenuation. \textit{Weeds} solves the radiative transfer equation, taking into account the finite angular resolution of the interferometer, the line opacity, line blending, and the continuum background. The spectroscopic predictions used to model the spectra mainly originate from the CDMS \citep[Cologne Database for Molecular Spectroscopy,][]{endres2016} and JPL \citep[Jet Propulsion Laboratory,][]{pearson2010} catalog. They are the same as in \citet{belloche2016, belloche2017} and \citet{muller2016}, except for ethanol, methanol, and methyl cyanide, for which we use new CDMS predictions. Each molecule is modeled separately, adjusting the following parameters: column density, rotational temperature, angular size of the emitting region (assumed to be Gaussian), velocity offset with respect to the systemic velocity of the source, and linewidth (full width at half maximum -- FWHM). For each molecule, a population diagram is plotted to derive the rotational temperature and 2D Gaussians are fitted to integrated intensity maps of unblended transitions to measure the size of the emitting region. 1D-Gaussian fits to emission lines are used to derive the linewidth and velocity offset. The column density is adjusted manually until a good fit to the data is obtained. Finally, the contribution from all species is added linearly to build up the complete synthetic spectrum. More details about the methods of analysis can be found in \citet{bonfand2017}.

\section{Physical properties of Sgr~B2(N)'s hot cores}
        \label{section-evolution-HCs}

In order to model the time-dependent chemical composition of Sgr~B2(N)'s sources, we need to determine their physical properties and characterize their time evolution, from the cold ambient cloud phase (that is prior to the free-fall collapse of the cloud) to the dense and warm hot core phase. In this section we derive the physical properties of the sources (\ce{H2} column density, mass, density, size, temperature, and luminosity) based on the observations. In particular, temperature and density will be used to derive the physical profiles used as inputs for the chemical modeling (Sect.~\ref{section-modelling}).

\subsection{\ce{H2} column densities and masses}
           \label{section-masses}

We derive \ce{H2} column densities from the dust thermal emission measured in the ALMA continuum emission maps toward each source, using the following equation \citep{bonfand2017}:
\begin{equation}
\label{eq-nh2}
N_{\rm H_2} = -\frac{1}{\mu_{\rm H_2}  m_{\rm H}  \kappa_{\nu}} \times \ln \left( 1 - \frac{S_{\nu}^{\rm beam}}{\Omega_{\rm beam}   B_{\nu}(T_d)}  \right)
\end{equation}
with $\mu_{\rm H_2}$ = 2.8 the mean molecular weight per hydrogen molecule \citep{kauffmann2008}, $m_{\rm H}$ the mass of atomic hydrogen, $\Omega_{\rm beam} =  \frac{\pi}{4 \ln 2} \times HPBW_{\rm max} \times HPBW_{\rm min}$ the solid angle of the synthesized ALMA beam, $B_{\nu}(T_{\rm d}$) the Planck function at the dust temperature $T_{\rm d}$ = 150~K (assuming $T_{\rm d}$ $\sim$ $T_{\rm g}$, the gas kinetic temperature, see Sect.~\ref{section-obs-constraints}), $\kappa_{\nu}$ the dust absorption coefficient per unit of mass density of gas (in cm$^2$~g$^{-1}$), and $S_{\nu}^{\rm beam}$ the peak flux density measured in Jy/beam toward each hot core in the ALMA continuum emission maps, after correction for the primary beam attenuation and the free-free contribution \citep{bonfand2017}. The dust mass opacity is given by the power law:
\begin{equation}
\label{eq-kappa}
\kappa_{\nu} = \kappa_0 \left( \frac{\nu}{\nu_0}  \right) ^{\beta},
\end{equation}
with $\kappa_0$ the reference dust mass absorption coefficient at the frequency $\nu_0$. In our previous analysis \citep{bonfand2017} we derived a dust emissivity exponent $\beta$ of 1.2~$\pm$~0.1 from the combined analysis of our ALMA data and the dust continuum emission map obtained with the submillimeter array (SMA) at 343~GHz \citep{qin2011}. According to the simulations of dust coagulation in cold dense cores conducted by \citet{ossenkopf1994}, a dust emissivity exponent of $\beta$~=~1.2 suggests an intermediate dust opacity spectrum between the models of dust grains without ice mantles and those with thin ice mantles (see Fig.~\ref{FIG-appendix-kappa-opacities}). Here we adopt the model with grains without ice mantle and for gas densities of 10$^6$~cm$^{-3}$, which corresponds to a dust emissivity exponent of $\beta$~=~1.3, with $\kappa_0 = 0.0199$~cm$^2$~g$^{-1}$ (of gas) at $\nu_0$~=~230~GHz (that is $\lambda_0 = 1.3$~mm, see Fig.~\ref{FIG-appendix-kappa-opacities}). 

Table~\ref{TAB-HCs-properties} shows the \ce{H2} column densities calculated at $T_{\rm d}$ = 150~K for Sgr~B2(N2-N5). In the case of Sgr~B2(N3), which is not detected in the ALMA continuum maps at 3~mm \citep{bonfand2017}, we use the peak flux density measured by \citet{sanchezmonge2017} at 242~GHz in ALMA data obtained at 0.4$\arcsec$ resolution.

For each hot core, we derive the gas mass of the envelope contained in the 1.6$\arcsec$ ALMA mean synthesized beam (or 0.4$\arcsec$ for Sgr~B2(N3)), from the \ce{H2} column density as follows \citep{hildebrand1983} :
\begin{equation}
\label{eq-mass}
M_{\rm g} (R) = N_{\rm H_2} \: \Omega_{\rm beam} \: \mu_{\rm H_2} \: m_{\rm H} \: D^2 
\end{equation}
with $D$ = 8.34~kpc the distance to Sgr~B2(N) from the Sun \citep{reid2014} and $R$ the radius of the mean ALMA synthesized beam: 
\begin{equation}
\label{eq-R-beam}
R = D \sqrt{\frac{\Omega_{\rm beam}}{\pi}} =  \frac{D \, \theta_b}{2 \sqrt{\ln 2}} \:  
\end{equation} 
with $\theta_{\rm b}$ = 1.6$\arcsec$ the mean synthesized beam of the ALMA observations (or 0.4$\arcsec$ for Sgr~B2(N3)). The resulting masses are given in Table~\ref{TAB-HCs-properties}.

\begin{table*}[!t]
\begin{center}
  \caption{\label{TAB-HCs-properties} Physical properties of four hot cores embedded in Sgr~B2(N).} 
  \vspace{-4mm}
  \setlength{\tabcolsep}{0.8mm}
  \begin{tabular}{lccrclccclll}
  \hline
Hot & $R$ \tablefootmark{a}  & $N_{\rm H_2}$\tablefootmark{b} & \multicolumn{1}{c}{$M_{\rm g}$ ($R$)\tablefootmark{c}} & $r_0$\tablefootmark{d} & \multicolumn{1}{c}{$T_0$\tablefootmark{e}} & $\rho_0$\tablefootmark{f} & $n_0$\tablefootmark{g}  & \multicolumn{1}{c}{$L_{\rm tot}$\tablefootmark{h}} & \multicolumn{1}{c}{$t_{\rm source}$\tablefootmark{i}} & \multicolumn{1}{c}{$M_*$\tablefootmark{j}} & $r_{\rm init}$\tablefootmark{k}   \\
  core         & (10$^3$ au)  & (10$^{24}$cm$^{-2}$) & \multicolumn{1}{c}{($\Msol$)}  & (10$^3$ au)  & \multicolumn{1}{c}{(K)} & (10$^{-19}$g cm$^{-3}$) & (10$^7$ cm$^{-3}$)  & (10$^5$ $\Lsol$)  & (10$^5$ yr)  & \multicolumn{1}{c}{($\Msol$)} & (10$^6$ au)  \\
  \hline
  \hline 
N2 &  8.01(0.50) & 1.42(0.39)   & 151(42) & 6.01(1.50)  & 150 ($^{+90}_{-20}$)     & 3.20(1.53) & 1.37(0.65)  &  2.63($^{+3.13}_{-1.58}$)  & 2.10($^{+2.10}_{-1.60}$)   & 42($^{+20}_{-15}$) & 1.65(0.67)  \\
N3\tablefootmark{*} &  2.00(1.25) & 0.90(0.17)   &   6(1)  & 2.00(0.50)  & 150($^{+20}_{-5})$  & 5.27(2.96)  & 2.25(1.26)  & 0.45($^{+0.16}_{-0.20}$)      & 0.20($^{+0.05}_{-0.06}$) & 18($^{+3}_{-6}$) & 0.77(0.55) \\
N4 &  8.01(0.50) & 0.41(0.04)   &  43(4)  & 5.01(1.50)  &  150($^{+40}_{-5})$ & 1.21(0.57) & 0.52(0.24)  & 3.92($^{+4.17}_{-2.62}$)   & 3.06($^{+2.34}_{-2.26}$)   & 51($^{+25}_{-22}$) & 0.72(0.52)  \\
N5 &  8.01(0.50) & 0.71(0.16)   & 72(17)  & 5.01(2.00)  &  150($^{+30}_{-5})$ & 2.09(1.31)  & 0.89(0.56)  & 2.82($^{+1.77}_{-1.52}$) & 2.25($^{+1.25}_{-1.45}$) & 44($^{+12}_{-14}$) & 1.04(0.78) \\
    \hline
\end{tabular}
\end{center}
\vspace{-4mm}
\tablefoot{The uncertainties are given in parentheses. 
\tablefoottext{a}{Radius of the mean synthesized beam of the ALMA data (Eq.~\ref{eq-R-beam})}
\tablefoottext{b}{Average peak H$_2$ column density for a mean synthesized beam radius $R$ (Eq.~\ref{eq-nh2}). The uncertainty corresponds to the standard deviation weighted by the error on the ALMA peak flux density, $S_{\nu}^{\rm beam}$, and on the correction factor for the free-free emission \citep{bonfand2017}.}
\tablefoottext{c}{Gas mass of the envelope contained in $R$ (Eq.~\ref{eq-mass}).}
\tablefoottext{d}{Radius of the COM emission region (Sect.~\ref{section-obs-constraints}).} 
\tablefoottext{e}{Excitation temperature derived from population diagrams (Sect.~\ref{section-obs-constraints}).}
\tablefoottext{f}{Dust mass density at the radius $r_0$ assuming a standard gas-to-dust mass ratio of 100 (Eq.~\ref{eq-rho0-3}).}
\tablefoottext{g}{Density of total hydrogen ($n_{\rm H}$~=~$n$(H)+2$n$(H$_2$)) at $r$~=~$r_0$ (Eq.~\ref{physical-profiles-eq6}).}
\tablefoottext{h}{Estimated current luminosity of the source (Sect.~\ref{section-radmc3d}).}
\tablefoottext{i}{Age of the source estimated from its current luminosity (Sect.~\ref{section-evolution-L-HCs}, Fig.~\ref{FIG-L-evolution}c).}
\tablefoottext{j}{Mass of the protostar derived from its current luminosity (Sect.~\ref{section-evolution-L-HCs}, Fig.~\ref{FIG-appendix-M-L-relation}).}
\tablefoottext{k}{Initial radius of the envelope of the source (Sect.~\ref{section-collapse-phase}).}
\tablefoottext{*}{In the case of Sgr~B2(N3), which is not detected in the EMoCA continuum data at 3~mm, we used the peak flux density reported by \citet{sanchezmonge2017}, measured at 1.3~mm in an ALMA synthesized beam of 0.4$\arcsec$ (that is $R$~=~2003~au), to derive the H$_2$ column density and mass.}
}

\end{table*}

\subsection{Source sizes and temperatures}
\label{section-obs-constraints}

In our previous analysis of Sgr~B2(N) we derived the size of the COM emission region, $\theta_{\rm s}$, for the hot cores Sgr~B2(N3-N5) \citep{bonfand2017}. For each source we selected in the observed spectra the transitions that have a high signal-to-noise ratio, are well reproduced by the LTE model, and are not severely contaminated by other species. We fitted 2D Gaussians to their integrated intensity maps to derive the emission size. We derived deconvolved sizes of 1.0 $\pm$ 0.3$\arcsec$ and 1.0 $\pm$ 0.4$\arcsec$ for Sgr~B2(N4) and N5 respectively, which are slightly smaller than the size of Sgr~B2(N2) \citep[$\sim$ 1.2$\arcsec$, with values ranging between 0.8$\arcsec$ and 1.5$\arcsec$][]{belloche2016, belloche2017}. Sgr~B2(N3) is more compact, with an emission size of 0.4$\arcsec$. For each hot core the radius of the emitting region, $r_0$, is computed using Eq.~\ref{eq-R-beam} and is given in Table~\ref{TAB-HCs-properties}. 

From the COM emission \citet{bonfand2017} also derived excitation temperatures by plotting population diagrams based on the transitions that are well detected and not severely contaminated by lines from other species. At high densities, where the LTE approximation is valid, the level populations can be described by a single excitation temperature, $T_{\rm rot}$. We found that $T_{\rm rot}$ varies from $\sim$~145~K to 190~K for Sgr~B2(N3-N5). The temperatures for Sgr~B2(N2) were determined by \citet{belloche2016, belloche2017} and \citet{muller2016}. We assume that the excitation temperature is equal to the kinetic temperature and we adopt a gas temperature $T_0$ of 150~K at the radius $r_0$ (see Table~\ref{TAB-HCs-properties}).

\subsection{Luminosity and temperature profile}   
           \label{section-radmc3d}

We use the radiative transfer code RADMC-3D (version 0.41), to compute dust temperature profiles in the envelopes of Sgr~B2(N2-N5), based on their current masses (Sect.~\ref{section-masses}) and temperatures (Sect.~\ref{section-obs-constraints}). RADMC-3D performs radiative transfer calculations using the Monte-Carlo method for a given dust distribution and computes the associated dust temperatures. Each source is modeled as a single protostar surrounded by a 1D spherically symmetric envelope with a radius of 10$^6$~au (that is 4.8~pc), divided into 5$\times$10$^4$ cells. The protostar is defined as a point-source (that is its radius is not taken into account) and we assume simple black body radiation by setting its effective temperature. For simplicity we adopt a single type of interstellar grains and do not include scattering. We use the dust opacities from \citet{ossenkopf1994} for dust without ice mantle and coagulated with gas densities of 10$^6$~cm$^{-3}$ as found to best fit Sgr~B2(N)'s dust properties (Sect.~\ref{section-masses}).

The dust mass density in the envelope of each source is assumed to follow a power-law distribution: 
\begin{equation}
\label{eq-dust-distribution}
\rho_{\rm d}(r) = \rho_0 \left( \frac{r}{r_0} \right)^{\alpha} ,
\end{equation}
with $\rho_0$ the dust mass density at the radius $r_0$. We assume $\alpha = -1.5$, holding for a free-falling envelope \citep{shu1977}. Assuming the following gas-to-dust mass ratio, $M_{\rm d} = M_{\rm g} / \chi_{\rm d}$ with $\chi_{\rm d}$~=~100, we can write: 
\begin{equation}
\label{eq-rho0-1}
M_{\rm g} (R) = \int_0^{\rm R} \frac{\mathrm{d}M_{\rm g} (r)}{\mathrm{d}r} \,  \mathrm{d}r = \int_0^{\rm R} 4 \pi \, \chi_{\rm d} \, r^2 \rho_{\rm d} (r) \,  \mathrm{d}r
\end{equation} 
with $M_{\rm g}(R)$ the gas mass inside of $R$ derived in Sect.~\ref{section-masses}. By replacing $\rho_d (r)$ by its expression (Eq.~\ref{eq-dust-distribution}) we obtain : 
\begin{equation}
\label{eq-rho0-2}
M_{\rm g} (R) = \int_0^{\rm R} 4 \pi \, \chi_{\rm d} \, r_0^{1.5} \rho_0 \, r^{0.5} \,  \mathrm{d}r \, .
\end{equation}
Finally, by integrating Eq.~\ref{eq-rho0-2} we obtain:
\begin{equation}
\label{eq-rho0-3}
\rho_0 = \frac{M_{\rm g} (R)}{\frac{8}{3} \pi \, \chi_{d} \, r_0^{1.5} \, R^{1.5}}  \, \, .
\end{equation}
The resulting dust mass densities, $\rho_0$, are given in Table~\ref{TAB-HCs-properties}. The dust mass density profiles derived in the envelope of Sgr~B2(N2-N5) using Eq.~\ref{eq-dust-distribution} is shown in Fig.~\ref{FIG-appendix-dust-profiles}. Because of the high densities reached in the envelopes, photon packages may get trapped inside optically thick regions, considerably slowing down the simulations. To prevent this effect, we use the Modified Random Walk Method \citep[MRW,][]{fleck1984} that allows RADMC-3D to predict where the photon package will go next and save computation time. 

\begin{figure}[!t]
   \begin{center}
    \includegraphics[width=\hsize]{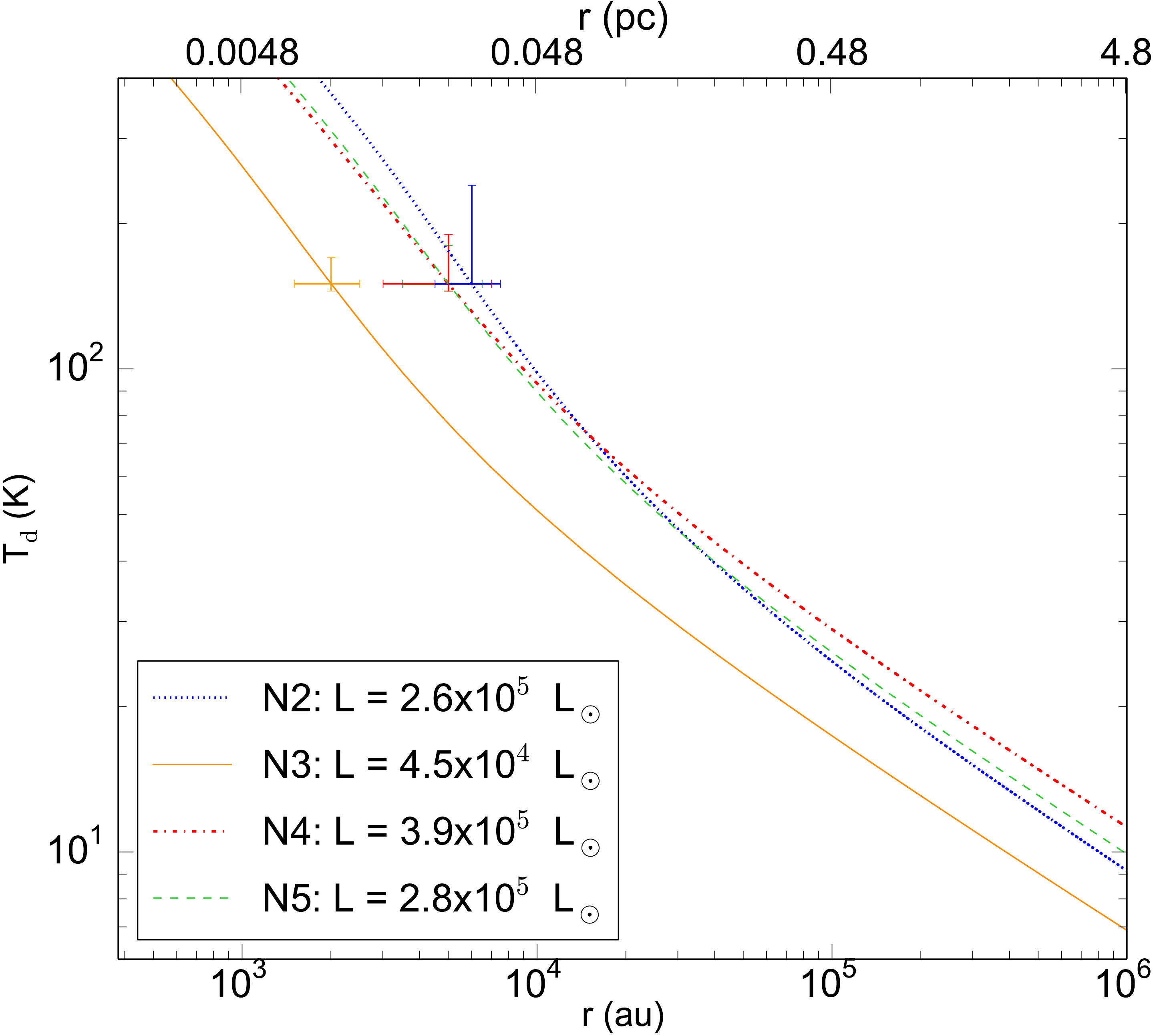} 
    \caption{\label{FIG-current-T-profiles} Dust temperature profiles computed for Sgr~B2(N2-N5) using RADMC-3D, assuming a dust mass density distribution proportional to $r^{-1.5}$ (Fig.~\ref{FIG-appendix-dust-profiles}). The observational constraint $\mathrm{T_{\rm d}}(r_0)$ = $T_0$ used to estimate the current luminosity of each source is plotted with errorbars (1$\sigma$).}
   \end{center}
\end{figure}

Finally, for each source the luminosity of its central protostar is adjusted until the calculated dust temperature in the envelope matches the observational constraint, $T_{\rm d}(r_0)$ = $T_0$. This gives us an estimate of the current luminosity of the sources. Figure~\ref{FIG-current-T-profiles} shows the dust temperature profiles computed by RADMC-3D in the envelope of the sources along with the total luminosities of their central protostars, which are also listed in Table~\ref{TAB-HCs-properties}.  

\subsection{Protostellar evolution}
           \label{section-evolution-L-HCs}

In Sect.~\ref{section-radmc3d} we estimated the current luminosity of Sgr~B2(N2-N5). In order to model the time-dependent chemical evolution of these sources, we need to determine the evolution of their protostellar properties, from the earlier stage of star formation, when a young protostar has just formed and started to warm up the surrounding gas. To this aim, we use published results from theoretical simulations of high-mass star formation.

\begin{figure*}[!t]
\begin{center}
       \includegraphics[width=\hsize]{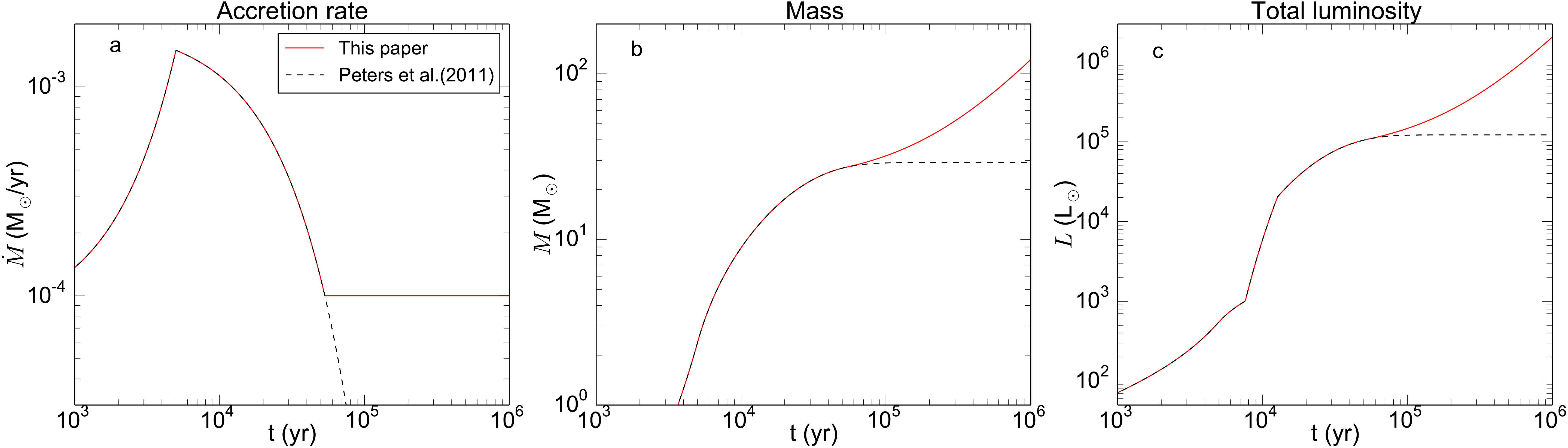} 
 \caption{\label{FIG-L-evolution}\textbf{a} Accretion rate evolution for a young high-mass protostar approximated from the RMHD simulations of \citet{peters2011} (dashed lines). \textbf{b} Time evolution of the protostar's mass derived from its accretion rate, $\dot{M}(t)$ (panel (a), dashed line). \textbf{c} Time evolution of the protostar's total luminosity derived from its mass, $M(t)$ (panel (b), dashed line), using the \citet{hosokawa2009} mass-luminosity relation (Fig.~\ref{FIG-appendix-M-L-relation}). In all panels the solid line shows the result obtained when keeping a constant accretion rate (10$^{-4}$~$\Msol$~yr$^{-1}$) after $\sim$~6$\times$10$^4$~yr.}
\end{center}
\end{figure*}

\citet{peters2011} report the results from the first 3D radiation-magnetohydrodynamical (RMHD) simulations of high-mass star formation including ionization feedback. They model the free-fall collapse of a magnetized rotating molecular cloud containing 1000~$\Msol$. The central core, with a gas density of $\rho = 1.27 \times 10^{-20}$~g cm$^{-3}$ within a radius of 0.5~pc, is surrounded by an envelope characterized by a density distribution $\propto$~$r^{-1.5}$. They assume an initial temperature of 30~K. After the first 20~kyr of simulation, many sink particles are formed, simulating a group of young stars all contributing to the radiative feedback (see their Fig.~1a, run E). In the initial accretion phase the first sink particle reaches a high accretion rate $> 10^{-3} \Msol$~yr$^{-1}$, which reduces to $\sim$~$10^{-4} \Msol$~yr$^{-1}$ after 0.7~Myr (see their Fig.~1b, run E). At the end of the simulation, the first sink particle is by far the most massive one. Although its accretion rate drops significantly when secondary sink particles form, it continues accreting material until the end of the simulation. Based on the accretion rate evolution of the most massive sink particle, without taking into account the influence of the formation of secondary particles, we approximate the relation $\dot{M}(t)$, characterizing the time-dependent evolution of the accretion rate for a young high-mass protostar. Figure~\ref{FIG-L-evolution}a (dashed line) shows the accretion rate as a function of time for a protostar that starts accreting material from its surrounding envelope at $t_0$~=~0. By integrating the relation $\dot{M}(t)$ we obtain the protostellar mass as a function of time, $M(t)$. Figure~\ref{FIG-L-evolution}b (dashed line) shows that after 10$^6$~yr the protostar reaches a final mass of about 30~$\Msol$.

In order to estimate the time-dependent evolution of Sgr~B2(N2-N5)'s luminosities, $L(t)$, from the relation $M(t)$ (Fig.~\ref{FIG-L-evolution}b), we use the mass-luminosity relation derived by \citet{hosokawa2009} for spherically accreting protostars with a constant accretion rate of $\dot{M} = 10^{-4} \Msol$~yr$^{-1}$ (see Fig.~\ref{FIG-appendix-M-L-relation}). The resulting total luminosity is plotted as a function of time in Fig.~\ref{FIG-L-evolution}c (dashed line). It shows that after 10$^6$~yr the protostar reaches a final luminosity of $\sim$~10$^5$~$\Lsol$, which is not high enough to reproduce the luminosities estimated for Sgr~B2(N2-N5) (up to 3.9$\times$10$^5$~$\Lsol$, Sect.~\ref{section-radmc3d}). In order to form more massive and thus more luminous protostars, we simply assume that the accretion rate does not drop below 10$^{-4} \Msol$~yr$^{-1}$ (Fig.~\ref{FIG-L-evolution}a, solid line). Using this simple assumption, protostars with masses $>$~10$^2$~$\Msol$ and luminosities up to $\sim$~2$\times$10$^6$~$\Lsol$ are formed after 10$^6$~yr (Figs.~\ref{FIG-L-evolution}b and c, solid lines). We use the relation $L(t)$ (Fig.~\ref{FIG-L-evolution}c) to estimate the age, $t_{\rm source}$, of Sgr~B2(N2-N5) based on their current luminosity (see Table~\ref{TAB-HCs-properties}).

The approach described in this section provides us with a simple model of the physical evolution of young high-mass protostars. Applied to Sgr~B2(N2-N5) it allows us to investigate the evolution of their chemical composition using astrochemical models. Our physical model is based on the simple assumption that in all sources, objects with similar final mass are formed, which might not be the case in reality (see discussion in Sect.~\ref{section-discussion-evolutionary-stage}). Our treatment is not intended to reflect the complexity of cloud-collapse dynamics but rather to provide a simple framework in which to model the behavior of the cloud chemistry under such conditions. Only a deeper analysis of the whole Sgr~B2 molecular cloud would provide a more accurate model for individual sources \citep[see for instance the 3D radiative transfer model of][which addresses the physical structure of Sgr B2 as a whole, from 45~pc down to 100 au]{schmiedeke2016}.

\subsection{The physical model}
           \label{section-physical-profiles}

Modeling the time-dependent chemical evolution of Sgr~B2(N2-N5) requires to know their physical structure and evolution over the whole star formation formation process, including the early cold phase preceding the hot core phase. However, the early initial stage of high-mass star formation, prior to the protostar's formation, and its timescale are still poorly known. Based on the fraction of star-forming versus quiescent clumps detected in the APEX Telescope Large Area Survey of the Galaxy (ATLASGAL), \citet{csengeri2014} estimated an upper limit of $\sim$~7.5$\pm$2.5$\times$10$^4$~yr for the quiescent deeply embedded phase ($\bar{n}$~$\sim$~4$\times$10$^5$~cm$^{-3}$) prior to the onset of free-fall collapse. \citet{wilcock2012} derived statistical lifetimes for starless infrared dark clouds (IRDCs) of $\sim$2.3$\times$10$^5$~yr from the Herschel Infrared Galactic Plane (HI-GAL) survey. This is a factor three longer than the prestellar phase described by \citet{csengeri2014} for objects with densities ranging from $\sim$~10$^4$ to 10$^5$~cm$^{-3}$. More recently, \citet{jeffreson2018} performed theoretical calculations to estimate the lifetimes of giant molecular clouds based on the time-scales derived for the large-scale dynamical processes that affect the ISM. For clouds located in the CMZ, at galactocentric radii from $\sim$~45 to 120 pc, they estimated median cloud lifetimes of 1.4 -- 3.9$\times$10$^6$~yr. 

\begin{figure*}[!t]
   \begin{center}
    \includegraphics[width=\hsize]{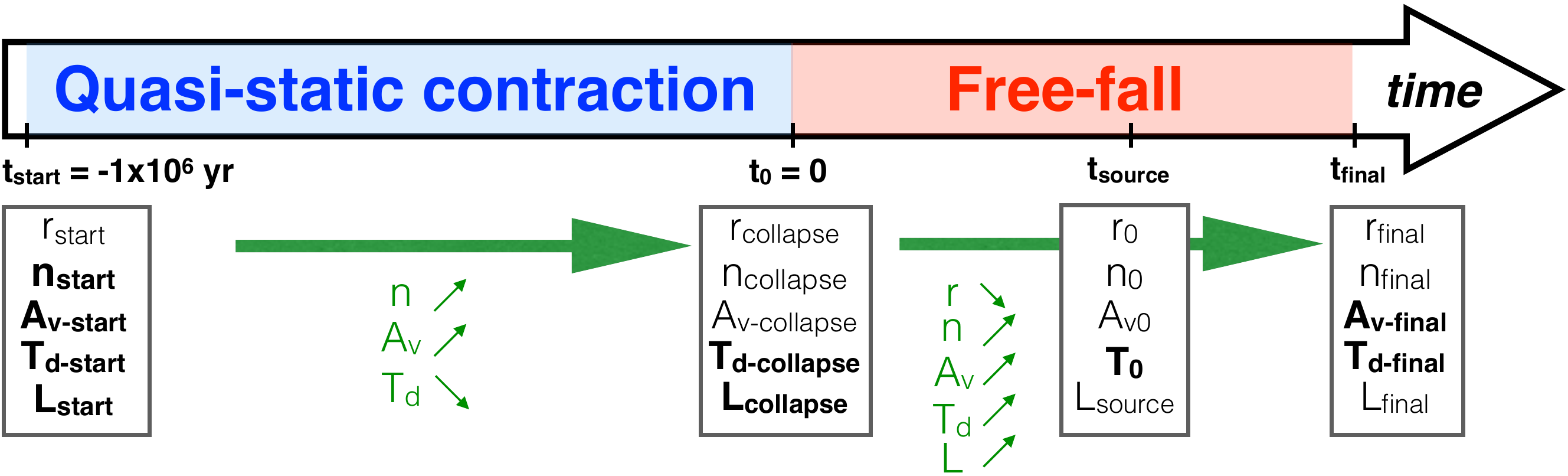} 
    \caption{\label{FIG-frise} Two-stage physical model assumed for the time evolution of Sgr~B2(N2-N5). The physical properties derived at the boundaries of each stage are shown in boxes (see also Table~\ref{TAB-profile-parameters}). The parameters that are common to all sources are highlighted in boldface. The evolution of the physical parameters during each stage is indicated in green with arrows. The evolution of $T_{\rm d}$ during the first stage depends on the adopted minimum dust temperature (see Sect.~\ref{section-precollapse-phase}).}
   \end{center}
\end{figure*}

We divide the early evolutionary phase of high-mass star formation into two distinct stages based on their physical conditions. The first stage traces the evolution of the chemistry for 1~Myr, which corresponds to an intermediate time between the IRDC starless phase derived by \citet{wilcock2012} and the molecular cloud lifetime determined by \citet{jeffreson2018}. This stage is characterized by low densities and low temperatures in the envelope of the source (see Sect.~\ref{section-precollapse-phase}), which undergoes quasi-static contraction leading to the formation a centrally peaked prestellar core. The second stage starts with the ignition of a young accreting protostar warming up its surrounding envelope. In this stage the envelope undergoes a free-fall collapse (see Sect.~\ref{section-collapse-phase}), starting from the conditions reached at the end of the first stage. The two stages of our physical model along with the associated timescales are schematized in Fig.~\ref{FIG-frise}. The physical properties characterizing each stage are computed in Sects.~\ref{section-precollapse-phase} and \ref{section-collapse-phase}.

\subsubsection{Quasi-static contraction}
              \label{section-precollapse-phase}

              The quasi-static contraction phase prior to the free-fall collapse comprises the evolution of density, visual extinction, and temperature in the envelope of Sgr~B2(N2-N5) from $t_{\rm start}$ = -1$\times$10$^6$~yr to $t_0$ = 0, corresponding to the formation of the stellar embryo and the onset of the free-fall collapse. For simplicity we assume that the radius does not change and remains constant throughout this phase (that is we look at the evolution of the density, visual extinction, and temperature of a given parcel of gas at a fixed radius, $r_{\rm start}$, see also Sect.~\ref{section-collapse-phase}). The quasi-static contraction phase starts at low density, with $n_{\rm H-start}$~=~3$\times$10$^3$~cm$^{-3}$ where $n_{\rm H}$ is the total hydrogen density (that is $n_{\rm H}$~=~$n$(H)+2$n$(H$_2$)), and low visual extinction, with $A_{\rm v-start}$~=~2~mag. These are the same initial condition as in \cite{garrod2013} and subsequent papers. We make the simple assumption that density and extinction at a given radius increase linearly as a function of time in log-log space (see Fig.~\ref{FIG-appendix-nH-Av-time} and Table~\ref{TAB-power-law-index}).


In order to characterize the behavior of the dust temperature during the cold, low density phase preceding the warming up of the envelope, we adopt the visual-extinction-dependent treatment presented in \citet{garrod2011}. Assuming that the rate of cooling of the dust is equal to the rate of radiative heating due to its exposure to a standard interstellar radiation field, ISRF~=~1~G$_0$ in units of the Draine field \citep{draine1978}, they found that for $A_{\rm v} < 10$~mag, the dust temperature can be described with a third-order polynomial: 
\begin{equation}
\label{physical-profiles-eqAv}
T_{\rm d} = 18.67 - 1.637 \, A_{\rm v} + 0.07518 \, A_{\rm v}^2 - 0.001495 \, A_{\rm v}^3 + 0.316 \; \rm K 
\end{equation}
Following \citet{garrod2011}, at higher visual extinctions ($A_{\rm v} > 10$~mag) we use the dust temperature profile given by \citet{zucconi2001}.

\begin{figure*}[!t]
   \begin{center}
    \includegraphics[width=\hsize]{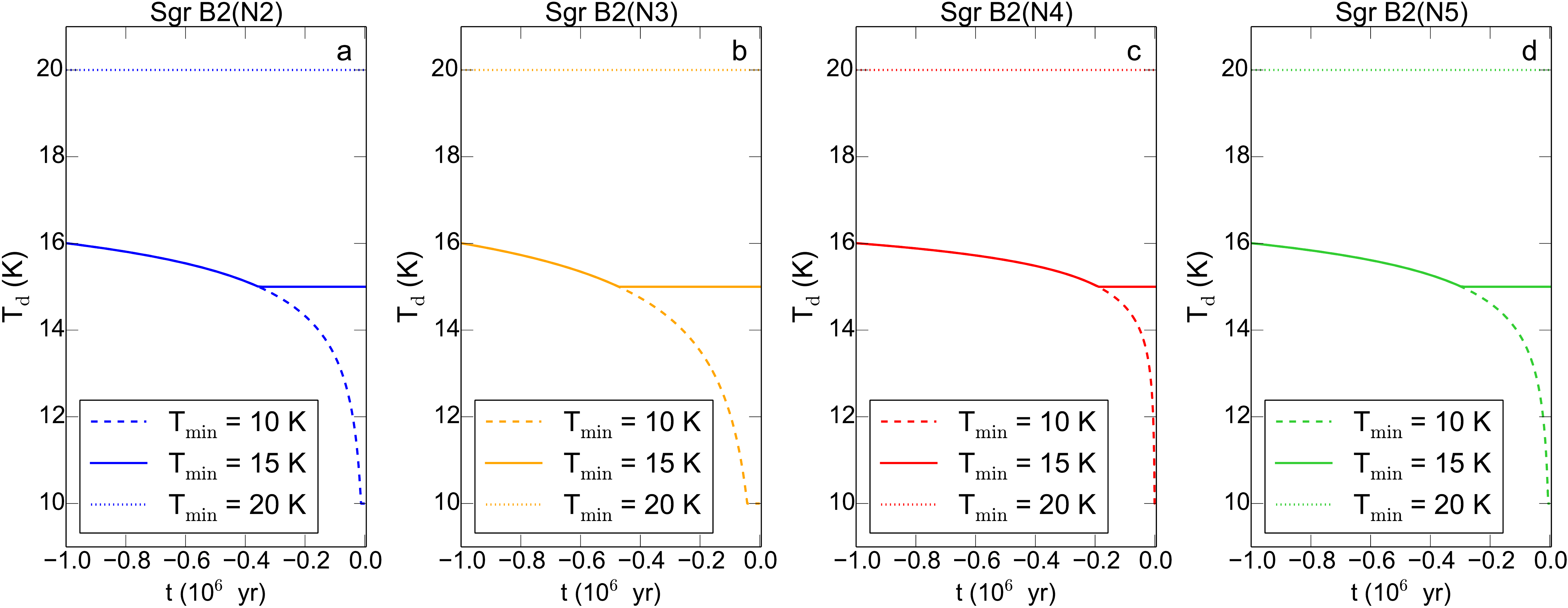} 
    \caption{\label{FIG-Td-time-pre-collapse} Dust temperature as a function of time during the quasi-static contraction phase (Eq.~\ref{physical-profiles-eqAv}). Temperatures are computed for a constant radius, $r$~=$r_{\rm start}$, and for three different $T_{\rm min}$.}
   \end{center}
\end{figure*}

For all sources, the quasi-static contraction phase starts with $T_{\rm d-start}$~=~16~K because $A_{\rm v-start}$~=~2~mag, and the dust temperature then decreases as the visual extinction increases. However, dust temperature measurements carried out at infrared wavelengths with the \textit{Herschel} Space Observatory report somewhat higher temperatures toward the GC region (20--28~K, see Sect.~\ref{section-discussion-Tmin}). Therefore, in order to keep the model physically meaningful and account for the higher dust temperatures expected toward the GC region, we define an arbitrary minimum temperature, $T_{\rm min}$, which represents the lowest temperature that is allowed in the chemical simulations (see also Sect.~\ref{section-chemical-models}). In Fig.~\ref{FIG-Td-time-pre-collapse} we plot the dust temperature as a function of time for Sgr~B2(N2-N5) with different $T_{\rm min}$. The gas kinetic temperature, $T_{\rm g}$, is held constant throughout the whole quasi-static contraction phase, with $T_{\rm g} = T_{\rm min}$.

The physical conditions (density, temperature, and visual extinction) achieved at the end of the quasi-static contraction phase are given for each source in Table~\ref{TAB-profile-parameters}. They represent the initial conditions for the free-fall collapse.

\begin{table}[!t]
\begin{center}
  \caption{\label{TAB-profile-parameters} Parameters of the physical models for Sgr~B2(N2-N5).} 
  \vspace{-2mm}
  \setlength{\tabcolsep}{1.3mm}
  \begin{tabular}{lcccc}
  \hline
   Parameters  & N2 & N3 & N4 & N5 \\ 
\hline
\hline
    & \multicolumn{4}{c}{$\mathbf{t_{\rm start}  = -1.0\times10^6 \: yr}$} \\
\hline
$r_{\rm start}$ (au) & 4.46(5) & 9.46(3) & 3.21(5) & 3.06(5) \\
$n_{\rm start}$ (cm$^{-3}$) & 3.00(3) & 3.00(3) & 3.00(3) & 3.00(3) \\
$A_{\rm v-start}$ (mag) & 2.0 & 2.0 & 2.0 & 2.0 \\
$T_{\rm d-start}$ \tablefootmark{*} (K) & 16.0 & 16.0 & 16.0 & 16.0  \\
$T_{\rm g-start}$ (K) & $T_{\rm min}$  & $T_{\rm min}$ & $T_{\rm min}$ & $T_{\rm min}$ \\
$L_{\rm start}$ ($\Lsol$) & 0 & 0 & 0 & 0 \\
\hline
    & \multicolumn{4}{c}{$\mathbf{t_0 = 0}$} \\
\hline
$r_{\rm collapse}$ (au) & $r_{\rm start}$ & $r_{\rm start}$  & $r_{\rm start}$  & $r_{\rm start}$\\
$n_{\rm collapse}$ (cm$^{-3}$) & 2.14(4) & 2.19(6) & 1.00(4) & 1.86(4) \\
$A_{\rm v-collapse}$ (mag) & 75.3 & 297 & 19.1 & 43.7 \\
$T_{\rm d-collapse}$ (K) & $T_{\rm min}$  & $T_{\rm min}$ & $T_{\rm min}$ & $T_{\rm min}$ \\
$T_{\rm g-collapse}$ (K) & $T_{\rm min}$  & $T_{\rm min}$ & $T_{\rm min}$ & $T_{\rm min}$ \\
$L_{\rm collapse}$ ($\Lsol$) & 0 & 0 & 0 & 0 \\
\hline
    &  \multicolumn{4}{c}{$\mathbf{t_{\rm final} }$}  \\
\hline
$t_{\rm final}$ (yr) & 2.10(5) & 2.06(4) & 3.07 (5) & 2.25(5) \\
$r_{\rm final}$ (au) & 1.38(3) & 5.70(2) & 1.32(3) & 1.42(3) \\
$n_{\rm final}$ (cm$^{-3}$) & 8.27(7) & 1.48(8) & 3.81(7) & 5.91(7) \\
$A_{\rm v-final}$ (mag) & 500  & 500 & 500 & 500 \\
$T_{\rm d-final}$ (K) &  400 &  400 &  400 &  400 \\
$T_{\rm g-final}$ (K) & 400 &  400 &  400 &  400 \\
$L_{\rm final}$ ($\Lsol$) & 2.63(5) & 4.71(4) & 3.92(5) & 2.82(5) \\
\hline
\end{tabular}
\end{center}
\vspace{-4mm}
\tablefoot{Physical properties derived or assumed for each source at three different stages of the high-mass star formation process (see also Fig.~\ref{FIG-frise}). $X(Y)$ means $X \times 10^Y$.
\tablefoottext{*}{$T_{\rm d-start}$ = 16~K is valid for the models with $T_{\rm min}$ = 10~K and 15~K. For $T_{\rm min}$ = 20~K, $T_{\rm d-start}$ = 20~K.}}
\end{table}

\subsubsection{Free-fall collapse}
              \label{section-collapse-phase}

In our physical models, the free-fall collapse phase starts with the ignition of a young high-mass protostar. As the newly formed protostar evolves, it starts heating up the surrounding gas and accretes material from its envelope. In order to characterize the evolution of the physical conditions in the collapsing envelope, we trace the trajectory of a parcel of gas gradually infalling toward the central protostar with the free-fall velocity
\begin{equation}
\label{physical-profiles-eq2}
v_{\rm ff} = \frac{dr}{dt} = \sqrt{2 G M_{\rm g}(r) \left( \frac{1}{r} - \frac{1}{r_{\rm start}}\right)}
\end{equation}
with M$_{\rm g}$(r) the mass of gas inside of $r$. The starting point, $r_{\rm start}$, of the parcel of gas is chosen such that it reaches $r_0$ at $t = t_{\rm source}$ (see Sect.~\ref{section-evolution-L-HCs}), partaking in the free-fall governed by the enclosed mass. The trajectory, $r(t)$, of a parcel of gas infalling through the envelope of Sgr~B2(N2-N5) with the free-fall speed $v_{\rm ff}$ is shown in Fig.~\ref{FIG-appendix-radius-time}.

We derive the gas density along the trajectory of the free-falling parcel of gas from the power-law profile:  
\begin{equation}
\label{physical-profiles-eq3}
n_{\rm H}(r) = n_0 \left( \frac{r}{r_0} \right)^{- 1.5} \, ,
\end{equation}
with $n_{\rm H}$ = 2~$n_{\rm H_2}$, the density of total hydrogen, and $n_0$ the reference density at $r_0$. For simplicity we neglect the time variation of $n_{\rm H}(r)$ during the free-fall collapse phase. From Eq.~\ref{eq-rho0-1}, we can write : 
\begin{equation}
\label{physical-profiles-eq4}
M_{\rm g} (R) = \int_{0}^{R} 4 \pi \, r^2 \mu_{\rm H_2} \, m_{\rm H} \, n_{\rm H_2}(r) \, \mathrm{d}r  ,
\end{equation}
which, in combination with Eq.~\ref{physical-profiles-eq3} yields:
\begin{equation}
\label{physical-profiles-eq6}
n_0 = \frac{M_{\rm g} (R)}{\frac{4}{3} \, \pi \, \mu_{\rm H_2} \, m_{\rm H} \, r_0^{1.5} \, R^{1.5}} \, .
\end{equation} 

The visual extinction along the trajectory of the free-falling parcel of gas is computed from the gas density as follows \citep{bohlin1978} :
\begin{equation}
\label{physical-profiles-eq7}
A_{\rm v}(r) = A_{\rm v-start} + \frac{3.1}{5.8 \times 10^{21}} \int_{r_{\rm init}}^{r} n_{\rm H}(r) \, \mathrm{d}r
\end{equation}
with $r_{\rm init}$ the initial radius of the source envelope computed such that $n_{\rm H}(r_{\rm init}) = n_{\rm start}$; thus $r_{\rm init} = r_0 \left( \frac{n_{\rm start}}{n_0}\right)^{- \frac{1}{1.5}}$ (see Table~\ref{TAB-HCs-properties}). During the free-fall collapse phase the visual extinction increases with density, until $A_{\rm v-max}$ = 500~mag. For simplicity we neglect the time variation of $A_{\rm v}(r)$ during the free-fall collapse. 

As the protostar's total luminosity rises, the temperatures in the envelope increase, progressively dominating the dust heating over the external radiation field. Based on the evolution of the luminosity over time, $L(t)$ (Sect.~\ref{section-masses}), we compute for each source dust temperature profiles for different luminosities, that is at different stages of star formation (see Fig.~\ref{FIG-appendix-interpolated-T-profiles}). Based on the relation $r(t)$ (Fig.~\ref{FIG-appendix-radius-time}) we derive $T_{\rm d}(r)$ the temperature evolution along the trajectory of a parcel of gas free-falling through the envelopes of Sgr~B2(N2-N5) during the collapse phase. The free-fall collapse phase stops when the temperature reaches $T_{\rm max}$~=~400~K. Given the high densities and high visual extinctions reached in the envelope of Sgr~B2(N2-N5), gas and dust temperatures are assumed to be well coupled during the free-fall collapse phase such that $T_{\rm g}\mathbf{(r)}$~=~$T_{\rm d}\mathbf{(r)}$.

The gas density, temperature, and visual extinction computed along the trajectory of the parcel of gas infalling through the envelopes of Sgr~B2(N2-N5) are plotted in Figs.~\ref{FIG-appendix-all-physical-profiles} and ~\ref{FIG-appendix-profiles-time}. The physical conditions (radius, density, temperature, and visual extinction) achieved at the end of the free-fall collapse phase are given for each source in Table~\ref{TAB-profile-parameters}.

\section{Chemical modeling}
        \label{section-modelling}

\subsection{A three-phase astrochemical code}
           \label{section-magickal}

In order to interpret the COM abundances observed toward Sgr~B2(N)'s hot cores (see Sect.~\ref{section-chemical-composition}), we use the astrochemical code MAGICKAL \citep[Model for Astrophysical Gas and Ice Chemical Kinetics and Layering,][]{garrod2013} to calculate chemical abundances in the envelopes of Sgr~B2(N2-N5). MAGICKAL is a single-point model that allows us to simulate in one stage the time-dependent physico-chemical evolution of a given source, from the early cold phase of the star-formation process to the warm up of the dense cores. The calculated chemical abundances characterize a range of distances from the cold extended region to the warmer innermost part of the protostellar collapsing envelope.

The chemical network is based on that of \citet{belloche2017}. It contains 1333 distinct chemical species of which neutrals can exist either as gas-phase, grain-surface, or ice-mantle species, while charged species are considered in the gas phase only. The grain surface is defined as the outermost layer of ice that resides on top of the other ice layers or directly onto the grain surface itself. The rest of the ice mantle, composed of the deeper ice layers below the ice surface, is treated as a separate phase. The network comprises 13374 chemical reactions and processes, which include gas-phase ion-molecule reactions and the related electronic recombination of the resulting protonated ions. Photodissociation and photoionization processes are also included, affecting both the gas-phase and grain-surface chemistry with UV photons from either the ambient radiation field or the CR-induced UV field. Gas-phase and ice-surface chemistry are coupled through the accretion of gas-phase material onto the grains, which may be released back into the gas phase via thermal or non-thermal desorption processes \citep[including chemical reactive desorption,][]{garrod2007}. Diffusion allows reactions within the mantle as well as exchange of chemical species between surface and mantle via swapping mechanisms. Details on the chemical processes and reactions rates can be found in \citet{garrodetal2008}, \citet{garrod2011}, and \citet{garrod2013}. 

The coupled gas-phase, dust-grain surface, and ice-mantle chemistry is solved using the modified rate-equation approach of \citet{garrod2008} to account for stochastic effects in the surface chemistry. In addition, the model uses the grain-surface back-diffusion correction of \citet{willis2017}, and assumes a single dust-grain size of 0.1 micron. 

The initial gas-phase material is assumed to be mostly composed of atoms and atomic ions, except for H$_2$. Initial abundances are taken from \citet{garrod2013}. As Sgr~B2(N)'s hot cores presumably originate from material with the same elemental abundances, the same initial abundances are used for all hot cores.

\subsection{The chemical models}
           \label{section-chemical-models}

In this section we present the grid of chemical models (see Table~\ref{TAB-our-models}) run to investigate the time-dependent chemical evolution of Sgr~B2(N2-N5) and explore the influence of environmental conditions on the calculated chemical abundances.

The first parameter we vary is the minimum dust temperature, $T_{\rm min}$, that can be reached in the simulations, in particular during the cold quasi-static contraction phase prior to the onset of the free-fall collapse (see also Sect.~\ref{section-precollapse-phase}). The minimum dust temperature adopted in the chemical simulations is critical because the production of the main ice constituents (\ce{H2O}, CO, \ce{CO2}, and \ce{CH3OH}), essential to the formation of COMs, is highly sensitive to the dust temperature. At low temperatures ($\sim$~10~K), mainly hydrogenation reactions are important on the surface of dust grains. As the temperature increases, the surface diffusion rates of heavier species become more important leading to a massive increase in the surface production of simple species by addition of radicals (OH, HCO, NH, \ce{NH2}, \ce{CH3}, \ce{CH2OH}, \ce{CH3O}). However, hydrogen atoms evaporate quickly and hydrogenation reactions become less efficient because the rate of desorption of H is much faster than the rate of reaction between H atoms. In order to explore the impact of the minimum dust temperature on the production of COMs we run ''low temperature'' models with $T_{\rm min}$~=~10~K, ''intermediate temperature'' models with $T_{\rm min}$~=~15~K, and finally, to better represent the higher temperatures expected toward the GC, ''high temperature'' models with $T_{\rm min}$~=~20~K (see Sect.~\ref{section-discussion-Tmin}). 

The second aspect explored in the chemical simulations is the impact of the CRIR on the production of COMs. We run chemical simulations using CRIRs ranging from the standard value $\zeta^{\rm H_2}_0$ = 1.3$\times$10$^{-17}$~s$^{-1}$ to 1000$\times \zeta^{\rm H_2}_0$~=~1.3$\times$10$^{-14}$~s$^{-1}$. Because of the uncertainties on the column-density dependence of the CRIR in dense regions (see Sect.~\ref{section-discussion-CR}), we assume for simplicity that no attenuation occurs in the envelope of the sources. Due to its location in the CMZ, Sgr~B2 is also known to be affected by X-ray emission \citep[see, e.g.,][]{terrier2018}. As the effects of X-rays are often considered to be similar to cosmic ray-related processes \citep[see, e.g.,][]{viti2003}, here the total ionization rate may be seen as a combination of both effects. 

For each hot core we run a grid of 18 models, one for each set of physical parameters (CRIR and $T_{\rm min}$, see Table~\ref{TAB-our-models}). The interstellar radiation field is held constant throughout the model grid, assuming a standard value of 1~G$_0$. The influence of the ISRF strength on the chemical simulations is discussed in Appendix~\ref{appendix-ISRF}.

\begin{table}[!t]
\begin{center}
  \caption{\label{TAB-our-models} Grid of chemical models.} 
  \setlength{\tabcolsep}{1.6mm}
  \begin{tabular}{l|lll}
    \hline
$\zeta^{\rm H_2}$ (s$^{-1}$) & \multicolumn{3}{c}{$T_{\rm min}$ (K)} \\[2mm]
               & \multicolumn{1}{c}{10}  & \multicolumn{1}{c}{15} & \multicolumn{1}{c}{20} \\
    \hline
    \hline 
  1.3$\times$10$^{-17}$ & T10-CR1    & \textbf{T15-CR1}    & T20-CR1    \\   
  1.3$\times$10$^{-16}$ & T10-CR10   & T15-CR10   & T20-CR10  \\ 
  6.5$\times$10$^{-16}$ & T10-CR50   & T15-CR50   & T20-CR50  \\ 
  1.3$\times$10$^{-15}$ & T10-CR100  & T15-CR100  & T20-CR100  \\
  6.5$\times$10$^{-15}$ & T10-CR500  & T15-CR500  & T20-CR500  \\
  1.3$\times$10$^{-14}$ & T10-CR1000 & T15-CR1000 & T20-CR1000 \\
    \hline
\end{tabular}
\end{center}
\vspace{-4mm}
\tablefoot{Our standard model, with $T_{\rm min}$~=~15~K and $\zeta^{\rm H_2}$~=~1.3$\times$10$^{-17}$~s$^{-1}$, is highlighted in boldface. It will be referred to as N2-T15-CR1, N3-T15-CR1, N4-T15-CR1, and N5-T15-CR1 for Sgr~B2(N2), N3, N4, and N5, respectively.}
\end{table}

\section{Results}
         \label{section-results}

\subsection{Observed chemical composition of the hot cores}
           \label{section-chemical-composition}

We use \textit{Weeds} as described in Sect.~\ref{section-modelling-obs} to derive the chemical composition of Sgr~B2(N3-N5) by modeling the continuum subtracted emission spectra observed toward each source. We present the results obtained for 11 COMs that include N-, O-, and S-bearing species, to provide a broad census of the chemical composition of the sources. The parameters of the best-fit LTE models are listed in Table~\ref{TAB-best-fit-parameters} along with the column densities derived for each molecule. The column densities derived for Sgr~B2(N2) are also given for comparison \citep{belloche2016, belloche2017, muller2016}. In the case of methyl cyanide (\ce{CH3CN}), we investigate its isotopologs $^{13}$\ce{CH3CN} and CH$_3^{13}$CN because the vibrational ground state transitions of \ce{CH3CN} are optically thick. We assume the isotopic ratio [\ce{CH3CN}]/[$^{13}$\ce{CH3CN}]~= [\ce{CH3CN}]/[CH$_3^{13}$CN]~=~21 derived by \citet{belloche2016} to obtain the column density of \ce{CH3CN}. 

The column densities derived for the 11 COMs in the four hot cores are plotted in Fig.~\ref{FIG-chemical-composition}a along with their uncertainties, estimated as described in Appendix~\ref{appendix-column-density-uncertainties}. In order to compute abundances relative to H$_2$ (Fig.~\ref{FIG-chemical-composition}b), we use the H$_2$ column densities derived for the compact region with warm dense gas from which the COM emission arises (see Table~\ref{TAB-appendix-H2}). The abundances relative to methanol (\ce{CH3OH}) are plotted in Fig.~\ref{FIG-chemical-composition}c. The overall chemical composition of Sgr~B2(N3) is relatively similar to that of Sgr~B2(N5), with at most a factor two of difference between the abundances (relative to \ce{CH3OH}) of all ten COMs. Sgr~B2(N4) shows higher abundances of O- and S-bearing species relative to \ce{CH3OH} than Sgr~B2(N3) and N5, except for \ce{NH2CHO}, but significantly lower abundances relatives to H$_2$. The chemical composition of Sgr~B2(N2) differs significantly from that of the other hot cores. For instance, \ce{C2H5CN} is about four times more abundant (with respect to \ce{CH3OH}) in Sgr~B2(N2) than toward the other hot cores; in addition, the ratio [\ce{C2H5CN}]/[\ce{C2H3CN}] is three times higher. The abundances relative to \ce{CH3OH} of the O-bearing species \ce{CH3OCHO} and \ce{CH3OCH3} are significantly lower toward Sgr~B2(N2) compared to the other sources, but similar to Sgr~B2(N3) in terms of abundances relative to H$_2$. Finally, \ce{NH2CHO} shows an astonishing ratio relative to \ce{CH3OH} about 15 times larger than toward Sgr~B2(N3-N5).

\begin{figure*}[!t]
   \resizebox{\hsize}{!}
   {\begin{tabular}{ccc}
       \includegraphics[width=\hsize]{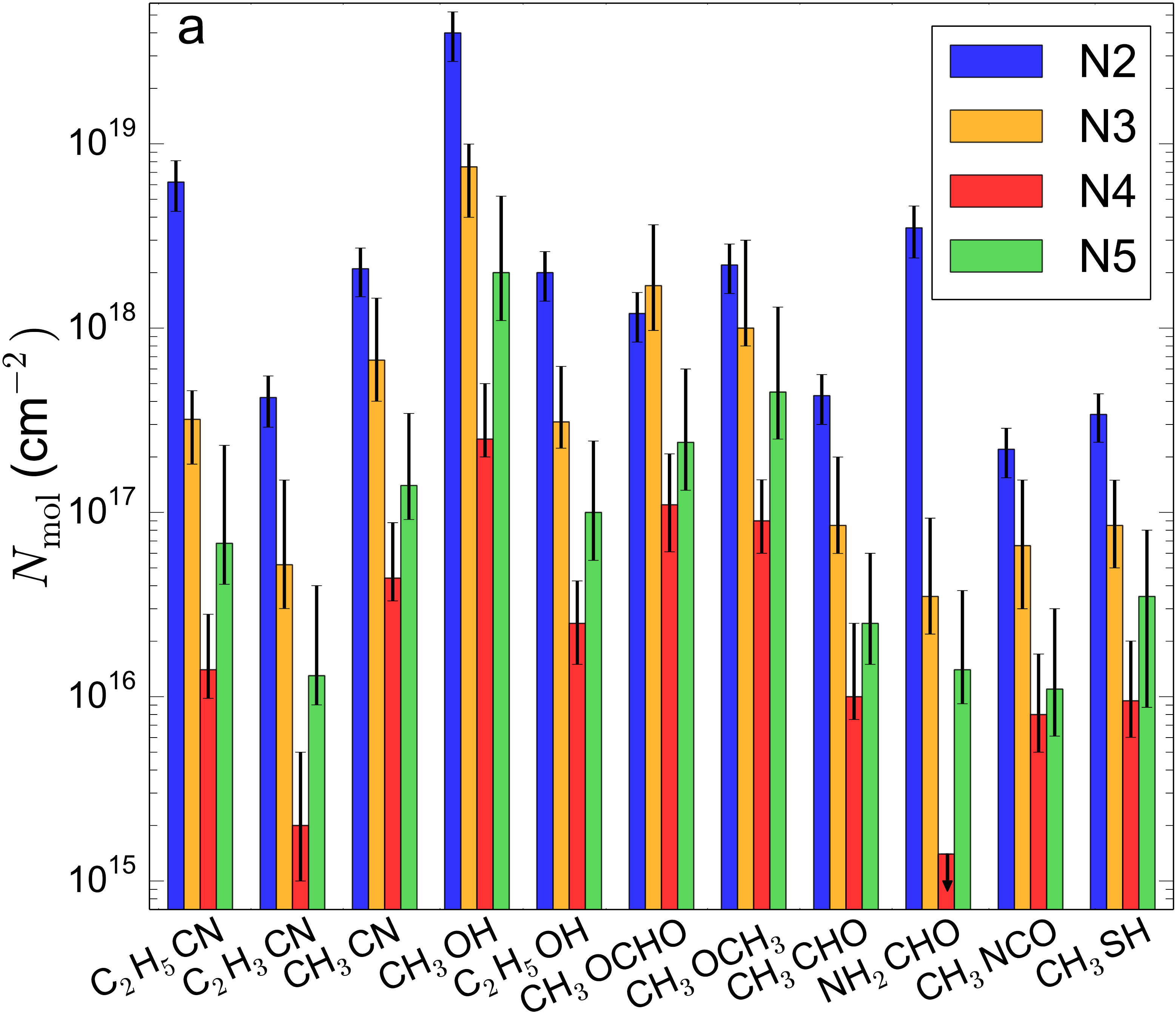}  &
       \includegraphics[width=\hsize]{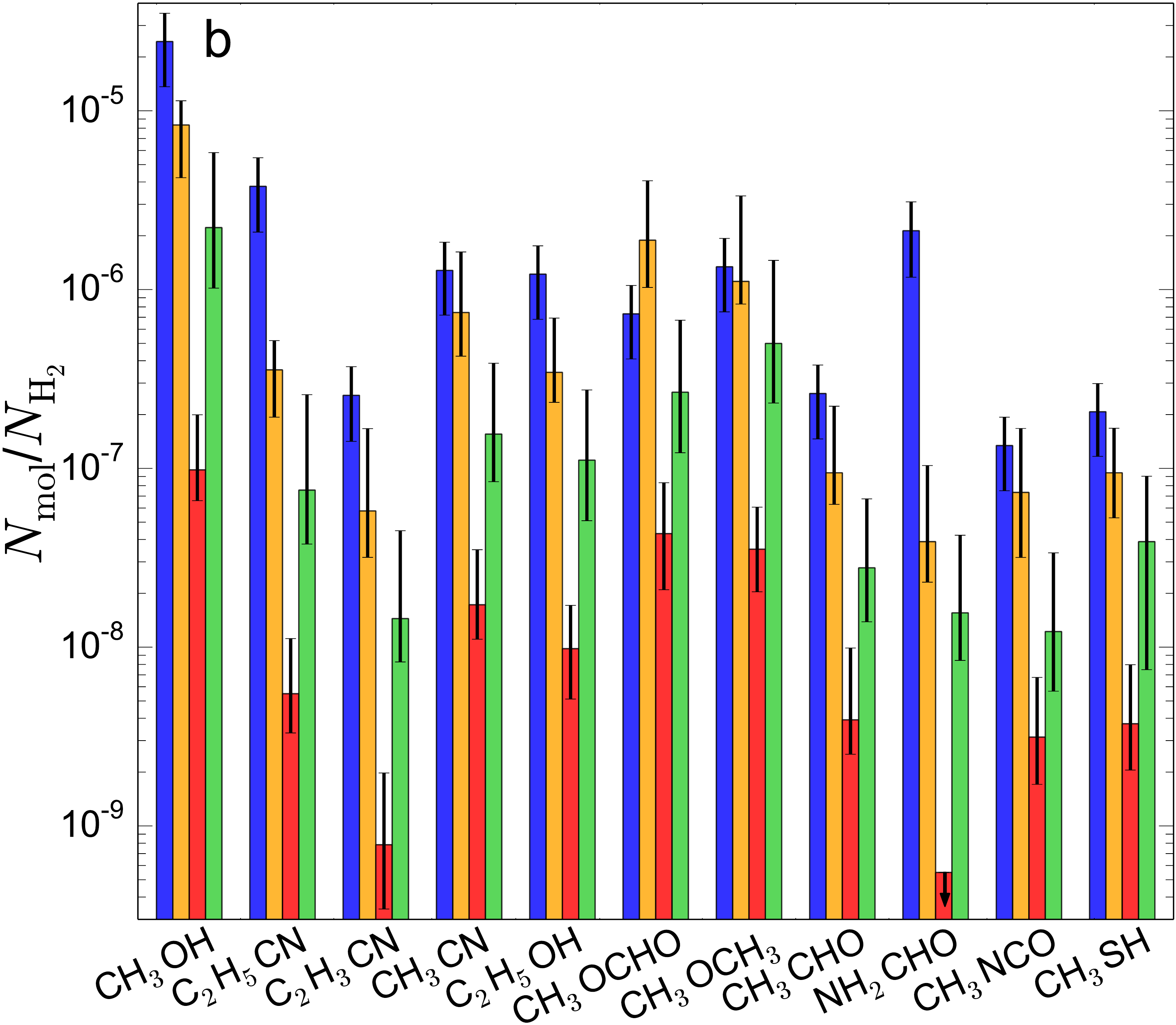}  &
       \includegraphics[width=\hsize]{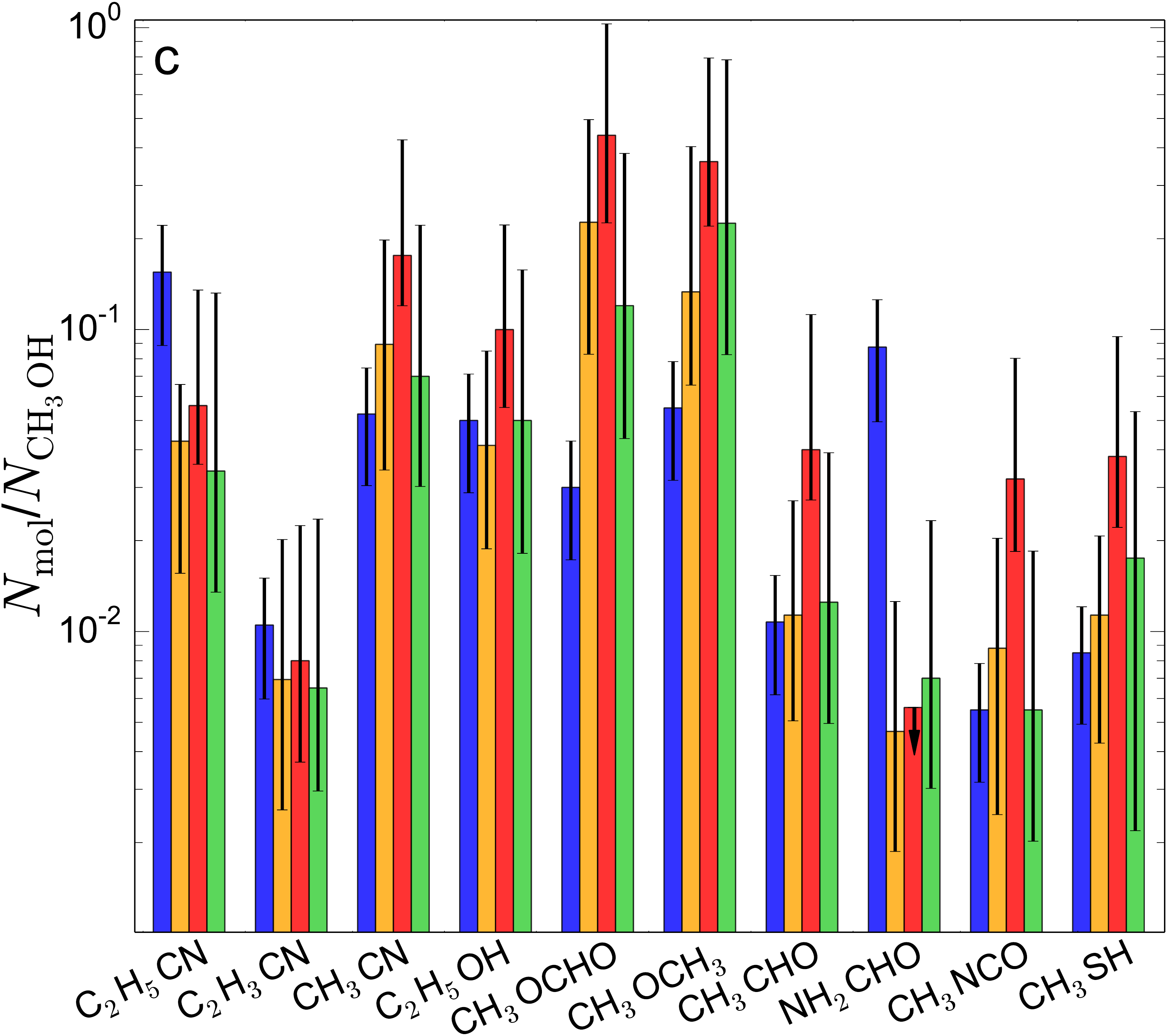} \\
    \end{tabular}}
  \caption{\label{FIG-chemical-composition} \textbf{a} Column densities of 11 COMs detected toward Sgr~B2(N2-N5) (see Table~\ref{TAB-best-fit-parameters}). The error bars show uncertainties on the derived column densities (1$\sigma$, see Sect.~\ref{appendix-column-density-uncertainties}). \textbf{b} Chemical abundances relative to H$_2$ of the COMs shown in panel (a). The error bars show uncertainties that also include the uncertainty on the H$_2$ column density (1$\sigma$, see Sect.~\ref{section-masses}). \textbf{c} Abundances relative to \ce{CH3OH} of the COMs shown in panel (a). The error bars represent 1$\sigma$ uncertainties on the derived ratios.} 
\end{figure*}

\subsection{Model results}
   \label{section-results-model}

   In order to interpret the results derived from the observations (Sect.~\ref{section-chemical-composition}) we use the astrochemical code MAGICKAL as described in Sect.~\ref{section-modelling} to compute time-dependent chemical abundances in the envelopes of Sgr~B2(N2-N5). In this section we present the results obtained for the standard models N(2-5)-T15-CR1. All abundances refer to fractional abundances with respect to the total hydrogen density ($n_{\rm H}$~=~2$n$(H$_2$)~+~$n$(H)) unless specified otherwise. The effects on the chemical abundances of variations in the minimum dust temperature reached in the cold prestellar phase (that is the quasi-static contraction phase) and the CRIR are discussed in Sects.~\ref{section-discussion-Tmin} and \ref{section-discussion-CR}, respectively. 

The final gas-phase fractional abundances calculated for the 11 investigated COMs in all models are given in Tables~\ref{TAB-appendix-model-results} and \ref{TAB-appendix-model-results2}.

\subsubsection{Building up ice mantles}
              \label{section-results-pre-collapse}

The 1~Myr cold phase prior to the free-fall collapse allows the gas-phase chemistry to quickly convert free atoms into stable simple molecules. Meanwhile, gas-phase material is accreted onto the interstellar dust grains to form ice mantles. The rate of growth of the ice mantles is not linear as a function of time. It is determined by the net rate of deposition of gas-phase material onto the grains and by the degree of surface coverage of bare grains. For the four standard models, N(2-5)-T15-CR1, an ice thickness of one layer is achieved at a time of 6.4$\times$10$^3$~yr into the simulation (that is $t$~=~-0.9936$\times$10$^6$~yr, see Fig.~\ref{FIG-appendix-growth-ice-mantle}). The rate of mantle deposition is relatively similar for models N2-T15-CR1, N4-T15-CR1, and N5-T15-CR1, over the whole quasi-static contraction phase. After 1~Myr (that is at $t_0$~=~0), about 110 ice layers have been formed on dust grains for the three models compared to the 161 layers formed in model N3-T15-CR1. This is due to the large densities computed at $r_{\rm start}$ in the envelope of Sgr~B2(N3) during the contraction phase (up to 2$\times$10$^6$~cm$^{-3}$, see Fig.~\ref{FIG-appendix-nH-Av-time}a). The accretion of material onto the dust grains is thus more efficient in model N3-T15-CR1, leading to rapid deposition of new ice layers. For all models, the net deposition rate continues to increase until the end of the quasi-static contraction phase, indicating that the depletion of gas-phase material has not reached completion yet.

\begin{figure*}[!t]
   \resizebox{\hsize}{!}
   {\begin{tabular}{ccc}
       \includegraphics[width=\hsize]{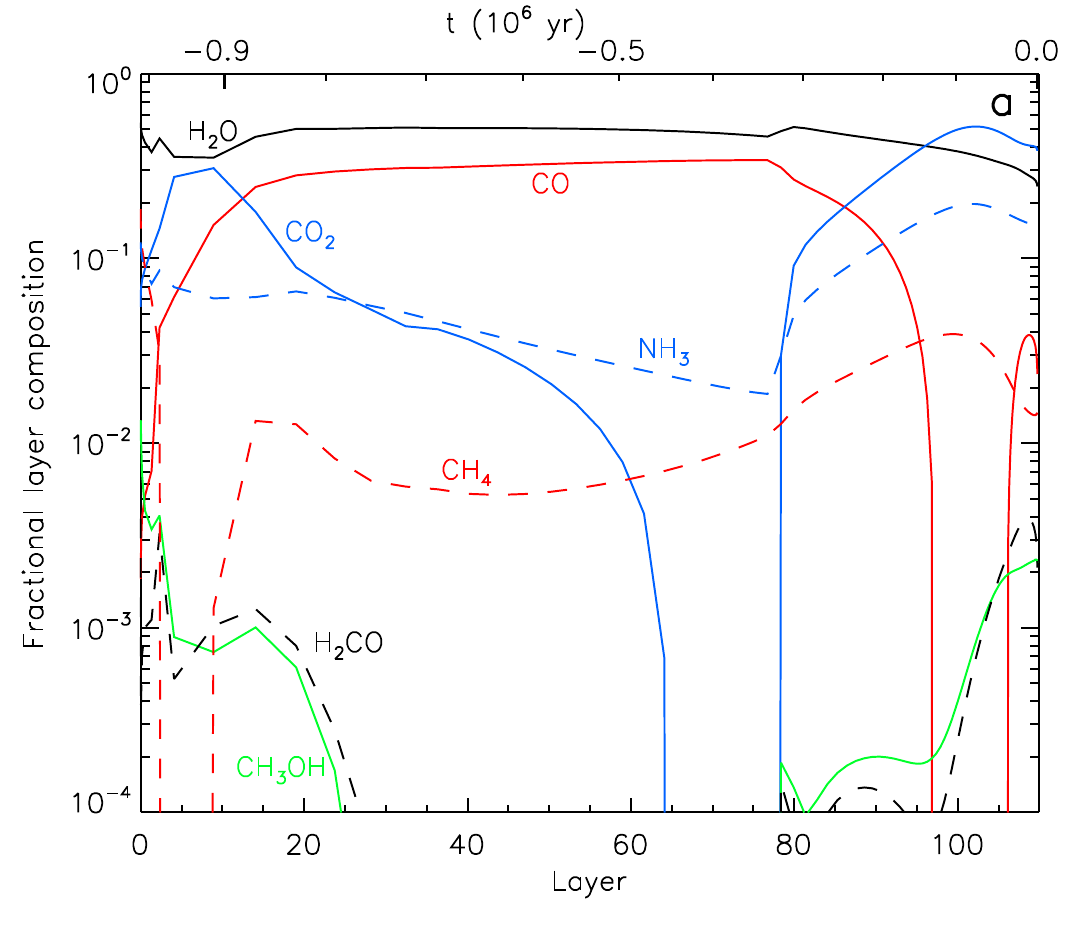}  &
       \includegraphics[width=\hsize]{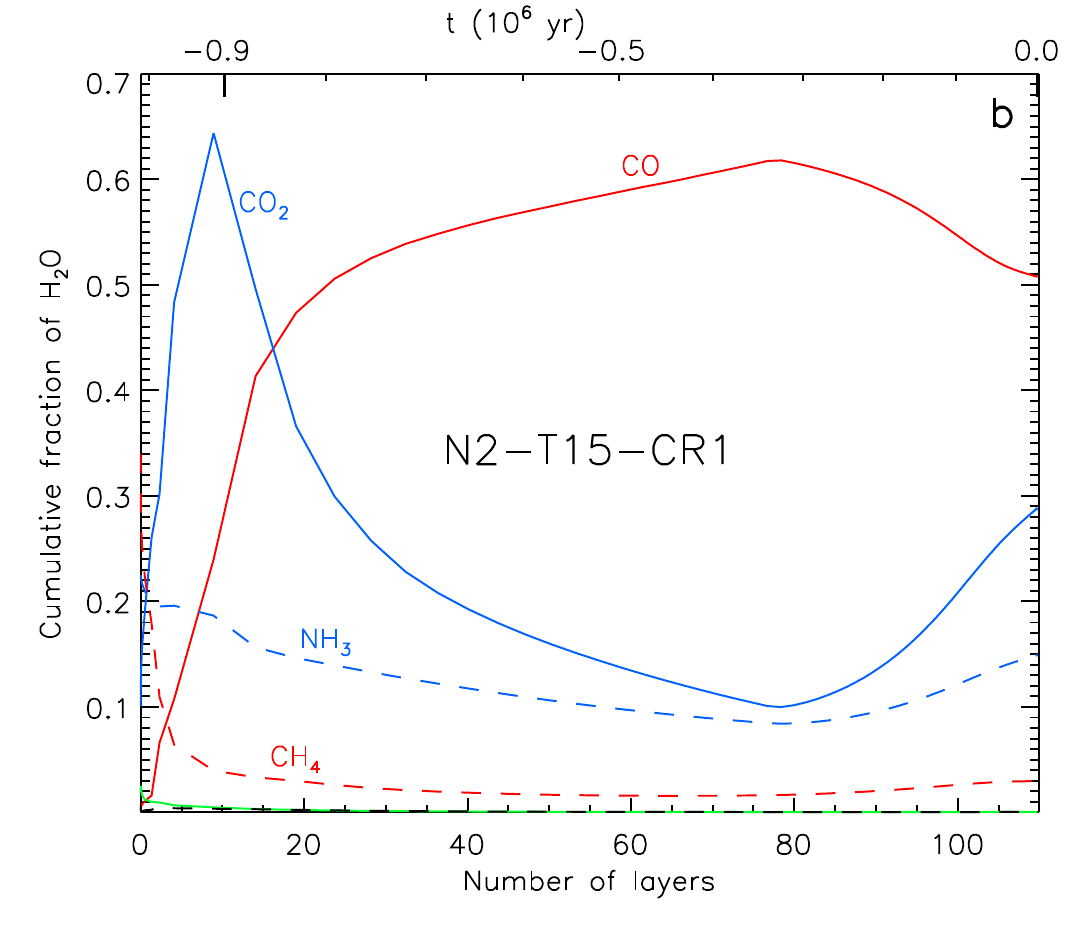}  &
       \includegraphics[width=\hsize]{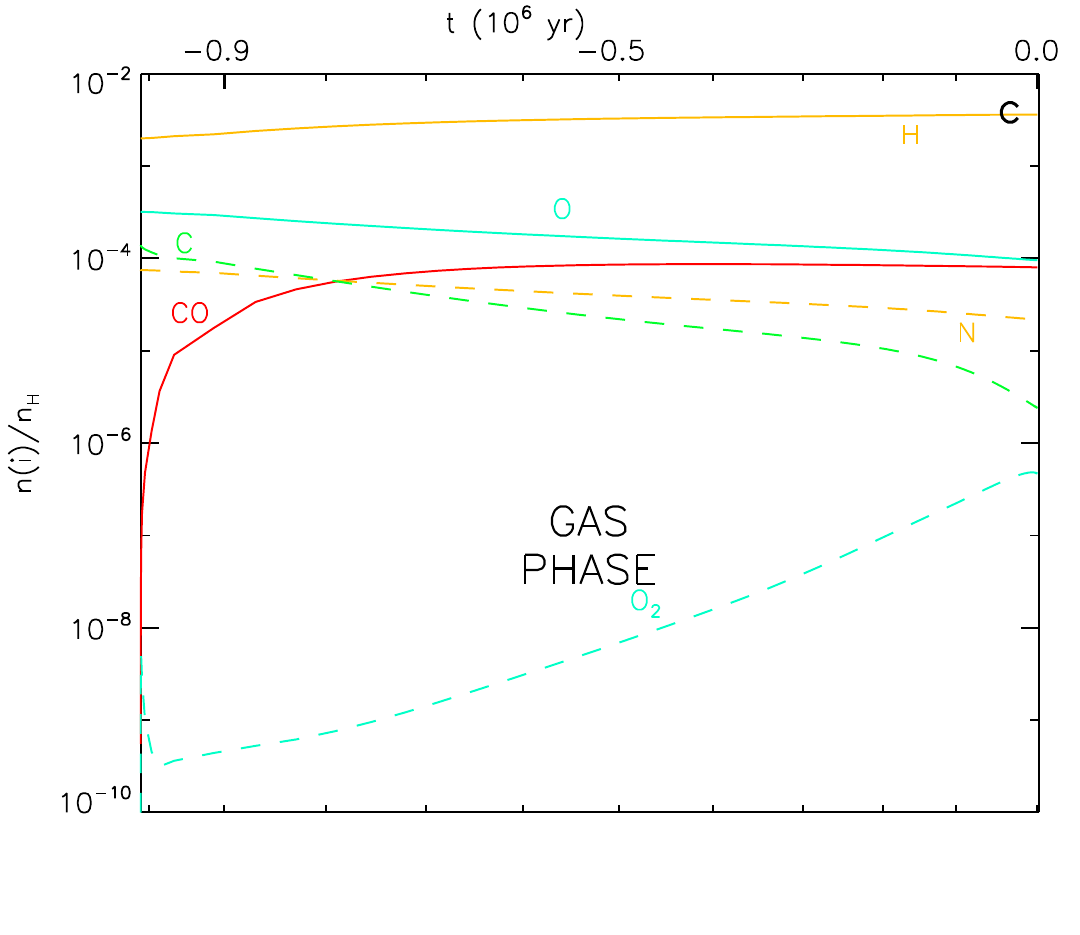} \\
       \includegraphics[width=\hsize]{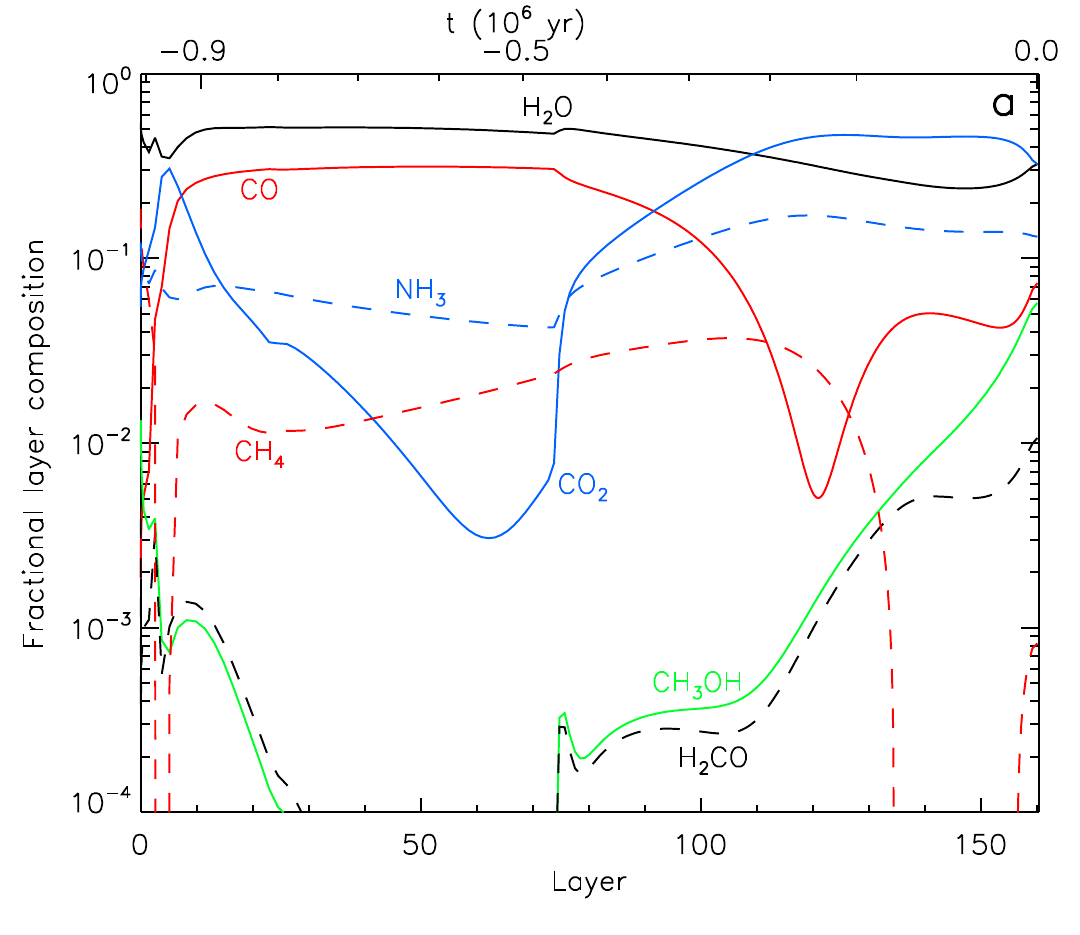}  &
       \includegraphics[width=\hsize]{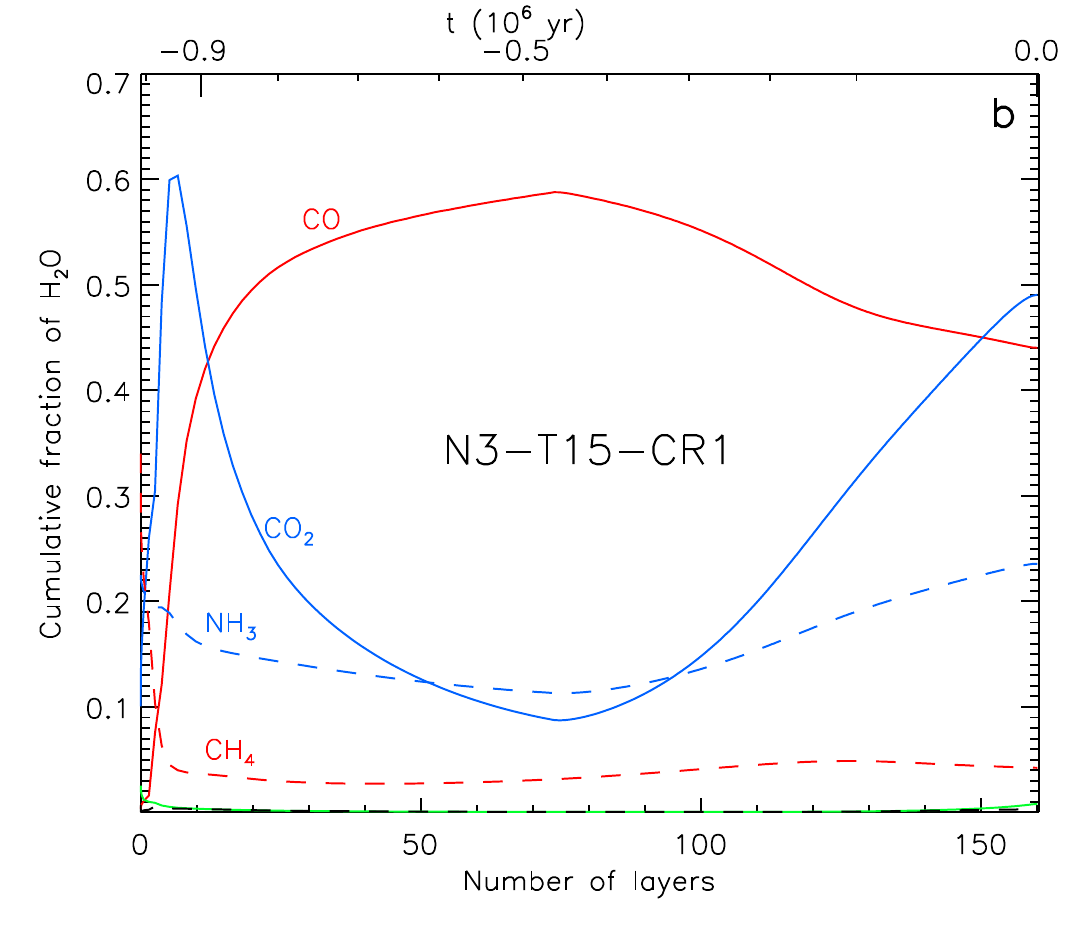}  &
       \includegraphics[width=\hsize]{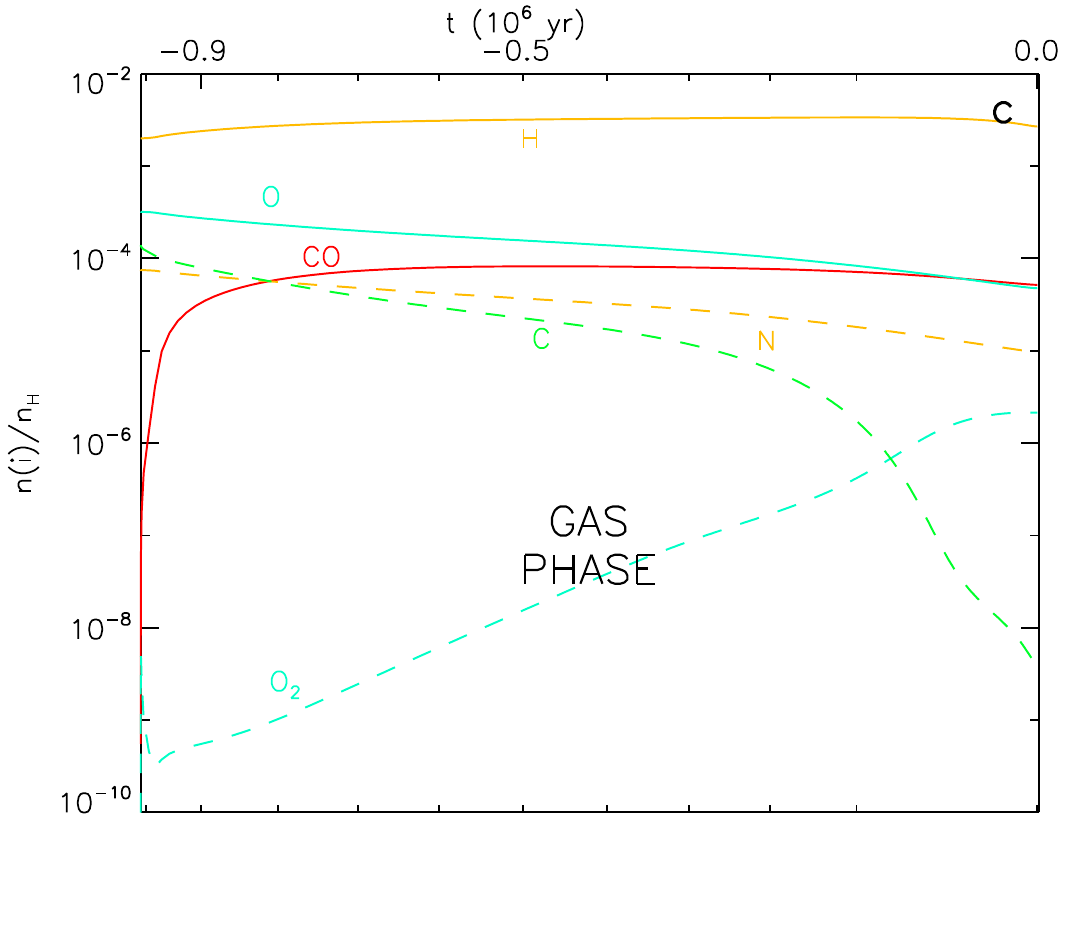} \\
       \includegraphics[width=\hsize]{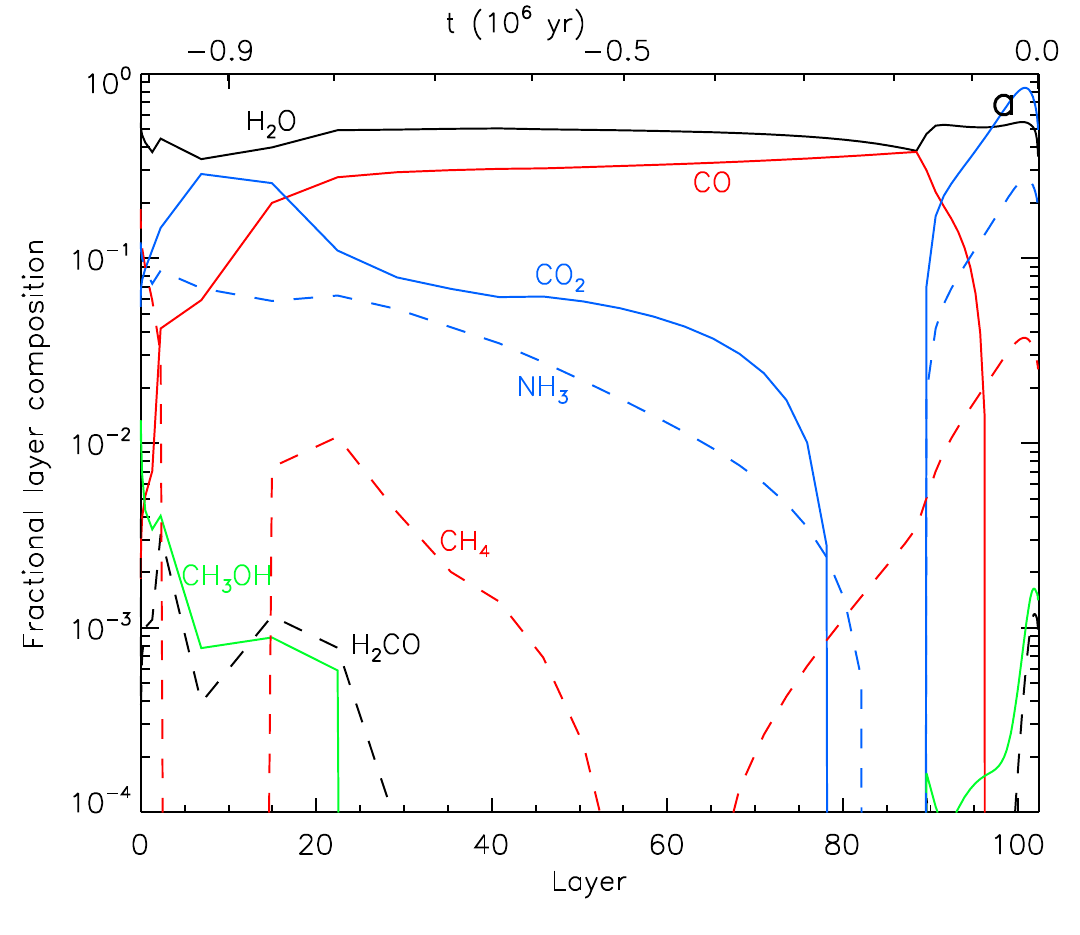}  &
       \includegraphics[width=\hsize]{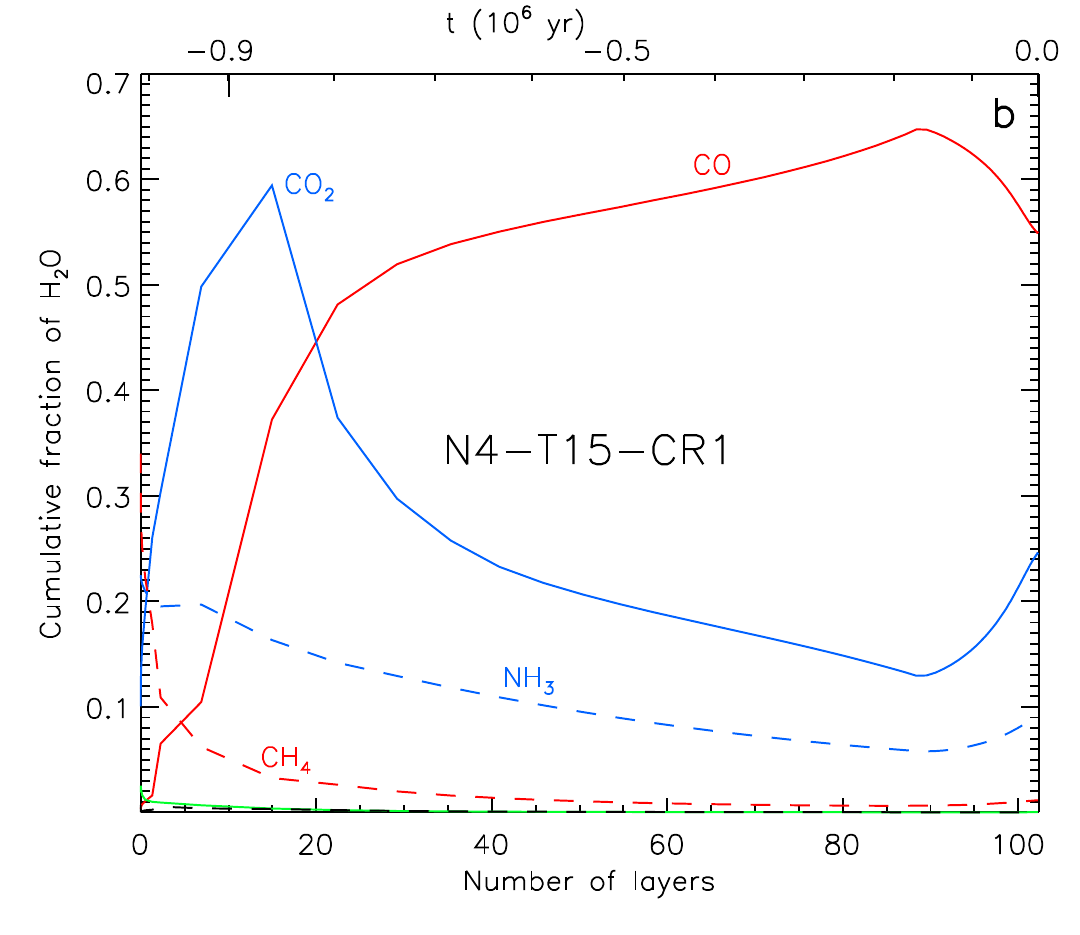}  &
       \includegraphics[width=\hsize]{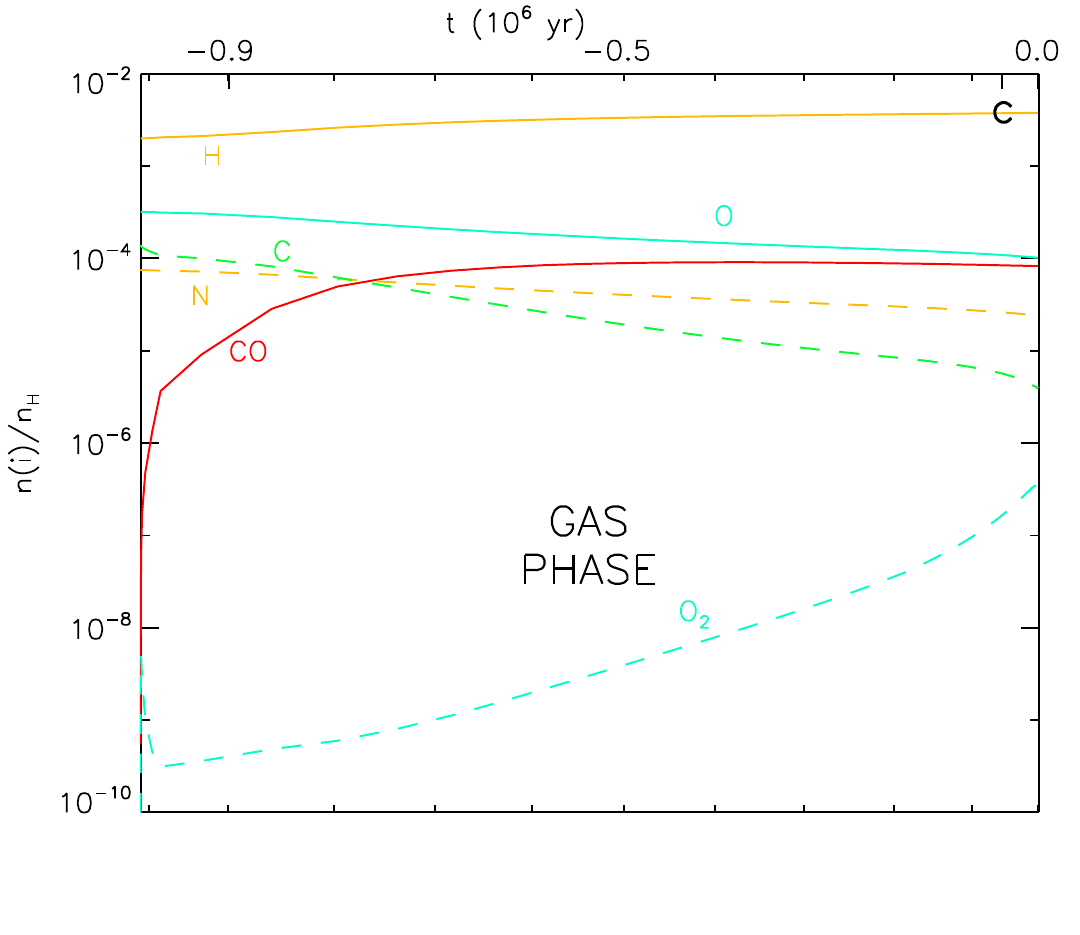} \\
       \includegraphics[width=\hsize]{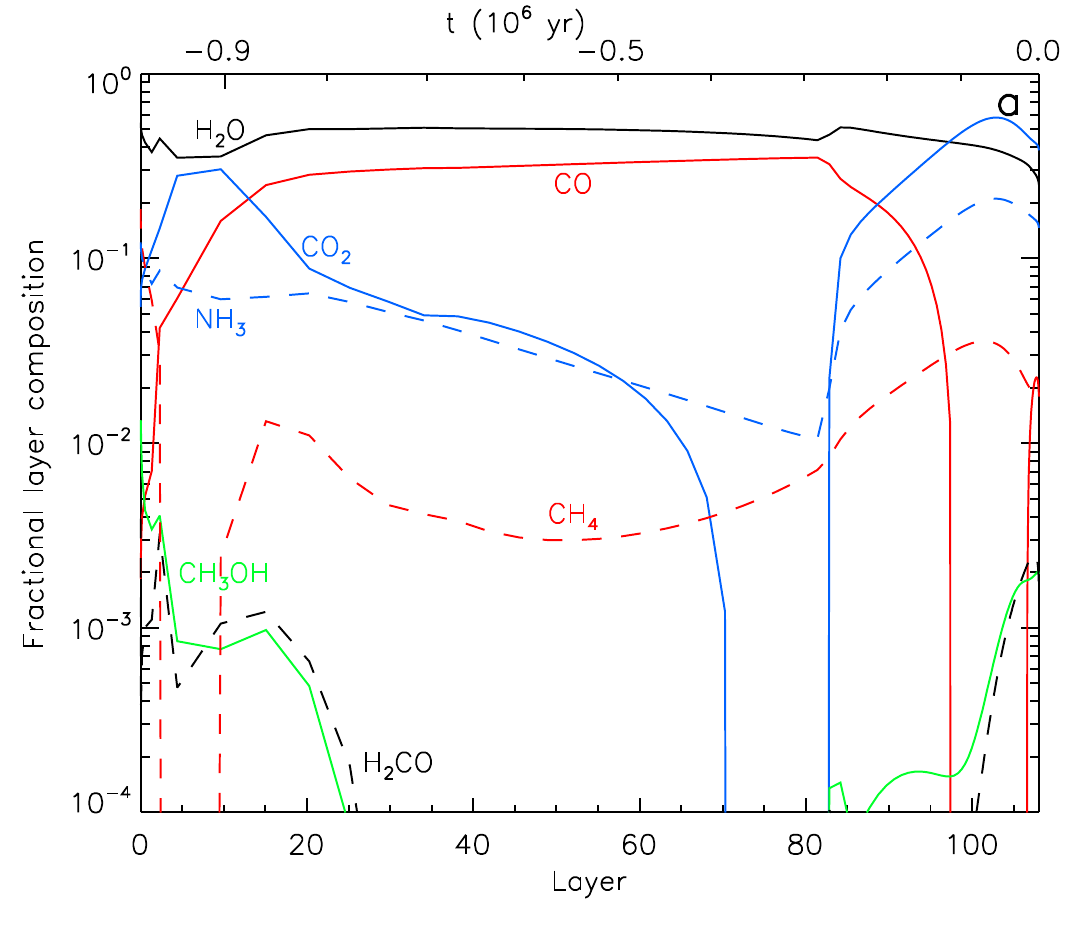}  &
       \includegraphics[width=\hsize]{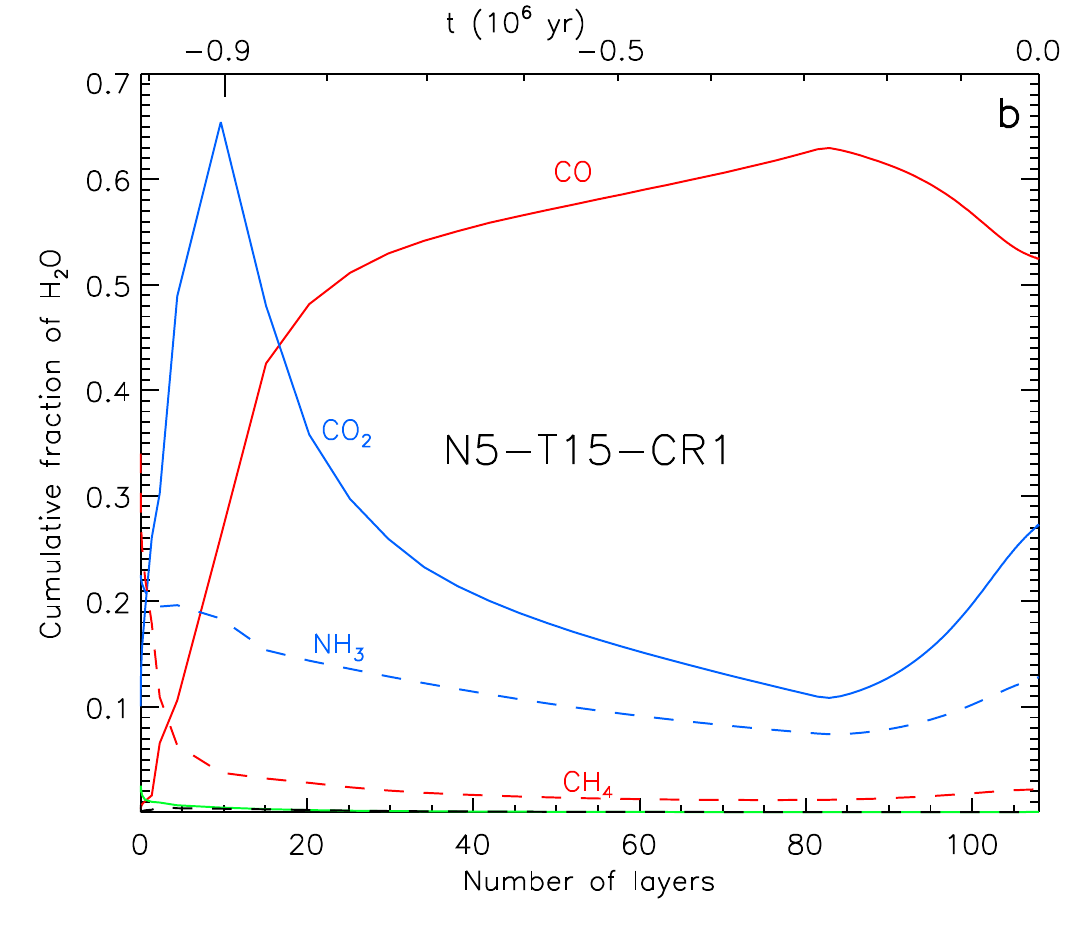}  &
       \includegraphics[width=\hsize]{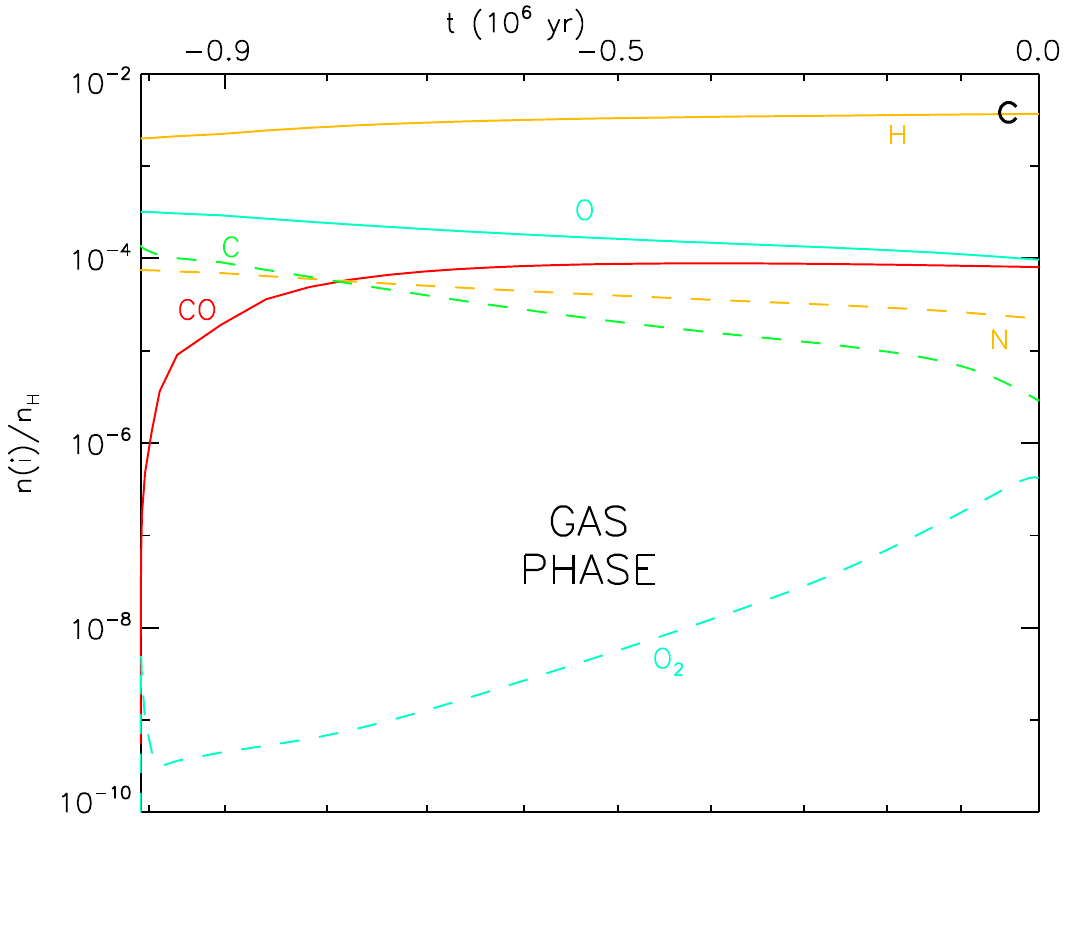} \\
    \end{tabular}}
    \caption{\label{FIG-ice-mantle-composition} Evolution of the ice-mantle and gas-phase chemistry during the cold quasi-static contraction phase. The results from the standard models N(2-5)-T15-CR1 are displayed from top to bottom. \textbf{a} Calculated fractional ice-mantle composition by ice layer. \textbf{b} Cumulative fractional composition of the ices with respect to water, summed over the grain mantle up to an ice thickness given on the abscissa. \textbf{c} Gas-phase fractional abundances (with respect to total hydrogen) of atomic and simple diatomic species as a function of time with the same time axis as panels (a) and (b) to facilitate the comparison.}
\end{figure*}

Figure~\ref{FIG-ice-mantle-composition}a shows the chemical composition of the grain mantles at each ice layer based on the surface ice that gets incorporated into the mantle. A timescale is indicated such that a time associated to a given ice layer represents the time at which the ice layer was deposited onto the dust grains. The chemical composition within the first 50 ice layers of the grain mantles is similar for the four standard models. Water is clearly the dominant ice constituent for all models, while CO and \ce{CO2} are also present in significant quantities (see also Fig.~\ref{FIG-ice-mantle-composition}b). Because of the relatively high dust temperatures ($T_{\rm min}$~=~15~K), the slow H-to-H$_2$ conversion on the grains results in high fractional abundances of atomic hydrogen in the gas phase over the whole quasi-static contraction phase (a few 10$^{-3}$, see Fig.~\ref{FIG-ice-mantle-composition}c). We find that the surface reaction OH~+~\ce{CH4}~$\rightarrow$~\ce{H2O}~+~\ce{CH3} is the main mechanism responsible for the formation of water ice, rather than the reaction of OH with atomic or molecular hydrogen expected at lower dust temperatures \citep[$T$~=~10~K, see, e.g.,][see also Sect.~\ref{section-discussion-Tmin}]{garrod2011}. At the beginning of the simulations, the rapid conversion of atomic carbon to CO in the gas phase causes a sharp decline in the abundance of \ce{CH4} ices, which in turn affects the abundance of water ice (Figs.~\ref{FIG-ice-mantle-composition}a and c). Due to the high dust temperature ($T_{\rm min}$~=~15~K) the gas-phase CO accreted onto the grains is mobile enough to react with OH radicals to form \ce{CO2} + H, limiting the fraction of OH going toward \ce{H2O}. The abundance of solid-phase \ce{CO2} reaches a peak within the first 10$^5$~yr of the simulation, representing 60--65~\% of the water abundance (Fig.~\ref{FIG-ice-mantle-composition}b). Then, due to the high accretion rate of gas-phase CO onto the dust grains, a rapid switch-over occurs and CO dominates over \ce{CO2}. \ce{H2CO} and \ce{CH3OH} form by the successive hydrogenation of surface CO.

Table~\ref{TAB-ice-composition} gives the chemical composition of the ice mantles at the end of the quasi-static contraction phase for the four standard models. The composition of ices observed in the line of sight of the field star Elias 16 \citep{whittet1998} and toward the high-mass protostar W33A \citep{gibb2000} are also given for comparison. Models N2-T15-CR1, N4-T15-CR1, and N5-T15-CR1 build up ices with similar chemical composition, also comparable (within a factor~$\sim$~2) to that of Elias~16. The ice composition in model N3-T15-CR1 differs from that of the other models, in particular with higher abundances of \ce{H2CO} and \ce{CH3OH} (see also Figure~\ref{FIG-ice-mantle-composition}a). These differences are due to the higher densities computed during the contraction phase at $r_{\rm start}$ in the more compact inner region of Sgr~B2(N3) compared to the other sources. High densities have an impact on the accretion rate of gas-phase species onto the dust grains, such that after 1~Myr simple species start to deplete onto dust grains in N3-T15-CR1 (Fig.~\ref{FIG-ice-mantle-composition}c).

\begin{table}[!t]
\begin{center}
  \caption{\label{TAB-ice-composition} Ice mantles composition.} 
  \vspace{-2mm}
  \setlength{\tabcolsep}{1.2mm}
  \begin{tabular}{lrrrrrr}
  \hline
        & \multicolumn{4}{c}{Standard models} & \multicolumn{2}{c}{Observations} \\ 
Species & N2 & N3 & N4 & N5 & Elias 16\tablefootmark{*} &  W33A\tablefootmark{*} \\
\hline
\hline
 \ce{H2O}   &  100 &  100 &  100 &  100 & 100     & 100 \\ 
 \ce{CO}    &   51 &   44 &   55 &   52 &   25    & 8 \\
 \ce{CO2}   &   29 &   49 &   25 &   27 &  18     & 13 \\
 \ce{NH3}   &   15 &   24 &    9 &   13 & $\leq$9 & 15 \\
 \ce{CH4}   &    3 &    4 &    1 &    2 &  \_     & 1.5 \\
 \ce{H2CO}  & 0.07 &  0.3 & 0.02 & 0.05 &  \_     & 6  \\
 \ce{CH3OH} & 0.07 &  0.8 & 0.03 & 0.05 & $<$3    & 18 \\
\hline
\end{tabular}
\end{center}
\vspace{-4mm}
\tablefoot{Abundances of the main ice constituents, summed over all the ice layers formed at the end of the quasi-static contraction phase. These results are for the standard models and are given in percentage of the water ice value.
\tablefoottext{*}{Observational values from \citet{gibb2000} and references therein.}}
\end{table}

\subsubsection{Production of COMs}
\label{section-formation-routes-COMs}

\begin{figure*}[!t]
 \hspace{0.12\linewidth}
 \begin{minipage}{0.4\linewidth}
   \includegraphics[scale=0.65]{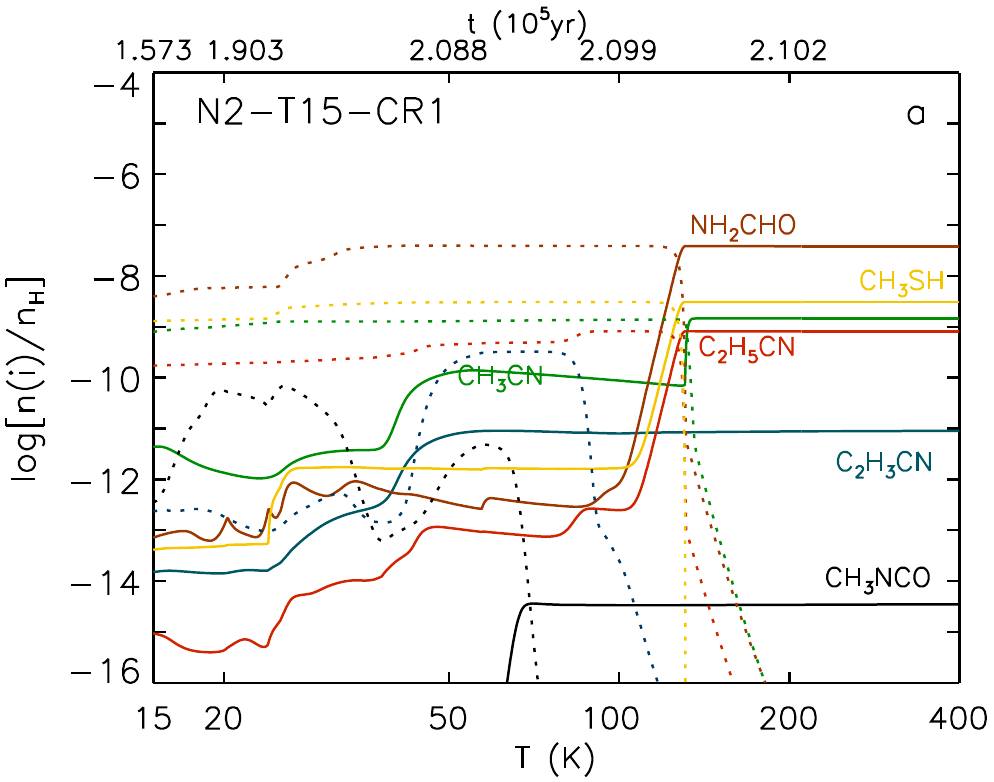} \\
   \includegraphics[scale=0.65]{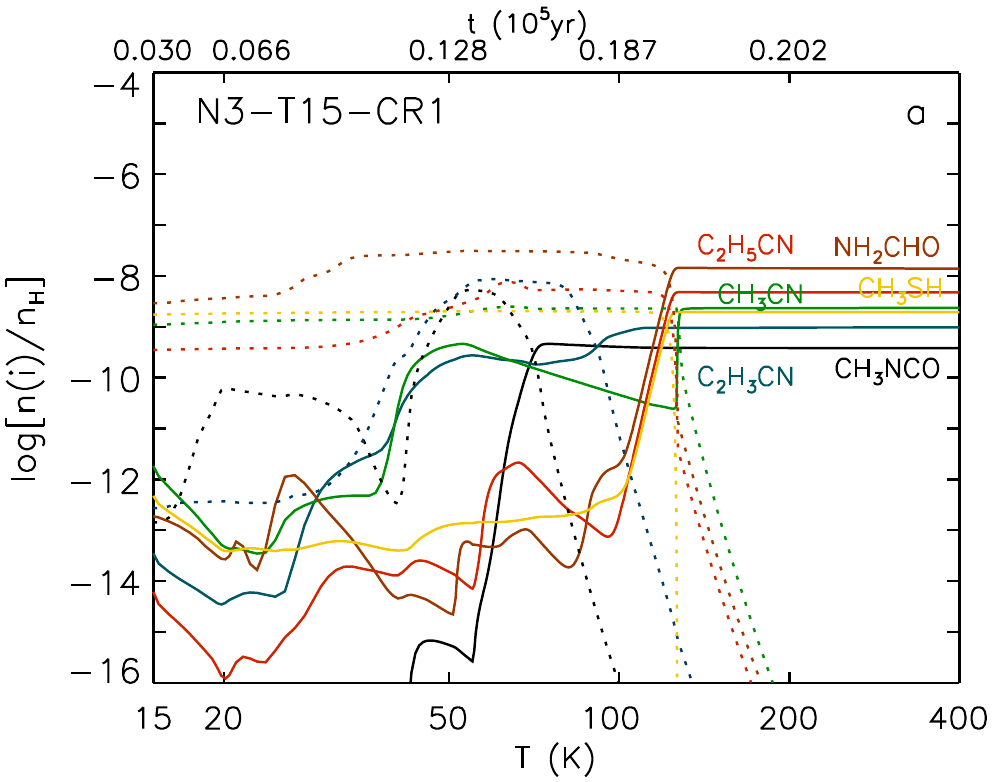} \\
   \includegraphics[scale=0.65]{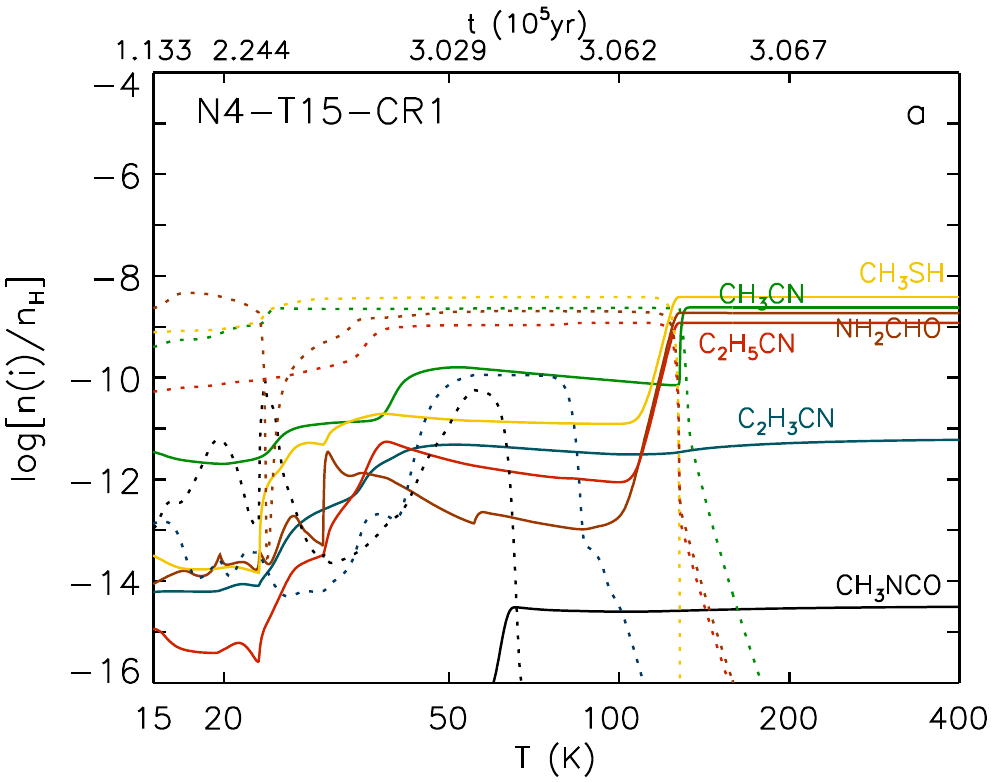} \\
   \includegraphics[scale=0.65]{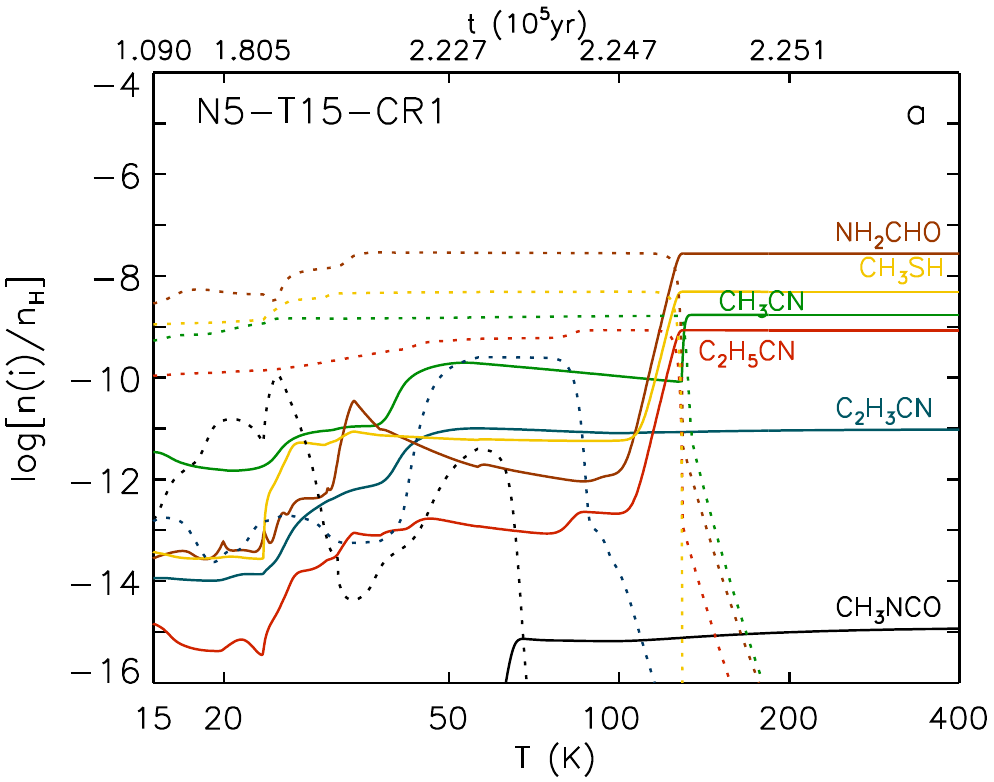} \\
 \end{minipage}
\hspace{-0.02\linewidth}
 \begin{minipage}{0.35\linewidth}
    \includegraphics[scale=0.65]{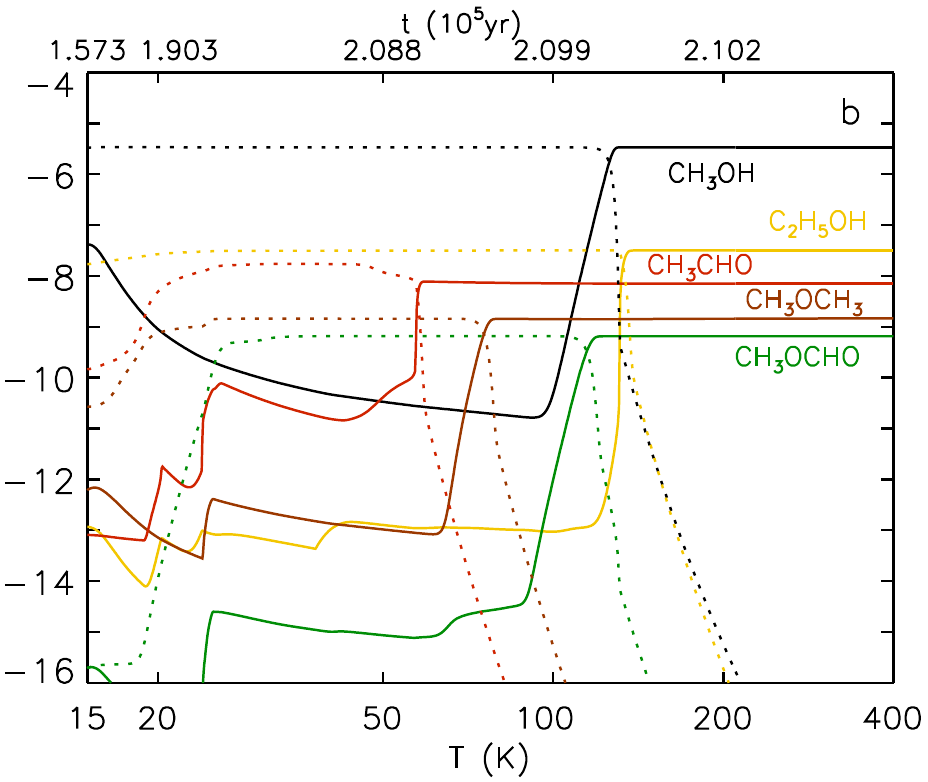} \\
    \includegraphics[scale=0.65]{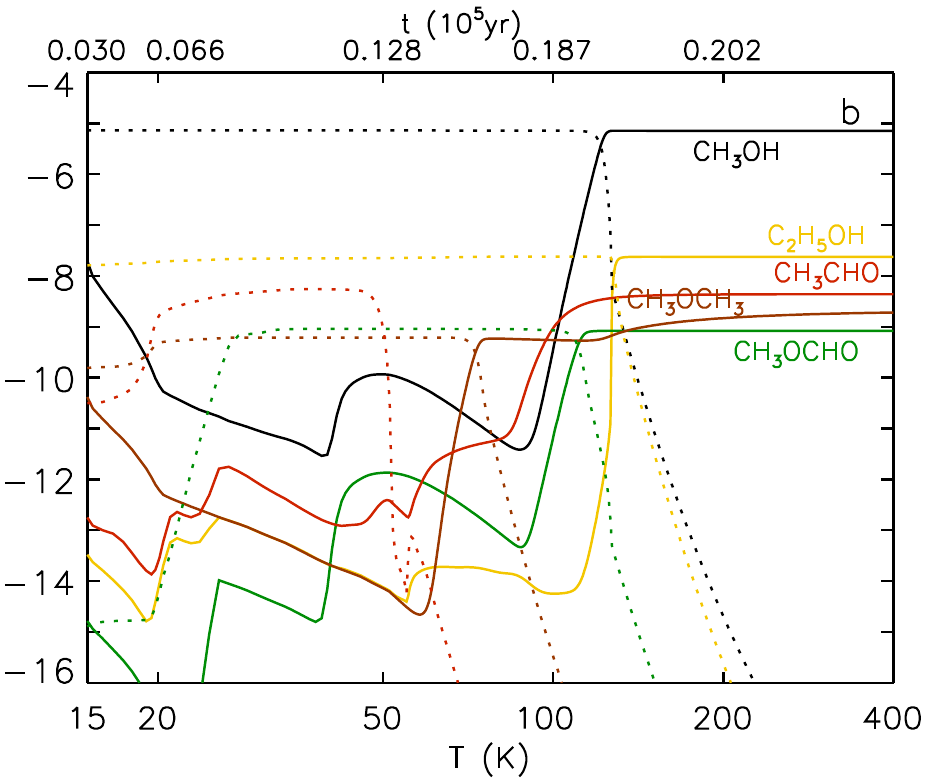} \\
    \includegraphics[scale=0.65]{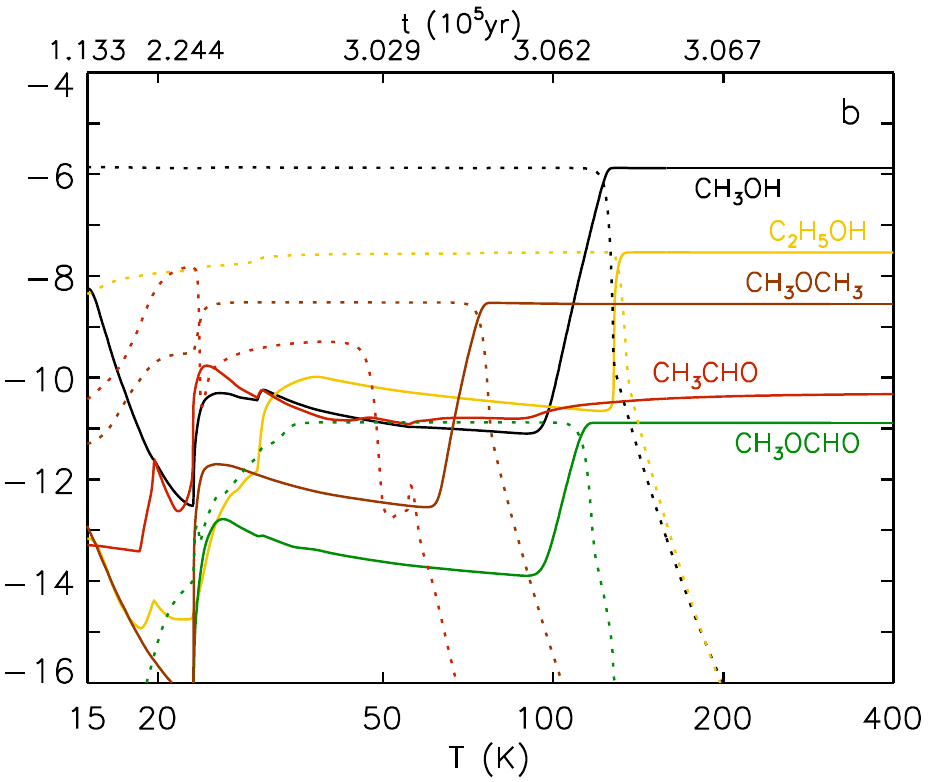} \\
    \includegraphics[scale=0.65]{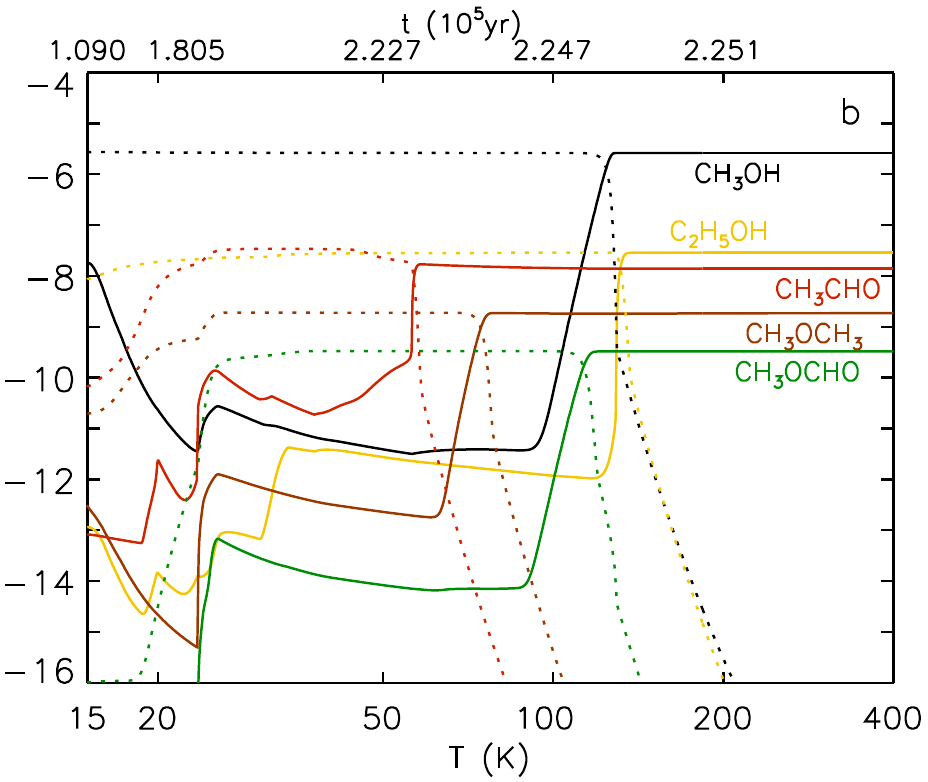} \\
 \end{minipage}
 \caption{\label{FIG-evolution-frac-abund} Calculated fractional abundances for 11 COMs detected toward Sgr~B2(N2-N5). Panels a show the N-bearing species \ce{C2H5CN}, \ce{C2H3CN}, \ce{CH3CN}, \ce{NH2CHO}, and \ce{CH3NCO}, as well as \ce{CH3SH}. Panels b show the O-bearing species \ce{CH3OH}, \ce{C2H5OH}, \ce{CH3OCHO}, \ce{CH3OCH3}, and \ce{CH3CHO}. The results for the standard models, N(2-5)-T15-CR1, are shown from top to bottom, as a function of the temperature in the envelope of the sources during the free-fall collapse phase. In each panel the timescale shown at the top is derived from the time-dependent evolution of the temperature in the envelopes of the sources (see Sect.~\ref{section-collapse-phase}). In each panel, the solid lines show the fractional abundances (with respect to total hydrogen) in the gas phase while the dotted lines show the abundances of the same species on the grains (ice surface+mantle).}
\end{figure*}

In this section we investigate the set of chemical reactions that drive the production of the 11 COMs listed in Table~\ref{TAB-best-fit-parameters}. All activation-energy barriers (when there is one) and binding energies may be found in \citet{garrod2013}, \citet{belloche2017}, and references therein. Figure~\ref{FIG-evolution-frac-abund} shows the results obtained for the standard models N(2-5)-T15-CR1 under free-fall collapse conditions. Each panel (a and b) shows the evolution of the fractional abundances of a subset of the sample of COMs. Panel (a) shows the N-bearing species \ce{C2H5CN}, \ce{C2H3CN}, \ce{CH3CN}, \ce{NH2CHO}, and \ce{CH3NCO}, as well as \ce{CH3SH}. Panel (b) shows the O-bearing species \ce{CH3OH}, \ce{C2H5OH}, \ce{CH3OCHO}, \ce{CH3OCH3}, and \ce{CH3CHO}. In each panel the solid lines indicate the gas-phase species while the dotted lines of the same color indicate the solid-phase species, combining both ice-surface and ice-mantle abundances. The calculated fractional abundances are plotted as a function of temperature in the envelope of the source (in log scale) for easy comparison between all models. A timescale is also indicated at the top of each plot. The figure shows that in most cases, the solid-phase abundances of the investigated COMs directly determine the final gas-phase abundances. This suggests that the gas-phase abundances observed in the warm envelopes of Sgr~B2(N2-N5) are dominated by the thermal desorption of the dust-grain ice mantles. In all four models, most COMs are already present on the grains with significant abundances before the warm-up phase ($T$~=~15~K). It is the case for the cyanides \ce{C2H5CN} and \ce{CH3CN}, as well as \ce{C2H5OH}, \ce{CH3OH}, and \ce{CH3SH}, indicating that their production mostly relies on the early cold chemistry. Other species still form efficiently on the grains up to $\sim$~50~K, in particular the O-bearing species \ce{CH3CHO}, \ce{CH3OCHO}, \ce{CH3OCH3}, \ce{NH2CHO}, and \ce{CH3NCO}, suggesting that their production relies on the warm-up stage chemistry.

The peak gas-phase fractional abundances of the ten investigated COMs are given in Table~\ref{TAB-peak-desorption-collapse} along with the temperatures at which they are achieved for each of the standard models. For comparison, we list also the fractional abundances reached at the end of the simulations (that is $T$~=~400~K). Most species reach their peak abundance right after desorption into the gas phase (except for \ce{C2H3CN} and \ce{CH3NCO}), then they quickly reach steady fractional abundances. The O-bearing species \ce{CH3OH}, \ce{C2H5OH}, \ce{NH2CHO}, \ce{CH3OCHO}, as well as \ce{CH3CN}, \ce{C2H5CN}, and \ce{CH3SH} desorb from the grains at high temperatures (120--147~K). The species which desorb at lower temperatures (via thermal or chemical desorption), or which are formed in the gas phase via ion-molecule reactions, are particularly exposed to gas-phase destruction through photodissociation by CR-induced UV photons or reactions with gas-phase ions.

There is little gas-phase formation of methanol; it is mostly formed on the grains via successive hydrogenation of the CO accreted from the gas phase: 
\begin{equation}
\ce{CO} \xrightarrow{\text{H}} \ce{HCO} \xrightarrow{\text{H}} \ce{H2CO} \xrightarrow{\text{H}} \ce{CH3O} \xrightarrow{\text{H}} \ce{CH3OH} .
\end{equation} 
The high gas-phase fractional abundance of \ce{CH3OH} at low temperature is mostly due to chemical desorption from the dust grains. Model N3-T15-CR1 shows a modest increase in the gas-phase abundance of \ce{CH3OH} around T~$\sim$~40~K, caused by the electronic recombination of HC(OH)O\ce{CH3}$^+$, product of the reaction between protonated methanol (\ce{CH3OH2}$^+$) and \ce{H2CO} when the latter is released abundantly into the gas phase.

Ethanol, \ce{C2H5OH}, is predominantly produced on the grains, mostly via the sequence:  
\begin{equation}
\label{chemical-reaction-c2h5oh}
\ce{CH3} \xrightarrow{\text{C}} \ce{C2H3} \xrightarrow{\text{H}} \ce{C2H4} \xrightarrow{\text{H}} \ce{C2H5} \xrightarrow{\text{O}} \ce{C2H5O} \xrightarrow{\text{H}} \ce{C2H5OH} ,
\end{equation}
followed by later desorption into the gas phase.

Formamide, \ce{NH2CHO}, is mainly produced on the grains via the successive hydrogenation of OCN accreted from the gas phase: 
\begin{equation}
\label{chemical-reaction-nh2cho}
\ce{OCN} \xrightarrow{\text{H}} \ce{HNCO} \xrightarrow{\text{H}} \ce{NH2CO} \xrightarrow{\text{H}}  \ce{NH2CHO} .
\end{equation}

The abundance of methyl cyanide, \ce{CH3CN}, mostly derives from the grain-surface hydrogenation reaction:
\begin{equation}
\label{chemical-reaction-ch3cn}
\ce{CH2CN} + \mathrm{H} \rightarrow \ce{CH3CN} ,
\end{equation}
where \ce{CH2CN} is formed via the grain-surface atomic-addition sequence: CN $\xrightarrow{\text{C}}$ \ce{C2N} $\xrightarrow{\text{H}}$ \ce{HC2N} $\xrightarrow{\text{H}}$ \ce{CH2CN}. At low temperatures ($T$~$<$~50~K), \ce{CH3CN} is also produced in the gas phase via the electronic recombination of protonated methyl cyanide (\ce{CH3CNH}$^+$), product of the ion-molecule reaction between \ce{CH3}$^+$ and \ce{HCN}. 

Methyl formate, \ce{CH3OCHO}, begins the free-fall collapse phase with modest abundances on the grains. It is formed more efficiently at later times ($T$~$\sim$~20--30~K), mainly via the grain-surface radical-radical addition reaction: 
\begin{equation}
\label{chemical-reaction-ch3ocho}
\ce{HCO} + \ce{CH3O} \rightarrow \ce{CH3OCHO} ,
\end{equation}
followed by later desorption into the gas phase. Model N3-T15-CR1 shows a modest increase in the gas-phase abundance of \ce{CH3OCHO} around 40~K, when (\ce{H2CO}) is abundantly released into the gas phase and reacts with \ce{CH3OH2}$^+$ to form HC(OH)O\ce{CH3}$^+$. Gas-phase \ce{CH3OCHO} is then produced via the electronic recombination of HC(OH)O\ce{CH3}$^+$. 

Methyl mercaptan, \ce{CH3SH}, is predominantly formed on the grains via the successive hydrogenation reactions: 
\begin{equation}
\label{chemical-reaction-ch3sh}
\ce{H2CS} \xrightarrow{\text{H}} \ce{CH2SH}/\ce{CH3S} \xrightarrow{\text{H}} \ce{CH3SH} .
\end{equation}
During the cold quasi-static contraction phase, \ce{CH3SH} forms via the successive hydrogenation of surface CS accreted from the gas phase. Above $\sim$~20~K, \ce{H2CS} is accreted onto the grains directly from the gas phase, where it is formed via the electronic recombination of \ce{H3CS}$^+$, deriving from the ion-molecule reaction S$^+$~+~\ce{CH4}.

Vinyl cyanide, \ce{C2H3CN},  is formed on dust grains through the successive hydrogenation of \ce{HC3N} accreted from the gas phase: 
\begin{equation}
\label{chemical-reaction-c2h3cn-grain}
\ce{HC3N} \xrightarrow{\text{H}} \ce{C2H2CN} \xrightarrow{\text{H}} \ce{C2H3CN}
\end{equation}
Then, as hydrogenation continues on the grain surfaces, \ce{C2H3CN} is quickly converted to \ce{C2H5CN} as follows:
\begin{equation}
\label{chemical-reaction-c2h5cn}
\ce{C2H3CN} \xrightarrow{\text{H}} \ce{C2H4CN} \xrightarrow{\text{H}} \ce{C2H5CN} .
\end{equation}
Ethyl cyanide, \ce{C2H5CN}, can thus maintain a steady abundance on the grains, dominating \ce{C2H3CN} at all times. Only a small fraction of solid-phase \ce{C2H3CN} contributes to the gas-phase abundance as it is quickly destroyed via a ion-molecule reaction right after desorption. At late times \ce{C2H3CN} may form in the gas phase via the reaction: 
\begin{equation}
\label{chemical-reaction-c2h3cn-gas}
\ce{C2H4} + \ce{CN} \rightarrow \ce{C2H3CN} + \ce{H} .
\end{equation}

Due to their low binding energies to water ices, \ce{CH3NCO} \citep[3575~K,][]{belloche2017}, \ce{CH3OCH3} \citep[3675~K,][]{garrod2013}, and \ce{CH3CHO} \citep[2275~K,][]{garrod2013} desorb at low temperatures (that is around 60--80~K). Methyl isocyanate, \ce{CH3NCO}, starts the free-fall collapse phase with modest abundances on the grains. It is more efficiently produced around 20~K through the radical-addition reaction:
\begin{equation}
\label{chemical-reaction-ch3nco}
\ce{CH3} + \ce{OCN} \rightarrow \ce{CH3NCO} .
\end{equation}
At low temperatures ($T$~$<$~50~K), \ce{CH3NCO} may be destroyed on the grains via hydrogenation reactions to form \ce{CH3NHCO}. In addition, \ce{CH3NCO} is quickly destroyed via gas-phase ion-molecule reactions right after desorption, leading to very low gas-phase fractional abundances at the end of the simulations ($<$~10$^{-14}$ with respect to total hydrogen, for all standard models but N3-T15-CR1). In model N3-T15-CR1, \ce{CH3NCO} is formed much more abundantly on the grains, mainly because of the high abundance of surface \ce{CH3} accreted from the gas phase. Therefore it cannot be destroyed completely once it is released into the gas phase, explaining the higher fractional abundances of gas-phase \ce{CH3NCO} observed in this model.   

Dimethyl ether, \ce{CH3OCH3}, is mostly formed on the surface of dust grains up to $T$~$\sim$~20--30~K, through the radical-addition reaction:
\begin{equation}
\label{chemical-reaction-ch3och3}
\ce{CH3} + \ce{CH3O} \rightarrow \ce{CH3OCH3} .
\end{equation}
In model N3-T15-CR1, the gas-phase fractional abundance of \ce{CH3OCH3} increases when \ce{CH3OH} desorbs from the dust grains around 130~K. This is due to the electronic recombination reaction \ce{CH3OCH4}$^+$ + e$^-$ $\rightarrow$ \ce{CH3OCH3} + H, where \ce{CH3OCH4}$^+$ is a product of the reaction between \ce{CH3OH} and \ce{CH3OH2}$^+$.

Finally, acetaldehyde, \ce{CH3CHO}, is formed on the grains up to $T$~$\sim$~30~K, via the surface reactions: 
\begin{equation}
\ce{CH2} + \ce{HCO} \rightarrow \ce{CH2CHO} \xrightarrow{\text{H}} \ce{CH3CHO} ,
\end{equation}
\begin{equation}
\ce{CH3} + \ce{HCO} \rightarrow \ce{CH3CHO} .
\end{equation}
Surface \ce{CH3CHO} may be destroyed either by reacting with \ce{CH3} to form \ce{CH4} + \ce{CH3CO}, or with \ce{NH2} to form \ce{NH3} + \ce{CH3CO}. In particular in models N3-T15-CR1 and N4-T15-CR1, only a small fraction of solid-phase \ce{CH3CHO} contributes to the gas-phase fractional abundance. At lower temperature ($T$~$\sim$~24~K), surface \ce{CH3CHO} is released into the gas phase via chemical desorption from the reaction \ce{CH2CHO} + H. Once in the gas phase, \ce{CH3CHO} becomes the dominant reaction partner for gas-phase ions, damping ionic abundances and thus limiting the destruction of other gas-phase species. In models N3-T15-CR1 and N4-T15-CR1, \ce{CH3CHO} is formed directly in the gas phase via the reaction \ce{C2H5} + \ce{O} $\rightarrow$ \ce{CH3CHO} + \ce{H}, when \ce{C2H5} desorbs from dust grains around 90~K.

\begin{table*}[!t]
\begin{center}
  \caption{\label{TAB-peak-desorption-collapse} Peak gas-phase abundances and associated temperatures for the standard models.} 
  \vspace{-4mm}
  \setlength{\tabcolsep}{1.2mm}
  \begin{tabular}{l|llr|llr|llr|llr}
\hline
 Species   &  \multicolumn{3}{c|}{N2-T15-CR1} & \multicolumn{3}{c|}{N3-T15-CR1}   & \multicolumn{3}{c|}{N4-T15-CR1}   & \multicolumn{3}{c}{N5-T15-CR1} \\
           &  final\tablefootmark{a} & peak\tablefootmark{b}     &    \multicolumn{1}{c|}{$T$\tablefootmark{c}}    &   final\tablefootmark{a} & peak\tablefootmark{b}        &   \multicolumn{1}{c|}{$T$\tablefootmark{c}}      &  final\tablefootmark{a} & peak\tablefootmark{b}   &  \multicolumn{1}{c|}{$T$\tablefootmark{c}}      & final\tablefootmark{a} & peak\tablefootmark{b}  &   \multicolumn{1}{c}{$T$\tablefootmark{c}}    \\
           & \multicolumn{2}{c}{($n[i]$/$n_{\rm H}$)} &  (K)    &  \multicolumn{2}{c}{($n[i]$/$n_{\rm H}$)}  &   (K)   &  \multicolumn{2}{c}{($n[i]$/$n_{\rm H}$)}  &   (K)  &  \multicolumn{2}{c}{($n[i]$/$n_{\rm H}$)}  &   (K)  \\
\hline
\hline
\ce{C2H5CN}  & 8.16(-10)  &   8.18(-10) &   131.4  & 4.70(-9)  & 4.81(-9)  &   127.1  & 1.19(-9)  &   1.20(-9)   &   128.5 & 8.53(-10)  &   8.56(-10)  &   129.7  \\
\ce{C2H3CN}  & 9.00(-12)  &   9.05(-12) &    61.4  & 9.80(-10) & 9.80(-10) &   400.0  & 6.01(-12) &   6.01(-12)  &   400.0 & 9.57(-12)  &   1.01(-11)  &    56.2  \\
\ce{CH3CN}   & 1.46(-9)   &   1.46(-9)  &   145.6  & 2.36(-9)  & 2.36(-9)  &   134.4  & 2.41(-9)  &   2.41(-9)  &   138.9 & 1.73(-9)   &   1.73(-9)   &   145.3  \\
\ce{CH3OH}   & 3.38(-6)   &   3.38(-6)  &   131.4  & 7.10(-6)  & 7.18(-6)  &   127.1  & 1.33(-6)  &   1.33(-6)   &   127.9 & 2.62(-6)   &   2.63(-6)   &   129.7  \\
\ce{C2H5OH}  & 3.18(-8)   &   3.19(-8)  &   147.3  & 2.38(-8)  & 2.39(-8)  &   143.8  & 2.90(-8)  &   2.91(-8)  &   140.0 & 2.87(-8)   &   2.87(-8)   &   143.4  \\
\ce{CH3OCHO} & 6.53(-10)  &   6.53(-10) &   129.6  & 8.32(-10) & 8.37(-10) &   126.3  & 1.30(-11) &   1.30(-11)  &   120.6 & 3.29(-10)  &   3.29(-10)  &   126.5  \\
\ce{CH3OCH3} & 1.44(-9)   &   1.44(-9)  &    80.1  & 1.92(-9)  & 1.92(-9)  &   400.0  & 2.82(-9)  &   2.97(-9)   &    77.6 & 1.86(-9)   &   1.87(-9)   &    78.9  \\
\ce{CH3CHO}  & 7.09(-9)   &   7.77(-9)  &    59.4  & 4.38(-9)  & 4.38(-9)  &   182.8  & 4.76(-11) &   1.72(-10)  &    24.4 & 1.40(-8)   &   1.70(-8)   &    58.4  \\
\ce{NH2CHO}  & 3.84(-8)   &   3.85(-8)  &   131.4  & 1.40(-8)  & 1.45(-8)  &   127.1  & 1.86(-9)  &   1.88(-9)   &   127.9 & 2.74(-8)   &   2.77(-8)   &   129.7  \\
\ce{CH3NCO}  & 3.49(-15)  &   3.62(-15) &    70.6  & 3.82(-10) & 4.62(-10) &    75.1  & 3.12(-15) &   3.12(-15)  &    67.0 & 1.17(-15)  &   1.17(-15)  &   400.0  \\
\ce{CH3SH }  & 3.08(-9)   &   3.08(-9)  &   131.4  & 1.94(-9)  & 1.95(-9)  &   126.3  & 3.85(-9)  &   3.86(-9)   &   127.9 & 4.90(-9)   &   4.91e(-9)  &   129.7  \\
\hline
\end{tabular}
\end{center}
\vspace{-4mm}
\tablefoot{$X(Y)$ means $X \times 10^Y$. 
\tablefoottext{a}{Gas-phase fractional abundances (with respect to total hydrogen) of the standard models at the end of the simulations.}
\tablefoottext{b}{Gas-phase peak fractional abundances (with respect to total hydrogen) reached during the free-fall collapse phase.}
\tablefoottext{c}{Temperature at which the peak fractional abundance is achieved.}
}
\end{table*}

\begin{figure}[!t]
   \begin{center}
    \includegraphics[width=\hsize]{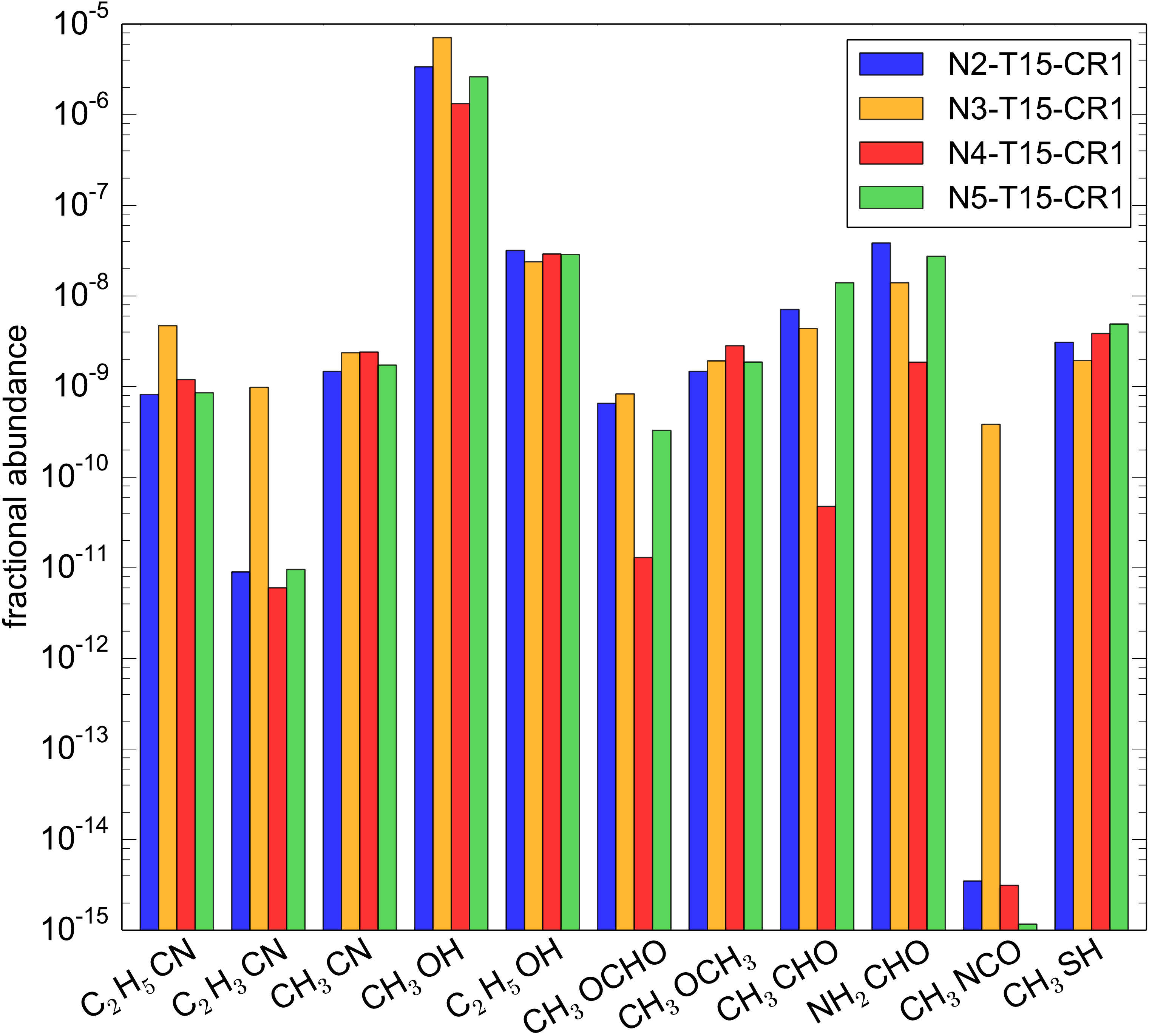} 
    \caption{\label{FIG-final-abundances-models} Fractional abundances (with respect to total hydrogen) calculated for the standard models at the end of the simulations (that is at $T$~=~400~K).}
   \end{center}
\end{figure}

The gas-phase fractional abundances of the 11 investigated COMs obtained at the end of the chemical simulations (that is at $T$~=~400~K) are given for the standard models in Table \ref{TAB-peak-desorption-collapse} and plotted in Fig.~\ref{FIG-final-abundances-models}. For each of the 11 COMs the abundances calculated in models N2-T15-CR1 and N5-T15-CR1 are relatively similar. Model N4-T15-CR1 shows significantly lower abundances of \ce{CH3CHO}, \ce{CH3OCHO}, and \ce{NH2CHO} than the other models. Model N3-T15-CR1 significantly differs from the other models. It shows higher abundances of cyanides, and has an astonishing [\ce{C2H5CN}]/[\ce{C2H3CN}] ratio $>$18 times lower than the other models. The final fractional abundance of \ce{CH3OH} is 2--5 times higher than in the other three models. Finally, N3-T15-CR1 shows an astonishing high abundance of \ce{CH3NCO}, more than five orders of magnitude higher than for the other models.

\section{Discussion}
        \label{section-disussion}

\subsection{Physical properties of Sgr~B2(N)'s hot cores}
\label{section-discussion-obs-NH2}

In Sect.~\ref{section-masses} we computed H$_2$ column densities from the dust continuum emission measured toward Sgr~B2(N2-N5) in the ALMA continuum emission maps. The results reported in Table~\ref{TAB-HCs-properties} show that at a given radius, Sgr~B2(N2) has the highest density, lying a factor $\sim$3 above N3 and N4 which have similar densities. The density computed for Sgr~B2(N5) lies in between that of the other sources. Given the uncertainties on the calculated densities (Table~\ref{TAB-HCs-properties}), the differences between the hot cores may be marginal.

The results discussed above slightly differ from our previous analysis of the same ALMA data \citep{bonfand2017}, where for the H$_2$ column density calculations we assumed a dust mass opacity coefficient ($\kappa_{\rm 100 \, GHz}$~=~5.9$\times$10$^{-3}$~cm$^2$~g$^{-1}$) $\sim$~1.1 times smaller than the value we use at the same frequency (100~GHz) in this paper (see Fig.~\ref{FIG-appendix-kappa-opacities}). In the case of Sgr~B2(N3), which is not detected in the ALMA continuum maps at 3~mm, we used the peak flux density given by \citet{sanchezmonge2017} (see their Table~1, source AN08) measured at 1.3~mm in an ALMA synthesized beam of 0.4$\arcsec$. \citet{sanchezmonge2017} derived a H$_2$ column density of 4.5$\times$10$^{24}$~cm$^{-2}$ (see their Table~3) from the integrated flux measured over a source size of 0.59$\arcsec$ \citep[see Table~1 of][]{sanchezmonge2017}. We note that they used a mean molecular weight of $\mu$~=~2.33, which corresponds to that of the mean free particle, so that their calculated value has to be divided by $\sim$1.2 to obtain the H$_2$ column density. This gives $N_{\rm H_2}$~=~3.7$\times$10$^{24}$~cm$^{-2}$, a factor $\sim$~4 higher than the value we derived in this paper. The difference is due to the different assumptions made to compute the H$_2$ column density. They assumed a lower dust temperature of 100~K and used a dust mass opacity coefficient ($\kappa_{\rm 230 \, GHz}$~=~0.011~cm$^2$~g$^{-1}$) $\sim$~1.8 times smaller than our value at the same frequency.

In Sects.~\ref{section-masses} and \ref{section-radmc3d} we used H$_2$ column densities to compute masses and densities, which are important inputs for the chemical modeling. One must keep in mind that the uncertainties in H$_2$ column densities may have an impact on the chemical model results.

\subsection{Influence of the minimum dust temperature on the chemistry}
\label{section-discussion-Tmin}

As mentioned in Sect.\ref{section-precollapse-phase}, dust temperature measurements carried out at infrared wavelengths with the \textit{Herschel} Space Observatory report somewhat higher dust temperatures toward the GC region than other typical regions forming high-mass stars. For instance, \citet{longmore2012} derived dust temperatures ranging from 19~K to 27~K from the center to the edge of the galactic center cloud G0.253+0.016 (also known as ''the Brick''). \citet{guzman2015} reported measurements of $T_{\rm d}$~=~20--28~K toward Sgr~B2, somewhat higher than the 17~K measured for dense cores embedded in the NGC~6334 molecular cloud lying at a distance of 1.3~kpc from the Sun \citep{chibueze2014}. Recently \citet{belloche2016} suggested that the low deuteration fractions of COMs observed toward Sgr~B2(N) could be explained by either an elemental abundance of deuterium toward the GC region lower than in the solar neighborhood, or higher temperatures during the prestellar phase that would have decreased the efficiency of the deuteration process. Deuteration levels of methanol similar to that of Sgr~B2(N) measured toward the aforementioned high-mass star-forming region NGC 6334I located at a distance of $\sim$ 7~kpc from the GC favor the second hypothesis \citep{bogelund2018}. Gas kinetic temperatures ranging from 50 K to more than 100 K were found on the basis of imaging of \ce{H2CO} transitions with the APEX 12~m telescope \citep{ao2013, ginsburg2016}. \citet{ott2014} showed that the kinetic temperature distribution of molecular clumps within the region between the supermassive black hole Sgr~A$^*$ and Sgr~B2 peaks at $\sim$~38~K based on a survey of \ce{NH3} conducted with the Australia Telescope Compact Array (ATCA). In comparison, lower temperatures have been derived from \ce{NH3} lines toward several massive clumps outside the GC region \citep[$T_{\rm g}$~=~10--30~K,][]{giannetti2013}.

In this section we explore the impact on the formation of COMs of varying arbitrarily the minimum dust temperature, $T_{\rm min}$, that can be reached in the chemical simulations, in particular during the cold quasi-static contraction phase, to account for the high dust temperatures observed toward the GC.  

\begin{figure*}[!t]
   \resizebox{\hsize}{!}
   {\begin{tabular}{cc}
       \includegraphics[width=\hsize]{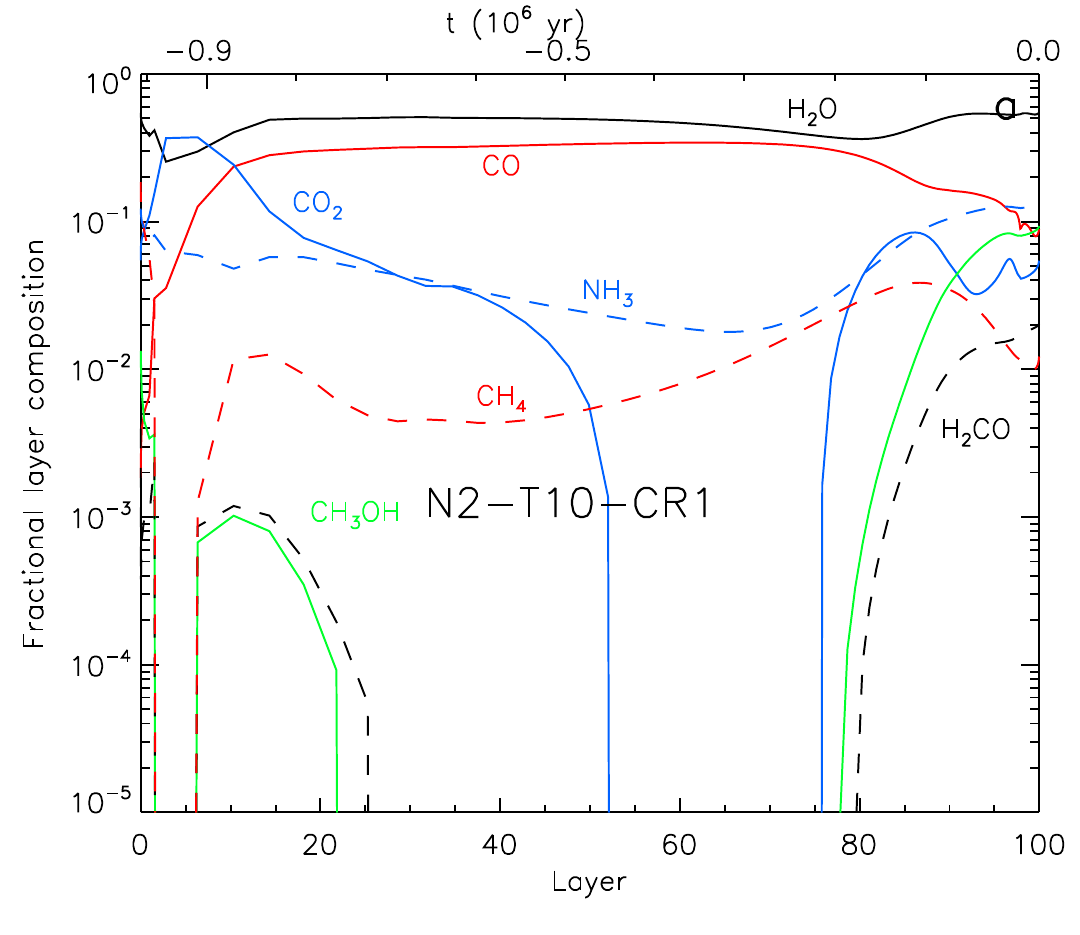} &
       \includegraphics[width=\hsize]{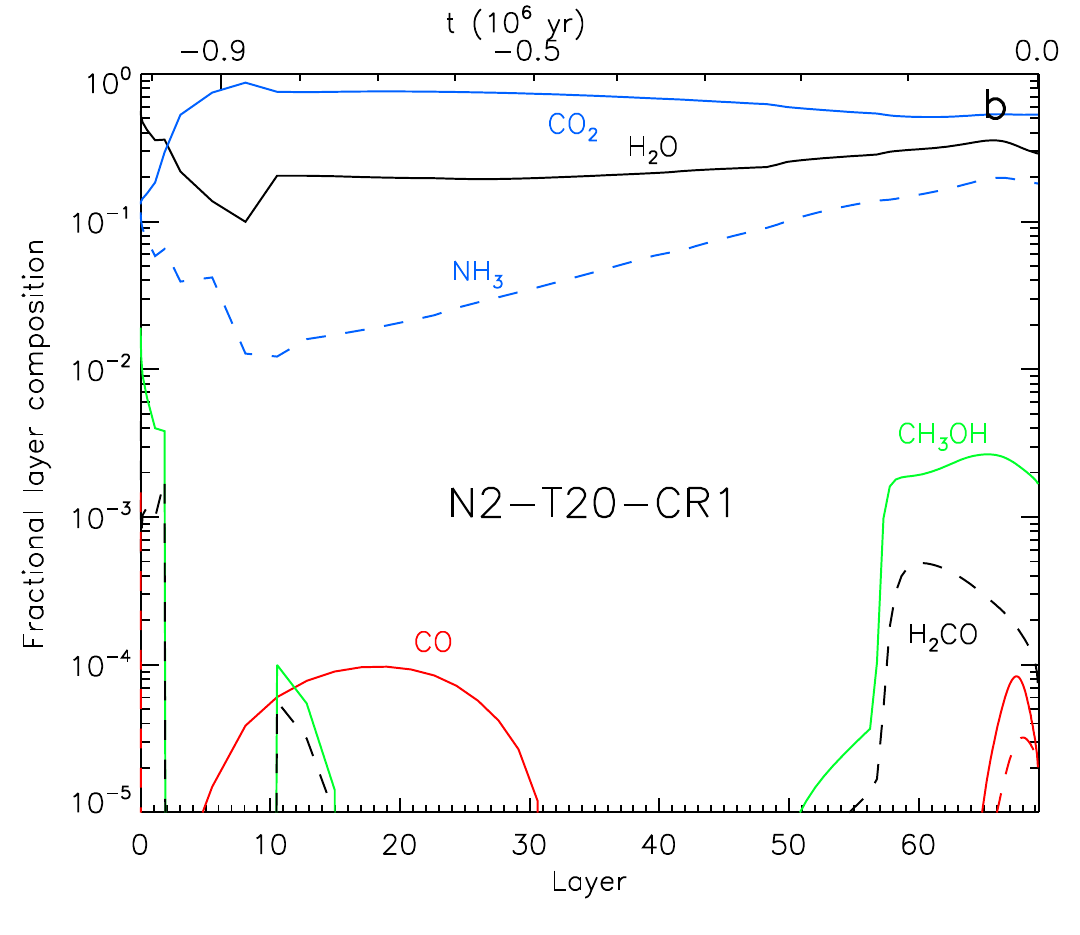}  \\
    \end{tabular}}
    \vspace{-3mm}
    \caption{\label{FIG-ice-mantle-T} \textbf{a} Composition of the ice-mantles built up during the cold quasi-static contraction phase prior to the free-fall collapse for the model N2-T10-CR1. \textbf{b} Same as a but for the model N2-T20-CR1.}
\end{figure*}

Figure~\ref{FIG-ice-mantle-T} shows the chemical composition of the ice mantles built up during the 1-Myr cold contraction phase for the ''low temperature'' model N2-T10-CR1 (panel a), and the ''high temperature'' model N2-T20-CR1 (panel b). Significant differences can be seen in the overall chemical composition of the ice mantles compared to the standard model (that is N2-T15-CR1, Fig.~\ref{FIG-ice-mantle-composition}a, top panel), indicating that the minimum dust temperature adopted in the chemical simulations is of critical importance. The ice mantle formed in the low-temperature model is enriched in CO and depleted in \ce{CO2} compared to the standard model. This is due to the less efficient CO-to-\ce{CO2} conversion when the dust temperature drops below $\sim$~14~K in the low-temperature model. At low temperature, H is converted more efficiently into \ce{H2} due to its lower thermal desorption rate, thus decreasing the abundance of gas phase atomic hydrogen compared to the standard model. Surface OH mostly reacts with \ce{H2} to form water ice rather than with CO to form \ce{CO2}. More CO is thus available to form \ce{H2CO} and \ce{CH3OH}, enhancing their abundances in the outer ice layers compared to the standard model. 

The high-temperature model produces thinner ice mantles than N2-T10/15-CR1, with fewer than 70 ice layers formed on dust grains after 1~Myr. The ice mantles are enriched in carbon dioxide as most surface CO and OH are destroyed to form \ce{CO2}, decreasing the amount of \ce{H2O}, \ce{H2CO}, and \ce{CH3OH} formed on the grains. The atomic carbon accreted to the grains mostly reacts with surface \ce{O2} instead of reacting with atomic hydrogen to form CH, which affects the abundance of \ce{CH4}. Furthermore, at high temperature reactions involving atomic hydrogen are less efficient because the H coverage of the grains is reduced due to rapid desorption, cutting down the formation of complex species.

\begin{figure*}[!t]
   \hspace{0.05\linewidth}
  \begin{minipage}{0.45\linewidth}
    \includegraphics[scale=0.26]{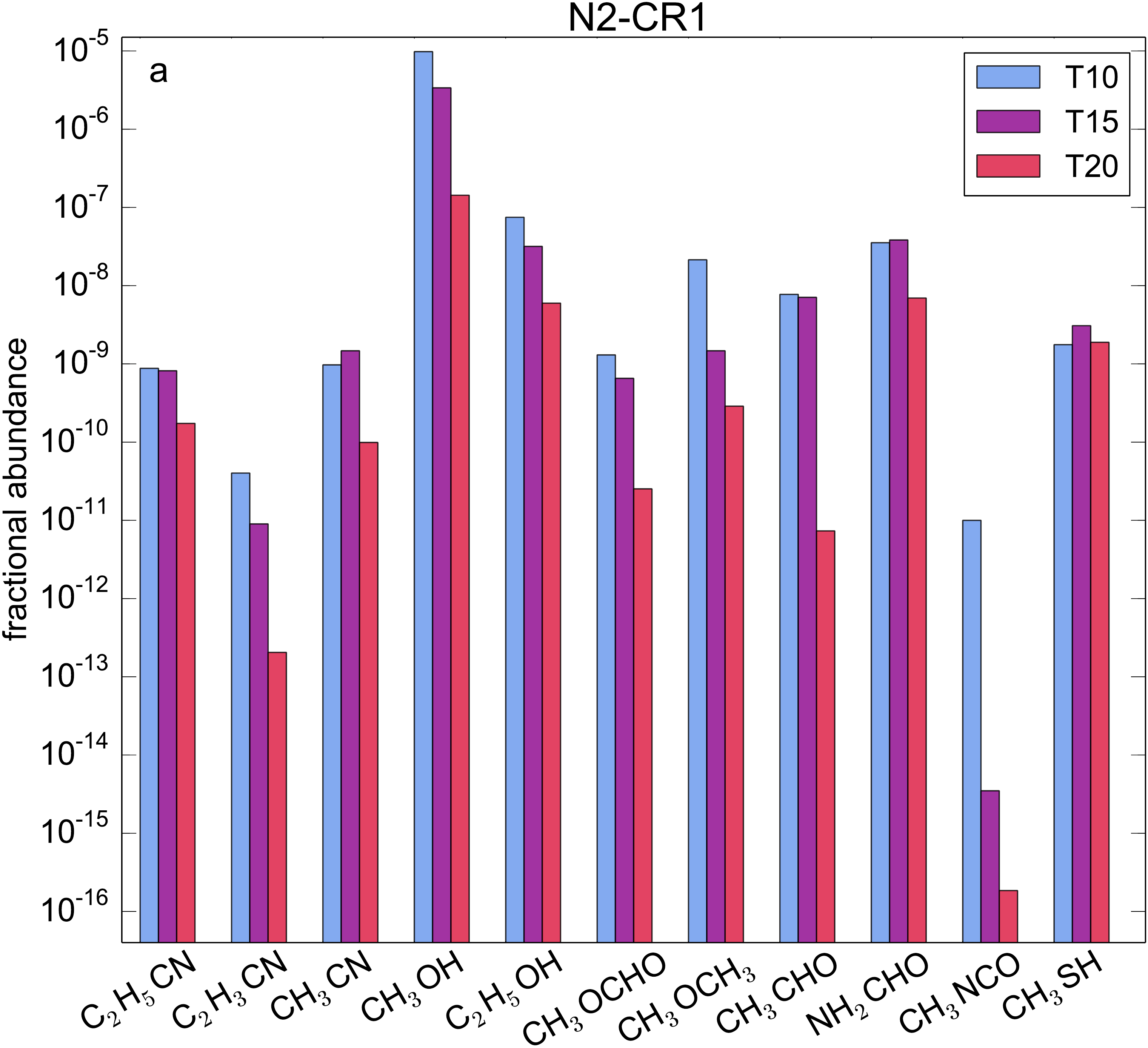} \\
    \includegraphics[scale=0.26]{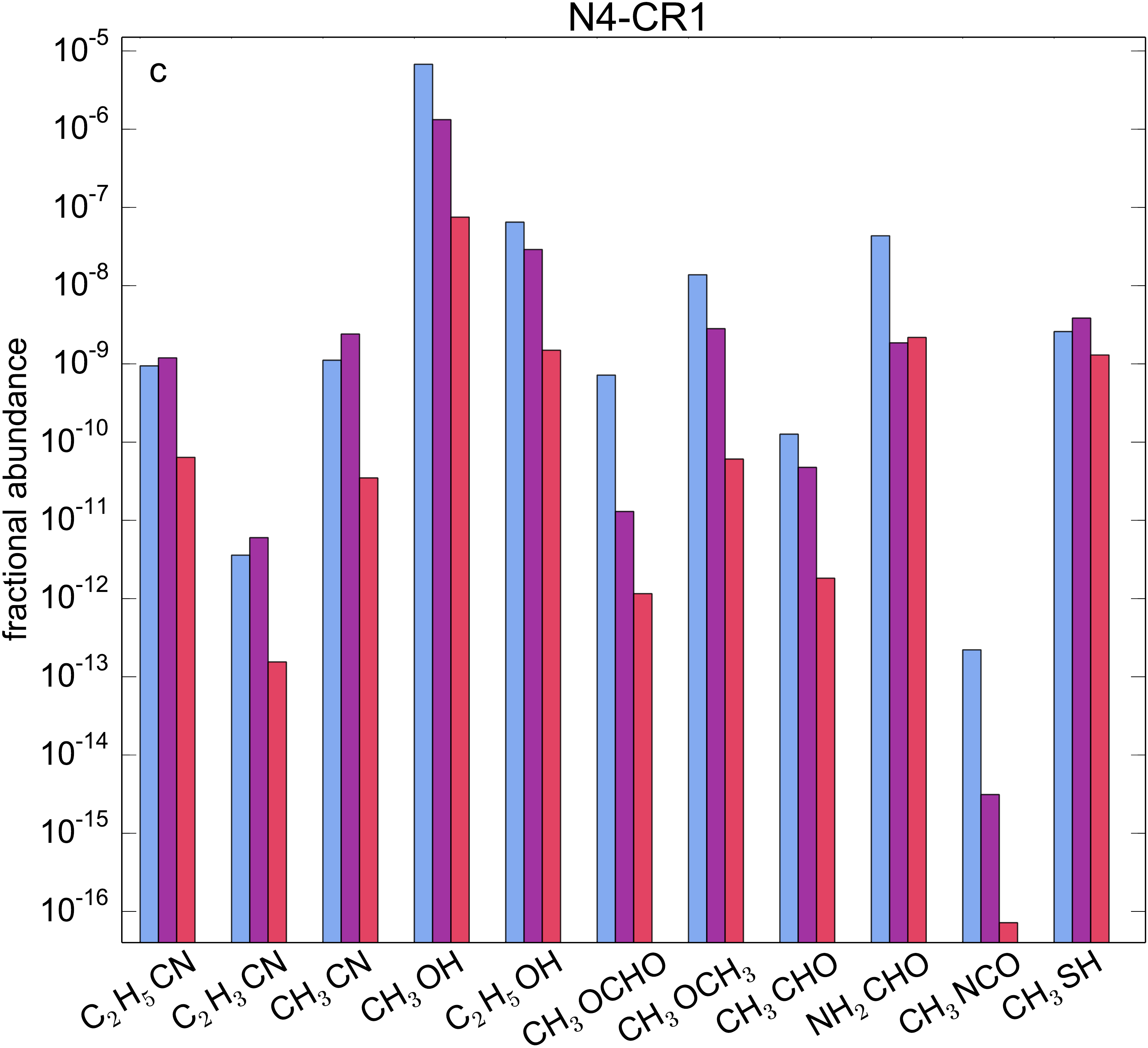} \\
 \end{minipage}
 \begin{minipage}{0.35\linewidth}    
    \includegraphics[scale=0.26]{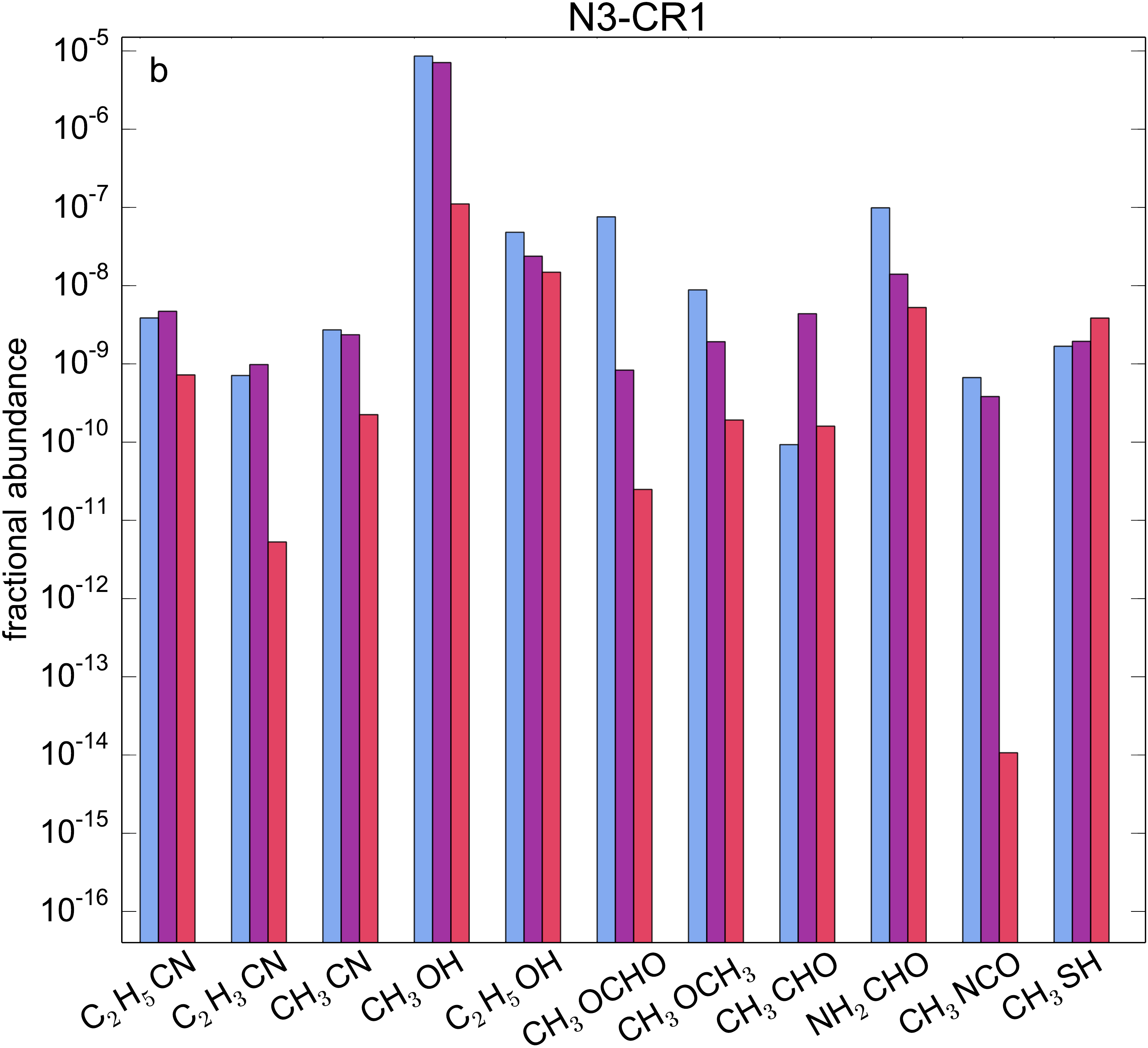}  \\
    \includegraphics[scale=0.26]{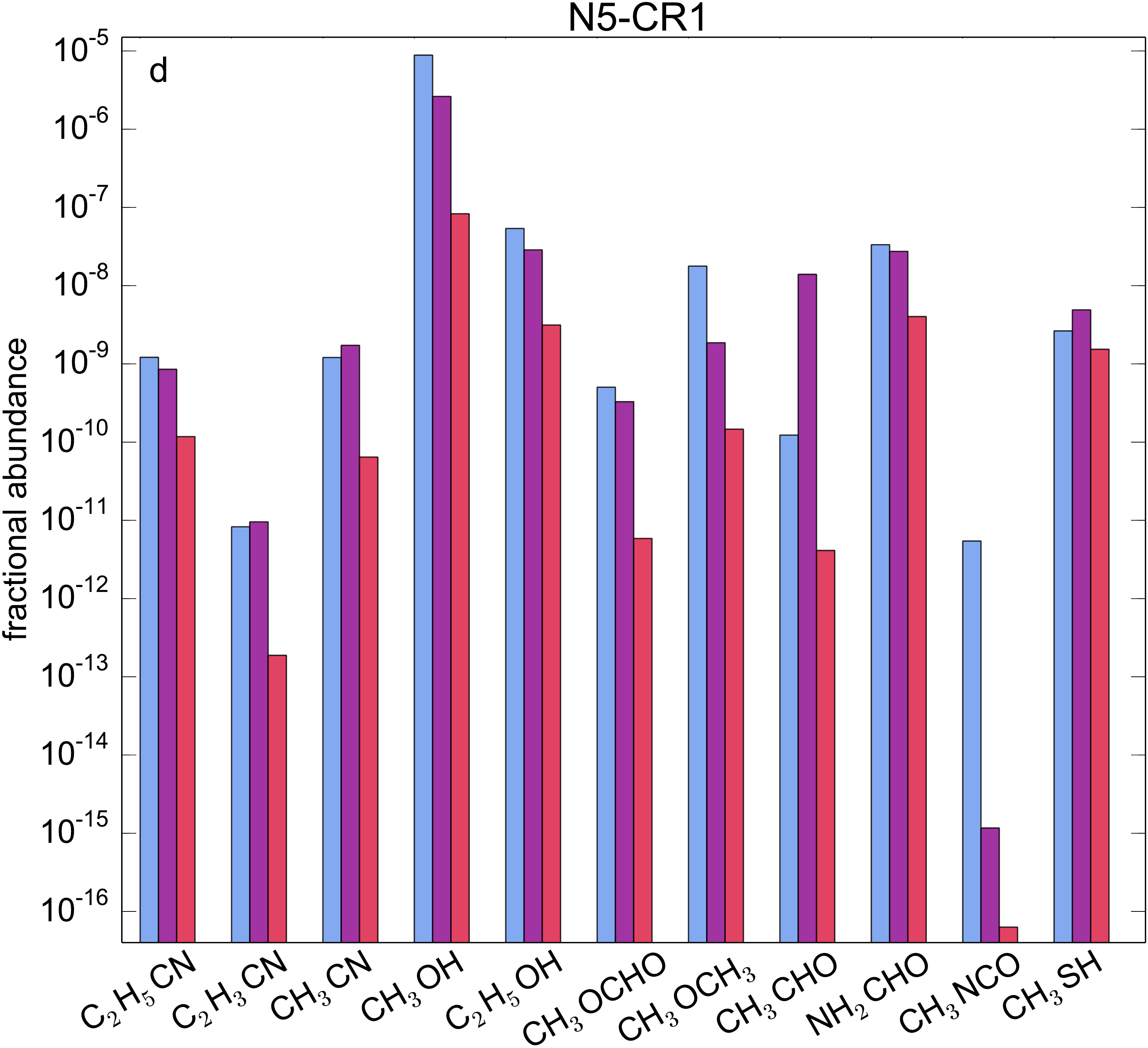} \\
     \end{minipage}
    \caption{\label{FIG-histos-T} Calculated fractional abundances (with respect to total hydrogen) of 11 COMs at the end of the simulations (that is $T$~=~400~K) for the models N2-CR1 (a), N3-CR1 (b), N4-CR1 (c), and N5-CR1 (d), assuming a standard CRIR and different $T_{\rm min}$: 10~K in blue, 15~K in purple, and 20~K in red (see Table~\ref{TAB-our-models}).}
\end{figure*}

The minimum dust temperature adopted in the chemical simulations also has an impact on the abundances of complex species at later times during the free-fall collapse phase, as the investigated COMs have been shown to be mainly produced on the grain surfaces (see Sect.~\ref{section-formation-routes-COMs}). Figure~\ref{FIG-histos-T} compares the final gas-phase fractional abundances of the low-, intermediate-, and high-temperature models (see also Figs.~\ref{FIG-appendix-collapse-Tmin-N2}--\ref{FIG-appendix-collapse-Tmin-N5}). In most cases, a higher $T_{\rm min}$ results in lower fractional abundances. In particular, the O-bearing species \ce{CH3OH}, \ce{CH3NCO}, \ce{CH3CHO}, \ce{CH3OCH3}, and \ce{CH3OCHO} are very sensitive to $T_{\rm min}$ with, for a given source, fractional abundances lower by more than two orders of magnitude in the high-temperature models compared to the low-temperature model. The fractional abundance of \ce{CH3NCO} varies by more than four orders of magnitude between the low- and high-temperature models for N2, N3, and N5. The fractional abundance of \ce{CH3SH} is barely sensitive to changes in $T_{\rm min}$, with at most a factor 3.2 of difference between the low-, intermediate-, and high-temperature models for all sources. This is due to its production mechanism, which relies on gas-phase \ce{H2CS} which sticks to the dust grains at $T$~$\sim$~20~K (see Sect.~\ref{section-formation-routes-COMs}).

A more accurate representation of the dust temperatures in the envelope of Sgr~B2(N2-N5), in particular in the outer parts exposed to external UV photons, requires to take into account the dust heating via the interstellar radiation field. The influence of the ISRF strength on the chemical model results is explored in Appendix~\ref{appendix-ISRF}. It shows that assuming a radiation field stronger than the standard ISRF value does not improve the agreement between calculated and observed abundances with respect to \ce{CH3OH} for Sgr~B2(N2-N5) (see discussion in Sect.~\ref{section-discussion-comparison-obs}).

Finally, a more realistic treatment of the grain-surface chemistry would require to consider a grain-size distribution rather than single dust-grain radius (0.1 micron) adopted in our simulations. \citet{pauly2016} investigated the effects of a broad grain-size distribution on the ice-surface chemistry by implementing a distribution of five distinct initial grain sizes to their dark cloud models, where the radii of each grain population increase as ice mantles form on the grains. These authors found that when using a uniform dust-grain temperature across all grain sizes the surface chemistry does not significantly differ from that of models assuming a single dust-grain radius. For chemical models with a more explicit treatment of the grain temperatures, in which $T_{\rm d}$ varies also with the grain-mantle growth of each individual grain-size population, \citet{pauly2016} found that the temperature difference plays a significant role in determining the total amount of ice formed and its composition for each grain-size population due to the temperature-sensitive nature of grain-surface reactions.

\subsection{Influence of the cosmic-ray ionization rate on the chemistry}
\label{section-discussion-CR}

\begin{figure*}[!t]
   \resizebox{\hsize}{!}
   {\begin{tabular}{ccc}
       \includegraphics[width=\hsize]{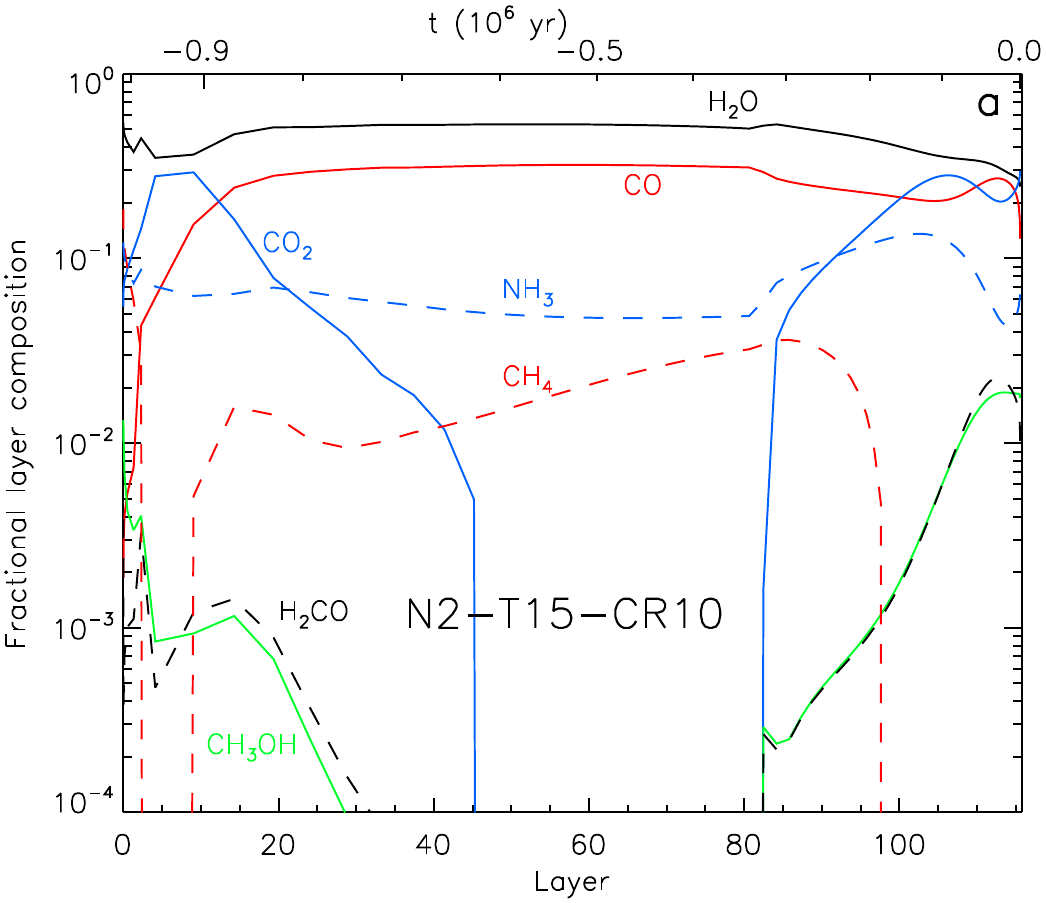} &
       \includegraphics[width=\hsize]{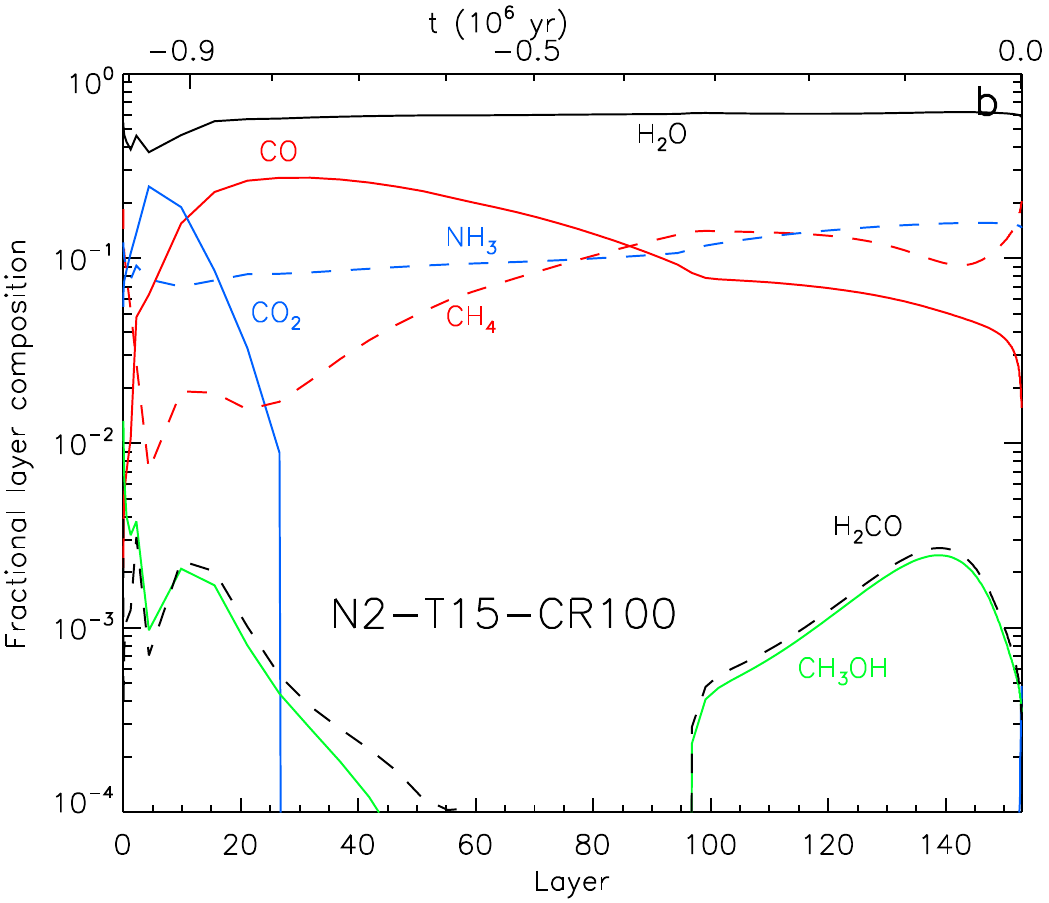} &
       \includegraphics[width=\hsize]{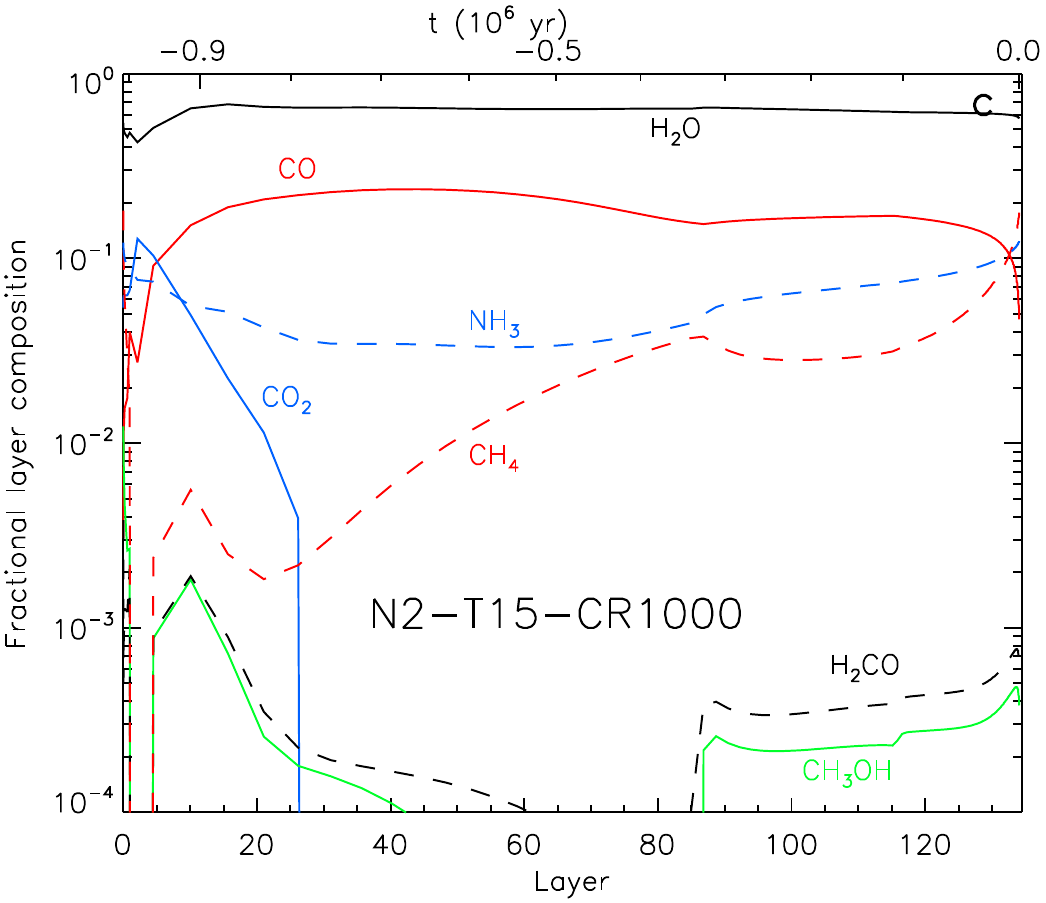} \\
    \end{tabular}}
    \caption{\label{FIG-ices-CR} Composition of the ice mantles built up during the cold quasi-static contraction phase for models N2-T15-CR10 \textbf{a}, N2-T15-CR100 \textbf{b}, and N2-T15-CR1000 \textbf{c}.}
\end{figure*}

In this section we explore the impact of varying $\zeta^{\rm H_2}$ on the formation of COMs. Figure~\ref{FIG-ices-CR} for instance shows the composition of the ice mantles built up during the cold phase prior to the free-fall collapse for Sgr~B2(N2), with $\zeta^{\rm H_2}$~=~10$\times \zeta^{\rm H_2}_0$ (N2-T15-CR10), 100$\times \zeta^{\rm H_2}_0$ (N2-T15-CR100) and 1000$\times \zeta^{\rm H_2}_0$ (N2-T15-CR1000). Panel a shows that the chemical composition of the ices exposed to a CRIR enhanced by a factor of 10 are relatively similar to that obtained for the standard model N2-T15-CR1 (Fig.~\ref{FIG-evolution-frac-abund}a, top panel). Bigger differences can be seen in the composition of the ice mantles for higher ionization rates. For instance in the models N2-T15-CR100 and N2-T15-CR1000, thicker ice mantles are formed on the grains (up to 153 ice layers for N2-T15-CR100 after 1~Myr, Fig.~\ref{FIG-ices-CR}b), than in the low-CRIR models ($\zeta^{\rm H_2}$~$\leq$~10$\times \zeta^{\rm H_2}_0$). At higher ionization rates, water is still the main ice constituent, but the grain mantles are strongly depleted in \ce{CO2} compared to the low-CRIR models, because most surface \ce{CO2} is photodissociated via CR-induced UV photons. The abundance of CO in the ice is also diminished because gas-phase CO reacts with He$^+$ before accretion onto dust grains. Gas-phase molecular hydrogen is rapidly dissociated, leading to higher abundances of atomic hydrogen in the gas phase compared to the standard models. Surface reactions involving atomic hydrogen are thus more efficient, as seen in model N2-T15-CR100 which produces ice mantles enriched in \ce{CH4} and \ce{NH3} compared to the other models. \ce{NH3} maintains a fairly stable abundance through the ice layers since the surface \ce{NH3} photodissociated to \ce{NH2} is quickly hydrogenated to form \ce{NH3} again. Finally, the ice mantles in the high CRIR models ($\zeta^{\rm H_2}$~$\geq$~100$\times \zeta^{\rm H_2}_0$) are significantly depleted in \ce{H2CO} and \ce{CH3OH}, which are efficiently destroyed by reacting with surface atomic hydrogen.

Besides their impact on the production of ice-mantle constitutents, cosmic rays also play an important role in gas-phase chemical processes, producing the ions that react with gas-phase species leading to their destruction. The results of models N2-T15, N3-T15, N4-T15, and N5-T15 with different CRIRs are plotted in Figs.~\ref{FIG-appendix-collapse-CR}, \ref{FIG-appendix-collapse-CR-N3}, \ref{FIG-appendix-collapse-CR-N4}, and \ref{FIG-appendix-collapse-CR-N5}, respectively. If we focus for instance on the results obtained for model N2-T15 with different CRIR, Fig.~\ref{FIG-appendix-collapse-CR} shows that higher gas-phase fractional abundances of \ce{C2H5OH} and \ce{CH3NCO} are obtained at the end of the simulation for a CRIR enhanced by a factor ten compared to the standard value. The final gas-phase fractional abundance of \ce{CH3CHO} obtained for N2-T15-CR10 is one order of magnitude lower than that of N2-T15-CR1. This is because of reactions with atomic hydrogen destroying \ce{CH3CHO} on the surface of dust grains. Only a small fraction of the \ce{CH3CHO} formed on the grains thus desorbs around 60~K and contributes to its final gas-phase abundance. The \ce{CH3CHO} present in the gas phase at lower temperatures (that is before thermal desorption) mainly derives from chemical desorption, but once in the gas phase it is quickly destroyed as it reacts with ions.

At higher CRIR, in the model N2-T15-CR100, \ce{C2H5OH} reaches a final gas-phase fractional abundance of about two orders of magnitude higher than in the standard model. \ce{CH3NCO} is formed approximately three orders of magnitude more abundantly than in the standard model although it is strongly destroyed by reacting with atomic hydrogen. Its final gas-phase abundance derives from the CR-induced photodissociation of \ce{CH3NHCO} which desorbs from the grains around $T$~$\sim$~90~K.

\begin{figure*}[!t]
   \resizebox{\hsize}{!}         
   {\begin{tabular}{cc}
       \includegraphics[width=\hsize]{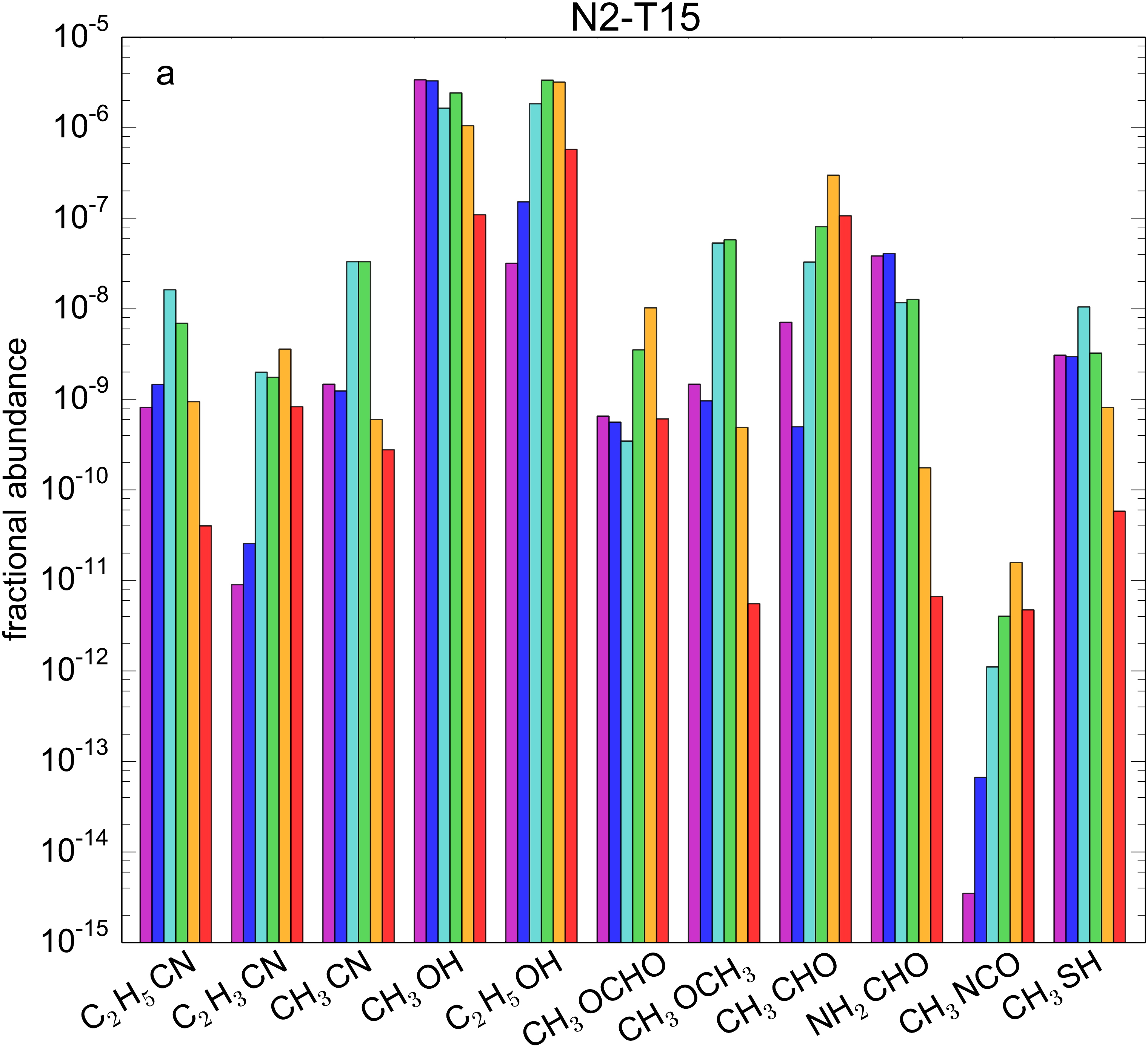} &
       \includegraphics[width=\hsize]{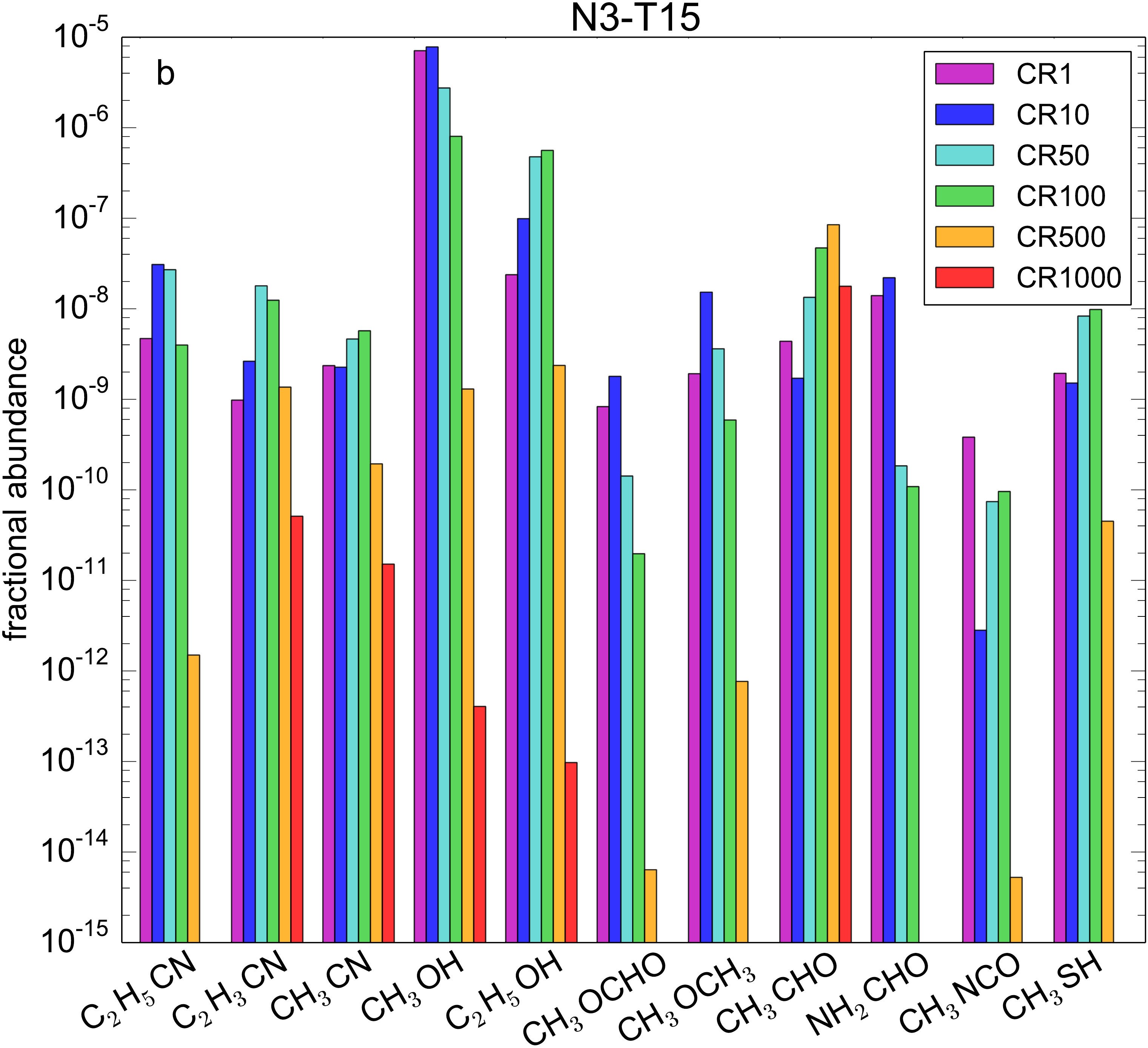}  \\
       \includegraphics[width=\hsize]{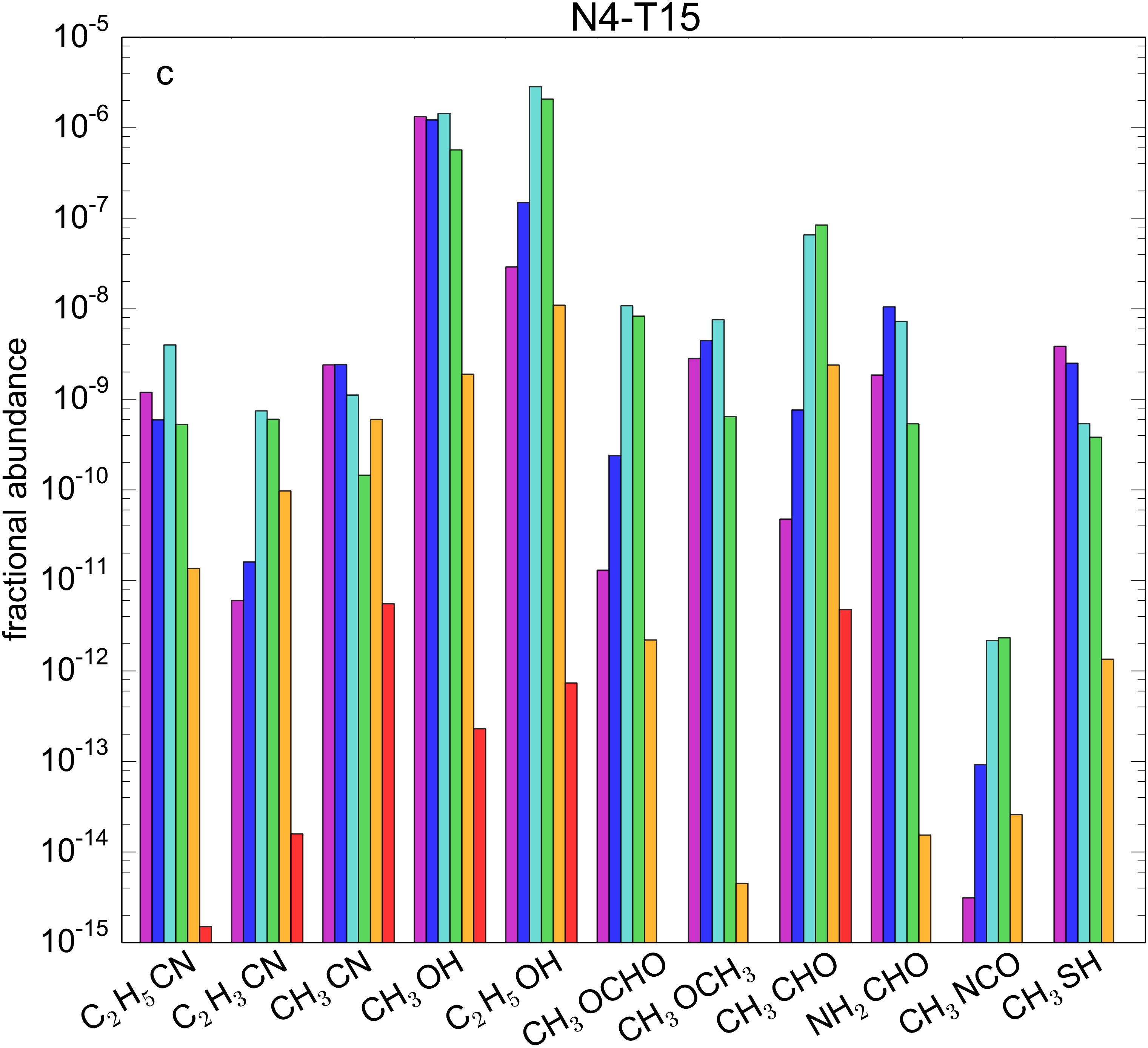} &
       \includegraphics[width=\hsize]{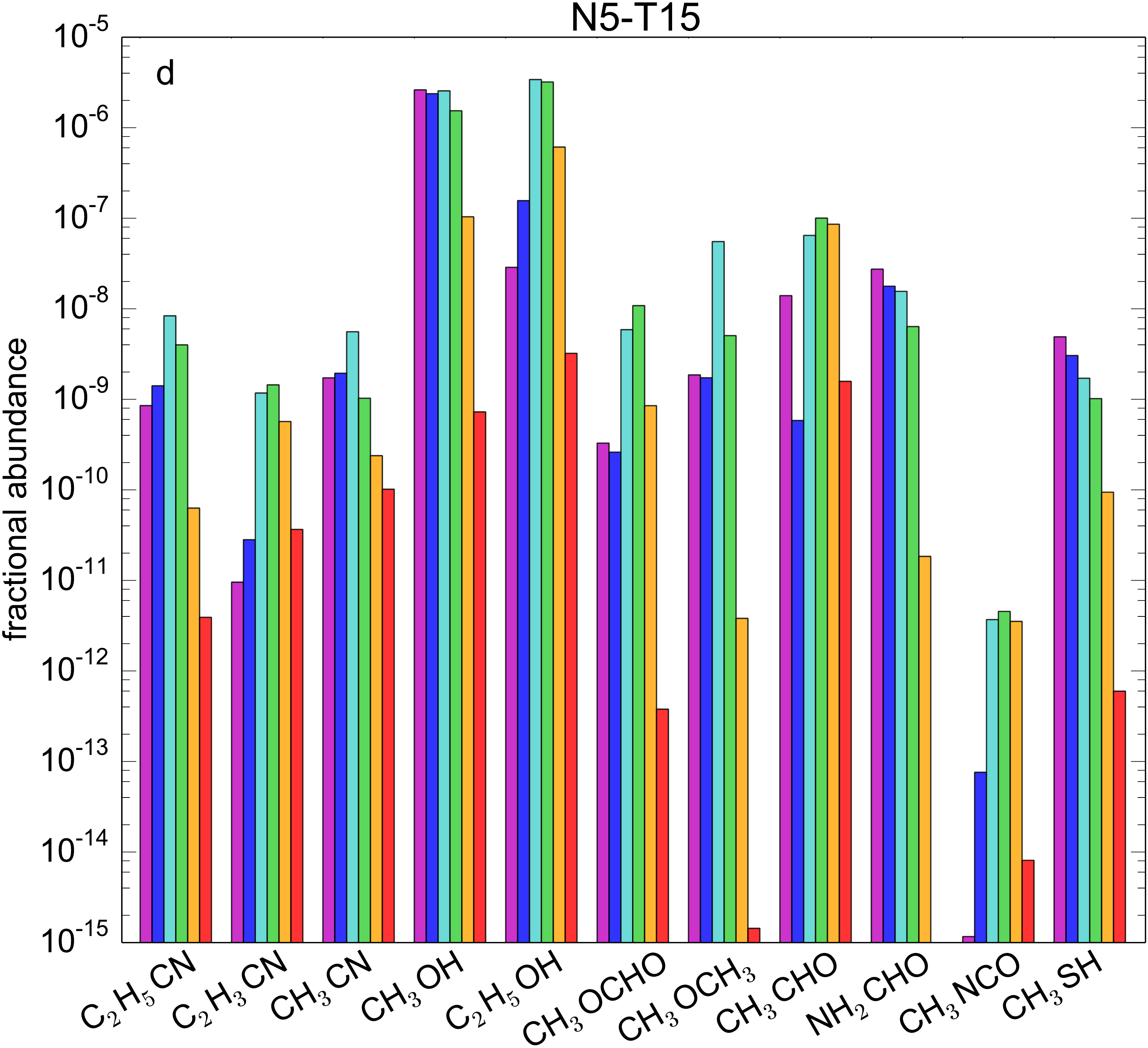} \\ 
    \end{tabular}}
    \caption{\label{FIG-histos-CR} Calculated gas-phase fractional abundances (with respect to total hydrogen) of 11 COMs obtained at the end of the simulations for model N2-T15 (a), N3-T15 (b), N4-T15 (c), and N5-T15 (d), with different CRIR values as shown in Table~\ref{TAB-our-models}. Species with fractional abundances lower than 10$^{-15}$ are not visible in the histograms.}
\end{figure*}

Finally in the model with the highest ionization rate, N2-T15-CR1000, all COMs are affected by cosmic rays (Fig.~\ref{FIG-appendix-collapse-CR}, bottom panel). In particular, the abundances of COMs with little to no gas-phase formation routes decrease right after desorption from the grain surfaces, leading to low final gas-phase fractional abundances (Fig.~\ref{FIG-histos-CR}). This is the case for \ce{C2H5OH}, \ce{CH3OH}, \ce{CH3OCHO}, \ce{CH3OCH3}, \ce{NH2CHO}, \ce{C2H5CN}, and \ce{CH3SH}, which are all destroyed in the gas phase as they react with one of the most abundant ions, \ce{H3O}$^+$, produced from water after it desorbs into the gas phase:
\begin{equation}   
\mathrm{H}_3^+ + \mathrm{H}_2\mathrm{O} \rightarrow  \mathrm{H}_3\mathrm{O}^+ 
\end{equation}
The gas-phase abundance of \ce{CH3OCH3} decreases by more than three orders of magnitude right after it desorbs from the dust grains. At later times, \ce{CH3OCH3} is formed in the gas phase via dissociative recombination of \ce{CH3OCH4}$^+$, a product of the ion-molecule reaction \ce{CH3OH2}$^+$ + \ce{CH3OH}, when \ce{CH3OH} desorbs from dust grains, around 130~K. The main contribution to the final gas-phase abundance of \ce{C2H3CN} and \ce{CH3CN} also comes from gas-phase reactions. \ce{C2H3CN} is mostly formed via the gas-phase reaction CN + \ce{C2H4} $\rightarrow$ \ce{C2H3CN} + H, when \ce{C2H4} desorbs from dust grains around 70~K, and the dissociative recombination of \ce{C2H6CN}$^+$. The later reaction is particularly efficient since \ce{C2H6CN}$^+$ is produced from the ion-molecule reaction \ce{H3O}$^+$ + \ce{C2H5CN}. The gas-phase abundance of \ce{C2H3CN} thus increases as \ce{C2H5CN} is destroyed. \ce{CH3CN} reaches its peak gas-phase abundance at early times, around 57~K, after which it is strongly destroyed by ion-molecule reactions. The main contribution to \ce{CH3CN}'s final gas-phase abundance comes from gas-phase recombination of \ce{CH3CNH}$^+$, product of the ion-molecule reaction \ce{CH3}$^+$ + HCN.

Figure~\ref{FIG-histos-CR} shows that the final gas-phase fractional abundances of the 11 investigated COMs are very sensitive to the CRIR for Sgr~B2(N2-N5). For instance, the abundances of cyanides rise as the CRIR increases up to 100$\times \zeta^{\rm H_2}_0$. For higher ionization rates ($\zeta_{\rm H_2}$~$\geq$~500$\times \zeta^{\rm H_2}_0$), the final gas-phase abundance of \ce{C2H5CN} decreases significantly, leading to [\ce{C2H5CN}]/[\ce{C2H3CN}]~$<$~1 in all models (see also Fig.~\ref{FIG-compare-obs-C2H5CN-C2H3CN}). The final fractional abundances calculated by models N3-T15 and N4-T15 are strongly diminished at high CRIR (1000$\times \zeta^{\rm H_2}_0$) compared to the low ionization-rate models, except for \ce{CH3CHO}. Model N2-T15-CR1000 shows significantly higher fractional abundances of \ce{C2H5OH}, \ce{CH3CHO}, and \ce{CH3NCO} compared to the standard model N2-T15-CR1. The impact of varying the CRIR on the calculated fractional abundances depends on the source, suggesting that the chemical differences discussed here may be exacerbated by the different physical properties of the cores.

\subsection{Comparison with observations}
\label{section-discussion-comparison-obs}

\begin{figure*}[!t]
   \hspace{0.02\linewidth}
   \begin{minipage}{0.32\linewidth}
     \includegraphics[scale=0.18]{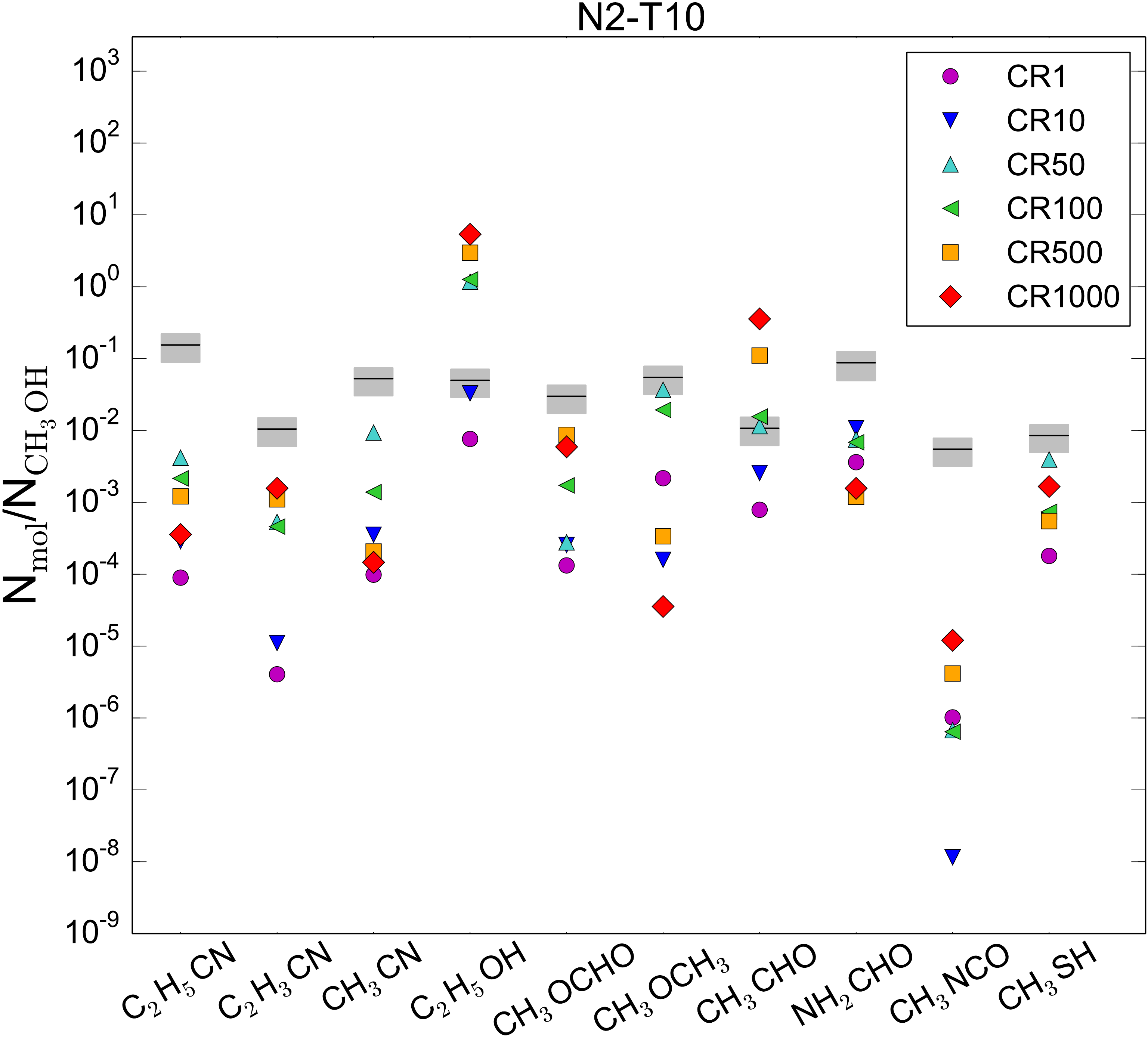}  \\
       \includegraphics[scale=0.18]{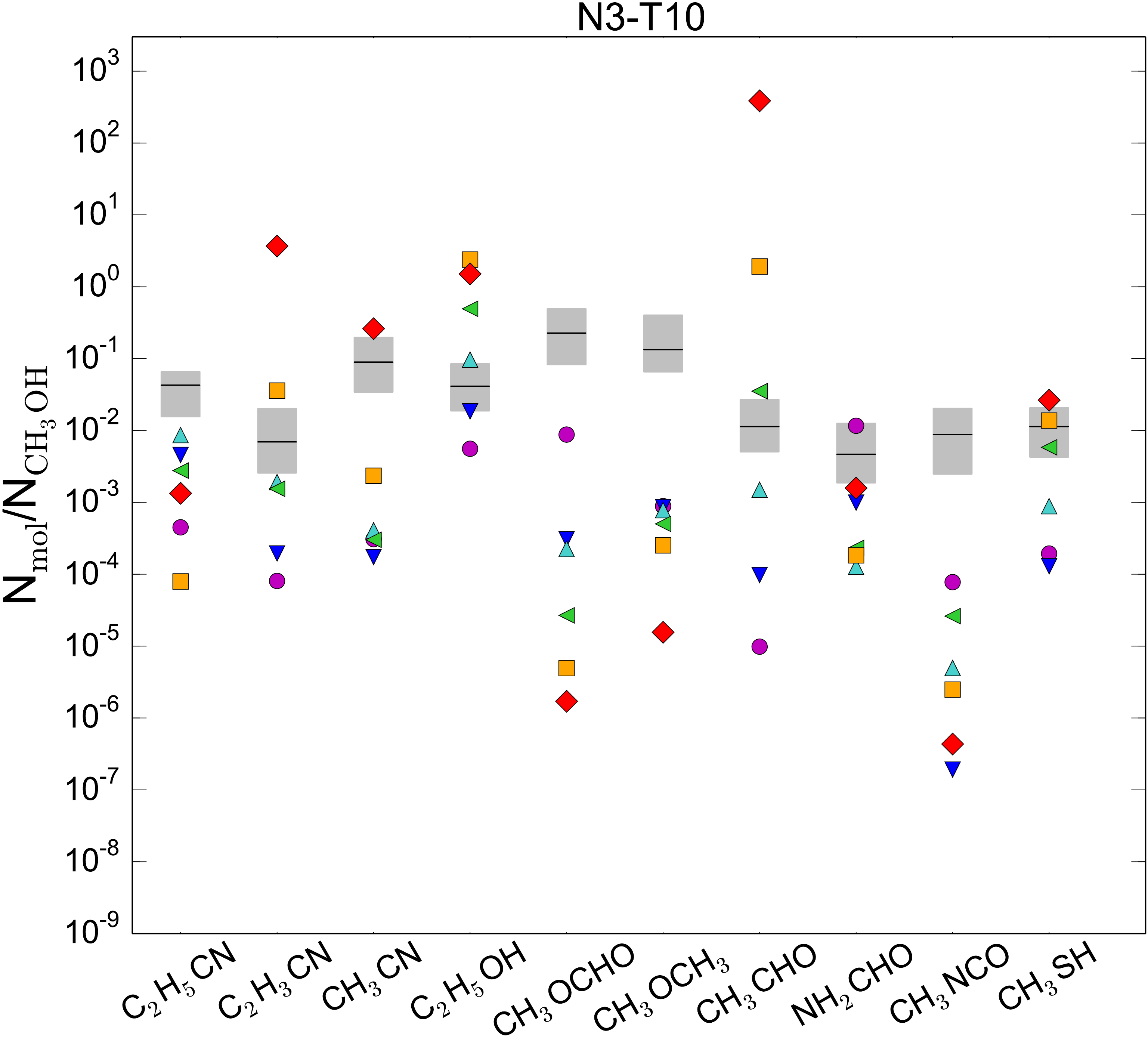} \\
       \includegraphics[scale=0.18]{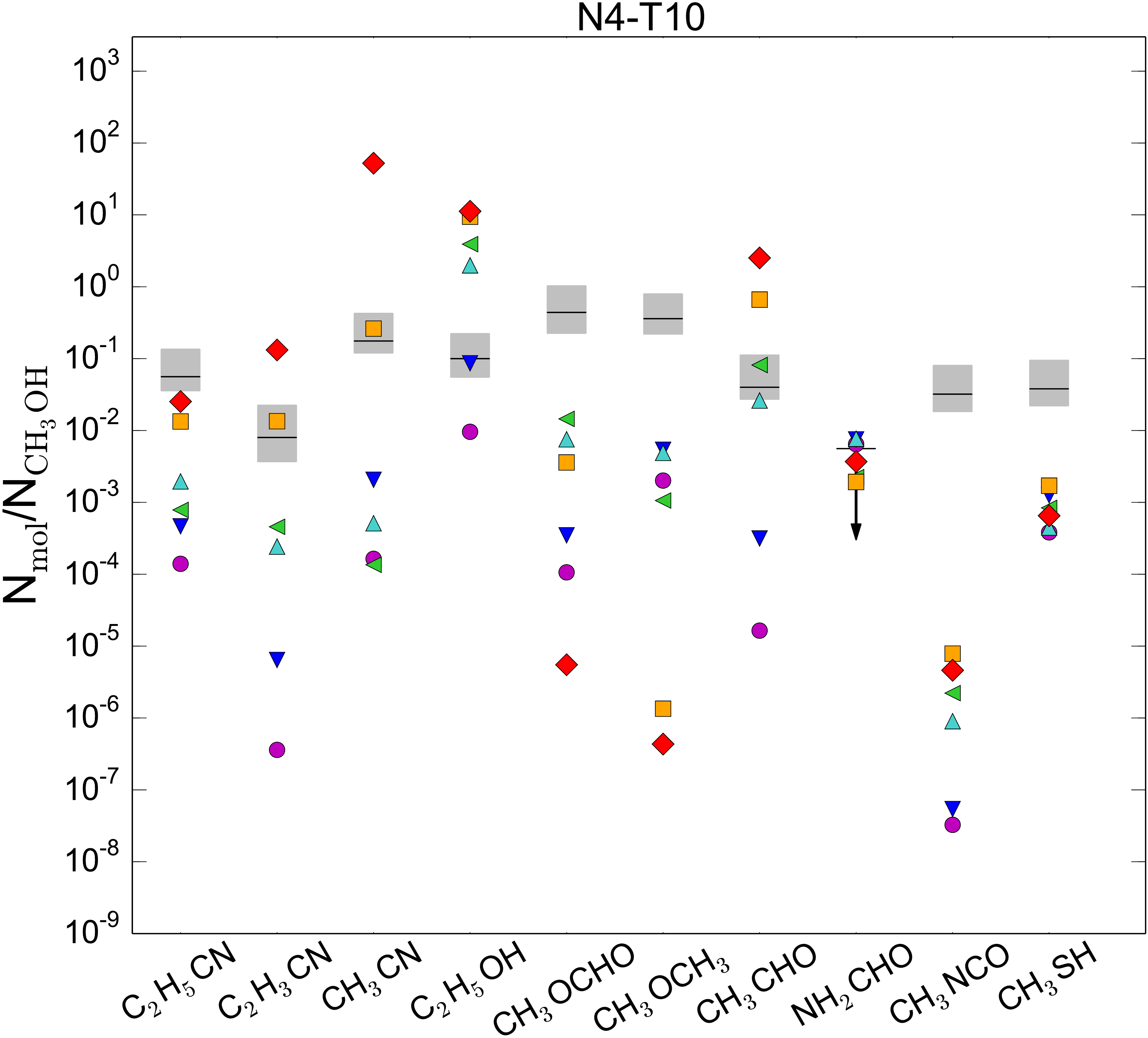} \\
       \includegraphics[scale=0.18]{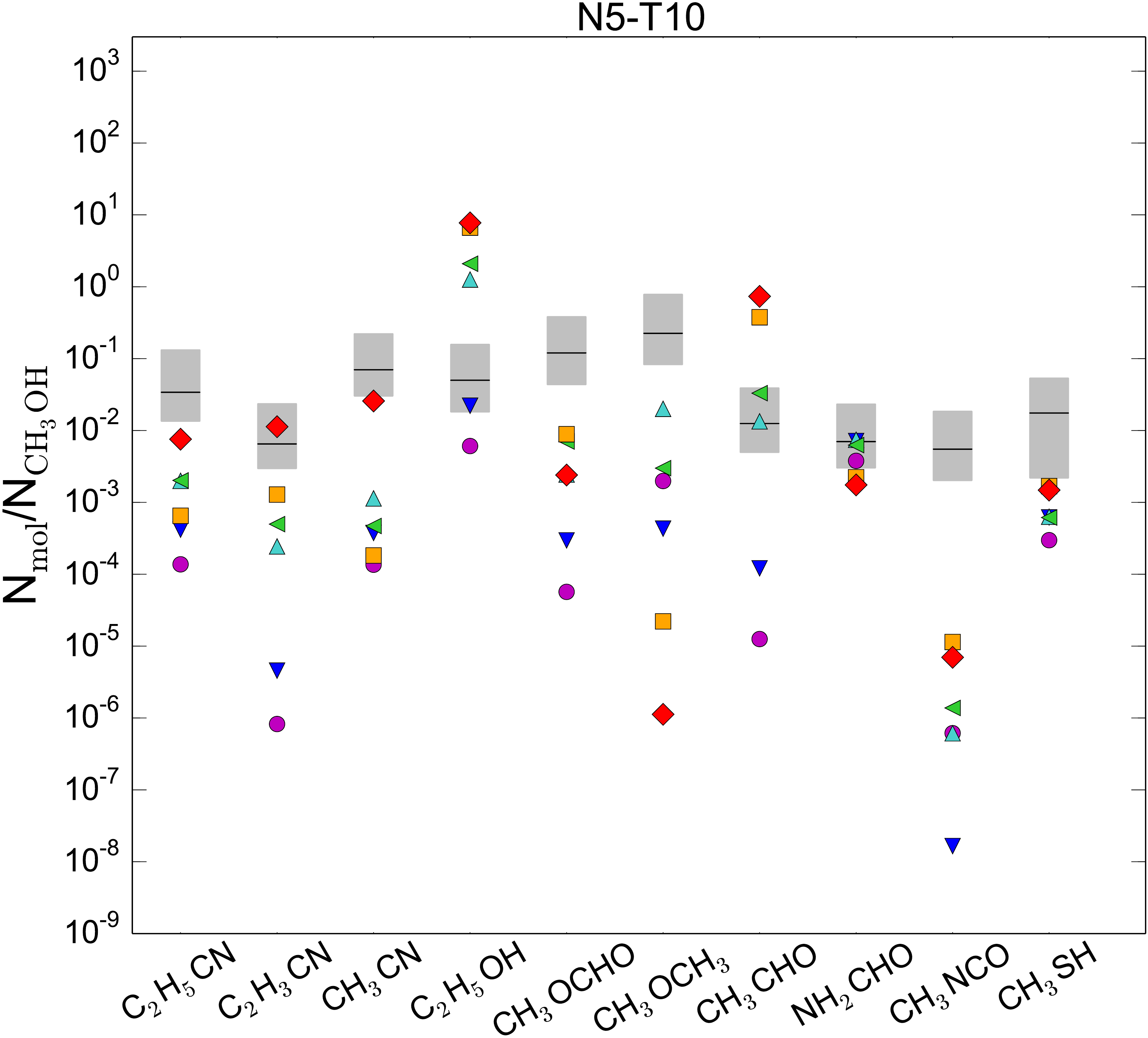} \\
   \end{minipage}
   \begin{minipage}{0.3\linewidth}
     \includegraphics[scale=0.18]{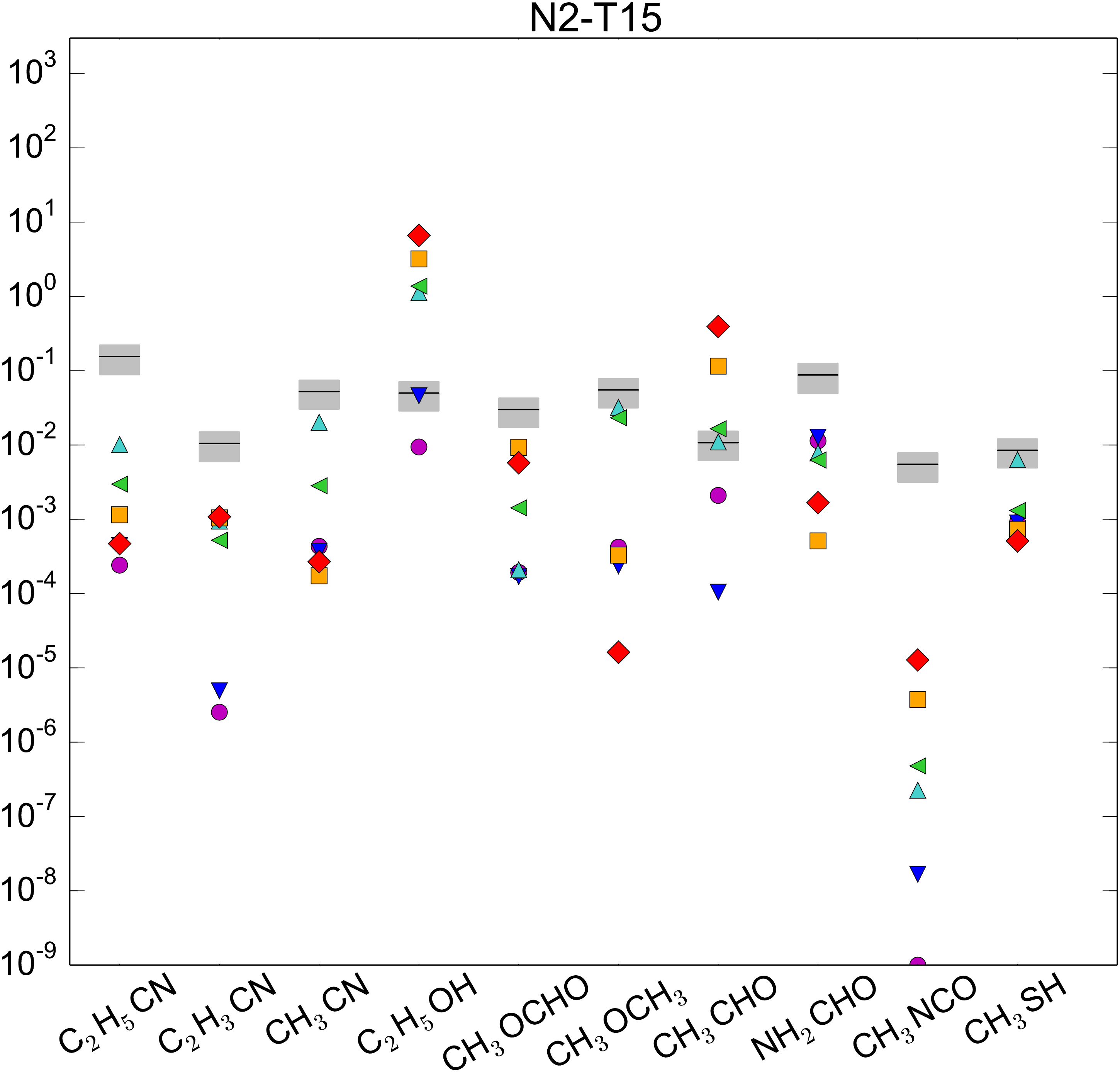}  \\ 
       \includegraphics[scale=0.18]{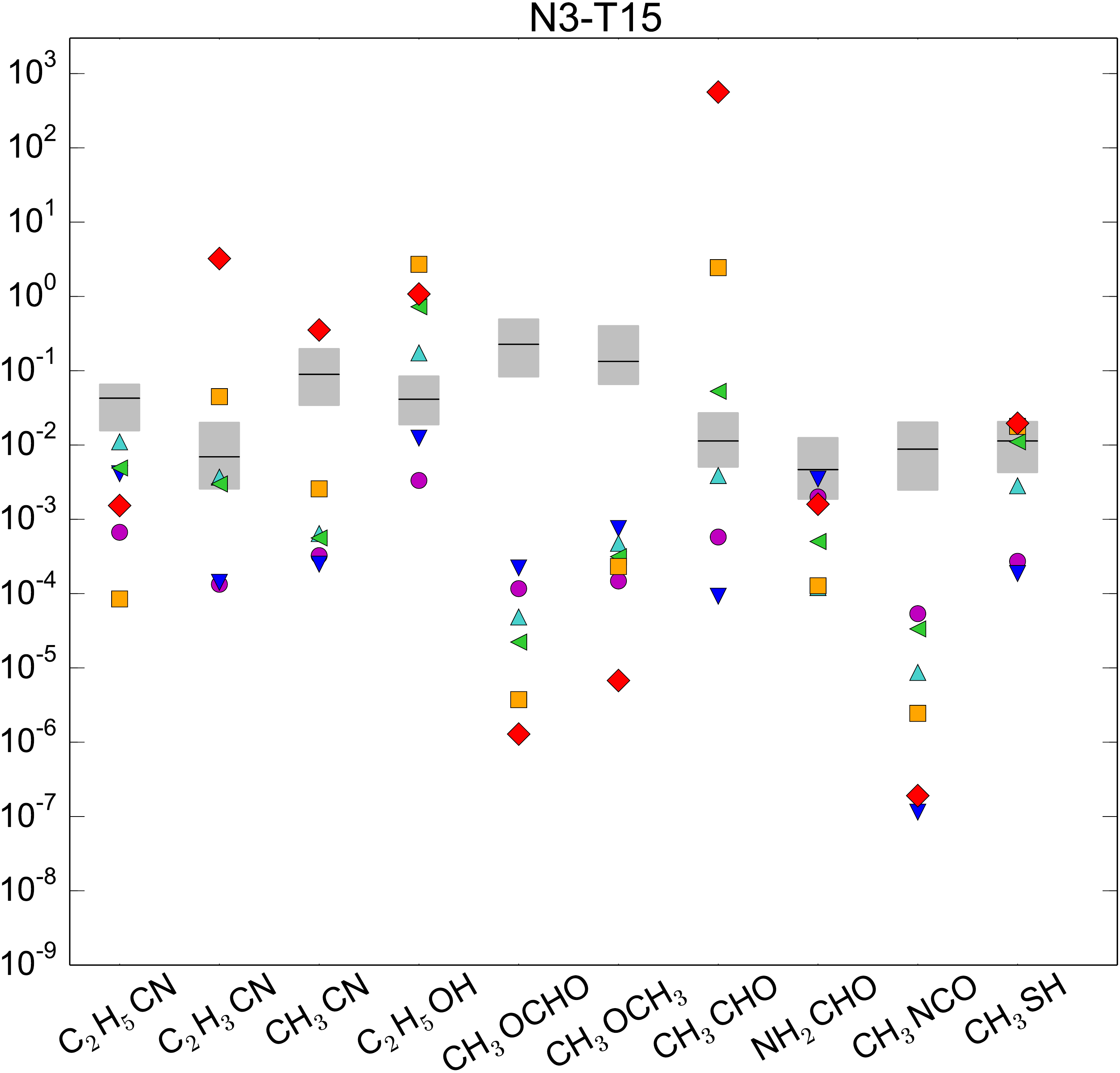} \\  
       \includegraphics[scale=0.18]{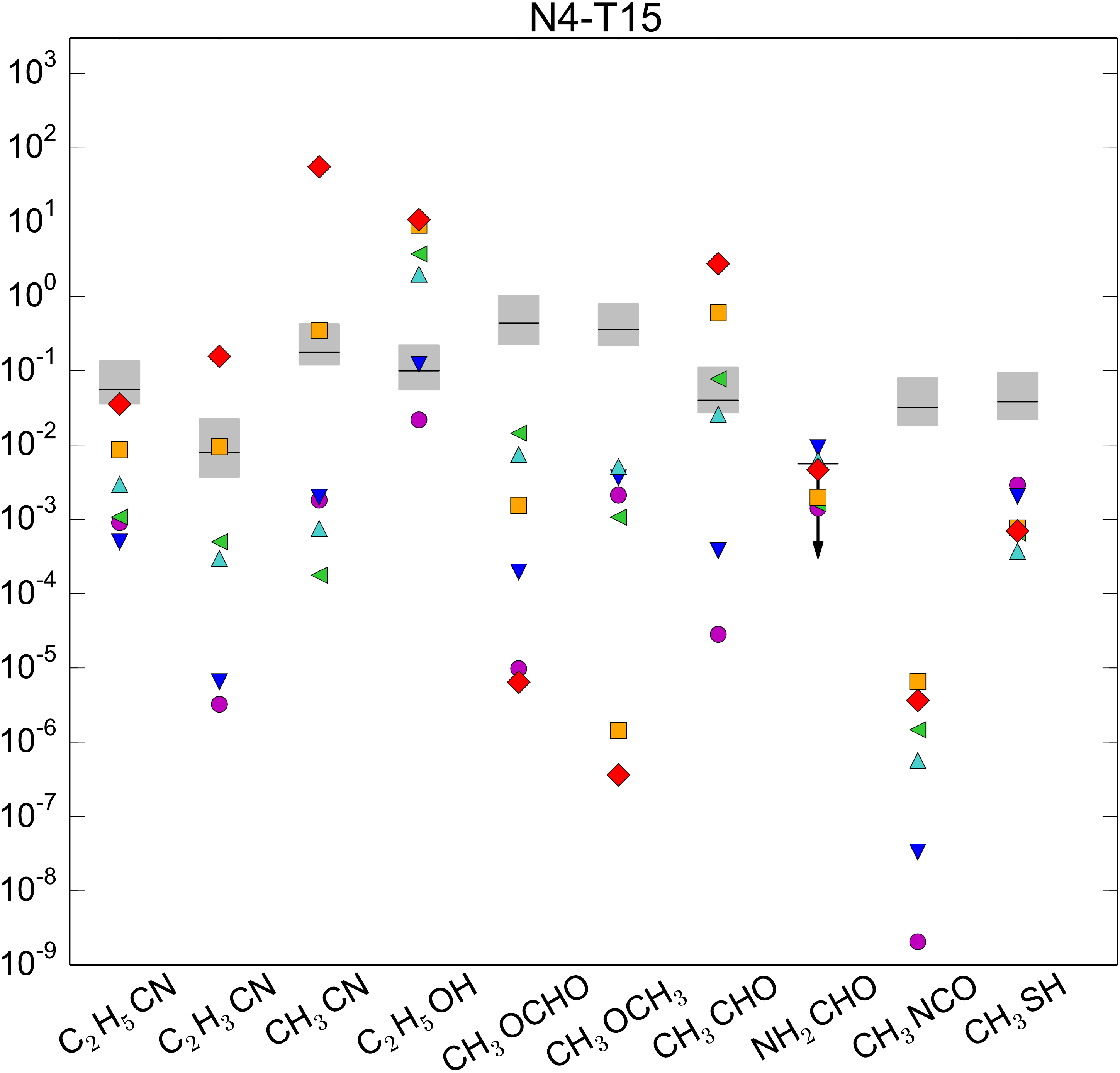}  \\ 
       \includegraphics[scale=0.18]{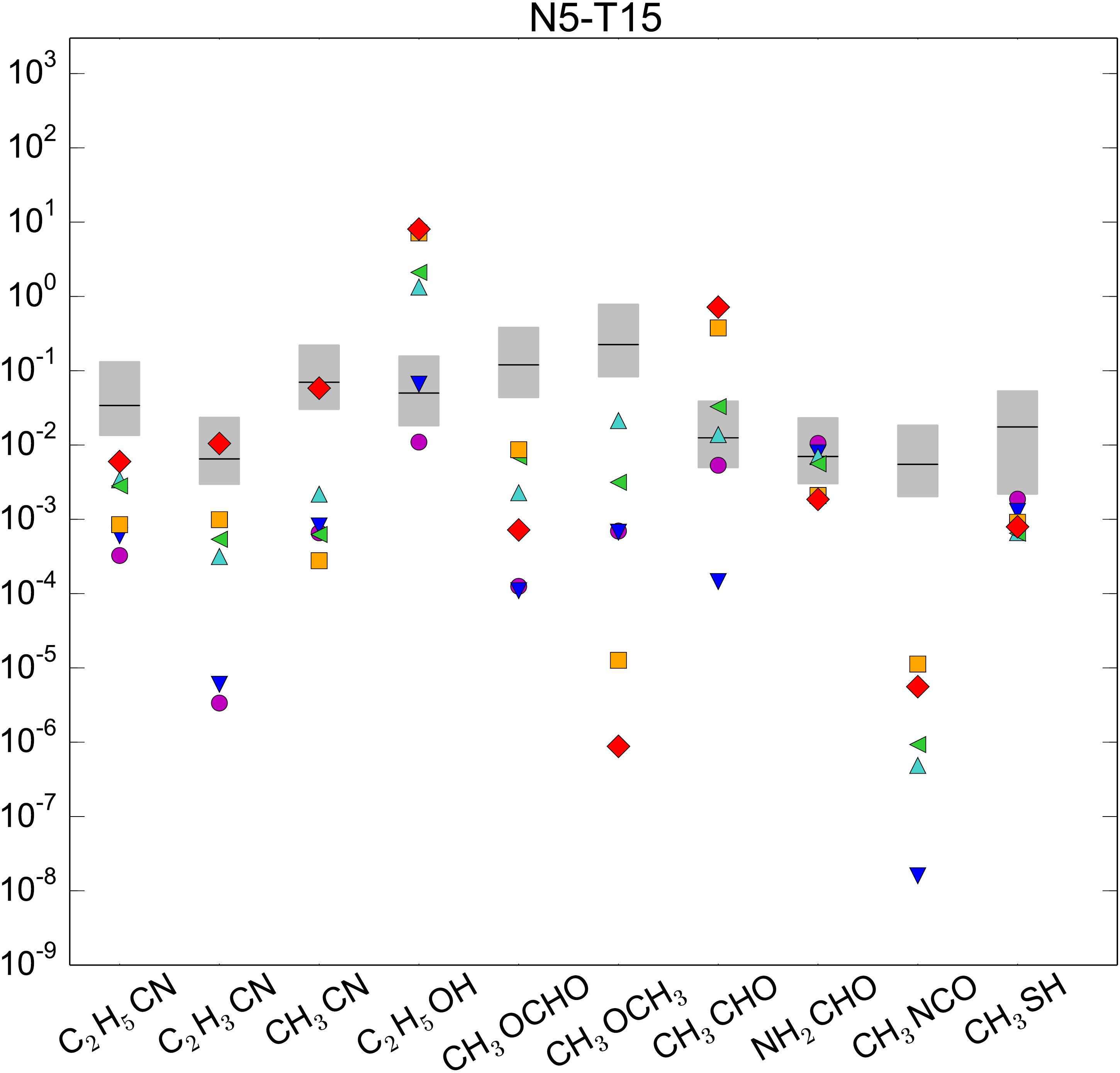}   \\
   \end{minipage}
   \begin{minipage}{0.30\linewidth}
       \includegraphics[scale=0.18]{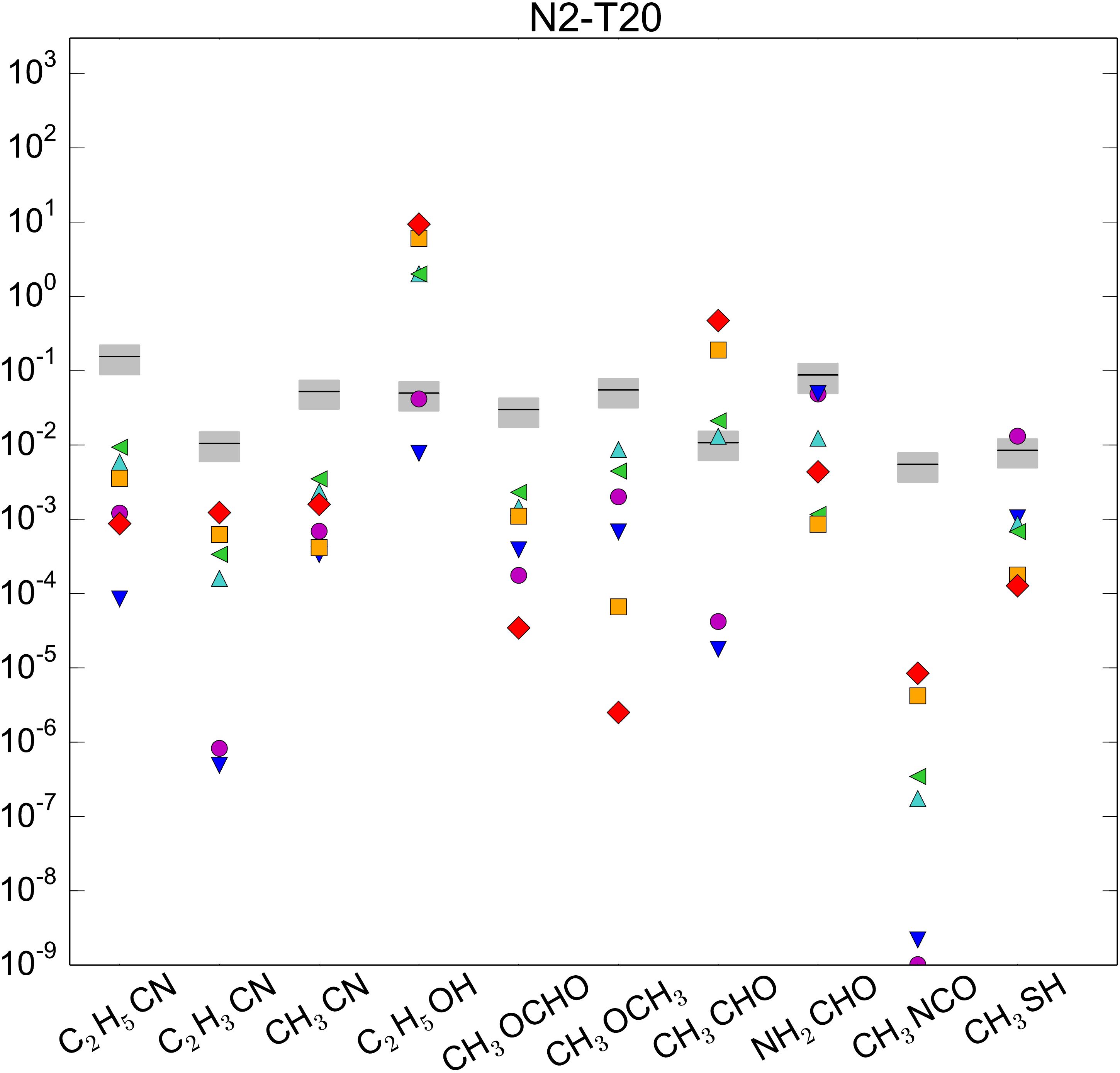}\\
       \includegraphics[scale=0.18]{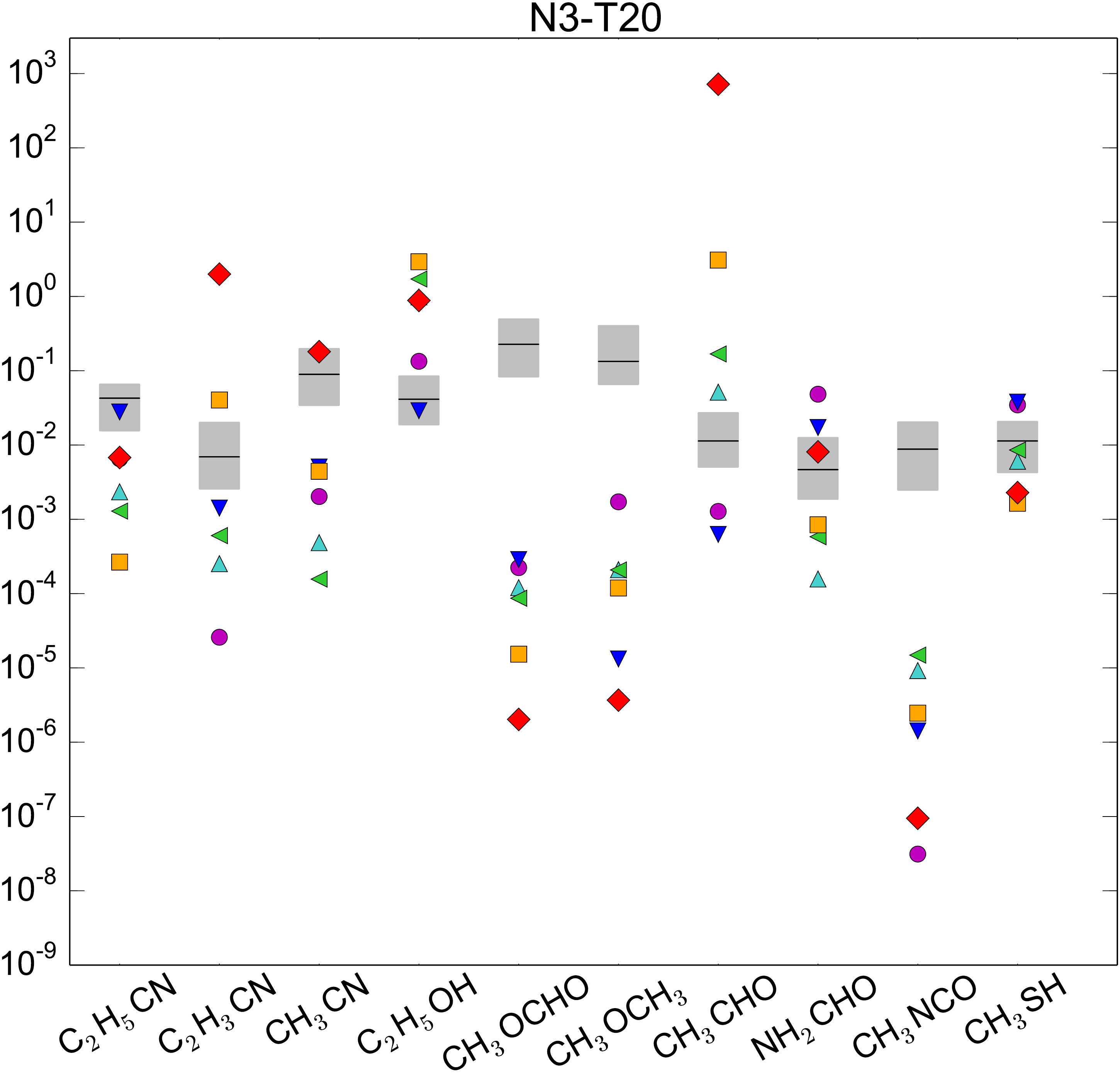}\\
       \includegraphics[scale=0.18]{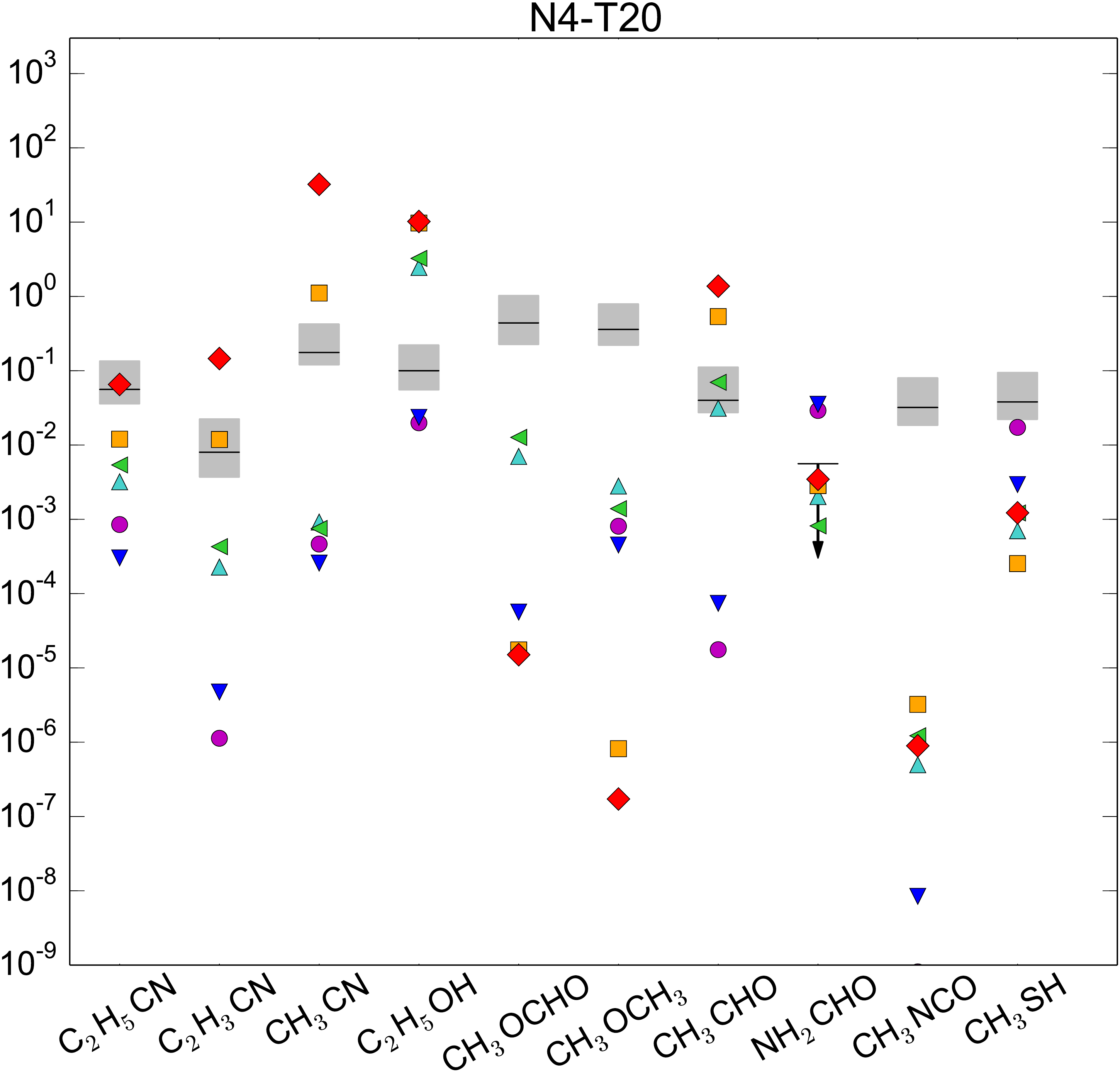}\\
       \includegraphics[scale=0.18]{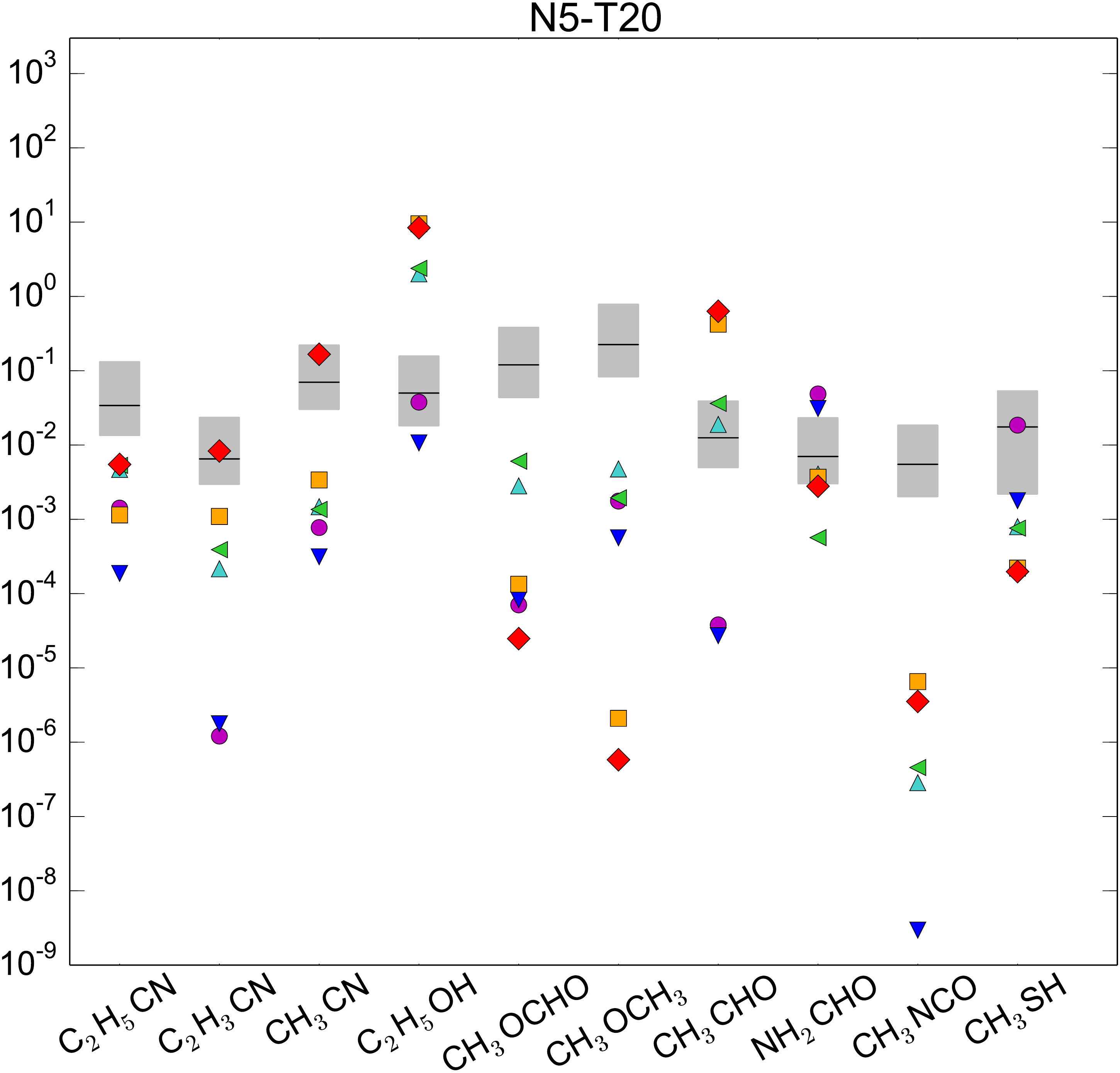}\\
        \end{minipage}  
    \caption{\label{FIG-comparison-obs} From top to bottom: abundances of ten COMs with respect to \ce{CH3OH} calculated at $T$~=~150~K by the models N2-N5, with different $T_{\rm min}$. In each panel the different symbols indicate the abundances calculated with different CRIR values. The horizontal black lines show the observed abundances relative to \ce{CH3OH}. The gray boxes represent the 1$\sigma$ uncertainties. Arrows indicate upper limits. The abundances relative to \ce{CH3OH} lower than 10$^{-9}$ are not visible in the plots.}
\end{figure*}

The gas-phase fractional abundances obtained at $T$~=~150~K (which corresponds to $T_0$, see Sect.~\ref{section-obs-constraints}) for the 11 investigated COMs in all models are given in Tables~\ref{TAB-appendix-model-results-150K} and \ref{TAB-appendix-model-results-150K2}. The results of all chemical models are also plotted in Fig.~\ref{FIG-comparison-obs} compared to the observed abundances with respect to \ce{CH3OH}. It shows that in most cases the chemical models tend to underestimate the observed abundances. In particular, this is the case for \ce{CH3OCHO}, for which the calculated abundances are lower than the observed ones toward all four sources. The abundance of \ce{CH3NCO} with respect to \ce{CH3OH} is also underestimated by all models, by at least two orders of magnitude compared to the observed abundances. In Sect.~\ref{section-formation-routes-COMs} we have seen that solid-phase \ce{CH3NCO} desorbs from the grains around 70~K, however, the measured rotational temperatures toward Sgr~B2(N2-N5) are higher than 140~K (Sect.~\ref{TAB-best-fit-parameters}). This suggests that some important chemical reactions involved in the production of \ce{CH3NCO} may be missing in the chemical network \citep[see, e.g.,][]{quenard2018}, or that the binding energy adopted for \ce{CH3NCO} in our chemical models \citep[3575~K,][]{belloche2017} is underestimated (see Belloche et al. subm. for models computed with a higher binding energy). For this reason we ignore \ce{CH3NCO} in the rest of the discussion.

In Fig.~\ref{FIG-comparison-obs} no error bars are plotted on the model results although it has been shown that calculated chemical abundances may be affected by rate coefficient uncertainties \citep[see, e.g.,][]{wakelam2005}. Because of the large number of parameters in the chemical code, a full analysis taking into account both observational and theoretical uncertainties \citep[see, e.g.,][]{wakelam2006} would be too difficult and is not attempted here. Besides, additional uncertainties affecting the model results derive from the lack of experimental and theoretical data on grain-surface processes. For instance, \citet{iqbal2018} found that the abundances of both gas-phase and surface species calculated under typical dark cloud conditions are affected by large errors due to the uncertainties in the diffusion energy of surface species, in particular for the molecules predominantly formed at the grain surface like \ce{CH3OH}. \citet{enriqueromero2016} presented quantum chemical computations showing that grain-surface radical-radical addition reactions do not necessarily lead to the formation of complex species because these radicals are trapped by the water-ice molecules with an orientation that favors other two-product reactions. This suggests that gas-phase reactions may play a more important role in the formation of COMs than currently assumed in chemical models. Recently \citet{skouteris2018a} computed rate coefficients for two gas-phase ion-molecule reactions proposed as main formation route for ce{CH3OCH3}: \ce{CH3OH} + \ce{CH3OH2}$^+$ $\rightarrow$ (CH$_3$)$_2$OH$^+$ + \ce{H2O} followed by the reaction of (CH$_3$)$_2$OH$^+$ with \ce{NH3} to form \ce{CH3OCH3} + \ce{NH4}$^+$. The latter reaction, not included in our chemical network, could help in improving the agreement between the calculated and observed [\ce{CH3OCH3}]/[\ce{CH3OH}] ratios, which differ by at least one order of magnitude in our models for N3-N5 (see Fig.~\ref{FIG-comparison-obs}).

\subsubsection{Constraining the physical parameters $T_{\rm min}$ and $\zeta^{H_2}$}
\label{section-discussion-constraining-parameters}

We use the method presented by \citet{garrod2007} to evaluate the level of confidence of each model with respect to the observations in order to determine which set of physical parameters ($T_{\rm min}$ and CRIR) best characterizes Sgr~B2(N)'s hot cores (see Appendix~\ref{appendix-level-confidence} for details about the method). We draw our attention first to the overall abundances of COMs with respect to H$_2$, and then examine in more detail the influence of $T_{\rm min}$ and CRIR on the chemical composition itself, that is the COM abundances relative to \ce{CH3OH} or \ce{CH3CN}.

\begin{figure*}[!t]
   \resizebox{\hsize}{!}
   {\begin{tabular}{ccc}
       \includegraphics[scale=0.45]{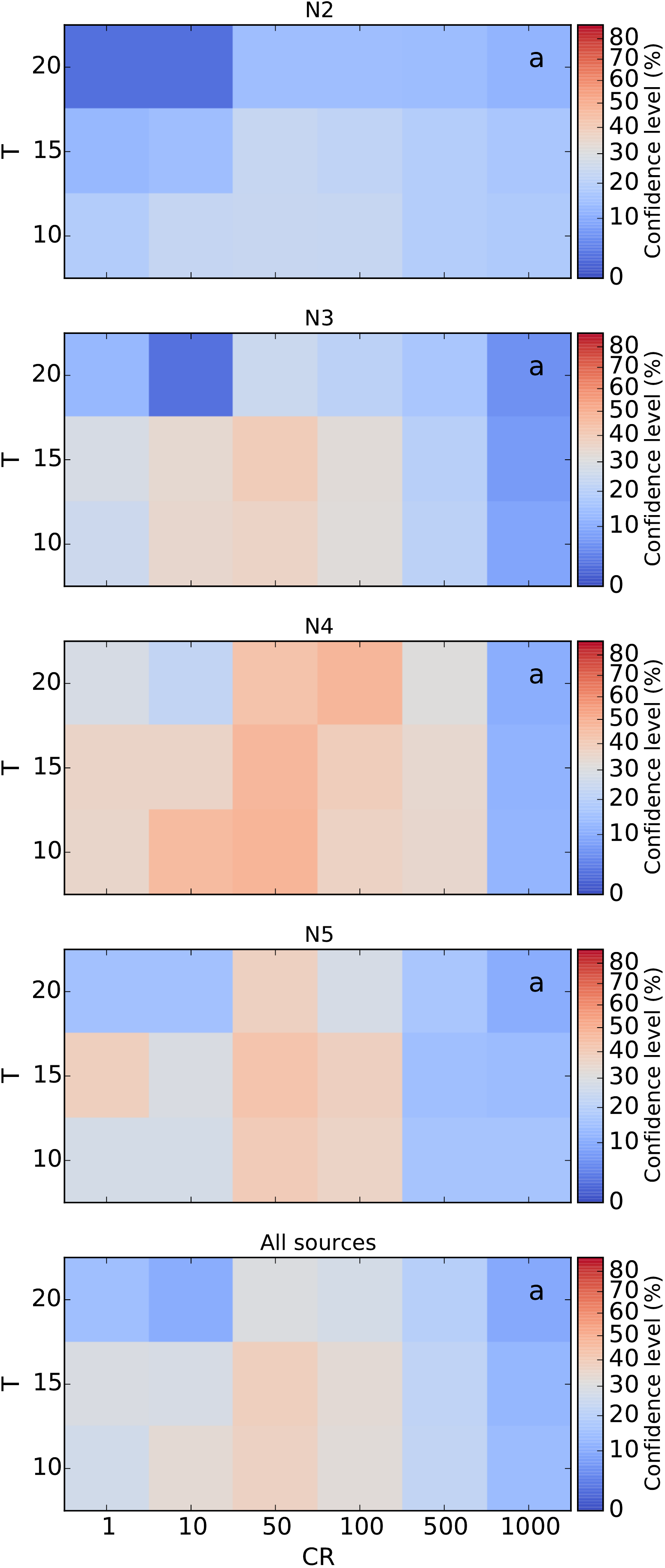} &
       \includegraphics[scale=0.45]{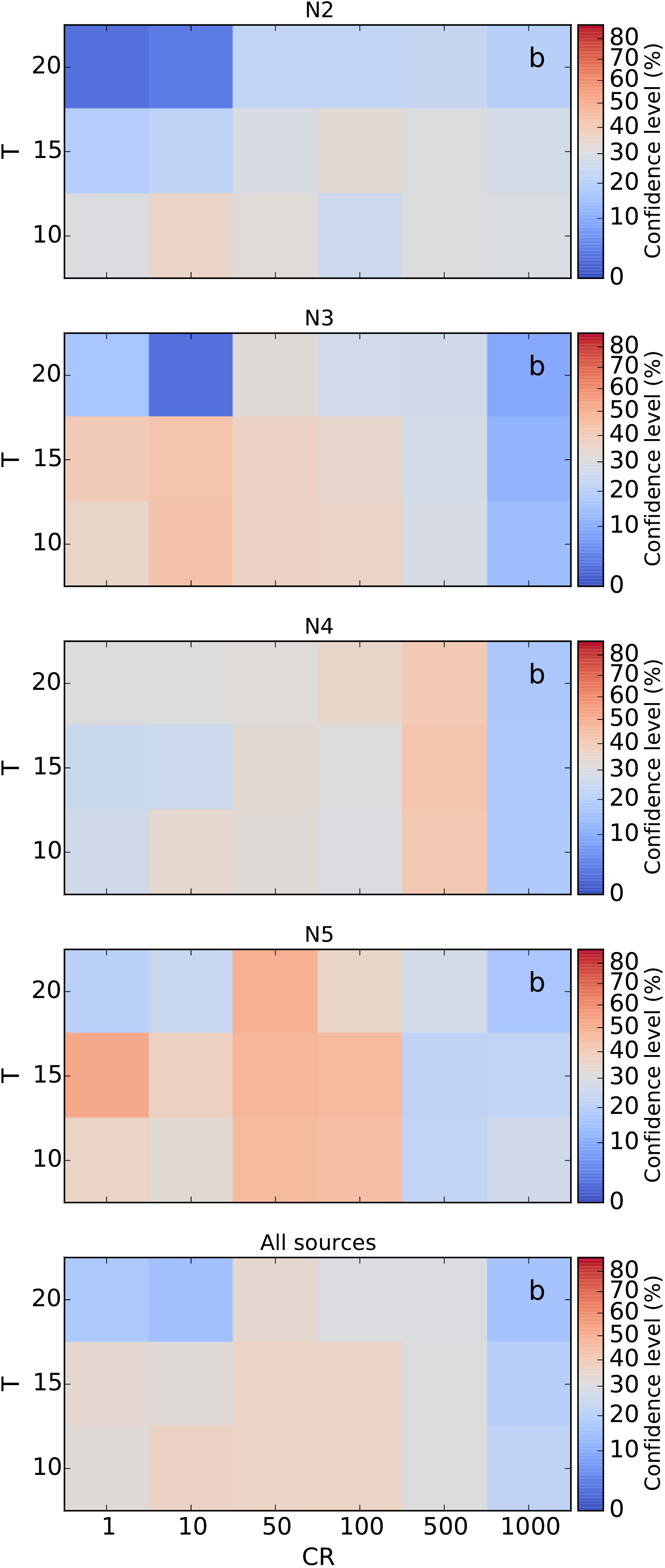} &
       \includegraphics[scale=0.45]{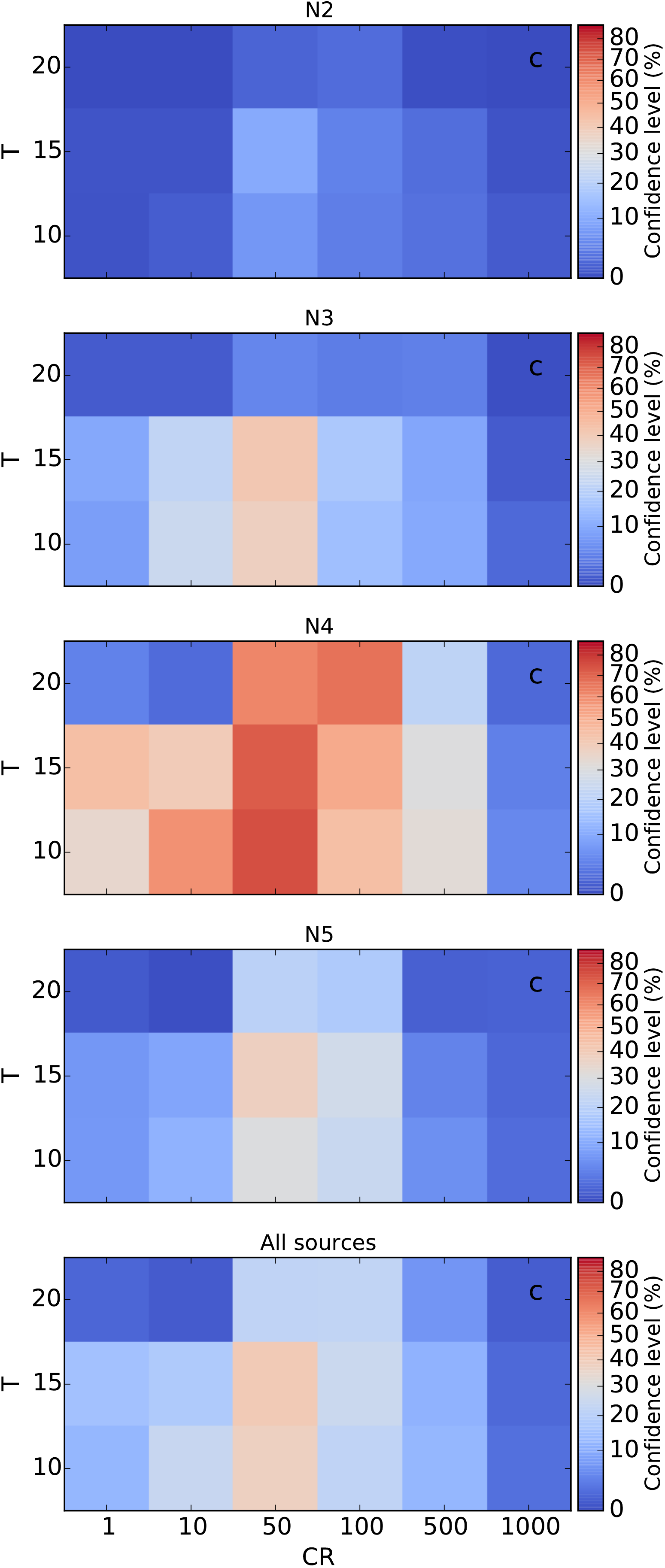} \\
    \end{tabular}}
    \caption{\label{FIG-confidence-level-H2} \textbf{a} Matrice of confidence levels of the models with respect to the observed abundances of ten COMs relative to \ce{H2}. The first four panels in this column show, from top to bottom, the results for Sgr~B2(N2), N3, N4, and N5, respectively. The bottom panel shows the average matrice for all four hot cores taken together. \textbf{b} Same as (a) but only for the O-bearing species \ce{CH3OH}, \ce{CH3OCHO}, \ce{CH3OCH3}, \ce{CH3CHO}, \ce{C2H5OH}, and \ce{NH2CHO}. \textbf{c} Same as (a) but only for the cyanides \ce{CH3CN}, \ce{C2H5CN}, and \ce{C2H3CN}.}
\end{figure*}

Figure~\ref{FIG-confidence-level-H2} shows the confidence levels for the ten investigated COMs (excluding \ce{CH3NCO}) relative to \ce{H2}, for Sgr~B2(N2-N5) taken individually and for all sources taken together. Panels a, b, and c show the confidence levels for all molecules, only the O-bearing molecules, and only the cyanides, respectively. The confidence level matrices differ from one source to the other, maybe reflecting the limits of our strong assumption that all sources share the same accretion history. Given that all sources should have been exposed to the same cosmic-ray flux and probably shared a similar thermal history during the prestellar phase, we focus our attention on the matrices corresponding to all sources taken together. The cyanides appear to be much more sensitive to $T_{\rm min}$ and CRIR than the O-bearing species. The cyanides constrain the CRIR value to be about 50 times the standard one, and the minimum temperature to be below 20 K. The best confidence levels for the O-bearing species are also obtained for these values, but the constraints are less sharp.

Given that a model with $T_{\rm min} =$~20~K still gives acceptable results for the O-bearing species, we run additional simulations with $T_{\rm min}$~=~25~K and 28~K for N2, with $\zeta^{\rm H_2}$~=~50$\times \zeta^{\rm H_2}_0$ (see Fig.~\ref{FIG-comparison-H2}a). Ignoring the cyanides, we obtain confidence levels of about 10\% and 5\% for the models N2-T25-CR50 and N2-T28-CR50, respectively. These values are much lower than the 23\% obtained for N2-T20-CR50. This results from the high dust temperatures ($T_{\rm min}$~$\geq$~25~K) preventing gas-phase atoms and simple molecular species such as CO to stick to the grains, damping the formation of ice mantles and thus the production of complex species. We expect the models of N3-N5 to behave in the same way and conclude that the COM abundances relative to H$_2$ in N2-N5 certainly exclude $T_{\rm min} >$ 20~K.

\begin{figure*}[!t]
   \resizebox{\hsize}{!}
   {\begin{tabular}{ccc}
       \includegraphics[scale=0.45]{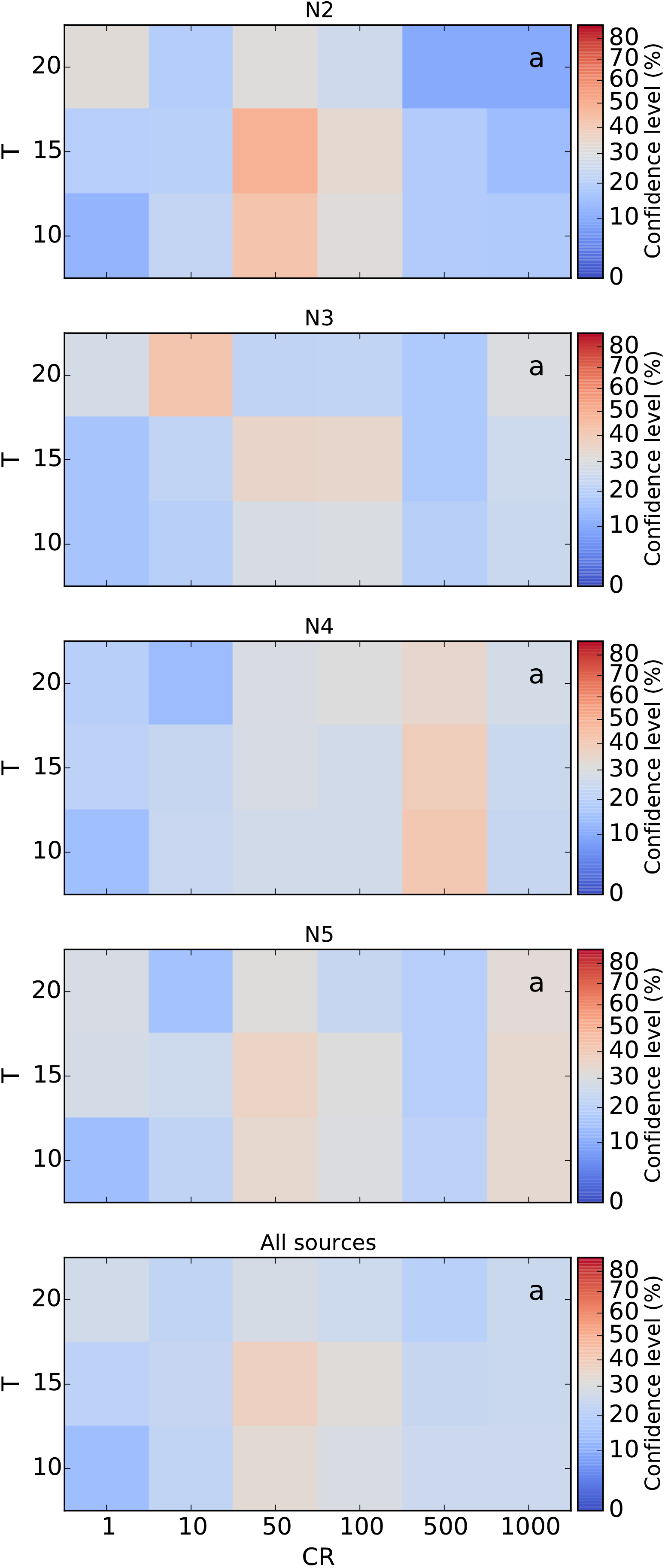} &
       \includegraphics[scale=0.45]{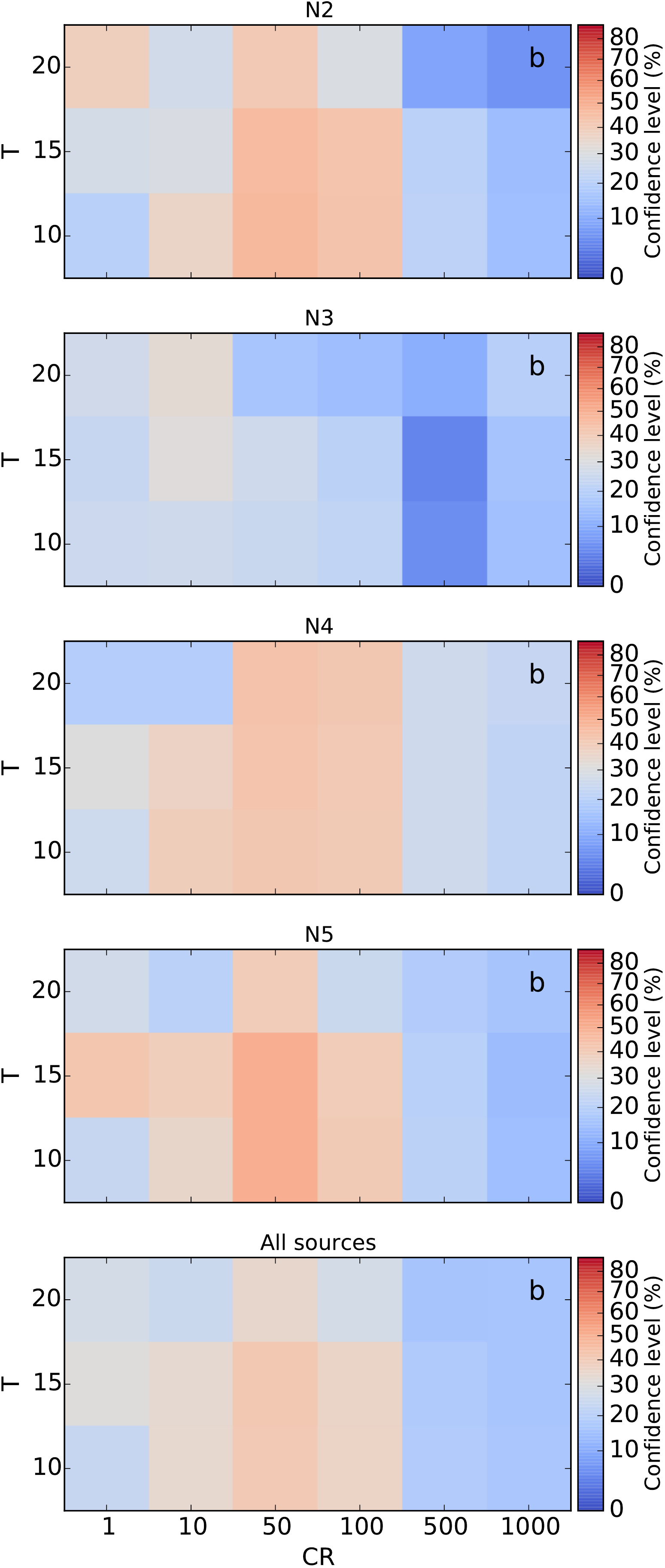} &
       \includegraphics[scale=0.45]{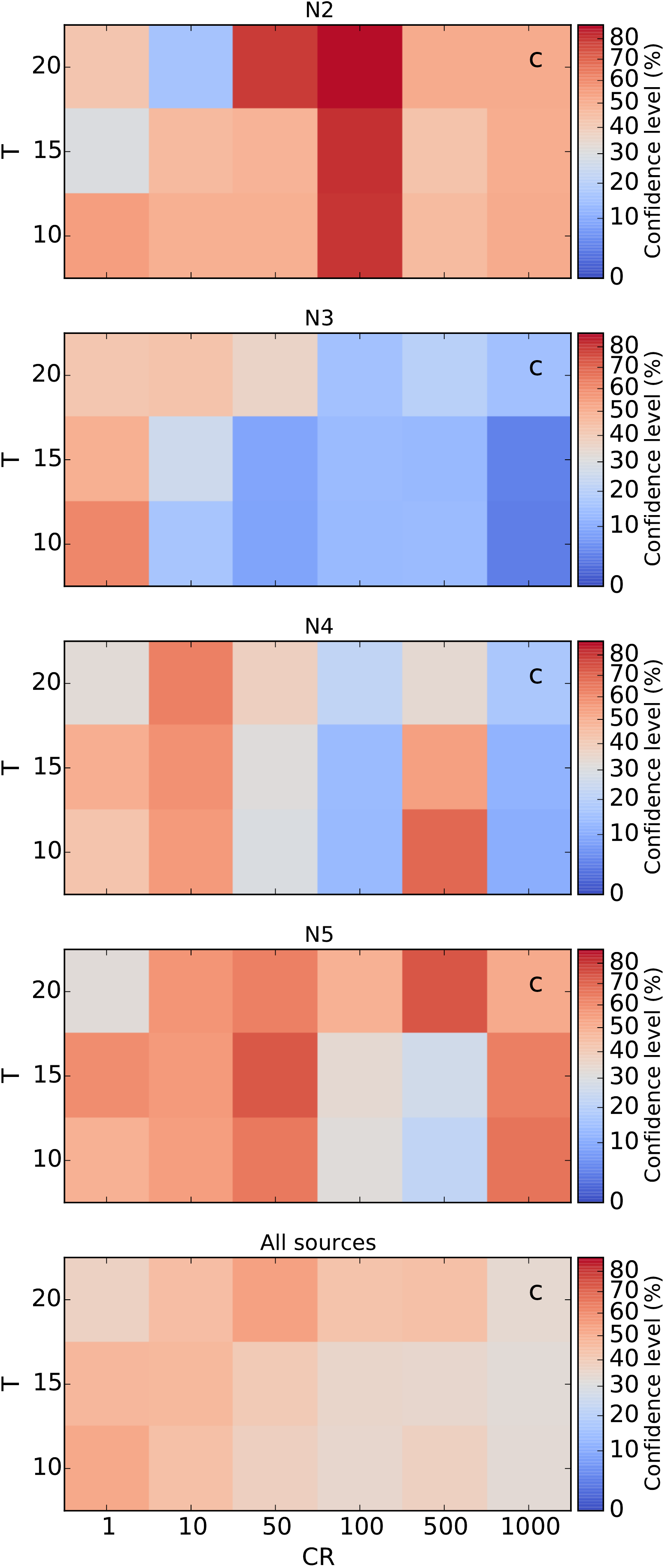} \\
    \end{tabular}}
    \caption{\label{FIG-confidence-level} \textbf{a} Matrice of confidence levels of the models with respect to the observed abundances of nine COMs relative to \ce{CH3OH}. The first four panels in this column show, from top to bottom, the results for Sgr~B2(N2), N3, N4, and N5, respectively. The bottom panel shows the average matrice for all four hot cores taken together. \textbf{b} Same as (a) but for the abundances relative to \ce{CH3OH} of the O-bearing species \ce{CH3OCHO}, \ce{CH3OCH3}, \ce{CH3CHO}, \ce{C2H5OH}, and \ce{NH2CHO} only. \textbf{c} Same as (a) but for the abundances relative to \ce{CH3CN} of the cyanides \ce{C2H5CN} and \ce{C2H3CN} only.}
\end{figure*}

\begin{figure*}[!h]
 \hspace{0.09\linewidth}
 \begin{minipage}{0.42\linewidth}
   \includegraphics[scale=0.3]{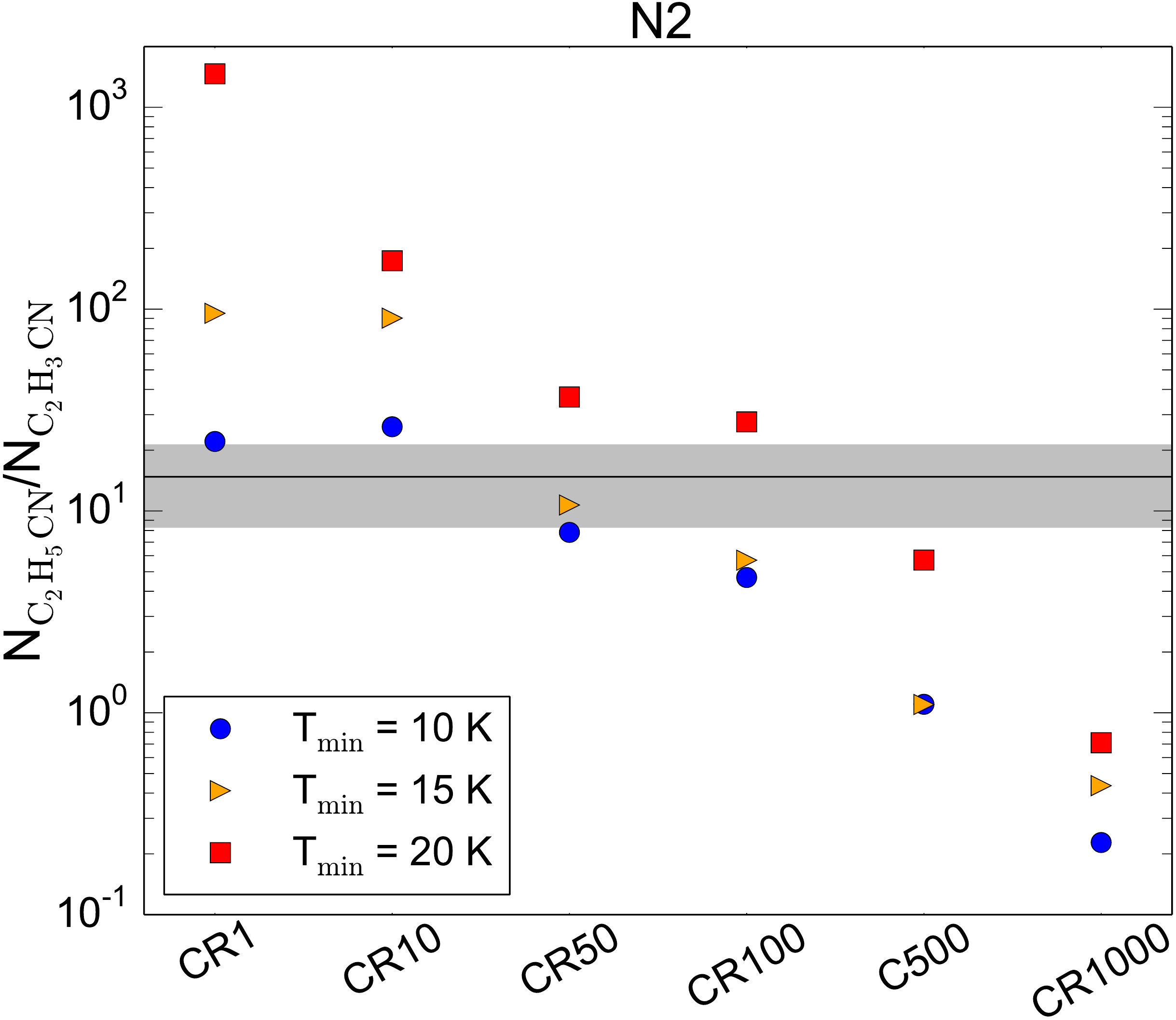} \\
   \includegraphics[scale=0.3]{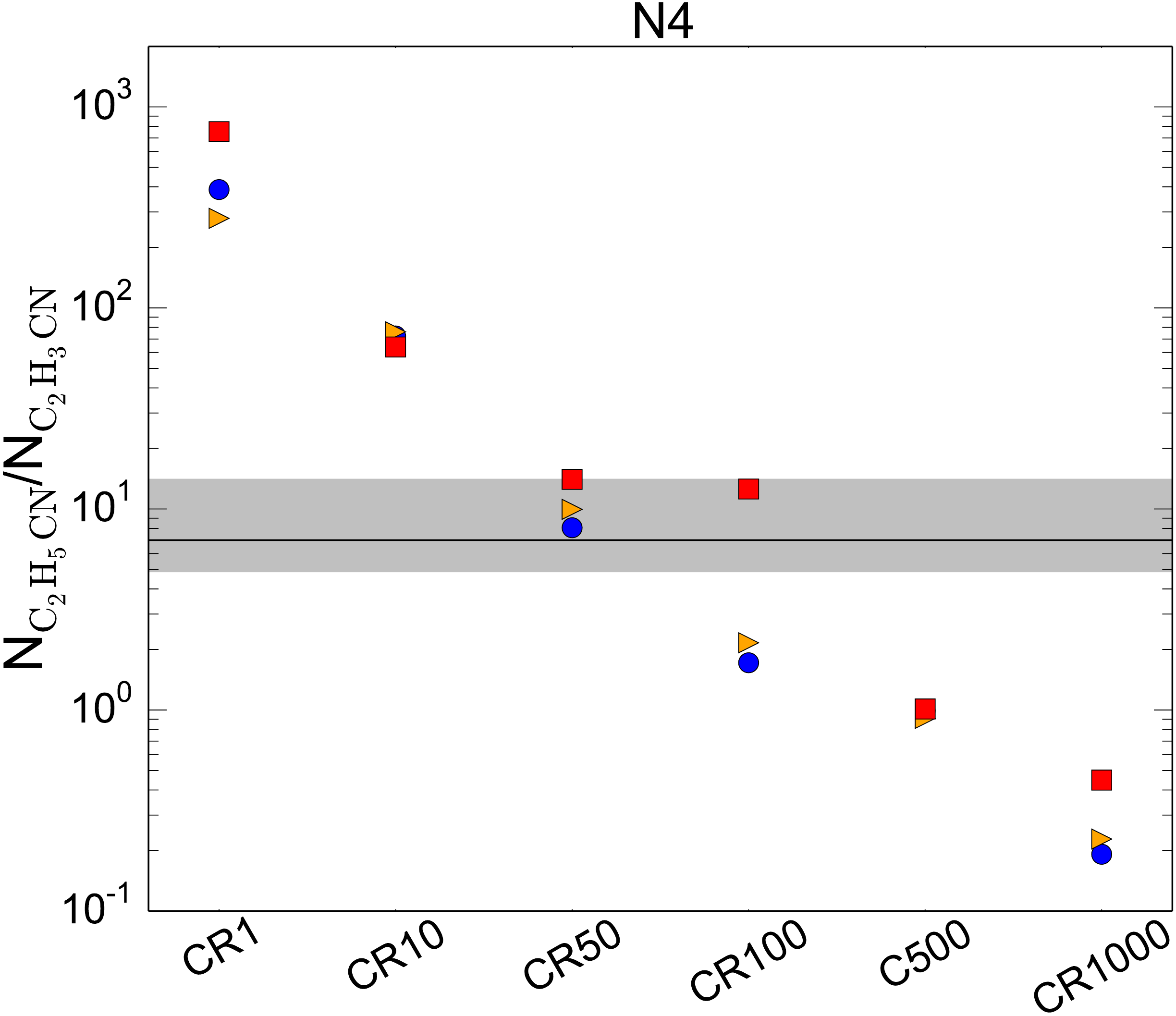} \\
  \end{minipage}
  \begin{minipage}{0.35\linewidth}
   \includegraphics[scale=0.3]{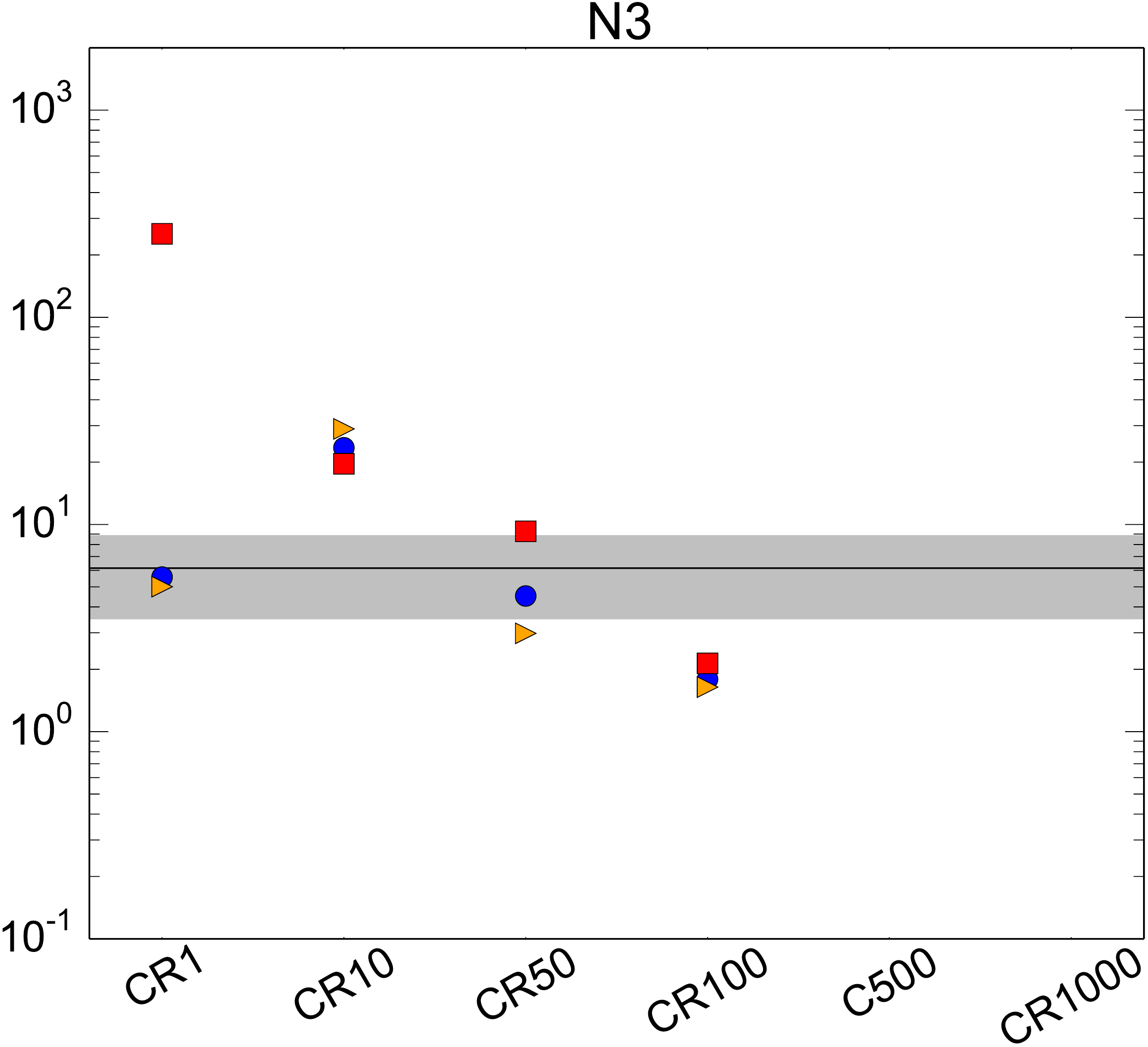} \\
   \includegraphics[scale=0.3]{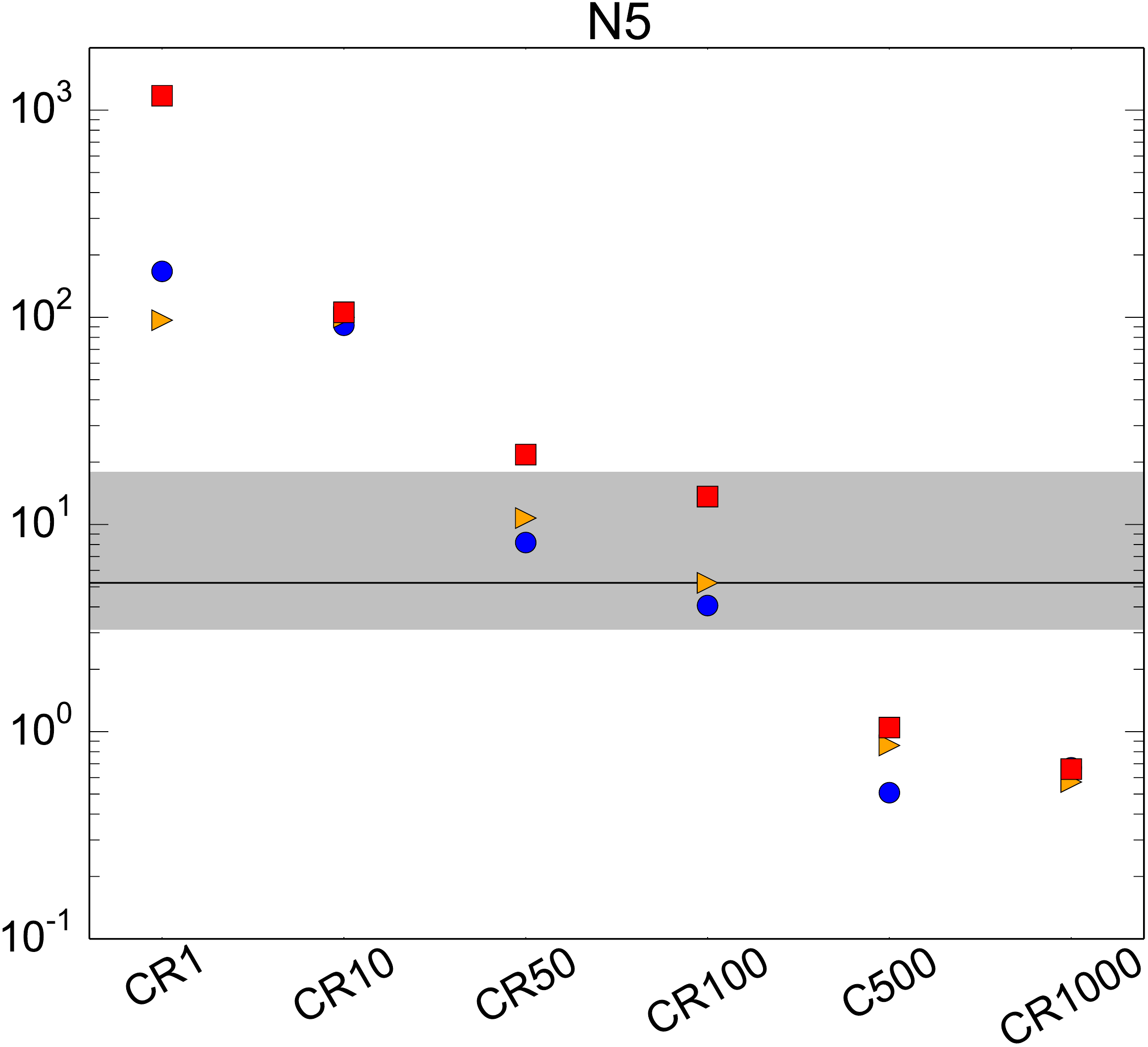} \\ 
\end{minipage}
    \caption{\label{FIG-compare-obs-C2H5CN-C2H3CN} Calculated fractional abundances of \ce{C2H5CN} with respect to \ce{C2H3CN} as a function of the CRIR. In each panel, the different symbols indicate the abundances calculated at $T$~=~150~K for different minimum dust temperatures and the horizontal black line shows the observed ratio. Uncertainties (1$\sigma$) are shown in gray. The [\ce{C2H5CN}]/[\ce{C2H3CN}] ratios obtained for models N3-CR500 ad N3-CR1000 ($<$10$^{-2}$) are not displayed.}
\end{figure*}

We now focus on the COM chemical composition itself and show in Fig.~\ref{FIG-confidence-level} the matrices of confidence levels computed for nine COMs relative to \ce{CH3OH} (panel a), the O-bearing molecules relative to \ce{CH3OH} (panel b), and the cyanides relative to \ce{CH3CN} (panel c). This refinement reveals that the O-bearing COM chemical composition relative to \ce{CH3OH} is sensitive to the CRIR value, with the best-fit value being 50 times the standard value, in agreement with the constraints obtained from the cyanide abundances relative to \ce{H2} (see Fig.~\ref{FIG-confidence-level-H2}). A minimum temperature below 20~K seems to be favored as well. The cyanide chemical composition with respect to \ce{CH3CN} is less sensitive to the two investigated parameters (bottom panel of Fig.~\ref{FIG-confidence-level}c). However, a more detailed investigation reveals that the abundance ratio of \ce{C2H5CN} to \ce{C2H3CN}, two species that are chemically linked (see Sect.~\ref{section-formation-routes-COMs}), is sensitive to the CRIR. Figure~\ref{FIG-compare-obs-C2H5CN-C2H3CN} shows that this ratio decreases as the CRIR increases in the four sources Sgr~B2(N2--N5). The observed ratios are best reproduced by the models with 50$\times \zeta^{\rm H_2}_0$~$\leq$~$\zeta^{\rm H_2}$~$\leq$100$\times \zeta^{\rm H_2}_0$, due to the increase in the gas-phase abundance of \ce{C2H3CN} at late times (that is high temperatures) in these models. This behaviour is consistent with what we observe, in particular toward Sgr~B2(N2) where the rotational temperature derived for \ce{C2H3CN} is higher than that of \ce{C2H5CN}. The flat gas-phase abundance profile of \ce{C2H3CN} at $T$~$>$~40~K in the standard models with lower CRIR (Fig.~\ref{FIG-evolution-frac-abund}) would imply that it traces more extended regions than \ce{C2H5CN}, which would be inconsistent with what is observed in Sgr~B2(N2).

Our analysis above shows that the chemical models with a CRIR enhanced by a factor 50 compared to the standard value best reproduce the observations for the four hot cores taken together. However, one must keep in mind that for simplicity our models assume that cosmic rays are not attenuated in the envelope of the sources, although the large visual extinctions of Sgr~B2(N2-N5) (see Figs.~\ref{FIG-appendix-nH-Av-time}b and \ref{FIG-appendix-all-physical-profiles}b) may actually lead to attenuation of the cosmic-ray flux. \citet{neufeld2017} showed, based on \ce{H2} and \ce{H3}$^+$ column densities measured toward diffuse clouds in the galactic disk, that the CRIR decreases with increasing \ce{H2} column density with a best-fit dependence scaling as $N$(\ce{H2})$^{-a}$ with $a$~=~0.92~$\pm$~0.32 for $N$(\ce{H2})~$\sim$~10$^{20}$ -- 10$^{22}$~cm$^{-2}$. This implies that the cosmic-ray flux is attenuated by a factor $\sim$~70--600 from the diffuse regions with $N$(\ce{H2})~=~10$^{21}$ cm$^{-2}$ to the dense gas with $N$(\ce{H2})~=~10$^{23}$--10$^{24}$ cm$^{-2}$. The attenuation of the cosmic-ray flux might have a significant impact on the chemical model results. For instance, \citet{rimmer2012} showed that a column-density-dependent CRIR improves the agreement between chemical model predictions and observations for the Horsehead nebula, compared to a standard model with constant a CRIR. For a more sophisticated treatment of the cosmic-ray driven chemistry in hot core models including cosmic-ray attenuation see, Willis et al. (in prep.).

 Moreover, our models do not take into account direct cosmic-ray bombardment of dust-grain ice mantles although laboratory experiments showed that it can trigger a rich chemistry that may lead to the formation of more complex species even at very low temperatures \citep{hudson2001, rothard2017}. Recently, \cite{shingledecker2018} used chemical models to investigate the effect of direct cosmic-ray collisions with dust grains on the solid-phase chemistry, including the formation of suprathermal species in the ices from collision with energetic particles. For instance, for a standard CRIR (10$^{-17}$~s$^{-1}$), they found that the abundance of \ce{CH3OCHO}, which is systematically underproduced in all our models (Fig.~\ref{FIG-comparison-obs}), is significantly enhanced in the gas phase, as well as on the grains compared to the standard chemical model (that is without cosmic-ray driven reactions).

As reported above, our models firmly exclude minimum dust temperatures of 25~K or higher during the prestellar phase of N2-N5. Such high temperatures would prevent the production of COMs at the level observed in these sources. Our models show that a value of 15~K, higher than the low values used in previous studies \citep[$\leq$10~K, see, e.g.,][]{muller2016, belloche2017, garrod2017}, still leads to an efficient production of COMs roughly consistent with the observations, while for 20~K signs of disagreement emerge between model and observations, especially for the cyanide abundances relative to H$_2$ and the O-bearing COM chemical composition relative to \ce{CH3OH}. A more accurate treatment of the dust temperatures in the envelope of  Sgr~B2(N2-N5) for different values of the ISRF does not modify the results much compared to our simple parametrization with an artificially-set minimum temperature (see Table~\ref{TAB-appendix-confidence-UV-ch3oh}). A value of 15~K is close to the lowest dust temperatures measured in the GC region (19--20~K, see Sect.~\ref{section-discussion-Tmin}). It could even be that models with $T_{\rm min} = $17--18~K produce COMs in amounts that are still consistent with the abundances measured in Sgr~B2(N)'s  hot cores. In addition, the coldest dust grains in the GC region may be masked by warmer outer layers and thus have been missed by \textit{Herschel} because of its limited angular resolution. Therefore, it is likely that the constraint we derived on $T_{\rm min}$ from the COM abundances is fully consistent with the thermal properties of the GC region.

\subsection{Chemical differentiation between N- and O-bearing complex molecules}
\label{section-discussion-chemical-differentiation}

Chemical differentiation between O- and N-bearing species has been observed toward several high-mass star-forming cores \citep{blake1987, wright1992, wyrowski1999, beuther2005, allen2017}. Recently, \citet{csengeri2018} investigated the young high-mass star-forming region G328.2551–0.5321 where \ce{CH3OH} and other O-bearing species show two distinct bright emission peaks spatially offset from the peak of the cyanides (see also Csengeri et al. subm.). The authors suggested that in this source, \ce{CH3OH} traces the accretion shocks resulting from the interaction between the collapsing envelope and the accretion disk, while the cyanide emission may trace the more compact hot region radiatively heated by the protostar. In this scenario, the O-bearing COM emission detected in the envelope of G328.2551–0.5321 arises from accretion shocks rather than from the region radiatively heated by the protostar itself. If such a mechanism is at play in Sgr~B2(N) at scales smaller than the beam of the EMoCA survey, then infering the luminosity of the protostars from the rotational temperature measured at the radius of the COM emission as we did in Sect.\ref{section-radmc3d} would be wrong. The thermal history derived for Sgr~B2(N2-N5) (see Sect.~\ref{section-collapse-phase}) would then be incorrect, and this could have a significant impact on the results of our chemical modeling. Observations at higher angular resolution that are attained in the ReMoCA survey (Belloche et al. subm.) are needed to investigate whether a chemical differentiation between N- and O-bearing species occurs in Sgr~B2(N)'s hot cores.

\subsection{Limitations on the physical models of Sgr~B2(N2-N5)}          
\label{section-discussion-evolutionary-stage}

In Sect.~\ref{section-evolution-L-HCs} we have estimated the age, $t_{\rm source}$, of each hot core based on the estimated current luminosities (Sect.~\ref{section-radmc3d}). To do so, we used the relation $L(t)$ obtained from RMHD simulations of high-mass star formation \citep[][see also Sect.~\ref{section-evolution-L-HCs}]{peters2011}, combined with the mass-luminosity relation given by \citet{hosokawa2009}. The estimated $t_{\rm source}$ values are unlikely to be accurate because our analysis assumes that the evolution of the accretion rate, $\dot{M}(t)$, is the same for all sources, although the density measured at a given radius toward Sgr~B2(N2) is a factor 3.2, 3.5, and 2.0 higher than that of N3, N4, and N5, respectively (Fig.~\ref{FIG-appendix-all-physical-profiles}b). In addition, the simulations of \citet{peters2011} were set up for a molecular cloud with a central core gas mass density of $\rho_{\rm g} = 1.27 \times 10^{-20}$~g~cm$^{-3}$ within a radius of 0.5~pc, which is about 10--40 times lower than the values we derive for Sgr~B2(N2-N5) at the same radius (Fig.~\ref{FIG-appendix-all-physical-profiles}b). Using the mass-luminosity relation provided by \citet{hosokawa2009} for a constant accretion rate of 10$^{-4}$~$\Msol$~yr$^{-1}$, the final mass of $\sim$~30~$\Msol$ obtained using the model from \citet{peters2011} leads to a final luminosity of $\sim$10$^5 \Lsol$, significantly lower than the values we derive for Sgr~B2(N2-N5) (Sect.~\ref{section-radmc3d}). 

Our simplified treatment, in which we assumed that the accretion rate does not drop below 10$^{-4} \Msol$~yr$^{-1}$ to form more massive and more luminous objects (Sect.~\ref{section-evolution-L-HCs}), results in significant differences in the time evolution of Sgr~B2(N3)'s physical parameters ($n_{\rm H}$, $A_{\rm v}$, and $T$) compared to the other three sources. Given its lower luminosity, N3 is apparently younger than the other sources (see Table~\ref{TAB-HCs-properties}). In the framework of our model, this results in a faster increase of the density and temperature along the trajectory of the modeled parcel of gas during the free-fall collapse phase of N3. Figure~\ref{FIG-comparison-H2} shows that our best models (that is with $\zeta^{H_2}$~=~50$\times \zeta^{H_2}_0$) better reproduce the chemical abundances (with respect to H$_2$) of \ce{C2H5OH} and \ce{CH3OH} in Sgr~B2(N3) compared to the other sources. However, the modeled abundances of \ce{CH3OCHO} and \ce{CH3OCH3} are 2--3 orders of magnitude lower than the abundances measured toward N3. This might suggest that the relations $\dot{M}(t)$, $M(t)$, and $L(t)$ we derived in Sect.~\ref{section-evolution-L-HCs} do not represent accurately the physical evolution of Sgr~B2(N2-N5). Furthermore, in our previous analysis of Sgr~B2(N)'s hot cores \citep[][]{bonfand2017}, we proposed a tentative evolutionary sequence in which N4 appears younger than N3, based on the molecular outflow signature detected toward N3, while N4 does not show any evidence of outflow yet. This is in contradiction with the present analysis in which we find that N3 is the youngest of the four investigated hot cores. This is another indication that the short evolution timescale we obtain here for N3 may not represent well the physical evolution of the source. However, the origin of some wing emission seen in the spectrum of N4 is not understood \citep[see discussion in Sect.~3.7 in][]{bonfand2017} and a search for a bipolar outflow in N4 should be carried out with more robust outflow tracers, such as CO, whose $J$=1--0 transition is not covered by the EMoCA survey.

We also point out that the ages we obtain for Sgr~B2(N2), N4, and N5 in Sect.~\ref{section-evolution-L-HCs} are about four times longer than the hot core lifetime we estimated in our previous analysis of Sgr~B2(N)'s hot core population \citep[$\sim$6$\times$10$^4$~yr,][]{bonfand2017} on the basis of a very simple statistical argument involving the number of hot cores and UCHII regions detected in the same area and assuming a lifetime of $\sim$10$^5$~yr for the UCHII phase \citep{peters2010}. This might suggest that the accretion rate, $\dot{M}(t)$, of these sources is/was actually higher than what we have assumed here.

In order to better represent the extreme physical conditions (densities/masses, luminosities) of Sgr~B2(N)'s hot cores and give an accurate estimate of their age, RMHD simulations treating specifically their different physical properties will be needed. In addition a more self-consistent treatment of the accretion rate, mass, and luminosity evolution (that is, taking into account the time-dependent evolution of the accretion rate to derive the mass-luminosity relation) is also required.

\section{Conclusions}

 We analyzed the results of the 3~mm imaging line survey EMoCA conducted with ALMA in its cycles 0 and 1 to characterize the hot core population embedded in Sgr~B2(N). We derived the chemical composition of the three hot cores Sgr~B2(N3-N5) focusing on 11 COMs. We investigated the time-dependent evolution of the protostellar properties and we derived the evolution of density, temperature, and visual extinction in the envelopes of Sgr~B2(N2-N5). We used the astrochemical code MAGICKAL to simulate the time-dependent chemical evolution of the sources based on their physical evolution. We investigated the impact of the cosmic rays and the minimum dust temperature reached during the cold prestellar phase on the chemical abundances calculated by the models. We compared the results to the observations to constrain the physical parameters which best represent the environmental conditions characterizing Sgr~B2(N) and the GC region. Our main results are the following:

\begin{enumerate}
\item{Sgr~B2(N3) and N5 share a similar chemical composition relative to methanol. The chemical composition of Sgr~B2(N2) differs significantly from the other hot cores Sgr~B2(N3-N5).}

\item{We derived gas densities of 1.4$\times$10$^7$~cm$^{-3}$ toward Sgr~B2(N2) at a radius of 6010~au ($\sim$0.03pc), while the values derived for N3, N4, and N5 at the same radius are a factor 3.2, 3.5, and 2.0 lower, respectively.}

\item{On the basis of the rotational temperature derived for the COMs and using the results of RMHD simulations of high-mass star formation, the current luminosities of the hot cores are estimated to be 2.6$\times$10$^5$~$\Lsol$, 4.5$\times$10$^4$~$\Lsol$, 3.9$\times$10$^5$~$\Lsol$, 2.8$\times$10$^5$~$\Lsol$ for Sgr~B2(N2), N3, N4, and N5, respectively. These results hold only if our assumption that the COM emission arises from the region radiatively heated by the protostar is valid.}

\item{The production of the cyanides \ce{C2H5CN} and \ce{CH3CN} as well as \ce{CH3OH} and \ce{C2H5OH} mostly relies on the early cold grain-surface chemistry ($T_{\rm d}$~$\leqslant$~15~K), while the O-bearing species \ce{CH3CHO}, \ce{CH3OCHO}, \ce{CH3OCH3} are predominantly formed during the warm-up phase of the protostellar evolution.}

\item{COMs form efficiently on grains during the prestellar phase of Sgr~B2(N2-N5) with minimum dust temperatures as high as 15~K. Minimum dust temperatures $\geqslant$25~K are firmly excluded by our chemical models.}

\item{The cosmic-ray ionization rate adopted in the chemical simulations is critical, in particular during the warm-up stage where cosmic rays may rapidly destroy COMs once they are released into the gas phase. The best match between chemical models and observations is obtained for $\zeta^{\rm H_2}$~=~7$\times$10$^{-16}$~s$^{-1}$, that is a CRIR enhanced by a factor 50 compared to the solar neighborhood value. This is somewhat lower than extreme values expected toward the diffuse medium in the GC region (1 -- 11$\times$10$^{-14}$~s$^{-1}$). This difference may reflect the attenuation of cosmic rays in denser gas.}

\end{enumerate}

The chemical composition calculated by our models for Sgr B2(N)'s hot cores strongly depends on their history since the initial stages in the prestellar phase. Our strong assumption about the accretion history being the same for all sources implies a much shorter timescale for Sgr~B2(N3) which has a significant impact on its calculated chemical composition and results in a worse agreement with the observations. Furthermore, our model in which the three other hot cores, N2, N4, and N5 are by far older than N3 is inconsistent with the tentative evolutionary sequence suggested in our previous analysis of Sgr~B2(N)'s hot cores based on their association with class II methanol masers, outflows, and UCHII regions \citep{bonfand2017}. This emphasizes the need for RMHD simulations tailored to each hot core in order to have a more realistic description of their history and improve the reliability of the chemical model results.


\begin{acknowledgements}
We would like to thank the anonymous referee for his/her comments that helped improving the paper. We thank Takeshi Hosokawa for providing us the results of their star-formation models in electronic form. MB thanks the University of Virginia for hosting her as a visiting student. This work was carried out within the Collaborative Research Centre 956, sub-project B3, funded by the Deutsche Forschungsgemeinschaft (DFG). This paper makes use of the following ALMA data: ADS/JAO.ALMA\#2011.0.00017.S, ADS/JAO.ALMA\#2012.1.00012.S. ALMA is a partnership of ESO (representing its member states), NSF (USA), and NINS (Japan), together with NRC (Canada), NSC and ASIAA (Taiwan), and KASI (Republic of Korea), in cooperation with the Republic of Chile. The Joint ALMA Observatory is operated by ESO, AUI/NRAO, and NAOJ. The interferometric data are available in the ALMA archive at https://almascience.eso.org/aq/. 
\end{acknowledgements}

\bibliographystyle{aa}
\bibliography{biblioPhD}

\clearpage

\begin{appendix}

\section{Uncertainties on the molecular column densities}
\label{appendix-column-density-uncertainties}

The column densities estimated in Sect.~\ref{section-chemical-composition} for Sgr~B2(N3-N5) (see also Table.~\ref{TAB-best-fit-parameters}) are based on simple assumptions, such as a single excitation temperature to characterize all transitions from a given molecule under the LTE approximation. In addition, because the hot core emission toward Sgr~B2(N3-N5) is resolved in our data for only a few species, especially toward Sgr~B2(N3) for which transitions from only two species could be used to derive the source size \citep{bonfand2017}, we assume for each of these three sources that all molecules are emitted from the same region characterized by a single size. This may not be true as it is shown in Sect.~\ref{section-formation-routes-COMs} that all investigated COMs do not desorb from the dust grains at the same temperature in the chemical models, suggesting that they may trace different regions.

In order to estimate the uncertainties on the column densities we varied for each species detected toward Sgr~B2(N3-N5) the emission size and rotational temperature within the uncertainties given by the 2D Gaussian fits to the integrated intensity maps and population diagrams, respectively. Comparing the column densities derived with our best-fit synthetic spectra (Table~\ref{TAB-best-fit-parameters}) with those obtained varying temperature and source size, we find that column densities vary at most by a factor four. In the case of Sgr~B2(N2) we simply assume uncertainties of 30\% on the molecular column densities of all investigated species.

\section{Results of the chemical simulations}
        \label{appendix-results-models}

\begin{figure}[!h]
   \begin{center}
    \includegraphics[width=\hsize]{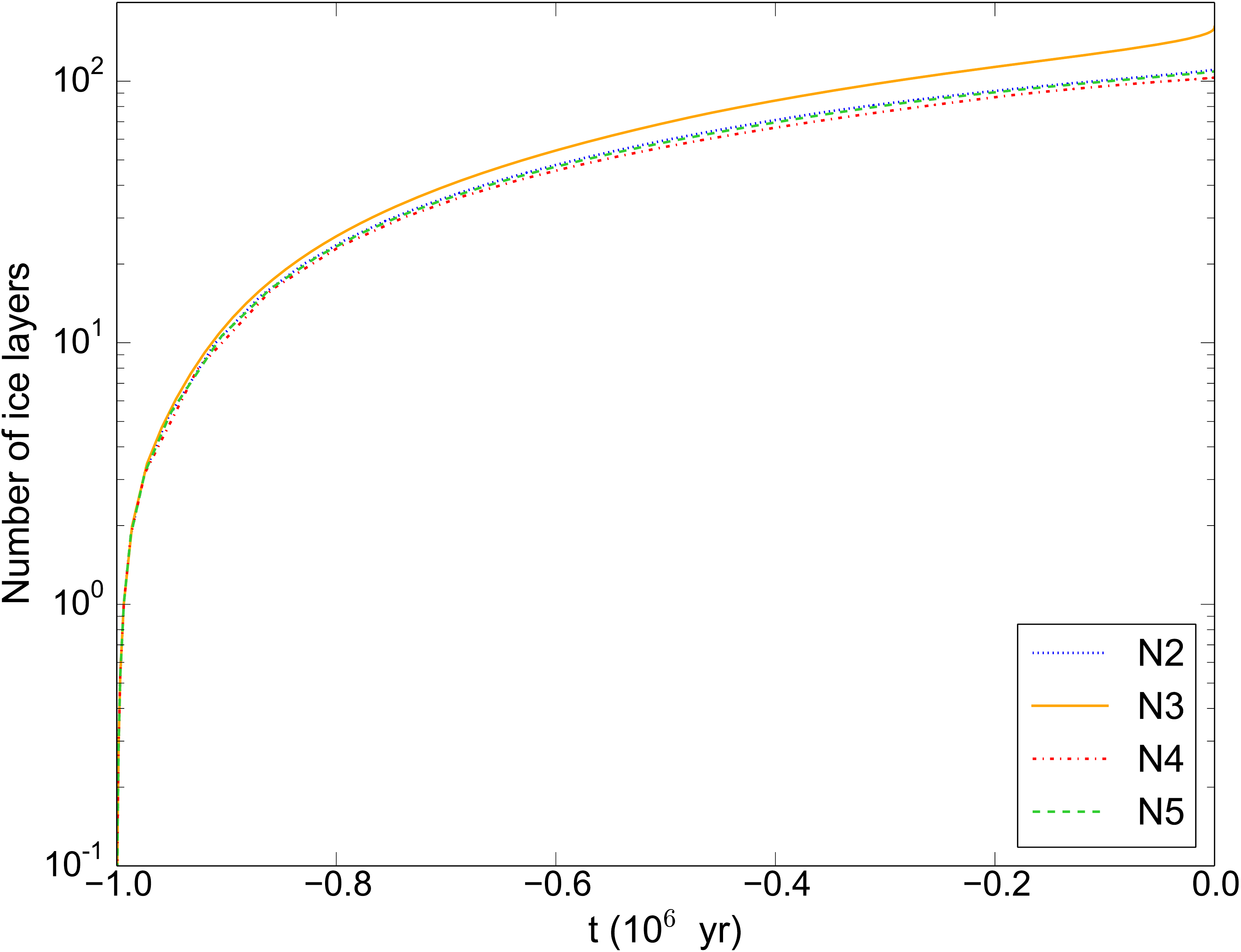} 
    \caption{\label{FIG-appendix-growth-ice-mantle} Growth of the dust-grain ice mantles over time during the quasi-static contraction phase for the standard models N(2-5)-T15-CR1.}
   \end{center}
\end{figure}

\begin{figure*}[!t]
 \hspace{0.03\linewidth}
 \begin{minipage}{0.5\linewidth}
   \includegraphics[scale=0.85]{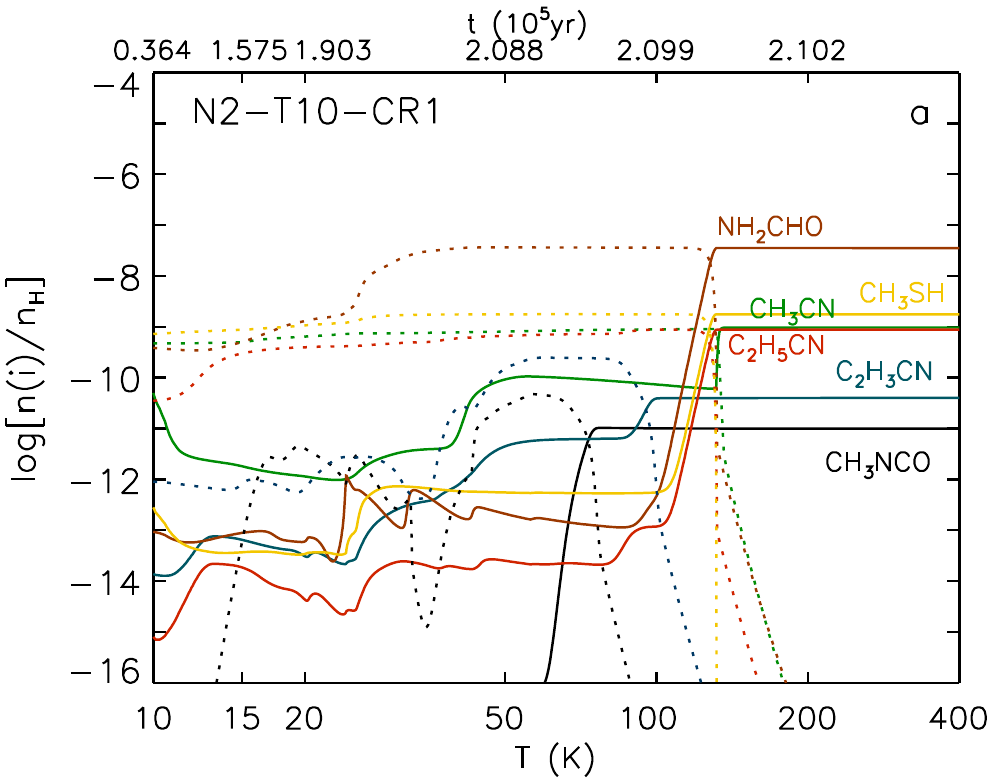} \\
   \includegraphics[scale=0.85]{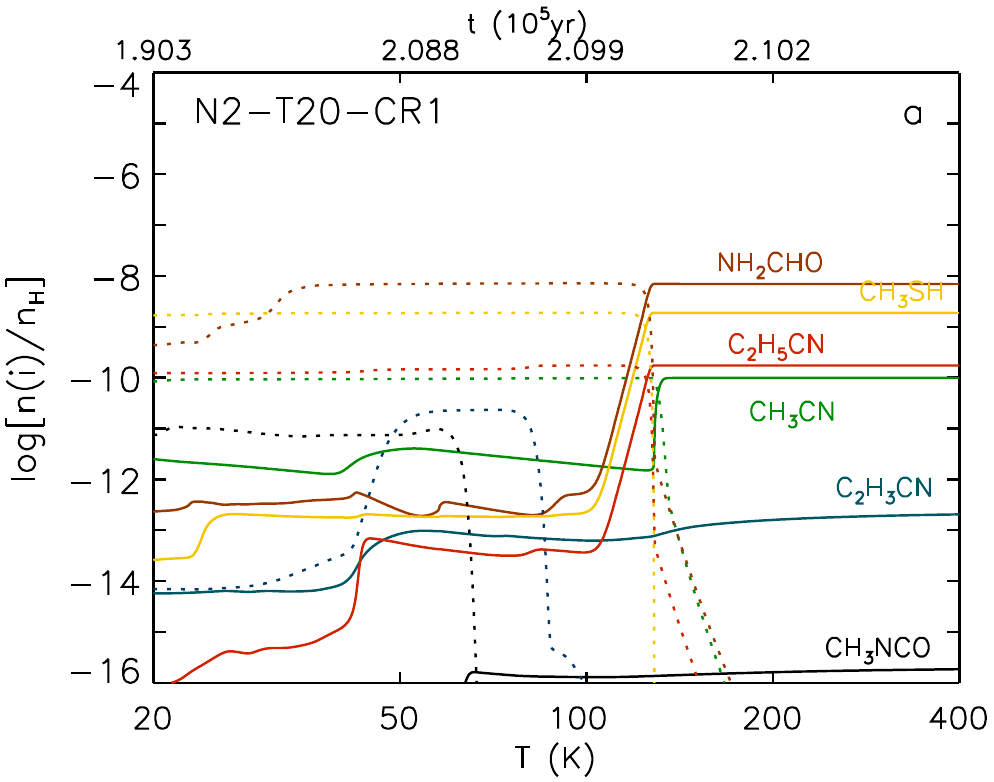} \\
 \end{minipage}
\hspace{-0.02\linewidth}
 \begin{minipage}{0.35\linewidth}
    \includegraphics[scale=0.85]{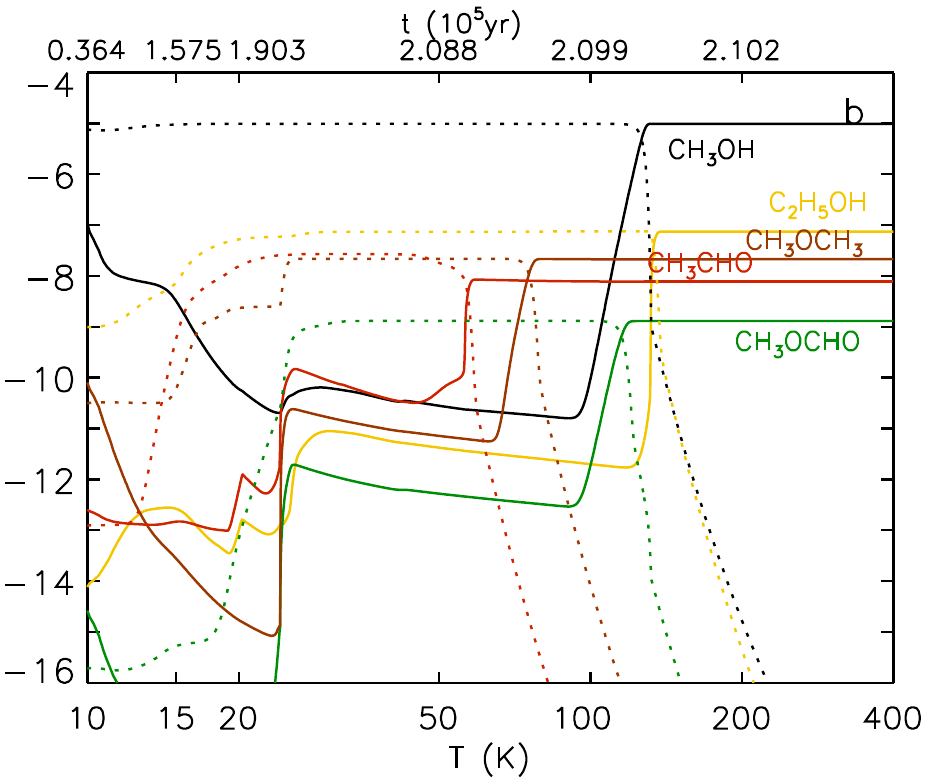} \\
    \includegraphics[scale=0.85]{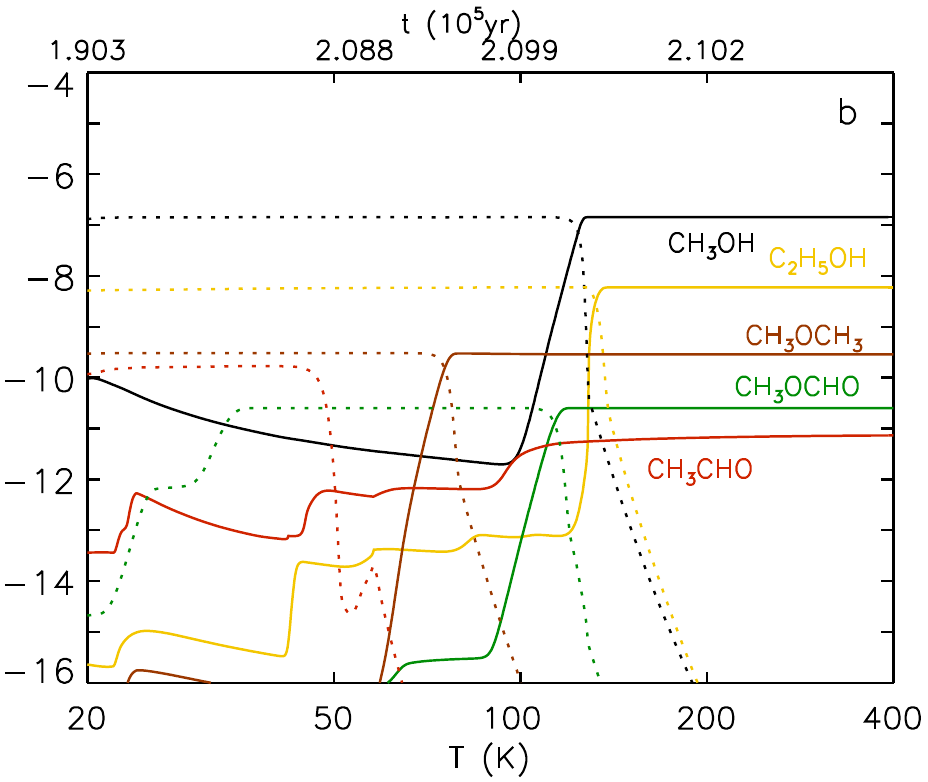} \\
 \end{minipage} 
\vspace{-5mm}
\caption{\label{FIG-appendix-collapse-Tmin-N2} Calculated fractional abundances of 11 COMs for models N2-T10-CR1 (top panels) and N2-T20-CR1 (bottom panels), plotted as a function of temperature in the envelope of Sgr~B2(N2) during the free-fall collapse phase. In each panel the timescale shown at the top is derived from the time-dependent evolution of the temperature along the trajectory of the infalling parcel of gas (see Sect.~\ref{section-collapse-phase}). In each panel, the solid lines show the fractional abundances (with respect to total hydrogen) in the gas phase while the dotted lines show the abundances of the same species on the grains (ice-surface+mantle).}
\end{figure*}

\begin{figure*}
 \hspace{0.03\linewidth}
 \begin{minipage}{0.5\linewidth}
   \includegraphics[scale=0.85]{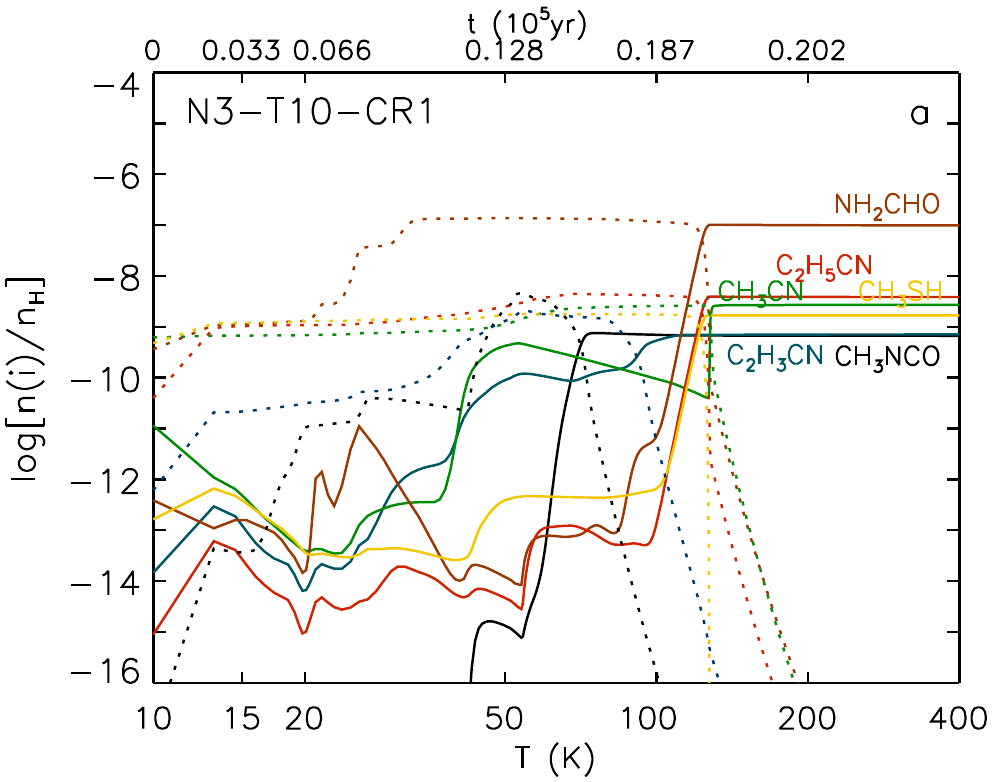} \\
   \includegraphics[scale=0.85]{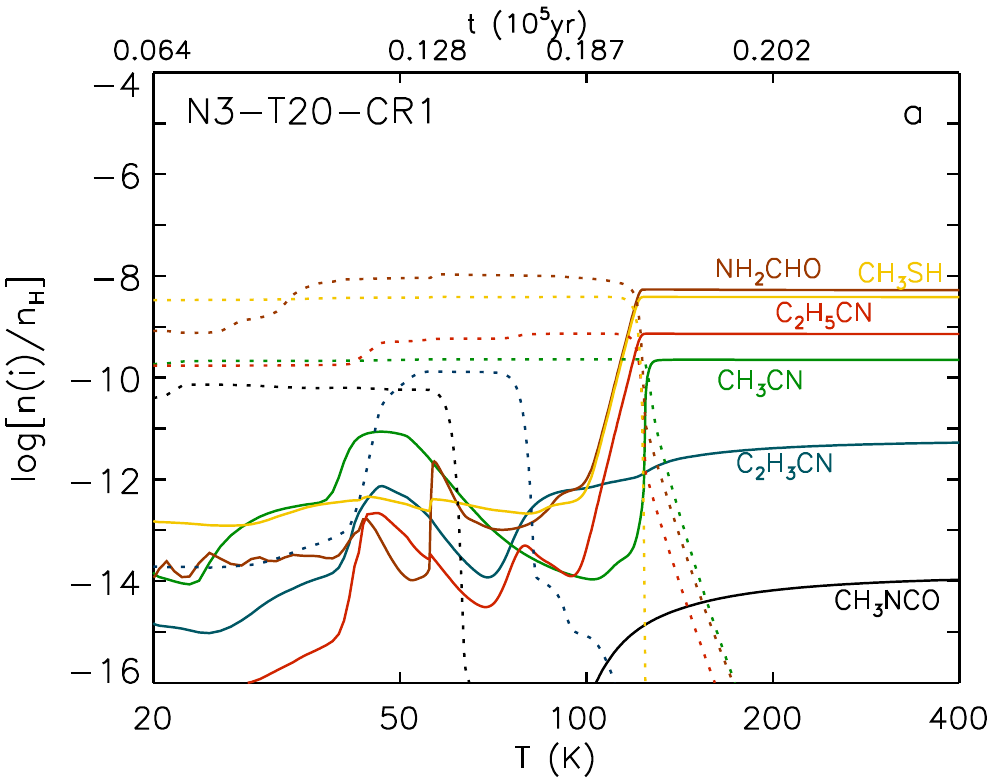} \\
 \end{minipage}
\hspace{-0.02\linewidth}
 \begin{minipage}{0.35\linewidth}
    \includegraphics[scale=0.85]{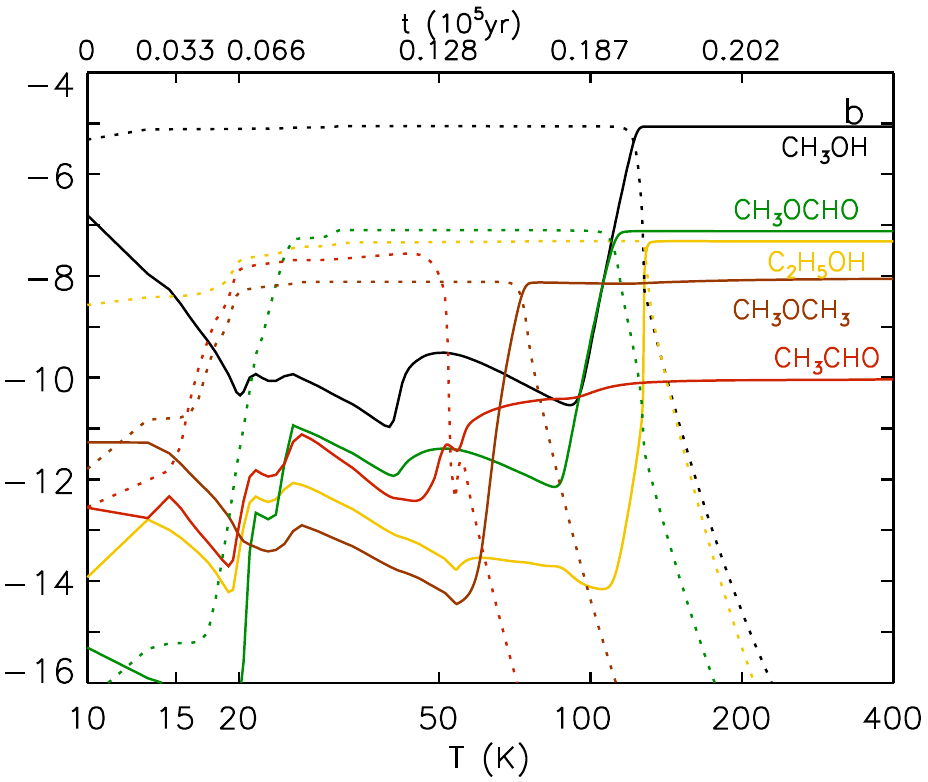} \\
    \includegraphics[scale=0.85]{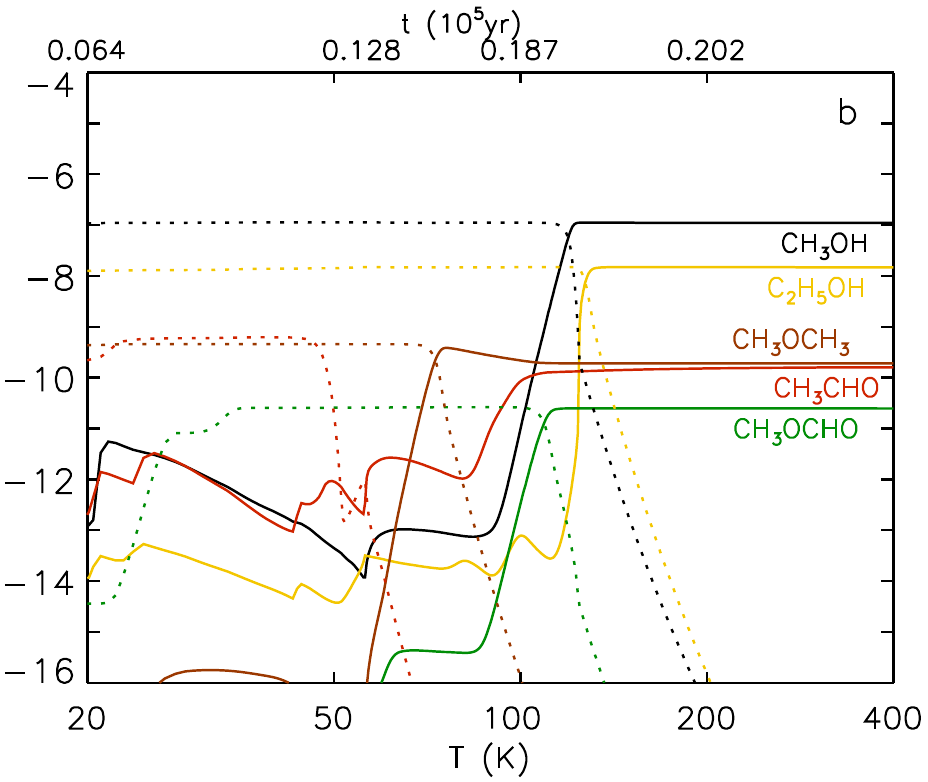} \\
 \end{minipage} 
\vspace{-5mm}
\caption{\label{FIG-appendix-collapse-Tmin-N3} Same as \ref{FIG-appendix-collapse-Tmin-N2} but for models N3-T10-CR1 (top panels) and N3-T20-CR1 (bottom panels).}
\end{figure*}

\begin{figure*}
 \hspace{0.03\linewidth}
 \begin{minipage}{0.5\linewidth}
   \includegraphics[scale=0.85]{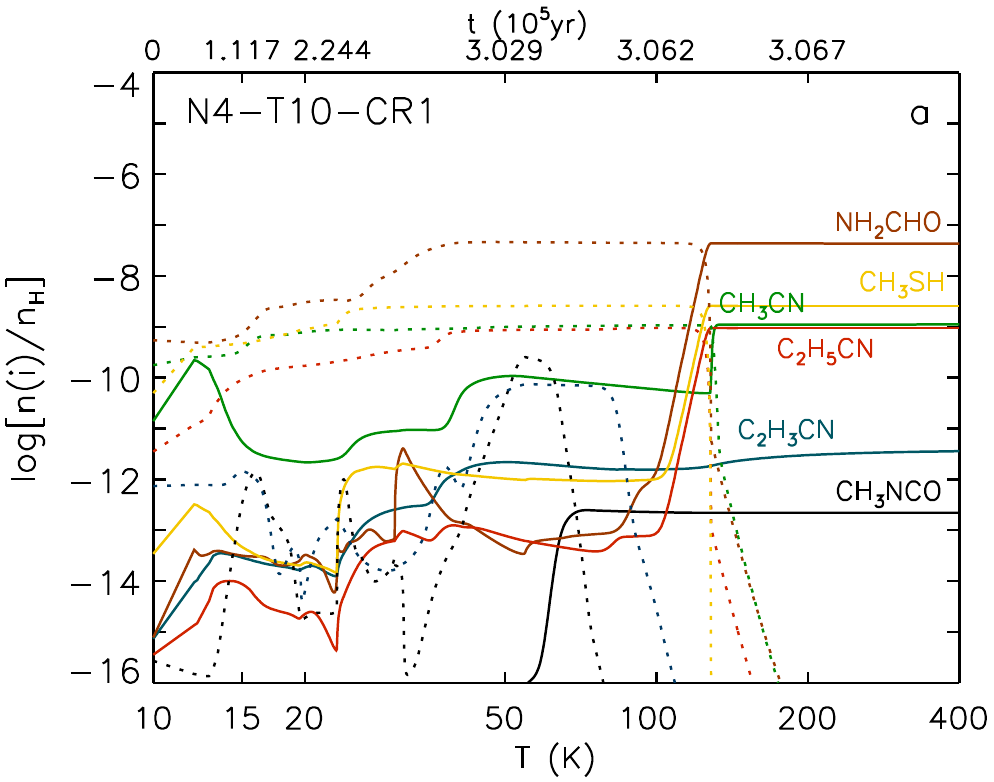} \\
   \includegraphics[scale=0.85]{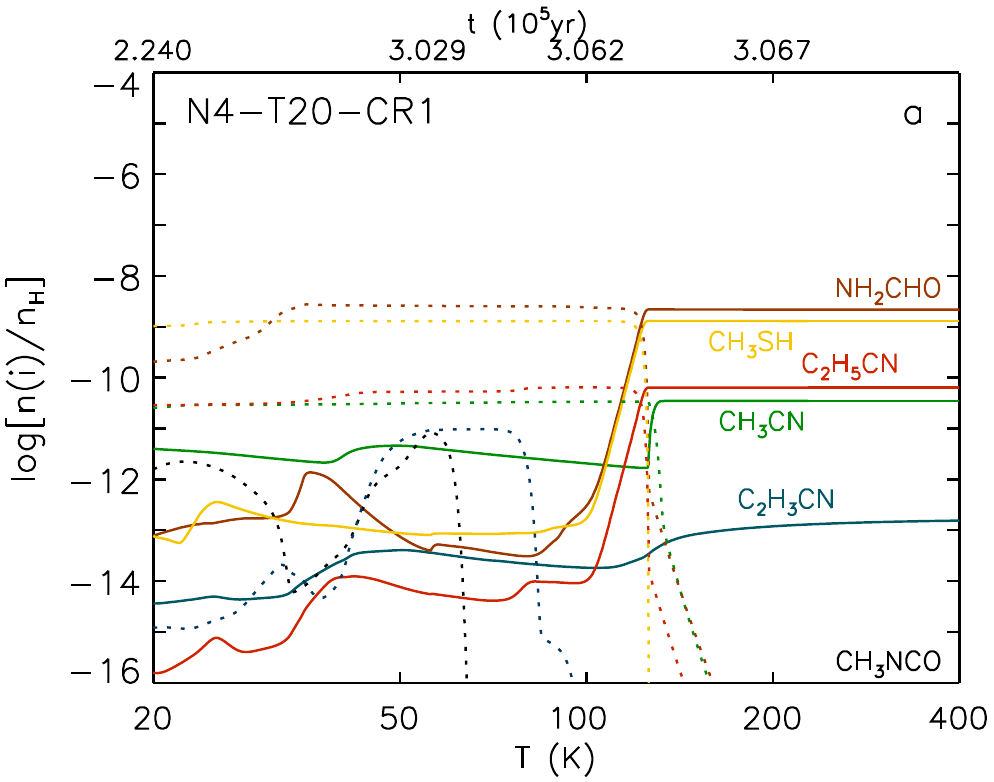} \\
 \end{minipage}
\hspace{-0.02\linewidth}
 \begin{minipage}{0.35\linewidth}
    \includegraphics[scale=0.85]{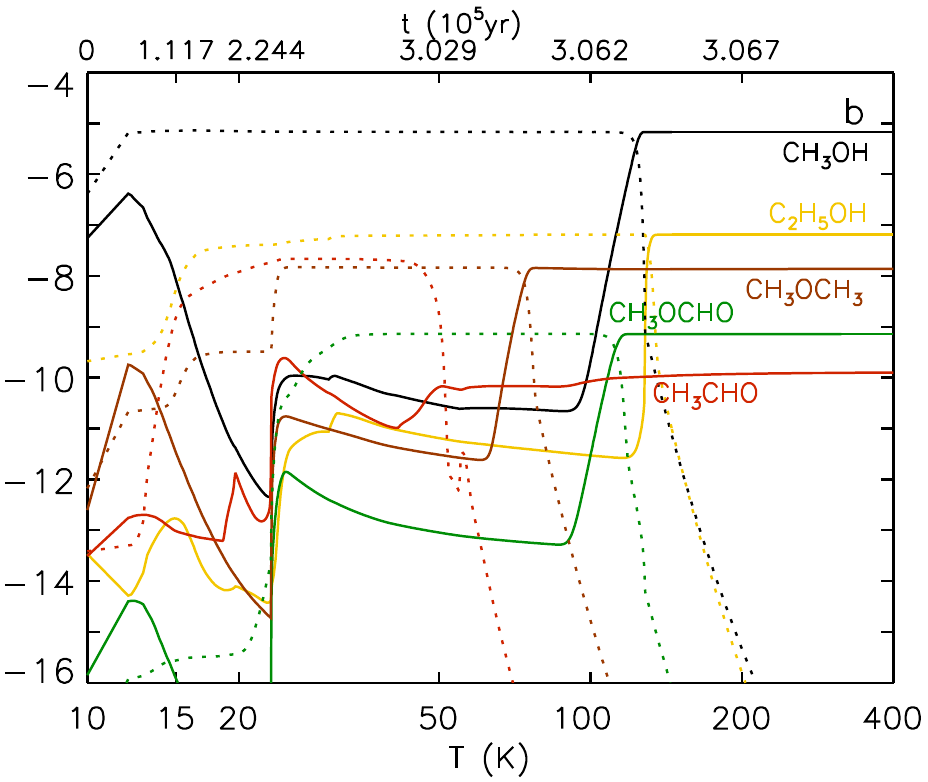} \\
    \includegraphics[scale=0.85]{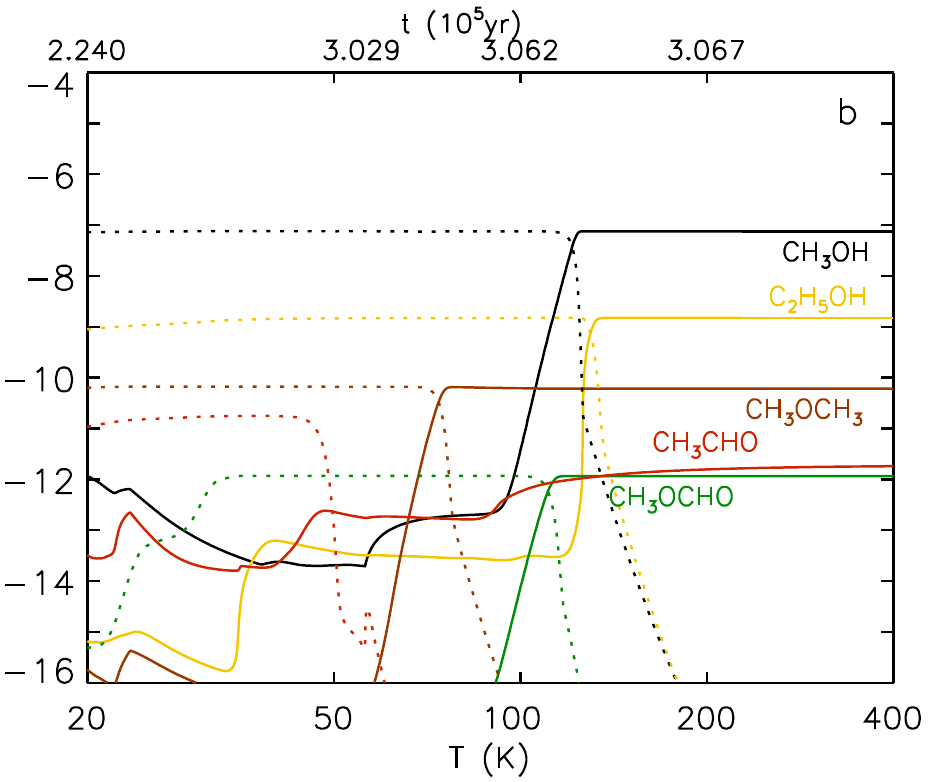} \\
 \end{minipage} 
\vspace{-5mm}
\caption{\label{FIG-appendix-collapse-Tmin-N4} Same as \ref{FIG-appendix-collapse-Tmin-N2} but for models N4-T10-CR1 (top panels) and N4-T20-CR1 (bottom panels).}
\end{figure*}

\begin{figure*}
 \hspace{0.03\linewidth}
 \begin{minipage}{0.5\linewidth}
   \includegraphics[scale=0.85]{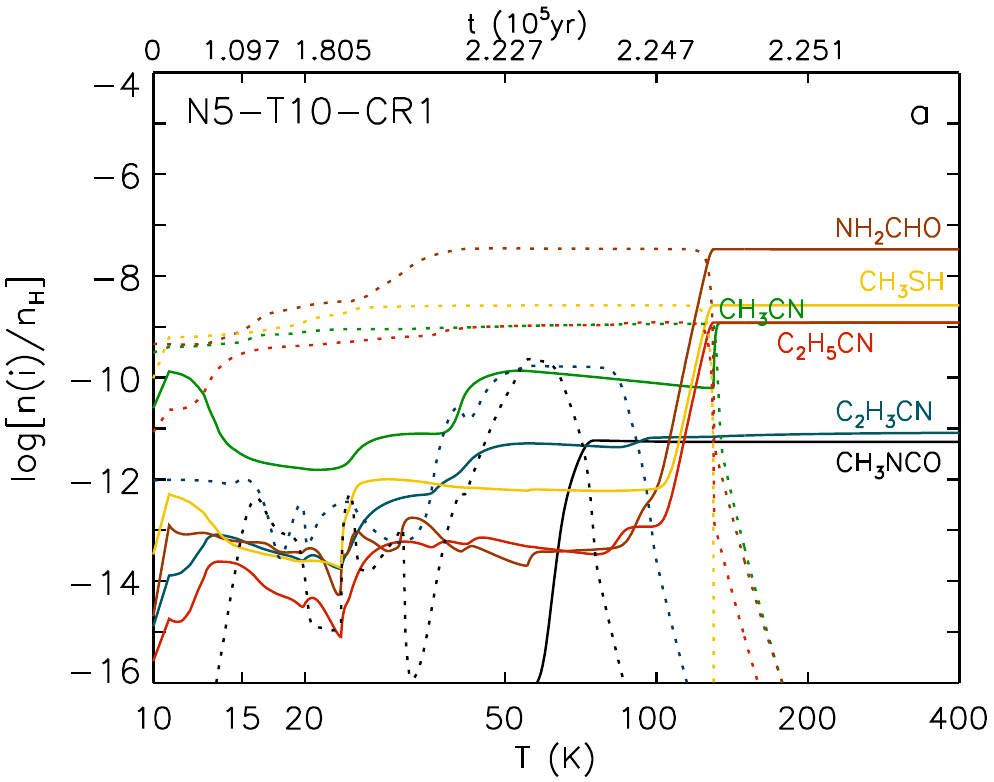} \\
   \includegraphics[scale=0.85]{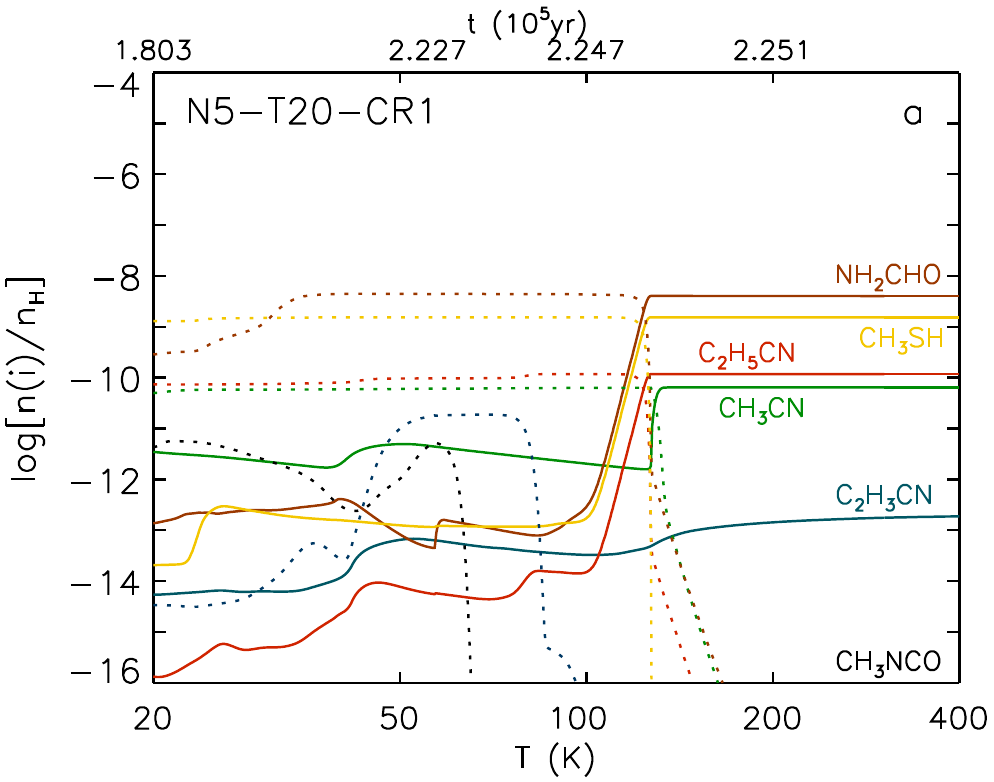} \\
 \end{minipage}
\hspace{-0.02\linewidth}
 \begin{minipage}{0.35\linewidth}
    \includegraphics[scale=0.85]{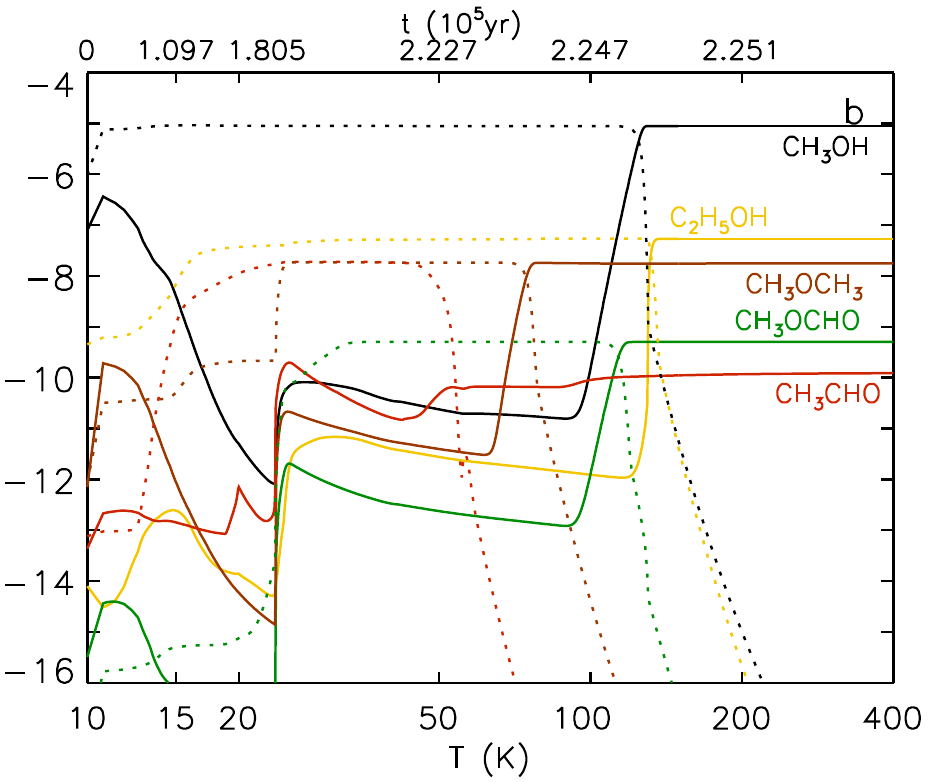} \\
    \includegraphics[scale=0.85]{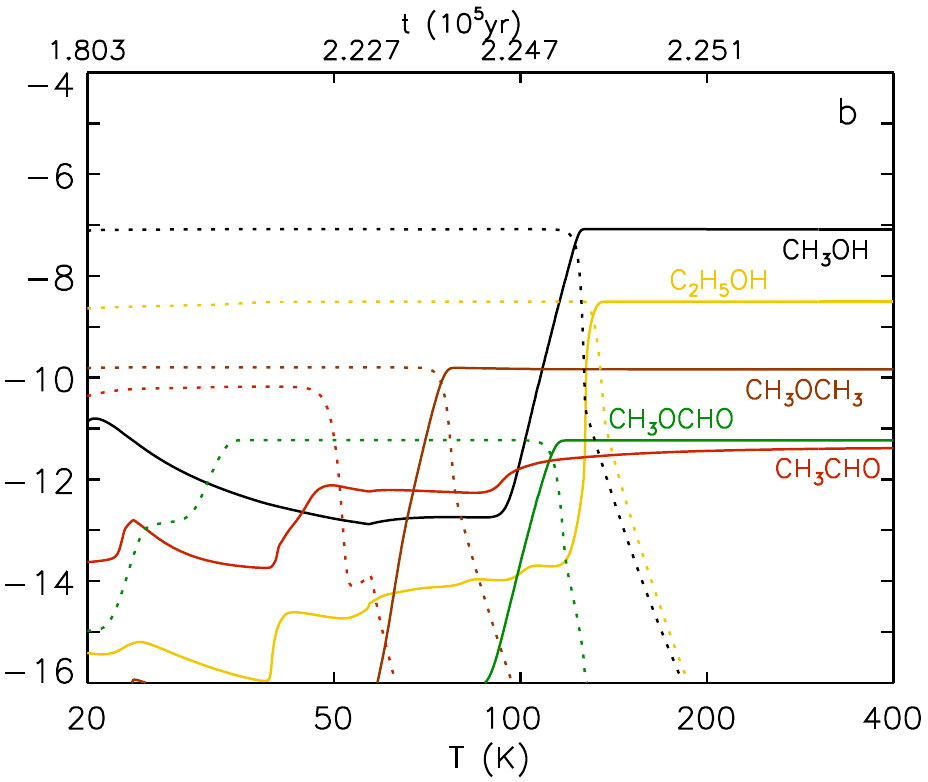} \\
 \end{minipage} 
\vspace{-5mm}
\caption{\label{FIG-appendix-collapse-Tmin-N5} Same as \ref{FIG-appendix-collapse-Tmin-N2} but for models N5-T10-CR1 (top panels) and N5-T20-CR1 (bottom panels).}
\end{figure*}

\begin{figure*}[!t]
 \hspace{0.03\linewidth}
 \begin{minipage}{0.5\linewidth}
   \includegraphics[scale=0.85]{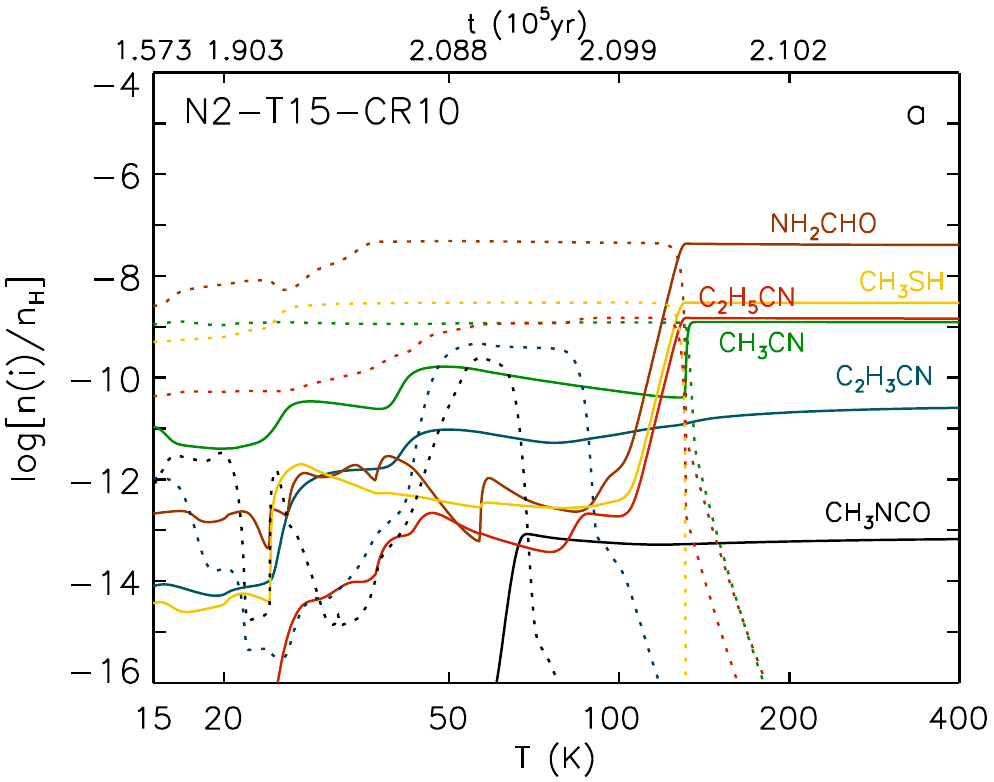} \\
   \includegraphics[scale=0.85]{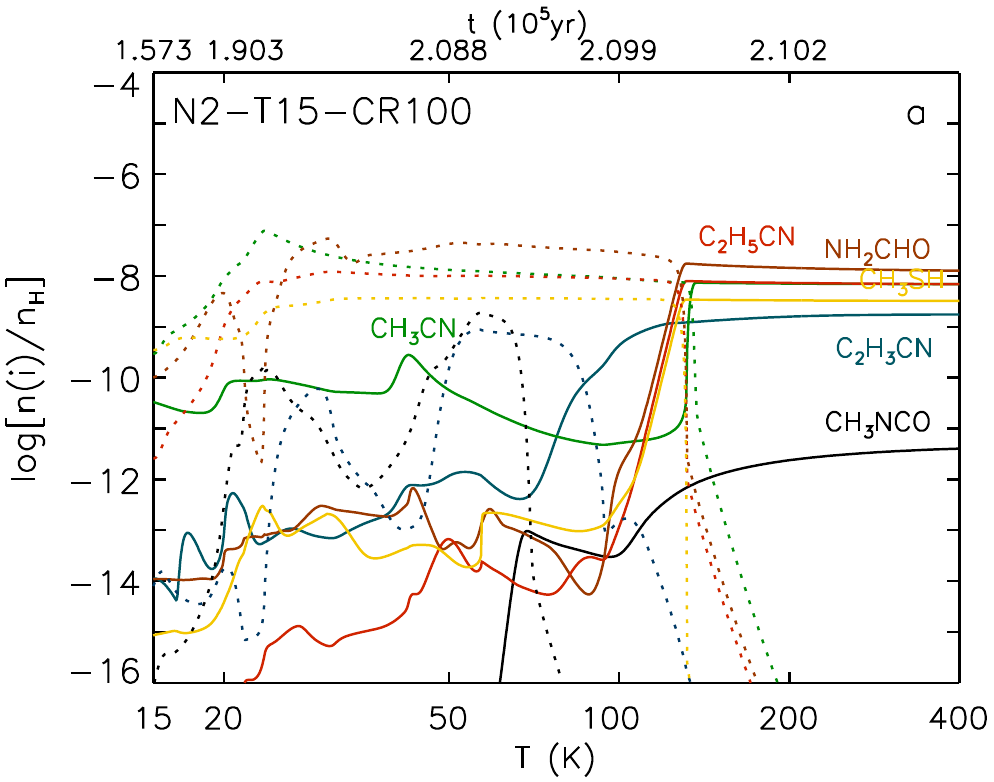} \\
   \includegraphics[scale=0.85]{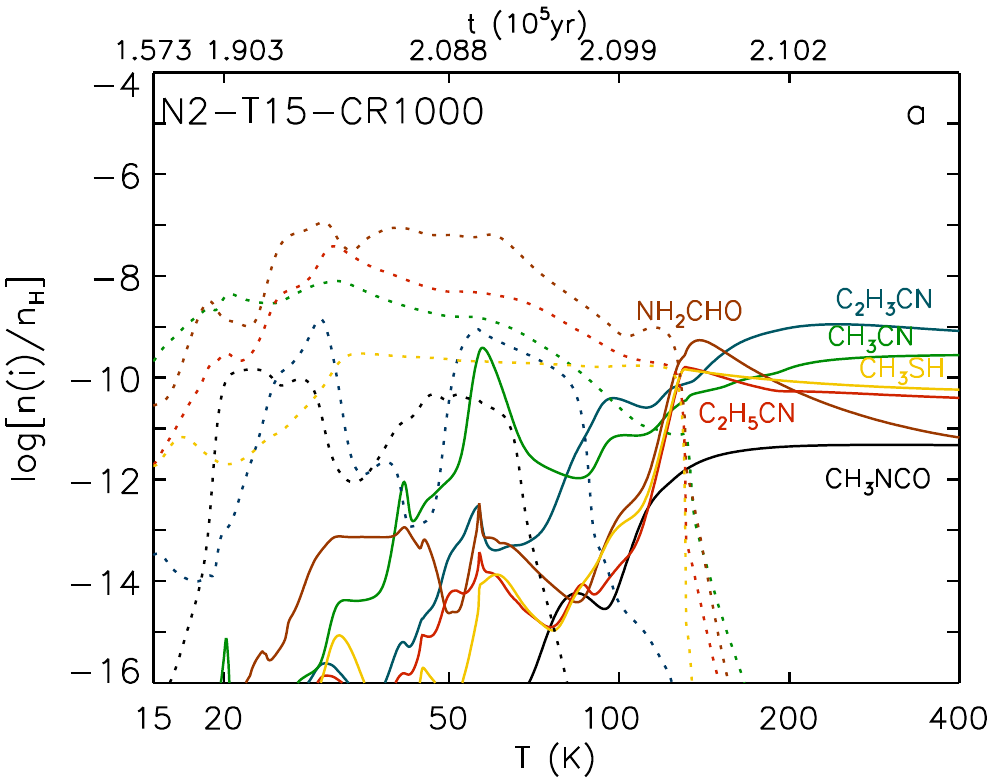} \\
 \end{minipage}
\hspace{-0.02\linewidth}
 \begin{minipage}{0.35\linewidth}
    \includegraphics[scale=0.85]{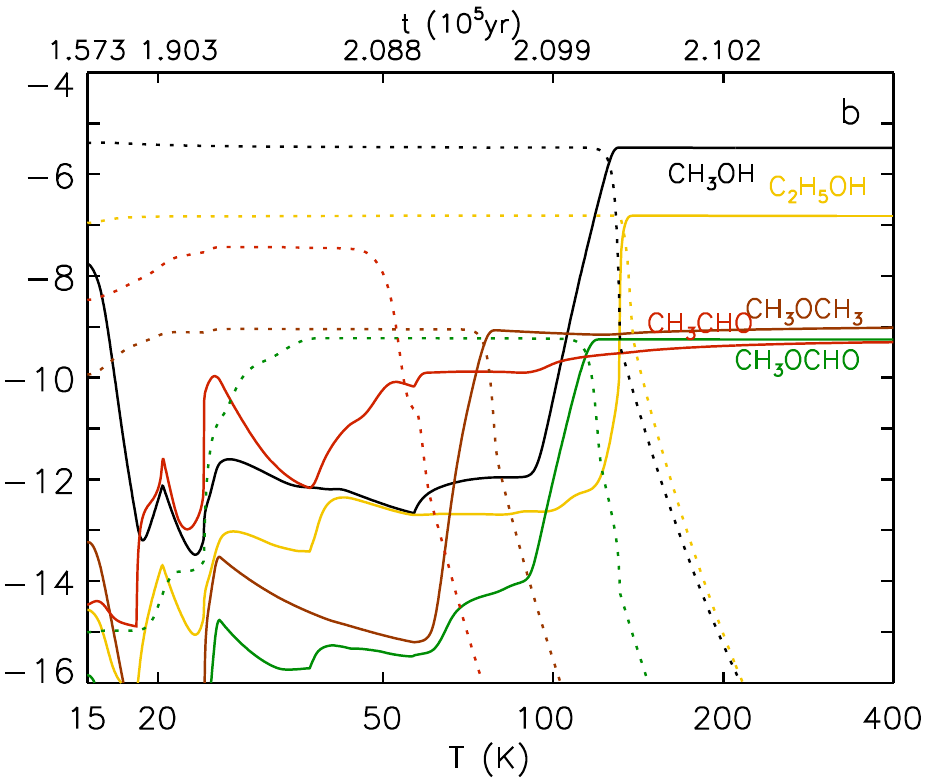} \\
    \includegraphics[scale=0.85]{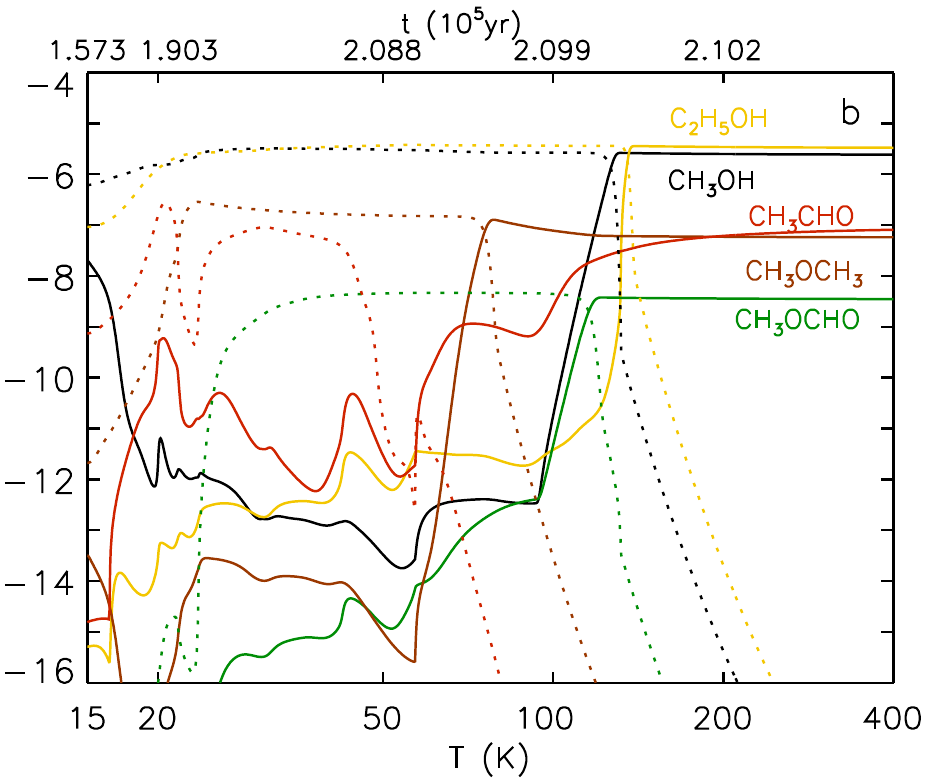} \\
    \includegraphics[scale=0.85]{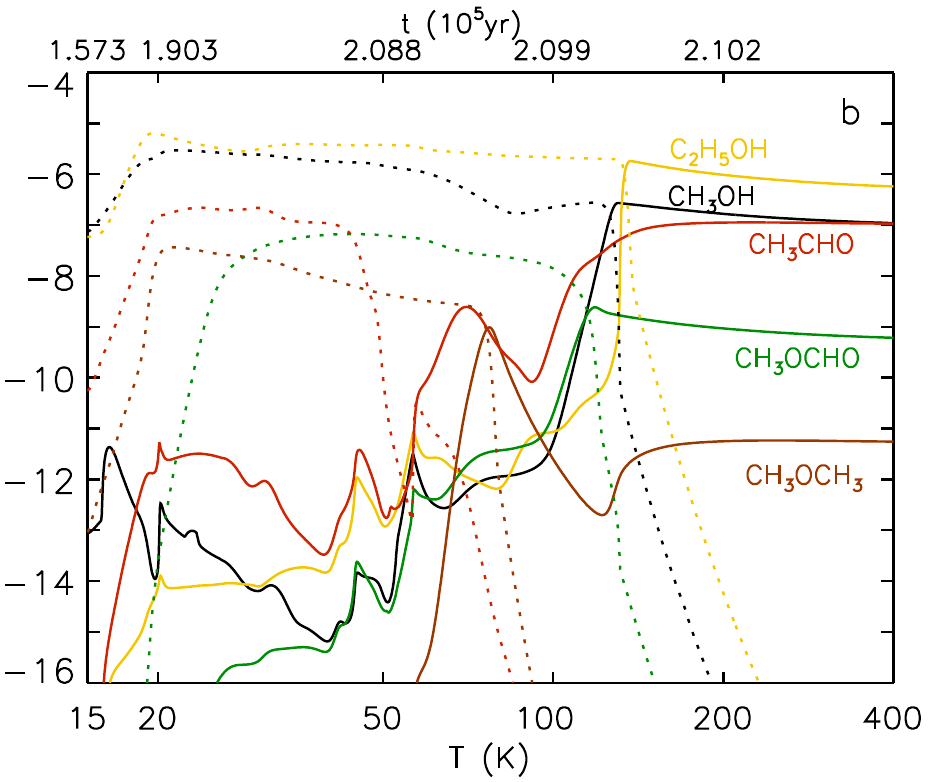} \\
 \end{minipage}
 \vspace{-5mm}
\caption{\label{FIG-appendix-collapse-CR} Same as \ref{FIG-appendix-collapse-Tmin-N2} but for models N2-T15-CR10 (top panels), N2-T15-CR100 (middle panels), and N2-T15-CR1000 (bottom panels).}
\end{figure*}

\begin{figure*}
 \hspace{0.03\linewidth}
 \begin{minipage}{0.5\linewidth}
   \includegraphics[scale=0.85]{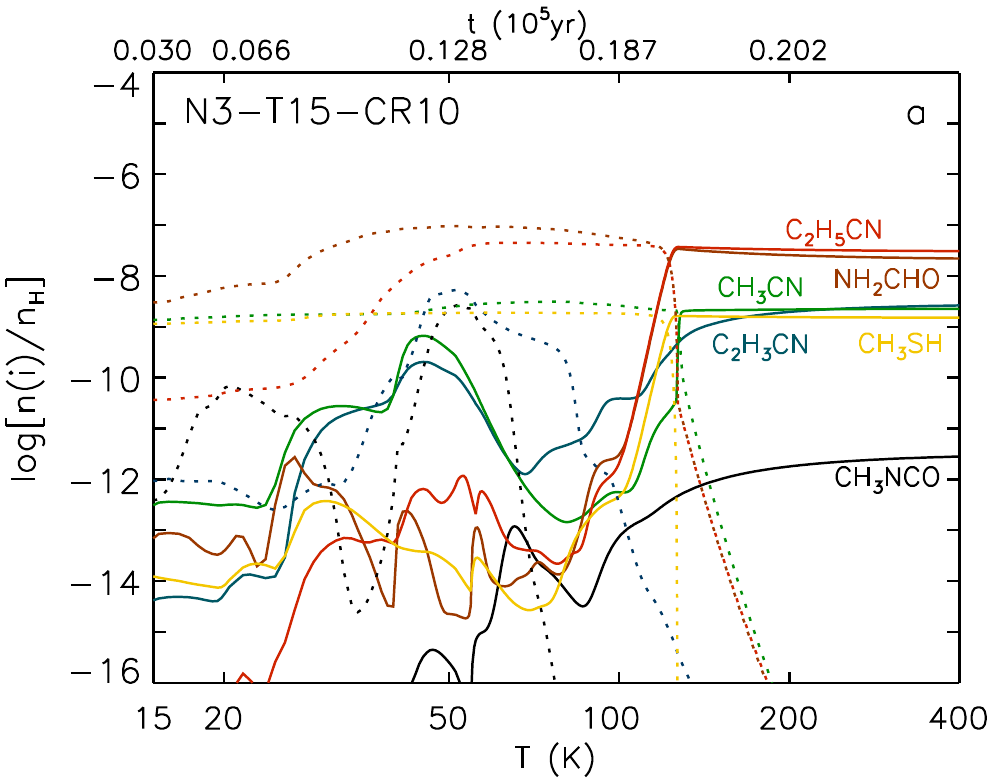} \\
   \includegraphics[scale=0.85]{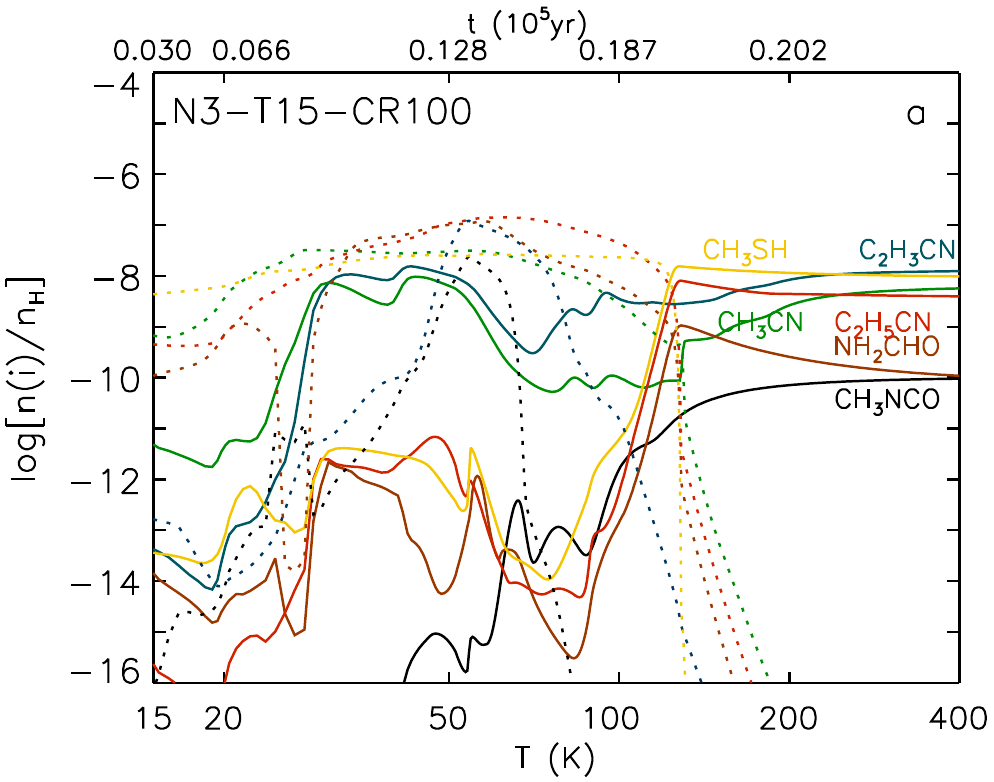} \\
   \includegraphics[scale=0.85]{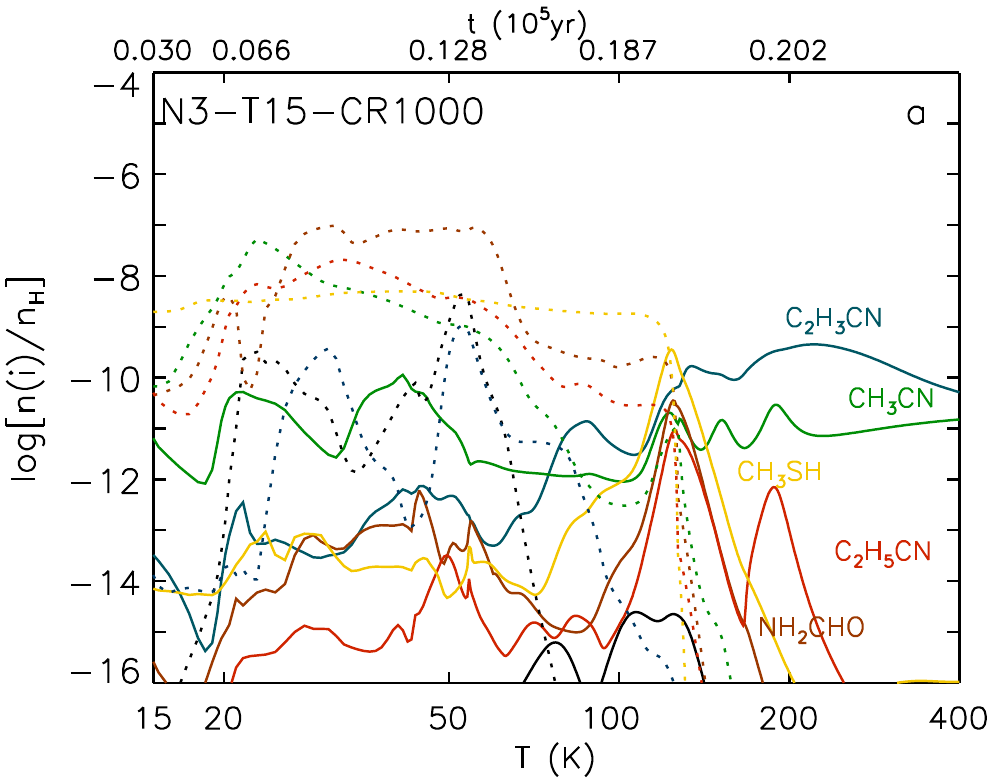} \\
 \end{minipage}
\hspace{-0.02\linewidth}
 \begin{minipage}{0.35\linewidth}
    \includegraphics[scale=0.85]{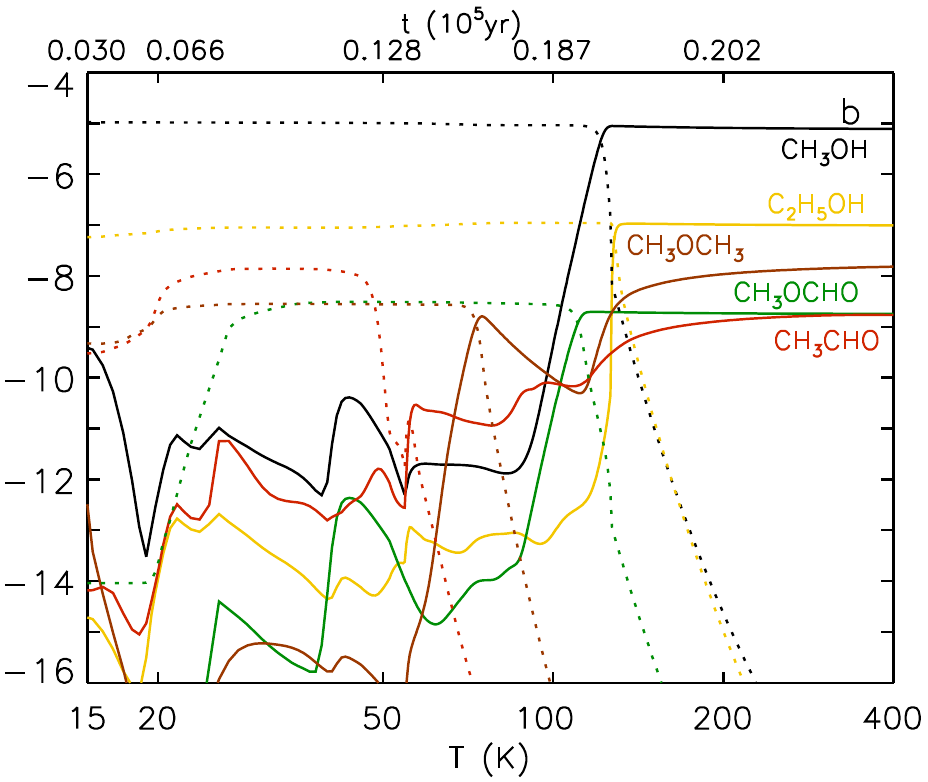} \\
    \includegraphics[scale=0.85]{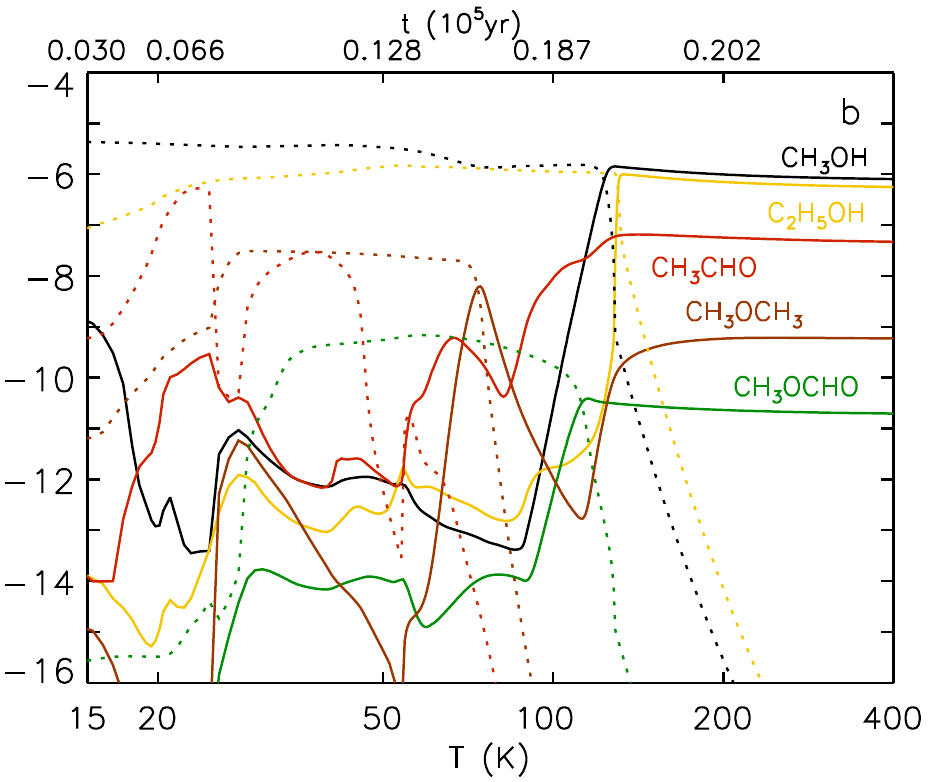} \\
    \includegraphics[scale=0.85]{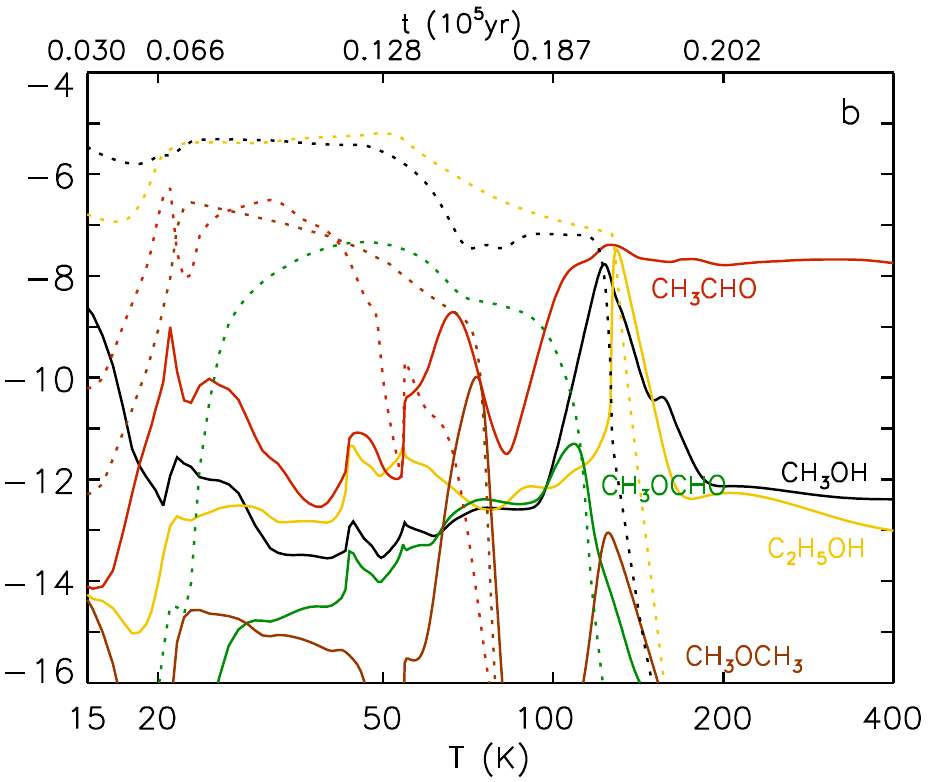} \\
 \end{minipage}
 \vspace{-5mm}
\caption{\label{FIG-appendix-collapse-CR-N3} Same as \ref{FIG-appendix-collapse-Tmin-N2} but for models N3-T15-CR10 (top panels), N3-T15-CR100 (middle panels), and N3-T15-CR1000 (bottom panels).}
\end{figure*}

\begin{figure*}
 \hspace{0.03\linewidth}
 \begin{minipage}{0.5\linewidth}
   \includegraphics[scale=0.85]{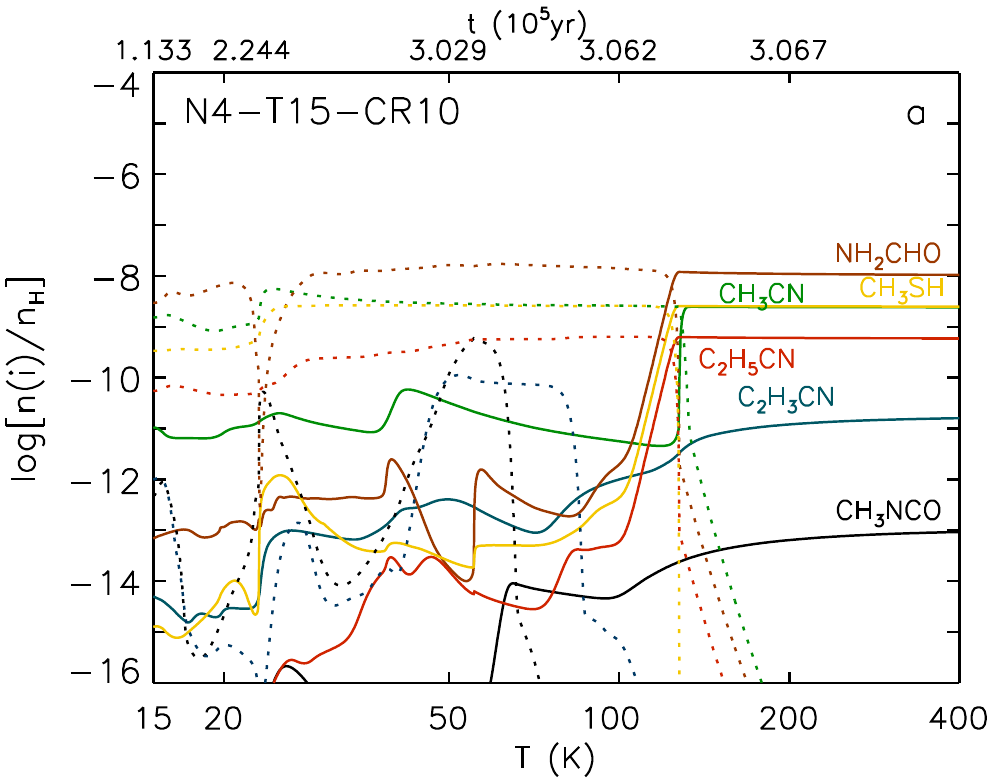} \\
   \includegraphics[scale=0.85]{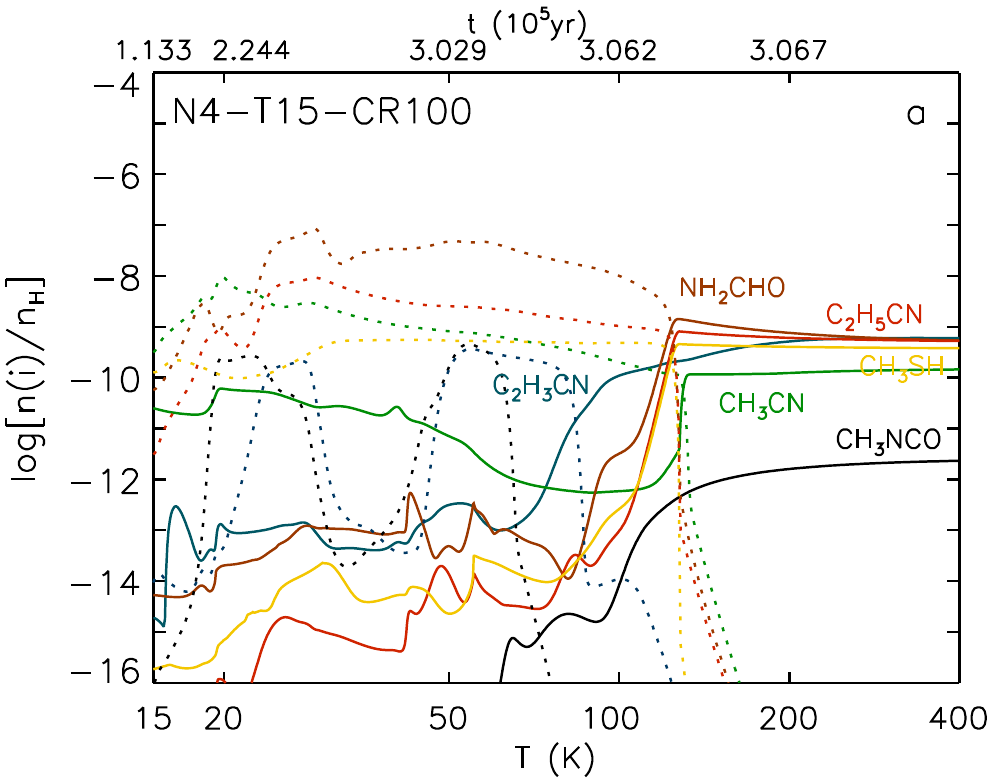} \\
   \includegraphics[scale=0.85]{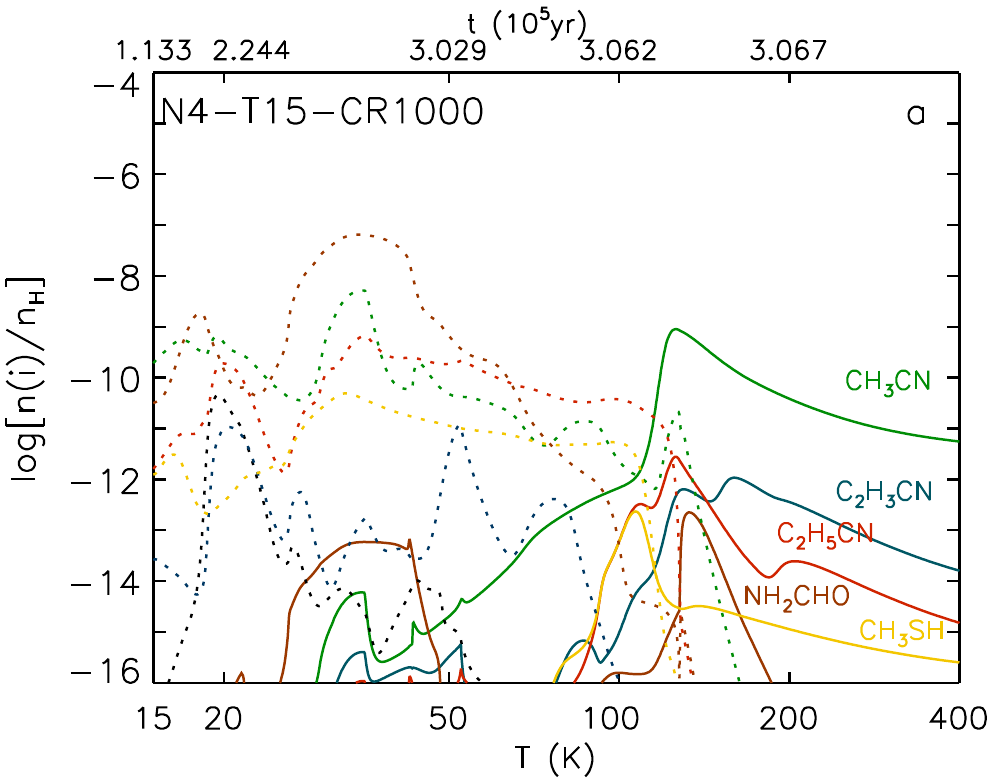} \\
 \end{minipage}
\hspace{-0.02\linewidth}
 \begin{minipage}{0.35\linewidth}
    \includegraphics[scale=0.85]{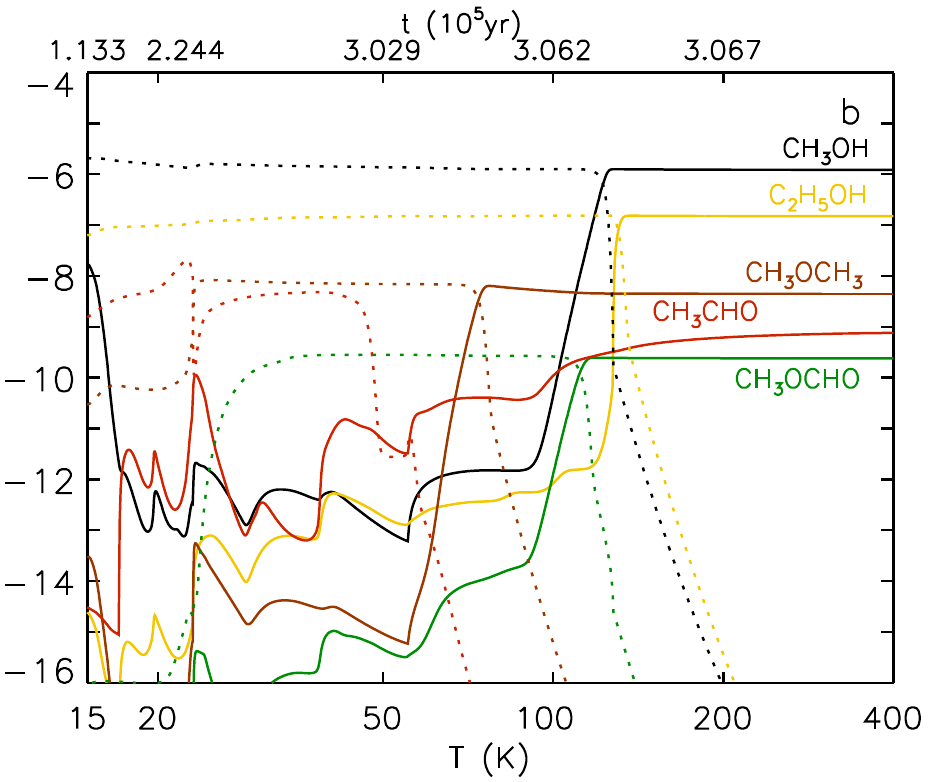} \\
    \includegraphics[scale=0.85]{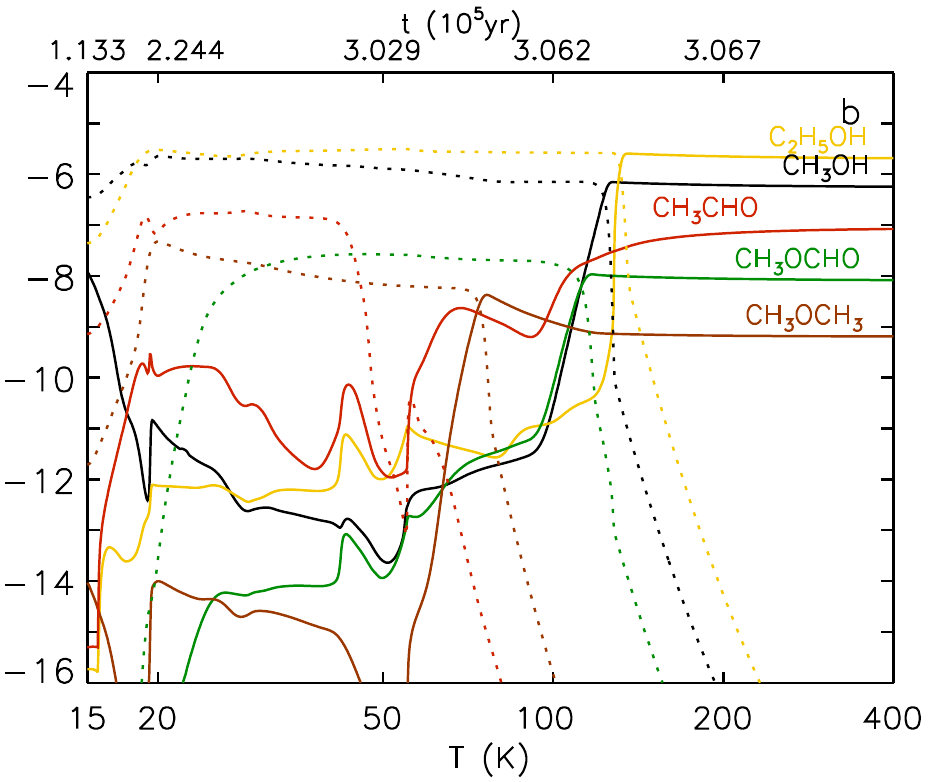} \\
    \includegraphics[scale=0.85]{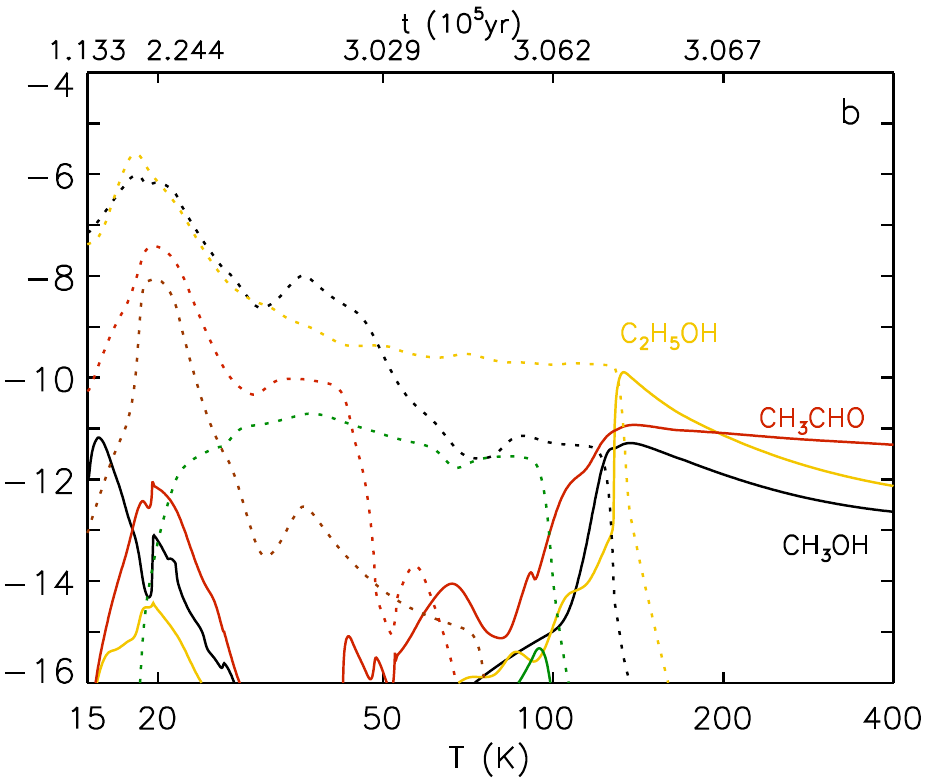} \\
 \end{minipage}
 \vspace{-5mm}
\caption{\label{FIG-appendix-collapse-CR-N4} Same as \ref{FIG-appendix-collapse-Tmin-N2} but for models N4-T15-CR10 (top panels), N4-T15-CR100 (middle panels), and N4-T15-CR1000 (bottom panels).}
\end{figure*}

\begin{figure*}
 \hspace{0.03\linewidth}
 \begin{minipage}{0.5\linewidth}
   \includegraphics[scale=0.85]{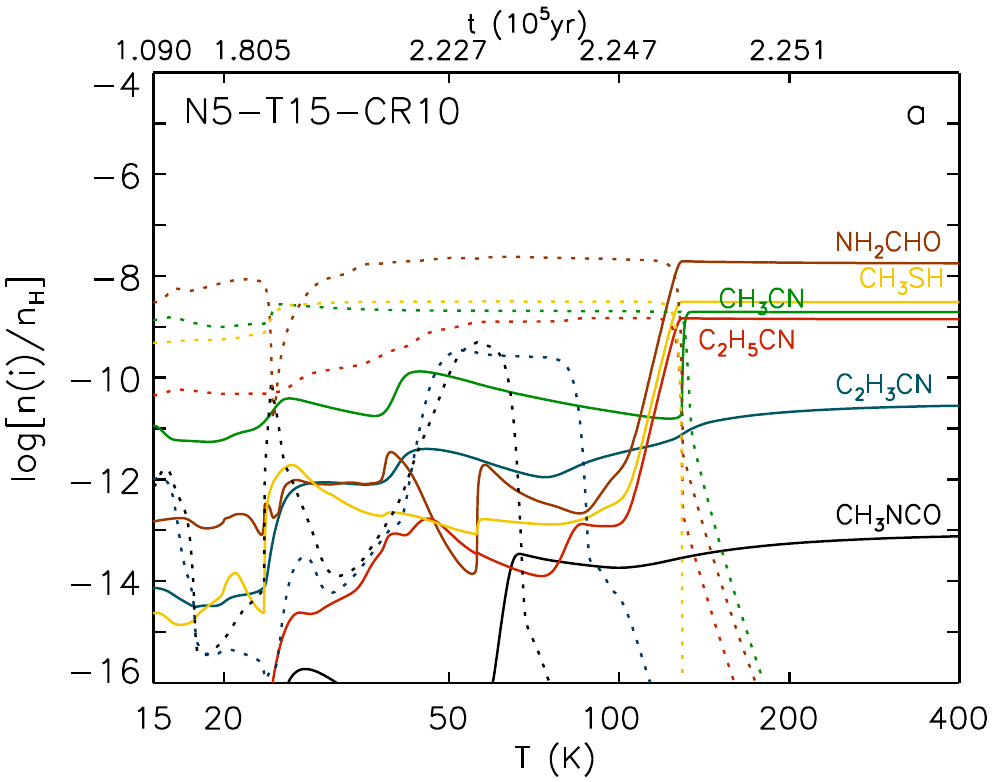} \\
   \includegraphics[scale=0.85]{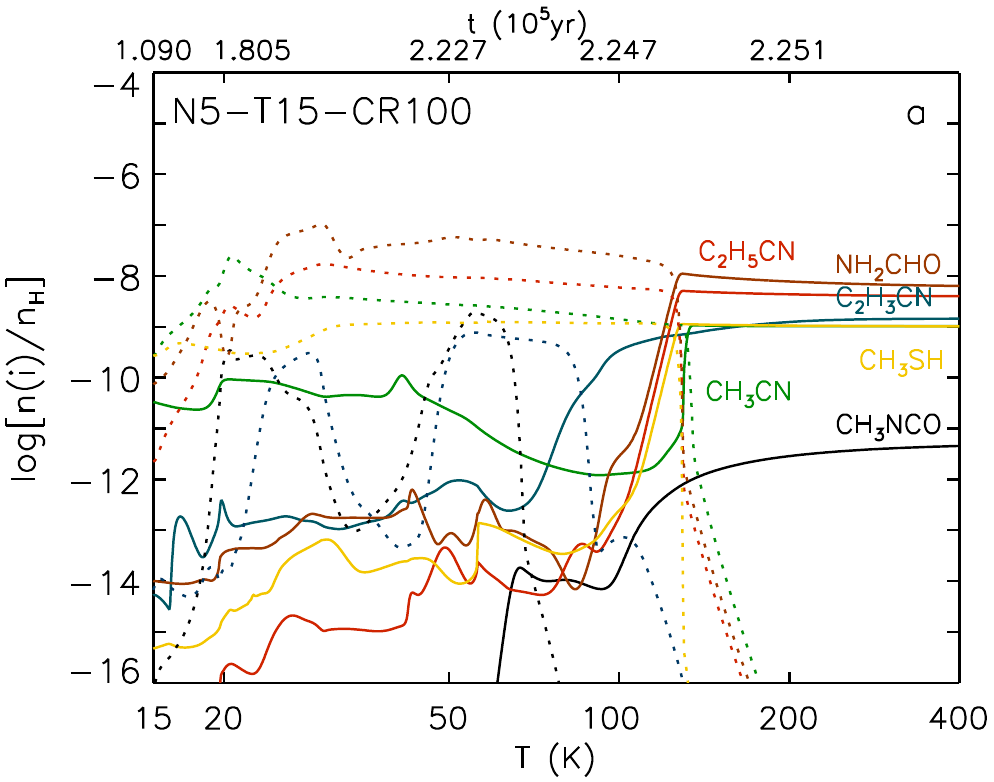} \\
   \includegraphics[scale=0.85]{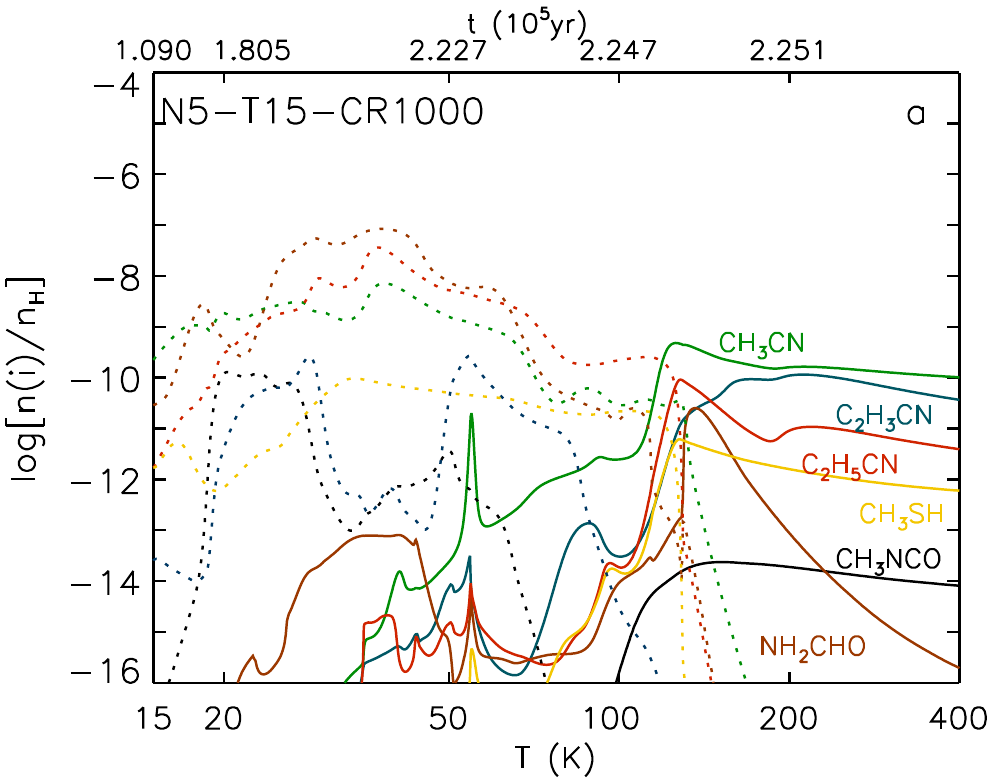} \\
 \end{minipage}
\hspace{-0.02\linewidth}
 \begin{minipage}{0.35\linewidth}
    \includegraphics[scale=0.85]{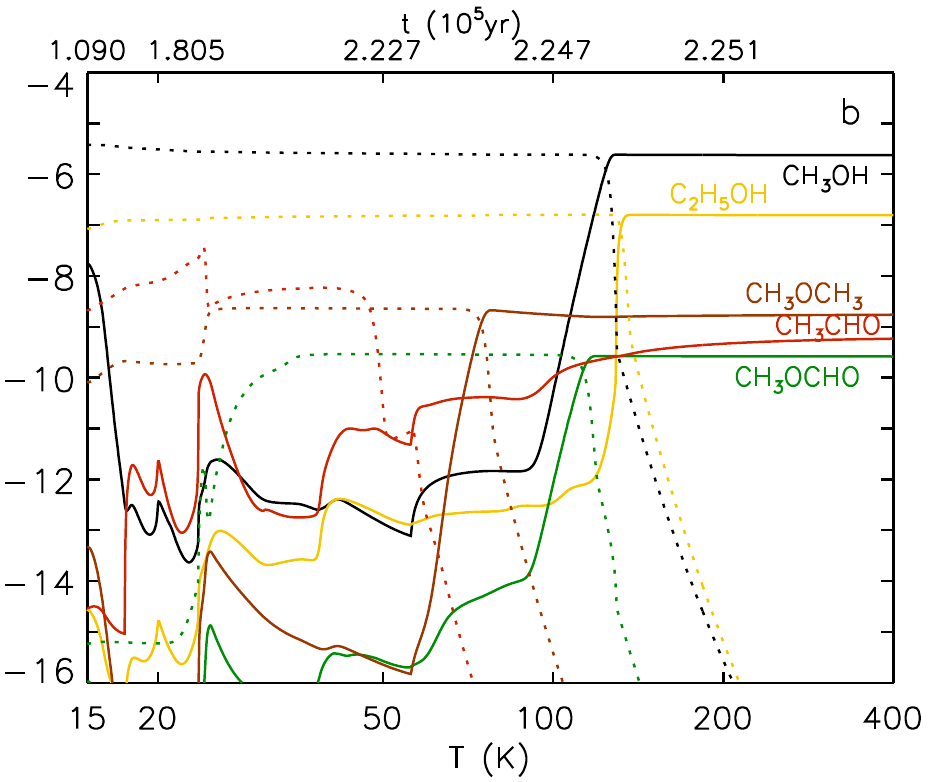} \\
    \includegraphics[scale=0.85]{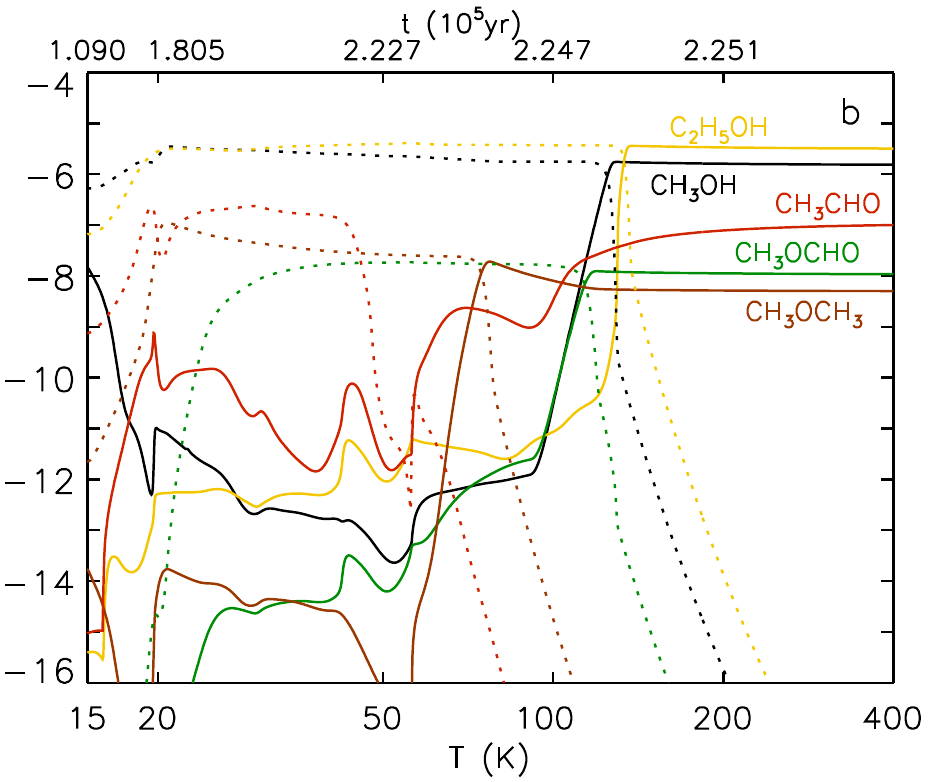} \\
    \includegraphics[scale=0.85]{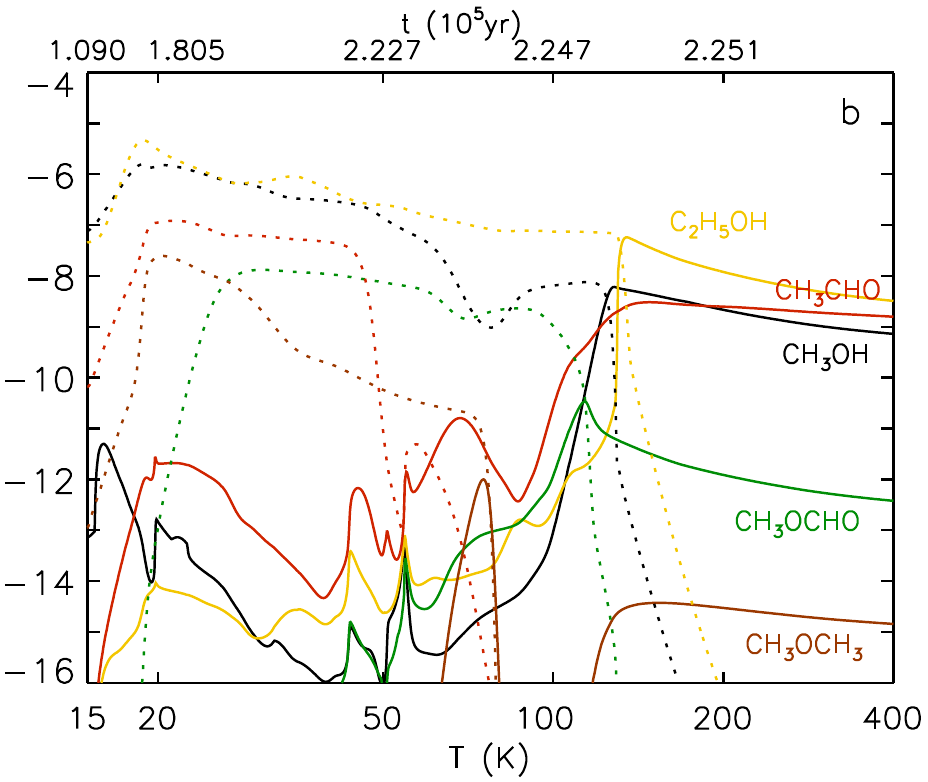} \\
 \end{minipage}
 \vspace{-5mm}
\caption{\label{FIG-appendix-collapse-CR-N5} Same as \ref{FIG-appendix-collapse-Tmin-N2} but for models N5-T15-CR10 (top panels), N5-T15-CR100 (middle panels), and N5-T15-CR1000 (bottom panels).}
\end{figure*}

Figure.~\ref{FIG-appendix-growth-ice-mantle} shows the growth of the dust-grain ice mantles calculated by the chemical models N(2-5)-T15-CR1 during the quasi-static contraction phase prior to the free-fall collapse. Figures~\ref{FIG-appendix-collapse-Tmin-N2}--\ref{FIG-appendix-collapse-Tmin-N5} and \ref{FIG-appendix-collapse-CR}--\ref{FIG-appendix-collapse-CR-N5} show the impact on the chemical model results for Sgr~B2(N2-N5) of varying the minimum dust temperature and the CRIR, respectively. Tables~\ref{TAB-appendix-model-results}--\ref{TAB-appendix-model-results2} and \ref{TAB-appendix-model-results-150K}--\ref{TAB-appendix-model-results-150K2} give the results of all the chemical models obtained at the end of the simulations ($T_{\rm max}$~=~400~K) and at $T_0$~=~150~K, respectively.



\begin{landscape}

\begin{table}
\begin{center}
  \caption{\label{TAB-appendix-model-results} Final gas-phas fractional abundances calculated by the chemical models.} 
    \vspace{-2mm}
  \setlength{\tabcolsep}{1.2mm}
  \begin{tabular}{llllllllllll}
 \hline
 Model  & \ce{C2H5CN} & \ce{C2H3CN} & \ce{CH3CN} & \ce{CH3OH} & \ce{C2H5OH} & \ce{CH3OCHO} & \ce{CH3OCH3} & \ce{CH3CHO} & \ce{NH2CHO} & \ce{CH3NCO} & \ce{CH3SH} \\ 
 \hline
 \hline
N2-T10-1 & 8.78(-10) & 4.02(-11) & 9.72(-10) & 9.81(-6) & 7.48(-8) & 1.30(-9) & 2.14(-8) & 7.74(-9) & 3.54(-8) & 9.99(-12) & 1.76(-9) \\
N2-T10-10 & 2.02(-9) & 9.25(-11) & 2.52(-9) & 7.07(-6) & 2.33(-7) & 1.81(-9) & 1.94(-9) & 1.84(-8) & 7.39(-8) & 1.32(-13) & 4.04(-9) \\
N2-T10-50 & 8.87(-9) & 1.45(-9) & 2.03(-8) & 2.18(-6) & 2.55(-6) & 6.09(-10) & 8.11(-8) & 4.70(-8) & 1.48(-8) & 3.13(-12) & 8.66(-9) \\
N2-T10-100 & 6.50(-9) & 2.04(-9) & 4.44(-9) & 3.17(-6) & 3.98(-6) & 5.53(-9) & 6.27(-8) & 9.95(-8) & 1.79(-8) & 6.49(-12) & 2.38(-9) \\
N2-T10-500 & 1.13(-9) & 4.25(-9) & 8.14(-10) & 1.18(-6) & 3.33(-6) & 1.07(-8) & 5.76(-10) & 3.21(-7) & 4.51(-10) & 1.88(-11) & 6.89(-10) \\
N2-T10-1000 & 5.57(-11) & 1.71(-9) & 3.67(-10) & 1.73(-7) & 7.74(-7) & 1.04(-9) & 1.51(-11) & 1.54(-7) & 1.25(-11) & 7.20(-12) & 3.08(-10) \\
            &            &           &            &           &            &            &            &            &            &             &           \\
N2-T15-1 & 8.16(-10) & 9.00(-12) & 1.47(-9) & 3.38(-6) & 3.18(-8) & 6.53(-10) & 1.47(-9) & 7.09(-9) & 3.84(-8) & 3.49(-15) & 3.08(-9) \\
N2-T15-10 & 1.46(-9) & 2.56(-11) & 1.24(-9) & 3.30(-6) & 1.52(-7) & 5.60(-10) & 9.62(-10) & 4.99(-10) & 4.08(-8) & 6.69(-14) & 2.96(-9) \\
N2-T15-50 & 1.63(-8) & 2.00(-9) & 3.31(-8) & 1.64(-6) & 1.85(-6) & 3.47(-10) & 5.33(-8) & 3.28(-8) & 1.17(-8) & 1.11(-12) & 1.05(-8) \\
N2-T15-100 & 6.91(-9) & 1.75(-9) & 6.89(-9) & 2.43(-6) & 3.36(-6) & 3.52(-9) & 5.78(-8) & 8.09(-8) & 1.27(-8) & 4.03(-12) & 3.24(-9) \\
N2-T15-500 & 9.44(-10) & 3.59(-9) & 6.01(-10) & 1.05(-6) & 3.19(-6) & 1.03(-8) & 4.90(-10) & 2.99(-7) & 1.76(-10) & 1.58(-11) & 8.13(-10 )\\
N2-T15-1000 & 4.01(-11) & 8.31(-10) & 2.78(-10) & 1.10(-7) & 5.76(-7) & 6.09(-10) & 5.53(-12) & 1.07(-7) & 6.64(-12) & 4.73(-12) & 5.81(-11) \\
            &            &           &            &           &            &            &            &            &            &             &           \\
N2-T20-1 & 1.73(-10) & 2.05(-13) & 9.90(-11) & 1.43(-7) & 5.97(-9) & 2.52(-11) & 2.88(-10) & 7.33(-12) & 6.95(-9) & 1.85(-16) & 1.89(-9) \\
N2-T20-10 & 2.63(-11) & 2.73(-13) & 1.05(-10) & 3.13(-7) & 2.42(-9) & 1.22(-10) & 2.14(-10) & 8.20(-12) & 1.50(-8) & 2.26(-15) & 3.29(-10) \\
N2-T20-50 & 6.60(-9) & 3.01(-10) & 2.69(-9) & 1.15(-6) & 2.37(-6) & 1.68(-9) & 1.01(-8) & 2.80(-8) & 1.31(-8) & 7.69(-13) & 9.95(-10) \\
N2-T20-100 & 8.19(-9) & 6.25(-10) & 3.21(-9) & 9.18(-7) & 1.83(-6) & 2.14(-9) & 4.17(-9) & 3.97(-8) & 8.95(-10) & 1.31(-12) & 6.35(-10) \\
N2-T20-500 & 2.75(-10) & 6.24(-10) & 3.53(-10) & 9.90(-8) & 5.38(-7) & 1.08(-10) & 8.58(-12) & 4.79(-8) & 1.82(-11) & 1.80(-12) & 1.80(-11) \\
N2-T20-1000 & 1.58(-11) & 1.78(-10) & 1.21(-10) & 1.73(-8) & 1.09(-7) & 4.89(-13) & 1.60(-13) & 1.85(-8) & 1.19(-12) & 4.93(-13) & 2.09(-12) \\
\hline
N3-T10-CR1 & 3.86(-9) & 7.10(-10) & 2.72(-9) & 8.61(-6) & 4.81(-8) & 7.58(-8) & 8.83(-9) & 9.29(-11) & 9.88(-8) & 6.69(-10) & 1.68(-9) \\
N3-T10-CR10 & 3.43(-8) & 3.63(-9) & 1.87(-9) & 7.83(-6) & 1.49(-7) & 2.50(-9) & 1.83(-8) & 8.95(-10) & 6.18(-9) & 4.59(-12) & 1.06(-9) \\
N3-T10-CR50 & 2.65(-8) & 1.38(-8) & 4.36(-9) & 3.56(-6) & 3.48(-7) & 8.65(-10) & 7.86(-9) & 6.22(-9) & 2.29(-10) & 5.77(-11) & 3.42(-9) \\
N3-T10-CR100 & 3.27(-9) & 1.07(-8) & 4.87(-9) & 1.19(-6) & 5.73(-7) & 3.55(-11) & 1.43(-9) & 5.04(-8) & 6.91(-11) & 1.14(-10) & 7.83(-9) \\
N3-T10-CR500 & 2.45(-12) & 2.07(-9) & 2.65(-10) & 2.66(-9) & 4.43(-9) & 1.79(-14) & 1.62(-12) & 8.20(-8) & 1.14(-17) & 1.19(-14) & 6.91(-11) \\
N3-T10-CR1000 & 1.36(-17) & 1.48(-10) & 2.57(-11) & 5.89(-13) & 1.63(-13) & 1.49(-18) & 5.28(-20) & 2.87(-8) & 6.93(-19) & 7.87(-24) & 2.32(-16) \\
            &            &           &            &           &            &            &            &            &            &             &           \\
N3-T15-CR1 & 4.70(-9) & 9.80(-10) & 2.36(-9) & 7.10(-6) & 2.38(-8) & 8.32(-10) & 1.92(-9) & 4.38(-9) & 1.40(-8) & 3.82(-10) & 1.94(-9) \\
N3-T15-CR10 & 3.09(-8) & 2.64(-9) & 2.27(-9) & 7.81(-6) & 9.91(-8) & 1.80(-9) & 1.53(-8) & 1.71(-9) & 2.21(-8) & 2.82(-12) & 1.51(-9) \\
N3-T15-CR50 & 2.70(-8) & 1.79(-8) & 4.64(-9) & 2.75(-6) & 4.77(-7) & 1.42(-10) & 3.62(-9) & 1.34(-8) & 1.84(-10) & 7.42(-11) & 8.31(-9) \\
N3-T15-CR100 & 3.98(-9) & 1.25(-8) & 5.72(-9) & 8.05(-7) & 5.62(-7) & 1.97(-11) & 5.92(-10) & 4.71(-8) & 1.09(-10) & 9.60(-11) & 9.83(-9) \\
N3-T15-CR500 & 1.50(-12) & 1.37(-9) & 1.94(-10) & 1.30(-9) & 2.37(-9) & 6.38(-15) & 7.67(-13) & 8.48(-8) & 1.67(-18) & 5.25(-15) & 4.51(-11) \\
N3-T15-CR1000 & 6.61(-19) & 5.12(-11) & 1.51(-11) & 4.06(-13) & 9.75(-14) & 1.18(-18) & 1.38(-20) & 1.76(-8) & 1.49(-19) & 1.13(-25) & 1.05(-16) \\
            &            &           &            &           &            &            &            &            &            &             &           \\
N3-T20-CR1 & 7.22(-10) & 5.29(-12) & 2.25(-10) & 1.11(-7) & 1.49(-8) & 2.48(-11) & 1.92(-10) & 1.60(-10) & 5.26(-9) & 1.07(-14) & 3.85(-9) \\
N3-T20-CR10 & 5.36(-10) & 5.20(-11) & 1.07(-10) & 2.07(-8) & 5.98(-10) & 6.00(-12) & 3.76(-13) & 2.44(-11) & 2.77(-10) & 1.06(-13) & 7.92(-10) \\
N3-T20-CR50 & 2.12(-9) & 9.24(-10) & 7.56(-10) & 1.06(-6) & 1.02(-6) & 1.34(-10) & 5.01(-10) & 1.17(-7) & 9.04(-11) & 3.82(-11) & 6.73(-9) \\
N3-T20-CR100 & 5.16(-10) & 1.44(-9) & 5.79(-10) & 5.26(-7) & 8.47(-7) & 4.90(-11) & 2.47(-10) & 2.05(-7) & 7.71(-11) & 3.03(-11) & 4.89(-9) \\
N3-T20-CR500 & 4.14(-14) & 1.96(-10) & 9.81(-11) & 1.21(-10) & 2.16(-10) & 2.14(-15) & 4.57(-14) & 3.30(-8) & 5.26(-18) & 4.03(-16) & 4.48(-13) \\
N3-T20-CR1000 & 4.02(-21) & 1.52(-11) & 2.45(-11) & 6.07(-13) & 4.21(-14) & 1.90(-18) & 4.20(-20) & 9.80(-9) & 1.73(-21) & 4.43(-27) & 6.70(-19) \\
\hline
\end{tabular}
\end{center}
\vspace{-6mm}
\tablefoot{Gas-phase fractional abundances (with respect to total hydrogen) reached at the end of the simulations (that is when the temperature reaches 400~K in the envelope of the sources). $X(Y)$ means $X \times 10^Y$.}
\end{table}
\end{landscape}

\begin{landscape}

\begin{table}
\begin{center}
  \caption{\label{TAB-appendix-model-results2} Table~\ref{TAB-appendix-model-results} continued.} 
  \setlength{\tabcolsep}{1.2mm}
  \begin{tabular}{lccccccccccc}
 \hline
 Model  & \ce{C2H5CN} & \ce{C2H3CN} & \ce{CH3CN} & \ce{CH3OH} & \ce{C2H5OH} & \ce{CH3OCHO} & \ce{CH3OCH3} & \ce{CH3CHO} & \ce{NH2CHO} & \ce{CH3NCO} & \ce{CH3SH} \\ 
 \hline
 \hline
N4-T10-CR1 & 9.43(-10) & 3.59(-12) & 1.11(-9) & 6.76(-6) & 6.49(-8) & 7.19(-10) & 1.38(-8) & 1.26(-10) & 4.34(-8) & 2.21(-13) & 2.59(-9) \\
N4-T10-CR10 & 1.20(-9) & 3.58(-11) & 5.42(-9) & 2.63(-6) & 2.26(-7) & 9.17(-10) & 1.44(-8) & 1.32(-9) & 1.84(-8) & 2.84(-13) & 3.31(-9) \\
N4-T10-CR50 & 2.54(-9) & 6.05(-10) & 7.44(-10) & 1.39(-6) & 2.73(-6) & 1.06(-8) & 6.97(-9) & 6.41(-8) & 8.42(-9) & 3.20(-12) & 6.30(-10) \\
N4-T10-CR100 & 3.36(-10) & 5.23(-10) & 1.10(-10) & 5.05(-7) & 1.92(-6) & 7.37(-9) & 5.70(-10) & 7.85(-8) & 6.14(-10) & 2.99(-12) & 4.31(-10) \\
N4-T10-CR500 & 1.55(-11) & 1.10(-10) & 4.84(-10) & 1.86(-9) & 1.09(-8) & 4.95(-12) & 4.23(-15) & 2.60(-9) & 1.19(-14) & 2.91(-14) & 2.77(-12) \\
N4-T10-CR1000 & 2.56(-15) & 2.26(-14) & 6.59(-12) & 2.91(-13) & 9.71(-13) & 1.12(-18) & 1.25(-19) & 5.53(-12) & 4.97(-22) & 1.18(-18) & 3.09(-16) \\
            &            &           &            &           &            &            &            &            &            &             &           \\
N4-T15-CR1 & 1.19(-9) & 6.01(-12) & 2.41(-9) & 1.33(-6) & 2.90(-8) & 1.30(-11) & 2.82(-9) & 4.76(-11) & 1.86(-9) & 3.12(-15) & 3.85(-9) \\
N4-T15-CR10 & 5.95(-10) & 1.60(-11) & 2.42(-9) & 1.22(-6) & 1.50(-7) & 2.39(-10) & 4.47(-9) & 7.63(-10) & 1.05(-8) & 9.25(-14) & 2.51(-9) \\
N4-T15-CR50 & 4.00(-9) & 7.47(-10) & 1.12(-9) & 1.44(-6) & 2.84(-6) & 1.08(-8) & 7.58(-9) & 6.56(-8) & 7.26(-9) & 2.17(-12) & 5.39(-10) \\
N4-T15-CR100 & 5.28(-10) & 6.02(-10) & 1.45(-10) & 5.70(-7) & 2.07(-6) & 8.30(-9) & 6.46(-10) & 8.42(-8) & 5.39(-10) & 2.33(-12) & 3.81(-10) \\
N4-T15-CR500 & 1.36(-11) & 9.76(-11) & 6.01(-10) & 1.89(-9) & 1.10(-8) & 2.20(-12) & 4.51(-15) & 2.40(-9) & 1.54(-14) & 2.59(-14) & 1.35(-12) \\
N4-T15-CR1000 & 1.50(-15) & 1.59(-14) & 5.53(-12) & 2.30(-13) & 7.38(-13) & 9.82(-19) & 8.18(-20) & 4.78(-12) & 4.65(-22) & 7.55(-19) & 2.51(-16) \\
            &            &           &            &           &            &            &            &            &            &             &           \\
N4-T20-CR1 & 6.38(-11) & 1.55(-13) & 3.49(-11) & 7.52(-8) & 1.49(-9) & 1.16(-12) & 6.07(-11) & 1.82(-12) & 2.18(-9) & 7.18(-17) & 1.30(-9) \\
N4-T20-CR10 & 2.34(-11) & 6.10(-13) & 2.04(-11) & 7.87(-8) & 1.86(-9) & 4.45(-12) & 3.58(-11) & 9.40(-12) & 2.62(-9) & 1.64(-15) & 2.32(-10) \\
N4-T20-CR50 & 1.89(-9) & 2.83(-10) & 5.96(-10) & 6.31(-7) & 1.53(-6) & 4.48(-9) & 1.81(-9) & 3.50(-8) & 1.02(-9) & 8.80(-13) & 4.48(-10) \\
N4-T20-CR100 & 2.48(-9) & 5.90(-10) & 5.07(-10) & 5.20(-7) & 1.65(-6) & 6.68(-9) & 7.56(-10) & 7.03(-8) & 2.72(-10) & 1.86(-12) & 6.45(-10) \\
N4-T20-CR500 & 4.89(-12) & 3.40(-11) & 4.76(-10) & 5.29(-10) & 3.14(-9) & 6.72(-15) & 5.97(-16) & 5.44(-10) & 5.45(-15) & 3.67(-15) & 1.67(-13) \\
N4-T20-CR1000 & 4.15(-16) & 6.84(-15) & 2.31(-12) & 1.58(-13) & 5.61(-13) & 1.25(-18) & 3.01(-20) & 1.32(-12) & 3.65(-22) & 1.80(-19) & 2.36(-16) \\
\hline
N5-T10-CR1 & 1.22(-9) & 8.26(-12) & 1.21(-9) & 8.85(-6) & 5.38(-8) & 5.05(-10) & 1.78(-8) & 1.23(-10) & 3.34(-8) & 5.44(-12) & 2.64(-9) \\
N5-T10-CR10 & 2.20(-9) & 4.76(-11) & 2.01(-9) & 5.33(-6) & 1.18(-7) & 1.58(-9) & 2.96(-9) & 8.35(-10) & 3.62(-8) & 1.52(-13) & 3.31(-9) \\
N5-T10-CR50 & 5.42(-9) & 1.00(-9) & 3.23(-9) & 2.82(-6) & 3.54(-6) & 6.97(-9) & 5.75(-8) & 6.91(-8) & 1.81(-8) & 4.82(-12) & 1.80(-9) \\
N5-T10-CR100 & 3.07(-9) & 1.46(-9) & 8.45(-10) & 1.66(-6) & 3.43(-6) & 1.19(-8) & 5.22(-9) & 1.09(-7) & 7.49(-9) & 6.99(-12) & 1.04(-9) \\
N5-T10-CR500 & 6.26(-11) & 8.44(-10) & 2.51(-10) & 1.25(-7) & 6.95(-7) & 1.07(-9) & 6.36(-12) & 1.03(-7) & 2.48(-11) & 4.31(-12) & 2.13(-10) \\
N5-T10-CR1000 & 5.03(-12) & 4.89(-11) & 1.18(-10) & 1.01(-9) & 4.44(-9) & 1.81(-12) & 2.69(-15) & 2.39(-9) & 2.79(-16) & 1.42(-14) & 1.45(-12) \\
            &            &           &            &           &            &            &            &            &            &             &           \\
N5-T15-CR1 & 8.53(-10) & 9.57(-12) & 1.73(-9) & 2.62(-6) & 2.87(-8) & 3.29(-10) & 1.86(-9) & 1.40(-8) & 2.74(-8) & 1.17(-15) & 4.90(-9) \\
N5-T15-CR10 & 1.41(-9) & 2.82(-11) & 1.94(-9) & 2.38(-6) & 1.57(-7) & 2.61(-10) & 1.73(-9) & 5.84(-10) & 1.78(-8) & 7.61(-14) & 3.05(-9) \\
N5-T15-CR50 & 8.37(-9) & 1.18(-9) & 5.58(-9) & 2.56(-6) & 3.41(-6) & 5.88(-9) & 5.53(-8) & 6.47(-8) & 1.56(-8) & 3.70(-12) & 1.71(-9) \\
N5-T15-CR100 & 4.00(-9) & 1.45(-9) & 1.03(-9) & 1.54(-6) & 3.20(-6) & 1.08(-8) & 5.06(-9) & 1.01(-7) & 6.37(-9) & 4.55(-12) & 1.02(-9) \\
N5-T15-CR500 & 6.29(-11) & 5.69(-10) & 2.39(-10) & 1.04(-7) & 6.12(-7) & 8.52(-10) & 3.81(-12) & 8.60(-8) & 1.85(-11) & 3.53(-12) & 9.43(-11) \\
N5-T15-CR1000 & 3.92(-12) & 3.66(-11) & 1.02(-10) & 7.26(-10) & 3.23(-9) & 3.78(-13) & 1.44(-15) & 1.58(-9) & 1.93(-16) & 8.11(-15) & 5.99(-13) \\
            &            &           &            &           &            &            &            &            &            &             &           \\
N5-T20-CR1 & 1.18(-10) & 1.88(-13) & 6.43(-11) & 8.31(-8) & 3.14(-9) & 5.86(-12) & 1.46(-10) & 4.12(-12) & 4.03(-9) & 6.30(-17) & 1.54(-9) \\
N5-T20-CR10 & 3.29(-11) & 5.47(-13) & 5.60(-11) & 1.78(-7) & 1.90(-9) & 1.48(-11) & 1.01(-10) & 7.78(-12) & 5.29(-9) & 1.52(-15) & 3.18(-10) \\
N5-T20-CR50 & 4.82(-9) & 3.96(-10) & 1.58(-9) & 1.07(-6) & 2.13(-6) & 3.03(-9) & 5.15(-9) & 3.71(-8) & 3.80(-9) & 1.01(-12) & 8.54(-10) \\
N5-T20-CR100 & 3.26(-9) & 5.71(-10) & 9.45(-10) & 6.67(-7) & 1.57(-6) & 4.10(-9) & 1.35(-9) & 4.90(-8) & 2.76(-10) & 1.08(-12) & 5.16(-10) \\
N5-T20-CR500 & 2.01(-11) & 1.57(-10) & 1.36(-10) & 1.88(-8) & 1.31(-7) & 2.12(-12) & 1.37(-13) & 1.64(-8) & 3.45(-12) & 3.66(-13) & 3.87(-12) \\
N5-T20-CR1000 & 1.99(-12) & 1.87(-11) & 7.24(-11) & 4.31(-10) & 1.95(-9) & 7.55(-15) & 5.14(-16) & 7.73(-10) & 2.02(-16) & 3.50(-15) & 1.27(-13) \\
\hline
\end{tabular}
\end{center}
\end{table}

\end{landscape}

\begin{landscape}

\begin{table}
\begin{center}
  \caption{\label{TAB-appendix-model-results-150K} Gas-phase fractional abundances at $T$~=~150~K.} 
    \vspace{-2mm}
  \setlength{\tabcolsep}{1.2mm}
  \begin{tabular}{llllllllllll}
 \hline
 Model  & \ce{C2H5CN} & \ce{C2H3CN} & \ce{CH3CN} & \ce{CH3OH} & \ce{C2H5OH} & \ce{CH3OCHO} & \ce{CH3OCH3} & \ce{CH3CHO} & \ce{NH2CHO} & \ce{CH3NCO} & \ce{CH3SH} \\ 
 \hline
 \hline
N2-T10-CR1 & 8.80(-10) & 3.98(-11) & 9.67(-10) & 9.82(-6) & 7.49(-8) & 1.30(-9) & 2.13(-8) & 7.74(-9) & 3.56(-8) & 1.00(-11) & 1.76(-9) \\
N2-T10-CR10 & 2.06(-9) & 7.90(-11) & 2.53(-9) & 7.16(-6) & 2.35(-7) & 1.83(-9) & 1.14(-9) & 1.83(-8) & 7.73(-8) & 8.24(-14) & 4.07(-9) \\
N2-T10-CR50 & 9.41(-9) & 1.20(-9) & 2.10(-8) & 2.25(-6) & 2.64(-6) & 6.25(-10) & 8.29(-8) & 2.59(-8) & 1.69(-8) & 1.53(-12) & 8.89(-9) \\
N2-T10-CR100 & 7.27(-9) & 1.55(-9) & 4.68(-9) & 3.37(-6) & 4.26(-6) & 5.81(-9) & 6.52(-8) & 5.26(-8) & 2.31(-8) & 2.17(-12) & 2.50(-9) \\
N2-T10-CR500 & 1.98(-9) & 1.80(-9) & 3.37(-10) & 1.63(-6) & 4.86(-6) & 1.42(-8) & 5.50(-10) & 1.80(-7) & 1.96(-9) & 6.77(-12) & 8.94(-10) \\
N2-T10-CR1000 & 1.30(-10) & 5.72(-10) & 5.34(-11) & 3.64(-7) & 1.96(-6) & 2.16(-9) & 1.30(-11) & 1.30(-7) & 5.71(-10) & 4.39(-12) & 6.06(-10) \\
             &            &           &            &           &            &            &            &            &            &             &           \\
N2-T15-CR1 & 8.17(-10) & 8.57(-12) & 1.46(-9) & 3.38(-6) & 3.19(-8) & 6.53(-10) & 1.43(-9) & 7.09(-9) & 3.85(-8) & 3.38(-15) & 3.08(-9) \\
N2-T15-CR10 & 1.48(-9) & 1.64(-11) & 1.25(-9) & 3.34(-6) & 1.54(-7) & 5.67(-10) & 7.88(-10) & 3.49(-10) & 4.26(-8) & 5.56(-14) & 2.98(-9) \\
N2-T15-CR50 & 1.72(-8) & 1.61(-9) & 3.42(-8) & 1.70(-6) & 1.91(-6) & 3.55(-10) & 5.44(-8) & 1.86(-8) & 1.33(-8) & 3.83(-13) & 1.07(-8) \\
N2-T15-CR100 & 7.68(-9) & 1.35(-9) & 7.30(-9) & 2.58(-6) & 3.55(-6) & 3.68(-9) & 6.01(-8) & 4.27(-8) & 1.62(-8) & 1.24(-12) & 3.39(-9) \\
N2-T15-CR500 & 1.67(-9) & 1.52(-9) & 2.52(-10) & 1.45(-6) & 4.63(-6) & 1.35(-8) & 4.78(-10) & 1.67(-7) & 7.42(-10) & 5.44(-12) & 1.06(-9) \\
N2-T15-CR1000 & 1.09(-10) & 2.50(-10) & 6.20(-11) & 2.31(-7) & 1.53(-6) & 1.33(-9) & 3.75(-12) & 9.09(-8) & 3.86(-10) & 2.95(-12) & 1.18(-10) \\
            &            &           &            &           &            &            &            &            &            &             &           \\
N2-T20-CR1 & 1.74(-10) & 1.19(-13) & 9.91(-11) & 1.43(-7) & 5.98(-9) & 2.53(-11) & 2.88(-10) & 6.03(-12) & 6.97(-9) & 1.46(-16) & 1.89(-9) \\
N2-T20-CR10 & 2.68(-11) & 1.54(-13) & 1.06(-10) & 3.16(-7) & 2.44(-9) & 1.23(-10) & 2.14(-10) & 5.63(-12) & 1.56(-8) & 6.89(-16) & 3.31(-10) \\
N2-T20-CR50 & 6.92(-9) & 1.88(-10) & 2.76(-9) & 1.18(-6) & 2.38(-6) & 1.71(-9) & 1.02(-8) & 1.54(-8) & 1.45(-8) & 2.06(-13) & 1.02(-9) \\
N2-T20-CR100 & 9.08(-9) & 3.28(-10) & 3.40(-9) & 9.71(-7) & 1.95(-6) & 2.24(-9) & 4.33(-9) & 2.05(-8) & 1.13(-9) & 3.35(-13) & 6.63(-10) \\
N2-T20-CR500 & 5.12(-10) & 8.96(-11) & 5.97(-11) & 1.44(-7) & 8.69(-7) & 1.58(-10) & 9.54(-12) & 2.72(-8) & 1.24(-10) & 6.06(-13) & 2.55(-11) \\
N2-T20-CR1000 & 3.27(-11) & 4.60(-11) & 5.94(-11) & 3.73(-8) & 3.50(-7) & 1.29(-12) & 9.37(-14) & 1.76(-8) & 1.62(-10) & 3.16(-13) & 4.76(-12) \\
\hline
N3-T10-CR1 & 3.90(-9) & 7.00(-10) & 2.68(-9) & 8.68(-6) & 4.83(-8) & 7.61(-8) & 7.69(-9) & 8.52(-11) & 1.01(-7) & 6.74(-10) & 1.68(-9) \\
N3-T10-CR10 & 3.97(-8) & 1.69(-9) & 1.51(-9) & 8.70(-6) & 1.61(-7) & 2.67(-9) & 7.47(-9) & 8.48(-10) & 8.62(-9) & 1.66(-12) & 1.13(-9) \\
N3-T10-CR50 & 3.99(-8) & 8.84(-9) & 1.89(-9) & 4.63(-6) & 4.48(-7) & 1.04(-9) & 3.60(-9) & 6.93(-9) & 5.86(-10) & 2.30(-11) & 4.10(-9) \\
N3-T10-CR100 & 5.20(-9) & 2.90(-9) & 5.67(-10) & 1.88(-6) & 9.28(-7) & 5.03(-11) & 9.43(-10) & 6.62(-8) & 4.37(-10) & 4.91(-11) & 1.10(-8) \\
N3-T10-CR500 & 5.59(-12) & 2.52(-9) & 1.65(-10) & 7.04(-8) & 1.68(-7) & 3.47(-13) & 1.77(-11) & 1.35(-7) & 1.29(-11) & 1.75(-13) & 9.67(-10) \\
N3-T10-CR1000 & 1.05(-13) & 2.89(-10) & 2.05(-11) & 7.86(-11) & 1.19(-10) & 1.34(-16) & 1.22(-15) & 3.03(-8) & 1.24(-13) & 3.41(-17) & 2.07(-12) \\
            &            &           &            &           &            &            &            &            &            &             &           \\
N3-T15-CR1 & 4.79(-9) & 9.56(-10) & 2.33(-9) & 7.16(-6) & 2.39(-8) & 8.35(-10) & 1.06(-9) & 4.14(-9) & 1.43(-8) & 3.85(-10) & 1.95(-9) \\
N3-T15-CR10 & 3.52(-8) & 1.22(-9) & 2.14(-9) & 8.57(-6) & 1.06(-7) & 1.90(-9) & 6.45(-9) & 7.91(-10) & 2.99(-8) & 9.81(-13) & 1.60(-9) \\
N3-T15-CR50 & 3.77(-8) & 1.27(-8) & 2.20(-9) & 3.42(-6) & 5.95(-7) & 1.66(-10) & 1.63(-9) & 1.33(-8) & 4.13(-10) & 2.97(-11) & 9.68(-9) \\
N3-T15-CR100 & 6.03(-9) & 3.67(-9) & 6.88(-10) & 1.23(-6) & 8.90(-7) & 2.73(-11) & 3.86(-10) & 6.48(-8) & 6.17(-10) & 4.11(-11) & 1.35(-8) \\
N3-T15-CR500 & 4.40(-12) & 2.33(-9) & 1.34(-10) & 5.19(-8) & 1.40(-7) & 1.94(-13) & 1.20(-11) & 1.27(-7) & 6.63(-12) & 1.27(-13) & 9.25(-10) \\
N3-T15-CR1000 & 5.61(-14) & 1.19(-10) & 1.30(-11) & 3.66(-11) & 3.95(-11) & 4.71(-17) & 2.48(-16) & 2.06(-8) & 5.85(-14) & 6.98(-18) & 7.19(-13) \\
            &            &           &            &           &            &            &            &            &            &             &           \\
N3-T20-CR1 & 7.30(-10) & 2.89(-12) & 2.26(-10) & 1.12(-7) & 1.49(-8) & 2.50(-11) & 1.92(-10) & 1.43(-10) & 5.39(-9) & 3.48(-15) & 3.87(-9) \\
N3-T20-CR10 & 6.19(-10) & 3.15(-11) & 1.12(-10) & 2.22(-8) & 6.45(-10) & 6.38(-12) & 2.93(-13) & 1.39(-11) & 3.82(-10) & 3.15(-14) & 8.39(-10) \\
N3-T20-CR50 & 3.06(-9) & 3.30(-10) & 6.34(-10) & 1.31(-6) & 1.29(-6) & 1.57(-10) & 2.74(-10) & 6.72(-8) & 2.05(-10) & 1.20(-11) & 7.90(-9) \\
N3-T20-CR100 & 1.04(-9) & 4.85(-10) & 1.26(-10) & 8.04(-7) & 1.38(-6) & 6.97(-11) & 1.67(-10) & 1.35(-7) & 4.69(-10) & 1.20(-11) & 6.88(-9) \\
N3-T20-CR500 & 5.04(-12) & 7.66(-10) & 8.39(-11) & 1.90(-8) & 5.56(-8) & 2.91(-13) & 2.26(-12) & 5.85(-8) & 1.60(-11) & 4.67(-14) & 3.12(-11) \\
N3-T20-CR1000 & 1.29(-13) & 3.80(-11) & 3.43(-12) & 1.90(-11) & 1.68(-11) & 3.86(-17) & 7.01(-17) & 1.37(-8) & 1.54(-13) & 1.80(-18) & 4.36(-14) \\
\hline
\end{tabular}
\end{center}
\vspace{-6mm}
\tablefoot{Gas-phase fractional abundances (with respect to total hydrogen) at $T$~=~150~K in the envelope of the sources. $X(Y)$ means $X \times 10^Y$.}
\end{table}
\end{landscape}

\begin{landscape}

\begin{table}
\begin{center}
  \caption{\label{TAB-appendix-model-results-150K2} Table~\ref{TAB-appendix-model-results-150K} continued.} 
  \setlength{\tabcolsep}{1.2mm}
  \begin{tabular}{lccccccccccc}
 \hline
 Model  & \ce{C2H5CN} & \ce{C2H3CN} & \ce{CH3CN} & \ce{CH3OH} & \ce{C2H5OH} & \ce{CH3OCHO} & \ce{CH3OCH3} & \ce{CH3CHO} & \ce{NH2CHO} & \ce{CH3NCO} & \ce{CH3SH} \\ 
 \hline
 \hline
N4-T10-CR1 & 9.46(-10) & 2.44(-12) & 1.11(-9) & 6.78(-6) & 6.50(-8) & 7.20(-10) & 1.36(-8) & 1.11(-10) & 4.38(-8) & 2.21(-13) & 2.59(-9) \\
N4-T10-CR10 & 1.25(-9) & 1.72(-11) & 5.53(-9) & 2.68(-6) & 2.31(-7) & 9.33(-10) & 1.45(-8) & 8.44(-10) & 2.02(-8) & 1.45(-13) & 3.37(-9) \\
N4-T10-CR50 & 2.90(-9) & 3.60(-10) & 7.61(-10) & 1.49(-6) & 2.95(-6) & 1.12(-8) & 7.26(-9) & 3.88(-8) & 1.14(-8) & 1.34(-12) & 6.65(-10) \\
N4-T10-CR100 & 4.61(-10) & 2.69(-10) & 7.95(-11) & 5.88(-7) & 2.30(-6) & 8.53(-9) & 6.24(-10) & 4.79(-8) & 1.32(-9) & 1.30(-12) & 4.93(-10) \\
N4-T10-CR500 & 6.74(-11) & 6.80(-11) & 1.32(-9) & 5.05(-9) & 4.76(-8) & 1.81(-11) & 6.80(-15) & 3.35(-9) & 9.70(-12) & 3.97(-14) & 8.61(-12) \\
N4-T10-CR1000 & 1.33(-13) & 6.91(-13) & 2.74(-10) & 5.24(-12) & 5.89(-11) & 2.89(-17) & 2.27(-18) & 1.32(-11) & 1.93(-14) & 2.42(-17) & 3.41(-15) \\
            &            &           &            &           &            &            &            &            &            &             &           \\
N4-T15-CR1 & 1.20(-9) & 4.30(-12) & 2.41(-9) & 1.33(-6) & 2.91(-8) & 1.30(-11) & 2.81(-9) & 3.76(-11) & 1.88(-9) & 2.74(-15) & 3.86(-9) \\
N4-T15-CR10 & 6.17(-10) & 8.11(-12) & 2.47(-9) & 1.25(-6) & 1.53(-7) & 2.43(-10) & 4.50(-9) & 4.71(-10) & 1.15(-8) & 4.14(-14) & 2.55(-9) \\
N4-T15-CR50 & 4.52(-9) & 4.54(-10) & 1.15(-9) & 1.53(-6) & 3.05(-6) & 1.14(-8) & 7.88(-9) & 3.95(-8) & 9.56(-9) & 8.65(-13) & 5.67(-10) \\
N4-T15-CR100 & 7.02(-10) & 3.25(-10) & 1.16(-10) & 6.54(-7) & 2.44(-6) & 9.43(-9) & 6.99(-10) & 5.07(-8) & 1.06(-9) & 9.62(-13) & 4.30(-10) \\
N4-T15-CR500 & 4.32(-11) & 4.76(-11) & 1.75(-9) & 5.02(-9) & 4.57(-8) & 7.73(-12) & 7.25(-15) & 3.03(-9) & 9.92(-12) & 3.30(-14) & 3.85(-12) \\
N4-T15-CR1000 & 1.38(-13) & 6.04(-13) & 2.15(-10) & 3.88(-12) & 4.19(-11) & 2.49(-17) & 1.41(-18) & 1.07(-11) & 1.79(-14) & 1.41(-17) & 2.70(-15) \\
            &            &           &            &           &            &            &            &            &            &             &           \\
N4-T20-CR1 & 6.40(-11) & 8.50(-14) & 3.49(-11) & 7.53(-8) & 1.50(-9) & 1.16(-12) & 6.08(-11) & 1.33(-12) & 2.20(-9) & 6.10(-17) & 1.30(-9) \\
N4-T20-CR10 & 2.42(-11) & 3.79(-13) & 2.06(-11) & 8.00(-8) & 1.89(-9) & 4.52(-12) & 3.61(-11) & 5.92(-12) & 2.83(-9) & 6.75(-16) & 2.35(-10) \\
N4-T20-CR50 & 2.15(-9) & 1.53(-10) & 6.19(-10) & 6.74(-7) & 1.65(-6) & 4.73(-9) & 1.89(-9) & 2.11(-8) & 1.36(-9) & 3.35(-13) & 4.73(-10) \\
N4-T20-CR100 & 3.16(-9) & 2.52(-10) & 4.42(-10) & 5.87(-7) & 1.91(-6) & 7.43(-9) & 8.16(-10) & 4.10(-8) & 4.79(-10) & 7.17(-13) & 7.15(-10) \\
N4-T20-CR500 & 1.66(-11) & 1.64(-11) & 1.54(-9) & 1.38(-9) & 1.35(-8) & 2.42(-14) & 1.13(-15) & 7.41(-10) & 3.93(-12) & 4.47(-15) & 3.51(-13) \\
N4-T20-CR1000 & 1.79(-13) & 3.98(-13) & 8.82(-11) & 2.73(-12) & 2.79(-11) & 4.11(-17) & 4.70(-19) & 3.76(-12) & 9.46(-15) & 2.44(-18) & 3.34(-15) \\
\hline
N5-T10-CR1 & 1.22(-9) & 7.31(-12) & 1.20(-9) & 8.87(-6) & 5.39(-8) & 5.04(-10) & 1.76(-8) & 1.11(-10) & 3.35(-8) & 5.45(-12) & 2.64(-9) \\
N5-T10-CR10 & 2.26(-9) & 2.47(-11) & 2.04(-9) & 5.42(-6) & 1.20(-7) & 1.60(-9) & 2.34(-9) & 6.56(-10) & 3.87(-8) & 8.97(-14) & 3.35(-9) \\
N5-T10-CR50 & 5.89(-9) & 7.19(-10) & 3.36(-9) & 2.95(-6) & 3.72(-6) & 7.22(-9) & 5.92(-8) & 3.96(-8) & 2.19(-8) & 1.81(-12) & 1.86(-9) \\
N5-T10-CR100 & 3.70(-9) & 9.09(-10) & 8.58(-10) & 1.83(-6) & 3.82(-6) & 1.28(-8) & 5.47(-9) & 6.04(-8) & 1.14(-8) & 2.52(-12) & 1.12(-9) \\
N5-T10-CR500 & 1.38(-10) & 2.72(-10) & 3.87(-11) & 2.12(-7) & 1.41(-6) & 1.88(-9) & 4.68(-12) & 7.97(-8) & 4.74(-10) & 2.42(-12) & 3.56(-10) \\
N5-T10-CR1000 & 4.49(-11) & 6.71(-11) & 1.53(-10) & 5.94(-9) & 4.60(-8) & 1.43(-11) & 6.67(-15) & 4.36(-9) & 1.04(-11) & 4.14(-14) & 8.79(-12) \\
            &            &           &            &           &            &            &            &            &            &             &           \\
N5-T15-CR1 & 8.55(-10) & 8.83(-12) & 1.73(-9) & 2.63(-6) & 2.87(-8) & 3.29(-10) & 1.83(-9) & 1.40(-8) & 2.76(-8) & 8.52(-16) & 4.91(-9) \\
N5-T15-CR10 & 1.45(-9) & 1.45(-11) & 1.97(-9) & 2.41(-6) & 1.59(-7) & 2.64(-10) & 1.62(-9) & 3.51(-10) & 1.89(-8) & 3.85(-14) & 3.09(-9) \\
N5-T15-CR50 & 9.05(-9) & 8.43(-10) & 5.82(-9) & 2.67(-6) & 3.57(-6) & 6.08(-9) & 5.69(-8) & 3.69(-8) & 1.86(-8) & 1.30(-12) & 1.77(-9) \\
N5-T15-CR100 & 4.74(-9) & 9.08(-10) & 1.05(-9) & 1.69(-6) & 3.55(-6) & 1.17(-8) & 5.30(-9) & 5.54(-8) & 9.43(-9) & 1.57(-12) & 1.10(-9) \\
N5-T15-CR500 & 1.48(-10) & 1.73(-10) & 4.84(-11) & 1.75(-7) & 1.25(-6) & 1.51(-9) & 2.21(-12) & 6.57(-8) & 3.65(-10) & 1.97(-12) & 1.59(-10) \\
N5-T15-CR1000 & 2.54(-11) & 4.44(-11) & 2.46(-10) & 4.23(-9) & 3.41(-8) & 3.04(-12) & 3.72(-15) & 3.04(-9) & 7.85(-12) & 2.37(-14) & 3.36(-12) \\
            &            &           &            &           &            &            &            &            &            &             &           \\
N5-T20-CR1 & 1.18(-10) & 1.00(-13) & 6.44(-11) & 8.32(-8) & 3.14(-9) & 5.86(-12) & 1.46(-10) & 3.16(-12) & 4.05(-9) & 4.89(-17) & 1.54(-9) \\
N5-T20-CR10 & 3.37(-11) & 3.19(-13) & 5.65(-11) & 1.80(-7) & 1.93(-9) & 1.49(-11) & 1.02(-10) & 4.90(-12) & 5.58(-9) & 5.33(-16) & 3.22(-10) \\
N5-T20-CR50 & 5.20(-9) & 2.39(-10) & 1.64(-9) & 1.11(-6) & 2.23(-6) & 3.13(-9) & 5.29(-9) & 2.10(-8) & 4.51(-9) & 3.16(-13) & 8.82(-10) \\
N5-T20-CR100 & 3.89(-9) & 2.86(-10) & 9.94(-10) & 7.32(-7) & 1.75(-6) & 4.43(-9) & 1.43(-9) & 2.66(-8) & 4.15(-10) & 3.34(-13) & 5.56(-10) \\
N5-T20-CR500 & 3.59(-11) & 3.43(-11) & 1.06(-10) & 3.14(-8) & 2.98(-7) & 4.20(-12) & 6.57(-14) & 1.33(-8) & 1.15(-10) & 2.05(-13) & 6.90(-12) \\
N5-T20-CR1000 & 1.32(-11) & 2.01(-11) & 4.02(-10) & 2.42(-9) & 2.03(-8) & 5.99(-14) & 1.40(-15) & 1.53(-9) & 6.73(-12) & 8.53(-15) & 4.77(-13) \\
\hline
\end{tabular}
\end{center}
\end{table}

\end{landscape}

\section{Confidence level of the models with respect to the observations}
\label{appendix-level-confidence}

In order to determine which set of physical parameters ($T_{\rm min}$ and CRIR) best characterizes Sgr~B2(N)'s hot cores (see Sect.~\ref{section-discussion-constraining-parameters}), we evaluate the success of each model in reproducing the observations using the method presented by \citet{garrod2007}. For each model, we compute a level of confidence, $\kappa_{\rm i}$, in the agreement between calculated and observed abundances (relative to \ce{H2}, \ce{CH3OH}, or \ce{CH3CN}). For each calculated abundance ratio, the level of confidence is given by:
\begin{equation}
\label{EQ-confidence-level}
\kappa_{\rm i} = erfc \left( \frac{| \, \mathrm{log}(R_{\rm model, i}) - \mathrm{log}(R_{\rm obs, i}) \, |}{\sqrt{2} \sigma} \right)
\end{equation} 
which computes the logarithmic distance of disagreement between the calculated abundance ratio ($R_{\rm model, i}$) and the observed one ($R_{\rm obs, i}$). We define the standard deviation $\sigma$~=~1 such that one standard deviation corresponds to one order of magnitude lower/higher than the observed abundance ratio. \textit{erfc} is the complementary error function (\textit{erfc}~=~1-\textit{erf}) such that $\kappa_{\rm i}$ ranges between 0 and 1. For instance, a calculated abundance ratio lying one order of magnitude lower/higher than the observed one has a confidence level $\kappa_{\rm i}$~=~31.7\% (and $\kappa_{\rm i}$~=~4.6\% for two orders of magnitude etc.). For each model we take the mean of the confidence levels obtained for the molecules in the sample to define the overall confidence level of the model with respect to the observations. The results are given in Tables~\ref{TAB-appendix-confidence-all-h2}--\ref{TAB-appendix-confidence-N-ch3cn}.

\begin{table}[!h]
\begin{center}
\caption{\label{TAB-appendix-confidence-all-h2} Mean confidence level (\%) for ten COMs with respect to H$_2$.} 
\vspace{-2mm}
\setlength{\tabcolsep}{1.2mm}
\begin{tabular}{lrrrrr}
\hline
Model & \multicolumn{1}{c}{N2}  & \multicolumn{1}{c}{N3} & \multicolumn{1}{c}{N4 \tablefootmark{*}} & \multicolumn{1}{c}{N5} & \multicolumn{1}{c}{All sources} \\
\hline
\hline
T10--CR1     & 18.8 & 25.3 & 35.2 & 27.8 & 26.8 \\
T10--CR10    & 23.5 & 34.6 & 46.1 & 27.9 & 33.0 \\
T10--CR50    & 24.0 & 36.4 & 48.2 & 39.9 & 37.1 \\
T10--CR100   & 23.6 & 31.6 & 37.0 & 36.4 & 32.1 \\
T10--CR500   & 19.1 & 21.0 & 34.5 & 16.2 & 22.7 \\
T10--CR1000  & 18.1 & 9.0 & 12.3 & 16.3 & 13.9 \\
\hline
T15--CR1     & 12.8 & 28.5 & 36.1 & 38.6 & 29.0 \\
T15--CR10    & 14.4 & 33.9 & 36.3 & 29.3 & 28.5 \\
T15--CR50    & 23.7 & 39.5 & 47.3 & 42.9 & 38.4 \\
T15--CR100   & 22.3 & 32.0 & 39.4 & 38.0 & 32.9 \\
T15--CR500   & 19.1 & 20.2 & 34.3 & 14.9 & 22.1 \\
T15--CR1000  & 16.7 & 7.4 & 11.8 & 14.0 & 12.5 \\
\hline
T20--CR1     & 2.1  & 12.9 & 28.4 & 15.4 & 14.7 \\
T20--CR10    & 2.1  & 2.2 & 22.6 & 15.4 & 10.6 \\
T20--CR50    & 14.5 & 25. & 43.0 & 37.8 & 30.1 \\
T20--CR100   & 14.5 & 21.0 & 47.8 & 28.0 & 27.8 \\
T20--CR500   & 14.3 & 16.9 & 31.0 & 16.7 & 19.7 \\
T20--CR1000  & 12.1 & 5.9 & 10.8 & 10.4 & 9.8 \\
T25--CR50    & 6.7  & \_       & \_      & \_      &   \_     \\
T28--CR50    & 3.3  &  \_       & \_      &  \_     &   \_     \\
\hline
\end{tabular}
\end{center}
\vspace{-4mm}
\tablefoot{Mean confidence level calculated for 10 COMs (excluding \ce{CH3NCO}) with respect to H$_2$ for each model. Results are given in percentages. \tablefoottext{*}{Because \ce{NH2CHO} is not detected toward N4 and we estimated only an upper limit to its molecular column density, a level of confidence of unity is attributed if the calculated ratio [\ce{NH2CHO}]/[\ce{H2}] is lower than the upper limit, otherwise it is treated normally.}}
\end{table}

\begin{table}[!h]
\begin{center}
\caption{\label{TAB-appendix-confidence-O-h2} Mean confidence level (\%) for O-bearing species with respect to H$_2$.} 
\vspace{-2mm}
\setlength{\tabcolsep}{1.2mm}
\begin{tabular}{lrrrrr}
\hline
Model & \multicolumn{1}{c}{N2}  & \multicolumn{1}{c}{N3} & \multicolumn{1}{c}{N4 \tablefootmark{*}} & \multicolumn{1}{c}{N5} & \multicolumn{1}{c}{All sources} \\
\hline
\hline
T10--CR1    & 29.9  & 35.8 & 26.4 & 36.4 & 32.1 \\
T10--CR10   & 36.2 & 43.4 & 34.3 & 33.3 & 36.8 \\
T10--CR50   & 31.8 & 36.8 & 32.0 & 46.3 & 36.7 \\
T10--CR100  & 35.8 & 36.4 & 29.7 & 44.9 & 36.7 \\
T10--CR500  & 30.1 & 28.6 & 41.1 & 22.8 & 30.7 \\
T10--CR1000 & 29.4  & 14.2 & 18.1 & 26.3 & 22.0 \\
\hline
T15--CR1    & 19.1 & 40.0 & 25.1 & 51.8 & 34.0 \\
T15--CR10   & 21.7 & 42.9 & 25.7 & 37.2 & 31.9 \\
T15--CR50   & 29.1 & 36.8 & 32.7 & 47.5 & 36.5 \\
T15--CR100  & 33.0 & 35.0 & 31.1 & 46.6 & 36.4 \\
T15--CR500  & 30.1 & 27.6 & 41.8 & 22.2 & 30.4 \\
T15--CR1000 & 27.7 & 12.0 & 17.7 & 22.7 & 20.0 \\
\hline
T20--CR1    & 2.1  & 16.6 & 30.7 & 20.8 & 17.6 \\
T20--CR10   & 3.3  & 2.1 & 30.7 & 24.4 & 15.1 \\
T20--CR50   & 22.8 & 32.1 & 31.6 & 49.6 & 34.0 \\
T20--CR100  & 22.8 & 26.6 & 35.2 & 35.6 & 30.0 \\
T20--CR500  & 23.8 & 26.2 & 40.9 & 27.3 & 29.6 \\
T20--CR1000 & 20.1 & 9.8 & 17.2 & 16.8 & 16.0 \\
T25--CR50   & 10.4 &   \_     &  \_    &  \_    &   \_    \\
T28--CR50   & 5.3  &  \_      &  \_    &   \_   &    \_   \\
\hline
\end{tabular}
\end{center}
\vspace{-4mm}
 \tablefoot{Mean confidence level calculated for the abundance ratios of the O-bearing species \ce{CH3OH}, \ce{CH3OCHO}, \ce{CH3OCH3}, \ce{CH3CHO}, \ce{C2H5OH}, and \ce{NH2CHO}, with respect to H$_2$ for each model. Results are given in percentages. \tablefoottext{*}{Because \ce{NH2CHO} is not detected toward N4 and we estimated only an upper limit to its molecular column density, a level of confidence of unity is attributed if the calculated ratio [\ce{NH2CHO}]/[\ce{H2}] is lower than the upper limit, otherwise it is treated normally.}}
\end{table}

\begin{table}[!h]
\begin{center}
\caption{\label{TAB-appendix-confidence-N-h2} Mean confidence level (\%) for cyanides with respect to H$_2$.} 
\vspace{-2mm}
\setlength{\tabcolsep}{1.2mm}
\begin{tabular}{lrrrrr}
\hline
Model & \multicolumn{1}{c}{N2}  & \multicolumn{1}{c}{N3} & \multicolumn{1}{c}{N4} & \multicolumn{1}{c}{N5} & \multicolumn{1}{c}{All sources} \\
\hline
\hline
T10--CR1    & 0.2 & 7.9 & 34.8 & 7.0 & 12.5 \\
T10--CR10   & 0.7 & 25.0 & 58.5 & 11.6 & 23.9 \\
T10--CR50   & 6.7 & 38.2 & 74.8 & 30.3 & 37.5 \\
T10--CR100  & 3.5 & 14.9 & 45.0 & 24.4 & 21.9 \\
T10--CR500  & 2.3 & 9.7 & 32.3 & 5.7 & 12.5 \\
T10--CR1000 & 0.6 & 1.5 & 4.7 & 1.8 & 2.1 \\
\hline
T15--CR1    & 0.3 & 9.4 & 45.0 & 6.7 & 15.4 \\
T15--CR10   & 0.3 & 22.5 & 39.9 & 9.1 & 18.0 \\
T15--CR50   & 10.1 & 41.6 & 72.0 & 38.2 & 40.5 \\
T15--CR100  & 3.9 & 17.3 & 51.7 & 26.6 & 24.9 \\
T15--CR500  & 1.9 & 9.2 & 30.6 & 4.1 & 11.4 \\
T15--CR1000 & 0.2 & 0.6 & 3.8 & 1.4 & 1.5 \\
\hline
T20--CR1    & 0.01 & 0.6 & 4.0 & 0.5 & 1.3 \\
T20--CR10   & 0.01 & 0.6 & 1.7 & 0.1 & 0.6 \\
T20--CR50   & 1.3 & 4.5 & 61.7 & 20.7 & 22.1 \\
T20--CR100  & 1.8 & 3.4 & 66.5 & 18.1 & 22.4 \\
T20--CR500  & 0.1 & 3.8 & 21.6 & 0.9 & 6.6 \\
T20--CR1000 & 0.02 & 0.1 & 1.6 & 1.1 & 0.7 \\
T25--CR50   & 0.5 &   \_     &   \_     &   \_     &    \_     \\
T28--CR50   & 1.4 &   \_       &   \_     &   \_     &    \_     \\
\hline
\end{tabular}
\end{center}
\vspace{-4mm}
\tablefoot{Mean confidence level calculated for the abundance of the cyanides \ce{CH3CN}, \ce{C2H5CN}, and \ce{C2H3CN} relative to \ce{H2} for each model. Results are given in percentages.}
\end{table}

\begin{table}[!h]
\begin{center}
\caption{\label{TAB-appendix-confidence-all-ch3oh} Mean confidence level (\%) for nine COMs with respect to \ce{CH3OH}.} 
\vspace{-2mm}
\setlength{\tabcolsep}{1.2mm}
\begin{tabular}{lrrrrr}
\hline
Model & \multicolumn{1}{c}{N2}  & \multicolumn{1}{c}{N3} & \multicolumn{1}{c}{N4 \tablefootmark{*}} & \multicolumn{1}{c}{N5} & \multicolumn{1}{c}{All sources} \\
\hline
\hline
T10--CR1    & 12.4 & 16.2 & 14.9 & 14.4 & 14.5 \\
T10--CR10   & 23.1 & 20.1 & 24.3 & 22.2 & 22.4 \\
T10--CR50   & 42.7 & 28.8 & 26.9 & 34.4 & 33.2 \\
T10--CR100  & 31.2 & 29.4 & 26.6 & 29.8 & 29.2 \\
T10--CR500  & 18.7 & 20.0 & 41.1 & 21.5 & 25.3 \\
T10--CR1000 & 18.2 & 25.0 & 24.0 & 33.5 & 25.2 \\
\hline
T15--CR1    & 19.6 & 16.3 & 21.3 & 27.9 & 21.3 \\
T15--CR10   & 20.4 & 22.8 & 23.8 & 25.8 & 23.2 \\
T15--CR50   & 49.1 & 35.8 & 28.5 & 36.7 & 37.5 \\
T15--CR100  & 33.9 & 34.9 & 27.0 & 30.4 & 31.5 \\
T15--CR500  & 18.7 & 17.8 & 39.1 & 19.6 & 23.8 \\
T15--CR1000 & 14.3 & 25.7 & 24.9 & 34.4 & 24.9 \\
\hline
T20--CR1    & 31.9 & 27.6 & 19.7 & 28.3 & 26.9 \\
T20--CR10   & 19.3 & 42.4 & 14.0 & 15.6 & 22.8 \\
T20--CR50   & 30.8 & 22.1 & 29.2 & 30.9 & 28.2 \\
T20--CR100  & 26.1 & 22.7 & 30.5 & 23.9 & 25.8 \\
T20--CR500  & 10.0 & 17.8 & 34.8 & 19.5 & 20.5 \\
T20--CR1000 & 9.9 & 29.8 & 27.6 & 32.3 & 24.9 \\
\hline
\end{tabular}
\end{center}
\vspace{-4mm}
\tablefoot{Same as \ref{TAB-appendix-confidence-all-h2} but for the abundances of nine COMs relative to \ce{CH3OH}.}
\end{table}

\begin{table}[!h]
\begin{center}
\caption{\label{TAB-appendix-confidence-O-ch3oh} Mean confidence level (\%) for O-bearing species with respect to \ce{CH3OH}.} 
\vspace{-2mm}
\setlength{\tabcolsep}{1.2mm}
\begin{tabular}{lrrrrrr}
\hline
Model & \multicolumn{1}{c}{N2}  & \multicolumn{1}{c}{N3} & \multicolumn{1}{c}{N4 \tablefootmark{*}} & \multicolumn{1}{c}{N5} & \multicolumn{1}{c}{All sources} \\
\hline
\hline
T10--CR1    & 20.3 & 25.3 & 25.7 & 23.9 & 23.8\\
T10--CR10   & 36.0 & 26.0 & 39.0 & 35.5 & 34.1 \\
T10--CR50   & 46.7 & 24.7 & 41.6 & 50.0 & 40.8 \\
T10--CR100  & 43.3 & 22.2 & 40.4 & 40.4 & 36.6 \\
T10--CR500  & 21.4 &  5.4 & 26.2 & 21.0 & 18.5 \\
T10--CR1000 & 14.7 & 15.1 & 22.3 & 14.8 & 16.7 \\
\hline
T15--CR1    & 27.7 & 23.8 & 30.7 & 41.9 & 31.0 \\
T15--CR10   & 29.2 & 31.3 & 37.0 & 38.7 & 34.0 \\
T15--CR50   & 46.2 & 26.0 & 42.9 & 50.2 & 41.3 \\
T15--CR100  & 43.0 & 21.2 & 40.8 & 39.6 & 36.1 \\
T15--CR500  & 20.8 &  4.3 & 26.1 & 20.5 & 17.9 \\
T15--CR1000 & 14.2 & 16.0 & 22.2 & 13.9 & 16.6 \\
\hline
T20--CR1    & 38.6 & 26.5 & 19.3 & 27.0 & 27.8 \\
T20--CR10   & 26.8 & 33.2 & 19.3 & 20.8 & 25.0 \\
T20--CR50   & 40.9 & 16.5 & 43.7 & 39.5 & 35.2 \\
T20--CR100  & 29.6 & 14.4 & 41.6 & 24.9 & 27.6 \\
T20--CR500  &  9.0 & 10.8 & 26.1 & 18.6 & 16.1 \\
T20--CR1000 &  6.4 & 19.9 & 23.4 & 16.1 & 16.4 \\
\hline
\end{tabular}
\end{center}
\vspace{-4mm}
 \tablefoot{Same as \ref{TAB-appendix-confidence-O-h2} but for the abundances of \ce{CH3OCHO}, \ce{CH3OCH3}, \ce{CH3CHO}, \ce{C2H5OH}, and \ce{NH2CHO} relative to \ce{CH3OH}.}
\end{table}

\begin{table}[!h]
\begin{center}
\caption{\label{TAB-appendix-confidence-N-ch3cn} Mean confidence level (\%) for cyanides with respect to \ce{CH3CN}.} 
\vspace{-2mm}
\setlength{\tabcolsep}{1.2mm}
\begin{tabular}{lrrrrr}
\hline
Model & \multicolumn{1}{c}{N2}  & \multicolumn{1}{c}{N3} & \multicolumn{1}{c}{N4} & \multicolumn{1}{c}{N5} & \multicolumn{1}{c}{All sources} \\
\hline
\hline
T10--CR1    & 55.1 & 61.3 & 42.9 & 49.3 & 52.1 \\
T10--CR10   & 49.8 & 16.4 & 56.3 & 54.8 & 44.3 \\
T10--CR50   & 50.0 &  8.8 & 29.5 & 64.7 & 38.2 \\
T10--CR100  & 80.3 & 13.4 & 13.4 & 31.7 & 34.7 \\
T10--CR500  & 46.0 & 13.6 & 69.2 & 22.3 & 37.8 \\
T10--CR1000 & 50.9 &  3.6 & 10.7 & 66.4 & 32.9 \\
\hline
T15--CR1    & 29.7 & 49.7 & 50.3 & 60.1 & 47.5 \\
T15--CR10   & 46.5 & 25.5 & 58.5 & 56.4 & 46.7 \\
T15--CR50   & 48.5 &  9.1 & 31.2 & 73.0 & 40.4 \\
T15--CR100  & 81.3 & 13.6 & 13.7 & 33.4 & 35.5 \\
T15--CR500  & 43.2 & 13.2 & 54.6 & 26.9 & 34.5 \\
T15--CR1000 & 50.7 &  4.0 & 11.7 & 63.7 & 32.5 \\
\hline
T20--CR1    & 42.4 & 42.0 & 32.5 & 32.0 & 37.2 \\
T20--CR10   & 15.9 & 43.3 & 63.2 & 57.6 & 45.0 \\
T20--CR50   & 79.2 & 36.2 & 38.0 & 63.0 & 54.1 \\
T20--CR100  & 85.9 & 15.3 & 22.4 & 49.4 & 43.3 \\
T20--CR500  & 51.2 & 20.3 & 33.6 & 73.2 & 44.6 \\
T20--CR1000 & 51.1 & 15.0 & 17.2 & 51.5 & 33.7 \\
\hline
\end{tabular}
\end{center}
\vspace{-4mm}
\tablefoot{Same as \ref{TAB-appendix-confidence-N-h2} but for the abundances of \ce{C2H5CN} and \ce{C2H3CN} relative to \ce{CH3CN}.}
\end{table}

\section{Influence of the external radiation field on the chemistry}
        \label{appendix-ISRF}

\begin{figure}[!h]
   \begin{center}
    \includegraphics[width=\hsize]{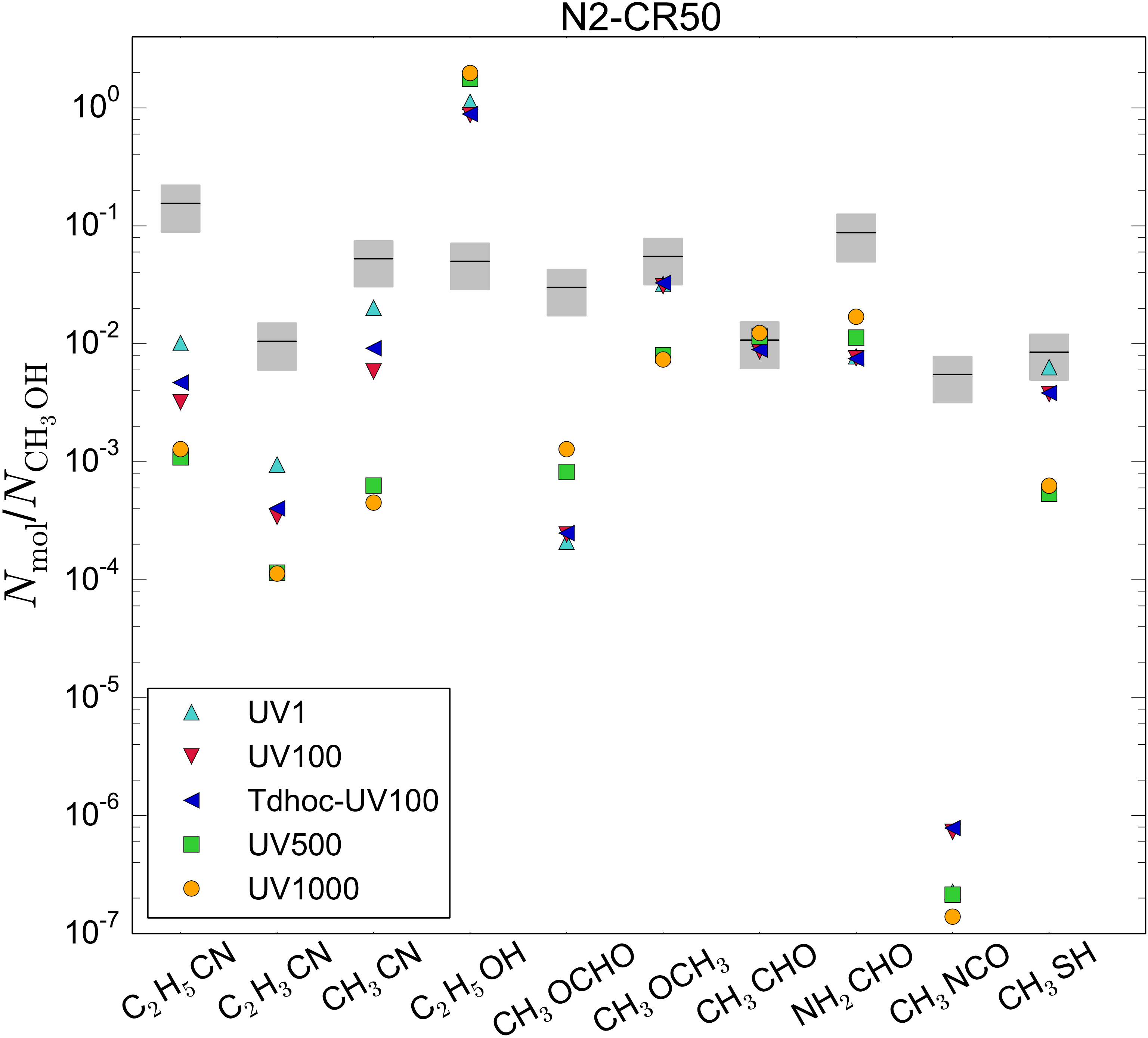} 
    \caption{\label{FIG-appendix-ISRF-obs} Chemical abundances with respect to \ce{CH3OH} of ten COMs calculated at $T$~=~150~K by the models N2-CR50. The different symbols indicate the abundances calculated with different $\chi_{\rm UV}$. Model Tdhoc-UV100 shows the results obtained using the dust temperature expression given by \citet{hocuk2017} with $\chi_{\rm UV}$~=~100. The horizontal black lines show the observed abundances relative to \ce{CH3OH}. The gray boxes represent the 1$\sigma$ uncertainties.}
   \end{center}
\end{figure}

Investigations carried out on the extreme environmental conditions characterizing the GC region suggest that the ambient radiation field may be enhanced by a factor $\sim$~500--1000 in the inner Galaxy \citep{lis2001, clark2013}, compared to the standard ISRF value (1~G$_0$) used in our simulations. A stronger ISRF may not have a big impact on the warm chemistry during the free-fall collapse phase, because external UV photons are rapidly absorbed as the visual extinction increases. However, it may be more critical during the preceding low-density phase, where dust grains are not efficiently shielded against external radiation. Recently, \citet{hocuk2017} derived a parametric expression of the dust temperature scalable with the interstellar radiation field strength. Based on their analysis we scale the dust temperature expression of \citet{garrod2011} (see Sect.~\ref{section-precollapse-phase}) as follows: $T_{\rm d}^{\rm UV}(A_{\rm v})$~=~$T_{\rm d}(A_{\rm v}) \, \chi_{\rm UV}^{1/5.9}$, where $\chi_{\rm UV}$ represents the strength of the UV field in units of the standard Draine field. In order to explore the impact of a stronger radiation field on the production of COMs, we run three additional chemical models, N2-UV100, N2-UV500, and N2-UV1000, with $\chi_{\rm UV}$= 100~G$_0$, 500~G$_0$, and 1000~G$_0$, respectively. No minimum threshold is fixed for the dust temperatures, which go as low as $\sim$14~K, 17~K, and 19~K in models N2-UV100, N2-UV500, and N2-UV1000, respectively. The gas temperature, $T_{\rm g}$, is held constant at 15~K during the whole quasi-static contraction phase. We adopt $\zeta_{\rm H_2}$~=~6.5$\times$10$^{-16}$~s$^{-1}$ as found to be the CRIR that best reproduces the observations toward Sgr~B2(N2-N5) (see Sect.~\ref{section-discussion-constraining-parameters}). Figure~\ref{FIG-appendix-ISRF-obs} compares the calculated gas-phase fractional abundances of ten COMs relative to \ce{CH3OH} for the models with strong UV fields to the observed abundances. The results of model N2-CR50-UV1, with a standard radiation field and a minimum dust temperature $T_{\rm min}$~=~15~K, are also shown for comparison. It shows that stronger UV fields increase the discrepancy between the calculated and observed abundances of cyanides with respect to \ce{CH3OH}, while the abundances of \ce{NH2CHO}, \ce{CH3CHO}, and \ce{C2H5OH} relative to \ce{CH3OH} are only slightly affected by the UV field strength. 

Table~\ref{TAB-appendix-confidence-UV-ch3oh} gives the mean confidence level of the models with the observations for different $\chi_{\rm UV}$. It shows that the highest confidence level for the abundances of nine COMs (excluding \ce{CH3NCO}) with respect to \ce{CH3OH} is obtained with the model assuming a standard ISRF (1~G$_0$). The confidence levels drop significantly for models with higher $\chi_{\rm UV}$. The opposite behavior is obtained when computing confidence levels with only the cyanides \ce{C2H3CN} and \ce{C2H5CN} with respect to \ce{CH3CN} (that is the agreement with observations increases with the UV field strength). Finally, all models have similar confidence levels when computed based only on the O-bearing species \ce{CH3OCHO}, \ce{CH3OCH3}, \ce{CH3CHO}, \ce{C2H5OH}, and \ce{NH2CHO}. This does not allow us to conclude strictly on which interstellar radiation field better characterizes Sgr~B2(N). This is consistent with the results we obtain by simply parametrizing the dust temperature profile with an artificially-set minimum temperature (see Sect.~\ref{section-discussion-constraining-parameters}).

\begin{table}[!h]
\begin{center}
\caption{\label{TAB-appendix-confidence-UV-ch3oh} Mean confidence level for high-UV field models.} 
\vspace{-2mm}
\setlength{\tabcolsep}{1.2mm}
\begin{tabular}{lccc}
\hline
Model &  \multicolumn{3}{c}{Confidence level (\%)} \\ 
  \cline{2-4}
                              &  All \tablefootmark{a} & O-bearing \tablefootmark{b} & N-bearing \tablefootmark{c} \\
\hline
\hline
N2-CR50-UV1 &  49.1 & 46.2 & 48.5 \\
N2-CR50-UV100 & 39.5 & 45.3 & 53.0 \\
N2-CR50-UV500 & 26.2 & 39.8 & 89.3 \\
N2-CR50-UV1000 & 27.5  & 41.8 & 95.5 \\
\hline
\end{tabular}
\end{center}
\vspace{-4mm}
\tablefoot{\tablefoottext{a}{Mean confidence level calculated for ten COMs with respect to \ce{CH3OH} for each model.} 
\tablefoottext{b}{Mean confidence level calculated for the abundances of the O-bearing species \ce{CH3OCHO}, \ce{CH3OCH3}, \ce{CH3CHO}, \ce{C2H5OH}, and \ce{NH2CHO}, with respect to \ce{CH3OH}.}
\tablefoottext{c}{Mean confidence level calculated for the abundance of the cyanides \ce{C2H5CN} and \ce{C2H3CN} relative to \ce{CH3CN}}}
\end{table}

As pointed out by \citet{hocuk2017}, the dust temperature expression of \citet{garrod2011} results in higher temperatures at low visual extinctions than the values computed using the parametric expression of \citet{hocuk2017} (see their Fig.~7.). For comparison, we plot in Fig.~\ref{FIG-appendix-ISRF-obs} the fractional abundances with respect to \ce{CH3OH} calculated by model N2-CR50-Tdhoc-UV100, which uses the dust temperature expression given by \cite{hocuk2017} (see their Eq.~8). Fig.~\ref{FIG-appendix-ISRF-obs} shows that the expression used to compute the dust temperature does not have a strong impact on the production of COMs, as there is at most a factor $\sim$1.6 of difference in the abundances with respect to \ce{CH3OH} calculated by models N2-CR50-Tdhoc-UV100 and N2-CR50-UV100.


\section{Other complementary tables}
        \label{appendix-tables}

\begin{table*}[!t]
\begin{center}
  \caption{\label{TAB-appendix-data-redu} Observational setups and noise level.} 
  \vspace{-2mm}
  \setlength{\tabcolsep}{1.1mm}
  \begin{tabular}{ccccccccccc}
  \hline
Setup & SPW \tablefootmark{a} & Frequency &  & & \multicolumn{6}{c}{Spectrum rms \tablefootmark{c}}  \\
  \cline{6-11} 
      &     & range     & \multicolumn{2}{c}{Channel map rms \tablefootmark{b}} & \multicolumn{2}{c}{N3} & \multicolumn{2}{c}{N4} & \multicolumn{2}{c}{N5} \\
      &     & (MHz)           & (mJy beam$^{-1}$) & (K) & (mJy beam$^{-1}$) & (K) & (mJy beam$^{-1}$) & (K) & (mJy beam$^{-1}$) & (K) \\
    \hline
\hline
 S1 & 0 &  84091--85966  & 3.0 & 0.16 &    2.6  & 0.14   & 2.9  & 0.15   &  4.1  & 0.22   \\
    & 1 &  85904--87779  & 2.7 & 0.14 &    2.7  & 0.14   & 2.6  & 0.14   &  3.9  & 0.21   \\
    & 2 &  96154--98029  & 3.0 & 0.16 &    2.9  & 0.15   & 3.1  & 0.16   &  4.5  & 0.23   \\
    & 3 &  97904--99779  & 3.1 & 0.16 &    3.4  & 0.18   & 2.9  & 0.15   &  3.8  & 0.20   \\
 S2 & 0 &  87729--89604  & 3.1 & 0.15 &    2.7  & 0.13   & 2.5  & 0.12   &  3.4  & 0.17   \\
    & 1 &  89554--91429  & 2.8 & 0.15 &    2.5  & 0.13   & 2.5  & 0.14   &  3.9  & 0.21   \\
    & 2 &  99728--101602 & 2.7 & 0.14 &    2.5  & 0.13   & 2.5  & 0.13   &  3.9  & 0.20   \\
    & 3 & 101552--103427 & 2.7 & 0.14 &    2.5  & 0.13   & 2.8  & 0.15   &  3.8  & 0.20   \\
 S3 & 0 &  91368--93242  & 3.4 & 0.12 &    2.7  & 0.09   & 2.9  & 0.10   &  5.0  & 0.17   \\
    & 1 &  93193--95067  & 3.1 & 0.10 &    3.3  & 0.11   & 2.4  & 0.08   &  4.7  & 0.16   \\
    & 2 & 103365--105239 & 3.4 & 0.11 &    3.0  & 0.10   & 3.0  & 0.10   &  5.2  & 0.36   \\
    & 3 & 105189--107064 & 3.6 & 0.12 &    3.1  & 0.11   & 3.4  & 0.12   &  5.7  & 0.19   \\
 S4 & 0 &  95021--96896  & 1.9 & 0.10 &    1.9  & 0.10   & 2.1  & 0.11   &  2.7  & 0.14   \\
    & 1 &  96846--98720  & 1.9 & 0.10 &    2.1  & 0.11   & 2.0  & 0.10   &  2.8  & 0.15   \\
    & 2 & 107019--108893 & 2.2 & 0.11 &    3.0  & 0.16   & 2.1  & 0.11   &  3.2  & 0.17   \\
    & 3 & 108843--110718 & 2.3 & 0.12 &    2.2  & 0.12   & 2.2  & 0.12   &  3.6  & 0.19   \\
 S5 & 0 &  98672--100546 & 2.8 & 0.14 &    2.9  & 0.14   & 2.5  & 0.12   &  3.8  & 0.19   \\
    & 1 & 100496--102370 & 2.7 & 0.13 &    2.5  & 0.13   & 2.3  & 0.12   &  3.9  & 0.19   \\
    & 2 & 110669--112543 & 3.5 & 0.17 &    3.2  & 0.15   & 3.7  & 0.18   &  4.9  & 0.23   \\
    & 3 & 112494--114368 & 4.9 & 0.24 &    4.0  & 0.20   & 4.0  & 0.20   &  5.8  & 0.29   \\
\hline
\end{tabular}
\end{center}
\vspace{-4mm}
\tablefoot{
\tablefoottext{a}{Spectral window.}
\tablefoottext{b}{Median noise level measured in the channel maps with the command \textit{go noise} in GREG in the continuum-subtracted datacubes \citep{belloche2016}.}
\tablefoottext{c}{Noise level measured in each spectral window of the new continuum-subtracted spectra obtained toward Sgr~B2(N3-N5) (Sect.~\ref{section-data-reduction}). The rms values are not corrected for the primary beam attenuation.}}
\end{table*}

\begin{table*}[!t]
\begin{center}
  \caption{\label{TAB-best-fit-parameters} Parameters of the best-fit LTE model.}
  \begin{tabular}{clrccccrc}
    \hline
Source &   \multicolumn{1}{c}{Species} & \multicolumn{1}{c}{$N_l$} \tablefootmark{a} & $N_{\rm mol}$ \tablefootmark{b} & $C_{\rm vib}$ \tablefootmark{c} &$T_{\rm rot}$ \tablefootmark{d}  & $D$ \tablefootmark{e} & \multicolumn{1}{c}{$v_{\rm off}$ \tablefootmark{f}} & $\Delta v$ \tablefootmark{g} \\
               &               &                                    & (cm$^{-2}$)                                   &  &  (K)                                        &   ($\arcsec$)                                      &  (km s$^{-1}$)                           &  (km s$^{-1}$)    \\
    \hline
    \hline
N3         & \ce{C2H5CN}, $\varv=0$   &  55  & 3.2$\times$10$^{17}$ & 1.54  & 170  &  0.4 & -0.3 & 5.5  \\
           & \ce{C2H3CN}, $\varv=0$   &  40  & 5.2$\times$10$^{16}$ & 1.00  & 150  &  0.4 & 0.0  & 7.0  \\
           & $^{13}$\ce{CH3CN}, $\varv=0$ & 8 & 3.2$\times$10$^{16}$ & 1.06  & 145  &  0.4 & 0.0  & 5.0 \\
           & \ce{CH3}$^{13}$CN, $\varv=0$ & 8 & 3.2$\times$10$^{16}$ & 1.06  & 145  &  0.4 & 0.0  & 5.0 \\
           & \ce{CH3OH}, $\varv_{\rm t}=0$    &  31  & 7.5$\times$10$^{18}$ & 1.00  & 170  &  0.4 & 0.0  & 5.0  \\
           & \ce{C2H5OH}, $\varv=0$   &  95  & 3.1$\times$10$^{17}$ & 1.24  & 145  &  0.4 & 1.0  & 4.0 \\
           & \ce{CH3OCHO}, $\varv=0$  &  133  & 1.7$\times$10$^{18}$ & 1.19  & 145  &  0.4 & 0.6  & 4.1  \\
           & \ce{CH3OCH3}, $\varv=0$  &  43  & 1.0$\times$10$^{18}$ & 1.00  & 145  &  0.4 & 0.0  & 5.0  \\
           & \ce{CH3CHO}, $\varv=0$   &  15  & 8.5$\times$10$^{16}$ & 1.00  & 145  &  0.4 & 0.0  & 4.7  \\
           & \ce{NH2CHO}, $\varv=0$   &  10  & 3.5$\times$10$^{16}$ & 1.09  & 145  &  0.4 & 0.5  & 5.6 \\
           & \ce{CH3NCO}, $\varv=0$   &  57  & 6.6$\times$10$^{16}$ & 1.00  & 145  &  0.4 & 0.5  & 4.0 \\
           & \ce{CH3SH}, $\varv=0$    &  7   & 8.5$\times$10$^{16}$ & 1.00  & 145  &  0.4 & 0.0  & 4.0 \\
           &  &  &  &  &  &  &  &  \\       
N4         & \ce{C2H5CN}, $\varv=0$   &  37  & 1.4$\times$10$^{16}$ & 1.35  & 145  &  1.0 & -0.5 & 5.5  \\
           & \ce{C2H3CN}, $\varv=0$   &  17   & 2.0$\times$10$^{15}$ & 1.00  & 145  &  1.0 & -0.6 & 4.5  \\
           & $^{13}$\ce{CH3CN}, $\varv=0$ & 7 & 2.1$\times$10$^{15}$ & 1.06  & 145  &  1.0 & 0.0  & 5.3 \\
           & \ce{CH3}$^{13}$CN, $\varv=0$ & 8 & 2.1$\times$10$^{15}$ & 1.06  & 145  &  1.0 & 0.0  & 5.3 \\
           & \ce{CH3OH}, $\varv_{\rm t}=0$ &  17  & 2.5$\times$10$^{17}$ & 1.00  & 190  &  1.0 & -0.6 & 5.0  \\
           & \ce{C2H5OH}, $\varv=0$   &  18  & 2.5$\times$10$^{16}$ & 1.24  & 145  &  1.0 & -0.3 & 3.5  \\ 
           & \ce{CH3OCHO}, $\varv=0$  &  62  & 1.1$\times$10$^{17}$ & 1.19  & 145  &  1.0 & 0.0  & 4.5  \\
           & \ce{CH3OCH3}, $\varv=0$  &  21  & 9.0$\times$10$^{16}$ & 1.00  & 145  &  1.0 & 0.0  & 5.0  \\
           & \ce{CH3CHO}, $\varv=0$   &  14  & 1.0$\times$10$^{16}$ & 1.00  & 145  & 1.0  & -0.5 & 4.0  \\
           & \ce{NH2CHO}, $\varv=0$   &  \_  & $<$1.4$\times$10$^{15}$ & 1.09  & 145  &  1.0  & 0.5  & 5.0 \\       
           & \ce{CH3NCO}, $\varv=0$   &  28  & 8.0$\times$10$^{15}$ & 1.00  & 145  & 1.0  & 0.0  & 4.0 \\
           & \ce{CH3SH}, $\varv=0$    & 2    & 9.5$\times$10$^{15}$ & 1.00  & 145  &  1.0 & 0.0  & 4.0 \\
           &  &  &  &  &  &  &  &  \\      
N5         & \ce{C2H5CN}, $\varv=0$   &  47  & 6.8$\times$10$^{16}$ & 1.35  & 145  & 1.0  & -0.5  & 5.5  \\    
           & \ce{C2H3CN}, $\varv=0$   &  34  & 1.3$\times$10$^{16}$ & 1.00  & 145  & 1.0  & -0.6  & 6.0  \\
           & $^{13}$\ce{CH3CN}, $\varv=0$ & 7 & 6.5$\times$10$^{15}$ & 1.06  & 145  & 1.0  & -0.6  & 4.6 \\
           & \ce{CH3}$^{13}$CN, $\varv=0$ & 8 & 6.5$\times$10$^{15}$ & 1.06  & 145  & 1.0  & -0.6  & 4.6 \\
           & \ce{CH3OH}, $\varv_{\rm t}=0$    &  29  & 2.0$\times$10$^{18}$ & 1.00  & 190  & 1.0  & -0.7  & 4.5  \\
           & \ce{C2H5OH}, $\varv=0$   &  74  & 1.0$\times$10$^{17}$ & 1.24  & 145  & 1.0  & -0.8  & 3.5 \\
           & \ce{CH3OCHO}, $\varv=0$  &  79  & 2.4$\times$10$^{17}$ & 1.19  & 145  & 1.0  & -0.5  & 3.5  \\
           & \ce{CH3OCH3}, $\varv=0$  &  53  & 4.5$\times$10$^{17}$ & 1.00  & 145  & 1.0  & 0.0  & 4.5  \\
           & \ce{CH3CHO}, $\varv=0$   &  16   & 2.5$\times$10$^{16}$ & 1.00  & 145  & 1.0  &  0.0  & 4.5  \\
           & \ce{NH2CHO}, $\varv=0$   &  10  & 1.4$\times$10$^{16}$ & 1.09  & 145  & 1.0  & -0.5  & 6.0 \\
           & \ce{CH3NCO}, $\varv=0$   &  14   & 1.1$\times$10$^{16}$ & 1.00  & 145  & 1.0  & -0.5  & 3.5 \\
           & \ce{CH3SH}, $\varv=0$    &  3   & 3.5$\times$10$^{16}$ & 1.00  & 145  & 1.0  & -0.5  & 5.0 \\
           &  &  &  &  &  &  &  &  \\    
 
N2  \tablefootmark{*}  & \ce{C2H5CN}, $\varv=0$ & 154 & 6.2$\times$10$^{18}$ & 1.38 & 150  &  1.2  & -0.8  & 5.0  \\
           & \ce{C2H3CN}, $\varv=0$   &  44   & 4.2$\times$10$^{17}$ & 1.00  & 200  & 1.1  & -0.6  & 6.0  \\
           & $^{13}$\ce{CH3CN}, $\varv=0$ & 8  & 9.9$\times$10$^{16}$ & 1.10  & 170  & 1.4  & -0.5  & 5.4 \\
           & \ce{CH3}$^{13}$CN, $\varv=0$ & 7  & 9.9$\times$10$^{16}$ & 1.10  & 170  & 1.4  & -0.5  & 5.4 \\
           & \ce{CH3OH}, $\varv_{\rm t}=0$    & 41    & 4.0$\times$10$^{19}$ & 1.00  & 160  & 1.4  & -0.5  & 5.4  \\
           & \ce{C2H5OH}, $\varv=0$   & 168   & 2.0$\times$10$^{18}$ & 1.24  & 150  & 1.5  & -0.4  & 4.7 \\
           & \ce{CH3OCHO}, $\varv=0$  & 90    & 1.2$\times$10$^{18}$ & 1.23  & 150  & 1.5  & -0.4  & 4.7  \\
           & \ce{CH3OCH3}, $\varv=0$  & 66   & 2.2$\times$10$^{18}$   & 1.00  & 130   & 1.6   & -0.5   & 5.0   \\
           & \ce{CH3CHO}, $\varv=0$     & 19  & 4.3$\times$10$^{17}$ & 1.00  & 150  &  1.2 &  0.0  & 5.6 \\
           &  \ce{NH2CHO}, $\varv=0$  &  30   & 3.5$\times$10$^{18}$ & 1.17  & 200  & 0.8  & 0.2   & 5.5  \\
           & \ce{CH3NCO}, $\varv=0$  &  60   & 2.2$\times$10$^{17}$ & 1.00  & 150  &  0.9 & -0.6  & 5.0  \\
           & \ce{CH3SH}, $\varv=0$    & 12    & 3.4$\times$10$^{17}$  & 1.00  & 180  & 1.4  & -0.5  & 5.4 \\
    \hline
\end{tabular}
\end{center}
\vspace{-4mm}
\tablefoot{\tablefoottext{a}{Number of lines detected above 3$\sigma$ (Table~\ref{TAB-appendix-data-redu}). One line may mean a group of transitions of the same molecule blended together.}
\tablefoottext{b}{Total column density of the molecule.}
\tablefoottext{c}{Correction factor applied to the column density to account for the contribution of vibrationally or torsionally excited states not included in the partition function.}
\tablefoottext{d}{Rotational temperature (Sect.~\ref{section-obs-constraints}).}
\tablefoottext{e}{Source diameter (FWHM, Sect~\ref{section-obs-constraints}) .}
\tablefoottext{f}{Velocity offset with respect to the assumed systemic velocity of the source: 74~km s$^{-1}$ for Sgr~B2(N3), 64~km s$^{-1}$ for N4, and 60~km s$^{-1}$ for N5 \citep{bonfand2017}.}
\tablefoottext{g}{Linewidth \citep[FWHM,][]{bonfand2017}.}
\tablefoottext{*}{Sgr~B2(N2)'s parameters are taken from \citet{belloche2016, belloche2017, muller2016} for all species, except for \ce{CH3CHO}, \ce{CH3OCH3} (Belloche, priv. comm.)}}
\end{table*} 

\begin{table}[!h]
  \caption{\label{TAB-appendix-H2} H$_2$ column densities used to derive chemical abundances.} 
\begin{center}
  \setlength{\tabcolsep}{1.5mm}
  \begin{tabular}{lcc}
    \hline
Source &  $\theta_s$\tablefootmark{a} & $N_{\rm H_2}$\tablefootmark{b}    \\  
       & ($\arcsec$)&   (10$^{24}$ cm$^{-2}$) \\
    \hline
    \hline 
N2 & 1.2(0.4) & 1.64(0.53)    \\
N3\tablefootmark{*} & 0.4(0.1) & 0.90(0.17)    \\
N4 & 0.7(0.1) & 2.55(0.66)    \\
N5 & 1.0(0.4) & 0.90(0.27)    \\
    \hline
\end{tabular}
\end{center} 
\tablefoot{Uncertainties are given in parentheses. 
\tablefoottext{a}{Sizes for which the new H$_2$ column densities are calculated as described in Sect.~3.8 of \citet{bonfand2017}.}
\tablefoottext{b}{H$_2$ column density calculated for the size indicated in the previous column. This value is used in Fig.~\ref{FIG-chemical-composition}b to plot chemical abundances relative to H$_2$. The uncertainties are calculated based on the uncertainties on the average H$_2$ column densities given in Table~\ref{TAB-HCs-properties} and on the source sizes.}
\tablefoottext{*}{In the case of N3 we use the H$_2$ column density given in Table~\ref{TAB-HCs-properties}, measured in the 1~mm ALMA continuum map at 0.4$\arcsec$ resolution (Table~\ref{TAB-HCs-properties}).}}
\end{table}

\begin{table}[!h]
\begin{center}
\caption{\label{TAB-power-law-index} Indexes of power-law function.}
\begin{tabular}{ccc}
\hline
 & \multicolumn{2}{c}{Index of power-law function} \\
Source & $n_{\rm H}(t)$ & $A_{\rm v}(t)$ \\
\hline
\hline
N2 & -0.17 & -0.32 \\
N3 & -0.57 & -0.43 \\
N4 & -0.10 & -0.20 \\
N5 & -0.16 & -0.27 \\
\hline
\end{tabular}
\end{center}
\tablefoot{Indexes of the power law functions, $n_{\rm H}(t)$ and $A_{\rm v}(t)$, plotted in Fig.~\ref{FIG-appendix-nH-Av-time}.}
\end{table}

\clearpage

\section{Other complementary figures}
        \label{appendix-figures}

\begin{figure}[!h]
   \begin{center}
    \includegraphics[width=\hsize]{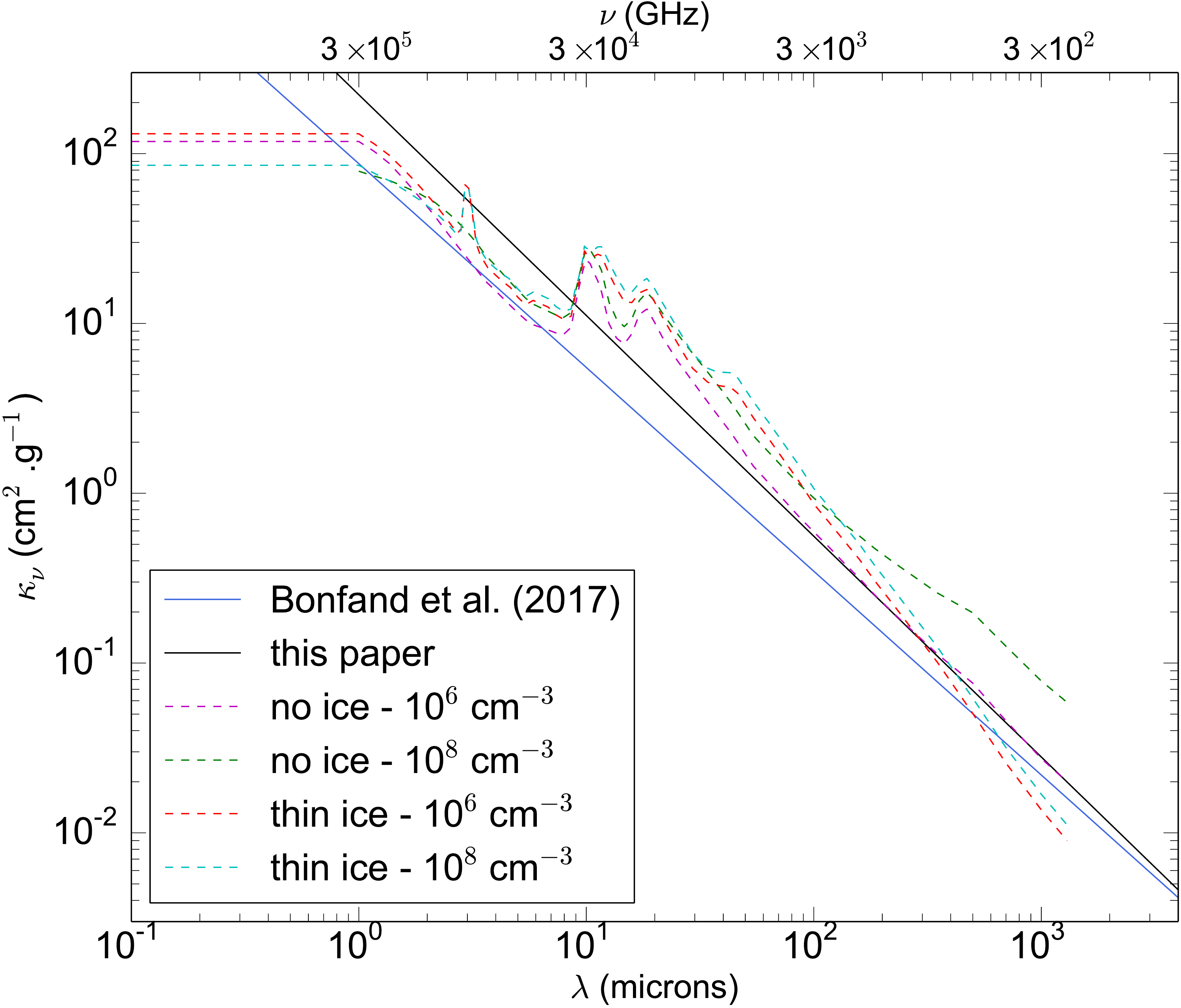} 
    \caption{\label{FIG-appendix-kappa-opacities} Dust mass opacity spectra for a standard gas-to-dust mass ratio of 100. The dashed lines show the results of \citet{ossenkopf1994}'s simulations after 10$^5$~yr of dust coagulation with gas densities of 10$^6$~cm$^{-3}$ and 10$^8$~cm$^{-3}$ and for dust grains without ice mantle and with thin ice mantle. The blue solid line shows the $\kappa_{\nu}$ (of gas) used to compute \ce{H2} column densities in \citet{bonfand2017}. The black solid line shows the $\kappa_{\nu}$ (of gas) used in this paper, computed with Eq.~\ref{eq-kappa}, and based on \citet{ossenkopf1994}'s model for grains without ice mantle and for gas densities of 10$^6$~cm$^{-3}$ (magenta dashed line).}
   \end{center}
\end{figure}

\begin{figure}[!h]
   \begin{center}
    \includegraphics[width=\hsize]{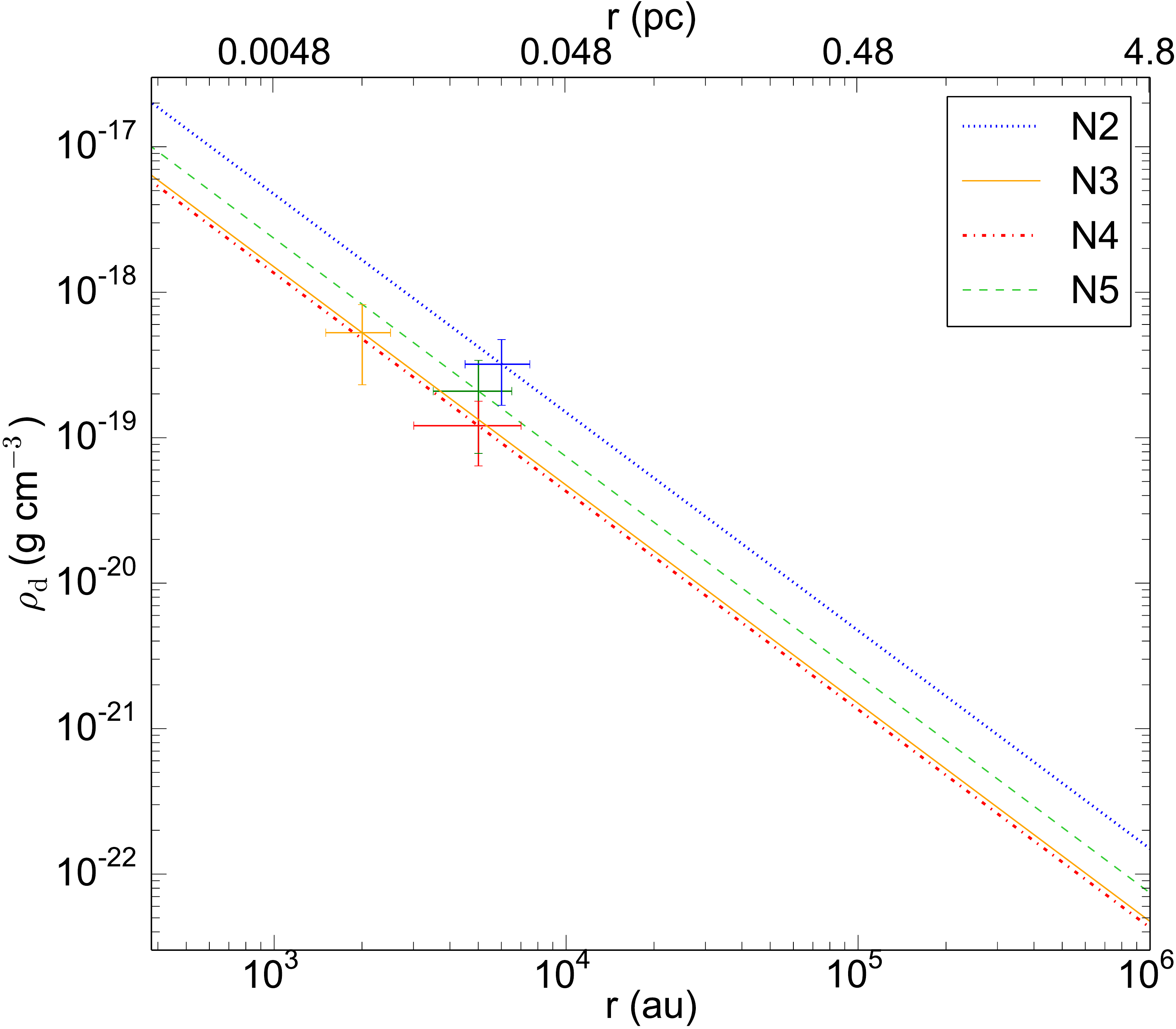} 
    \caption{\label{FIG-appendix-dust-profiles} Dust mass density profiles derived for Sgr~B2(N2-N5) assuming $\rho_d$ $\propto$ $r^{-1.5}$ \citep{shu1977}. For each source the observational constraint, $\rho$($r_0$)~=~$\rho_0$ is plotted with errorbars (1$\sigma$).}
   \end{center}
\end{figure}

\begin{figure}[!h]
   \begin{center}
    \includegraphics[width=\hsize]{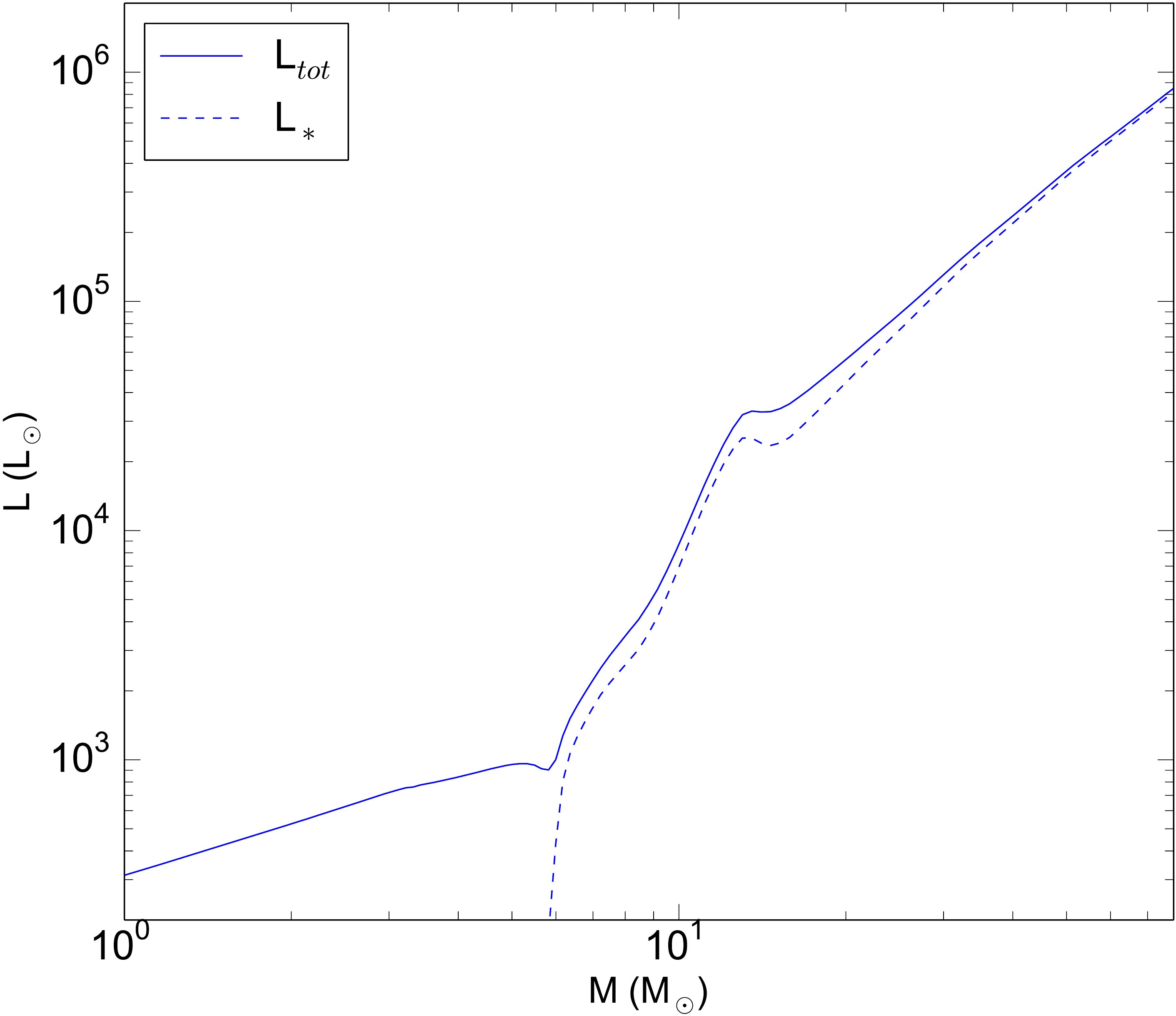} 
    \caption{\label{FIG-appendix-M-L-relation} Evolution of the total luminosity (L$_*$ + L$_{\rm acc}$) for a young high-mass protostar assuming a constant accretion rate of $\dot{M}$ = 10$^{-4}$~$\Msol$~yr$^{-1}$ \citep[Hosokawa, priv. comm., based on][]{hosokawa2009}.}
   \end{center}
\end{figure}

\begin{figure}[!h]
   \begin{center}
    \includegraphics[width=\hsize]{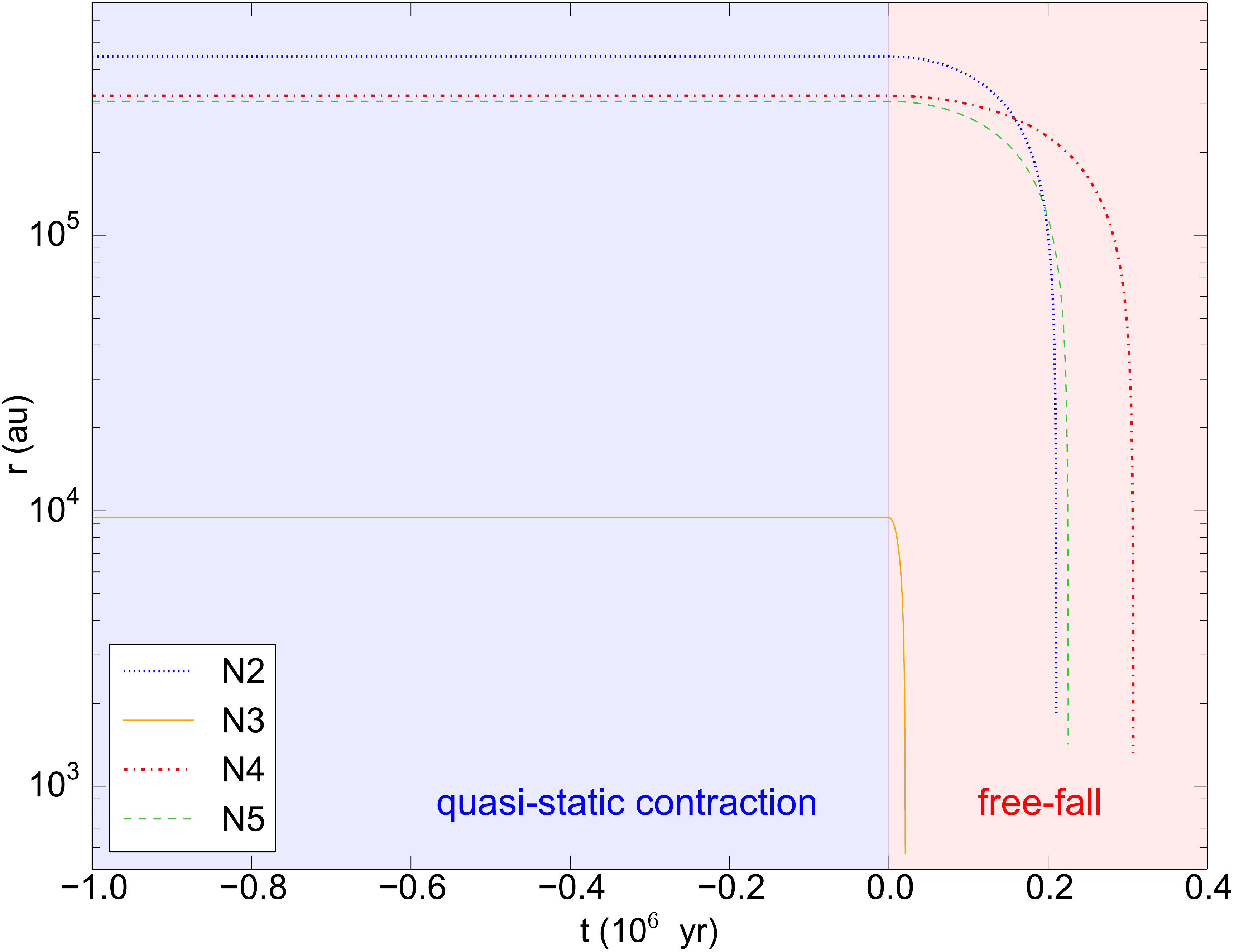} 
    \caption{\label{FIG-appendix-radius-time} Trajectory of a parcel of gas as function of time in the envelopes of Sgr~B2(N2-N5). For each source the parcel of gas is held at a constant radius, $r_{\rm start}$ (see Table~\ref{TAB-profile-parameters}), during the whole quasi-static contraction phase. During the free-fall collapse phase, the parcel of gas gradually falls toward the central protostar with the free-fall speed $v_{\rm ff}$ (Eq.~\ref{physical-profiles-eq2}).}
   \end{center}
\end{figure}

\begin{figure}[!h]
   \begin{center}
    \includegraphics[width=\hsize]{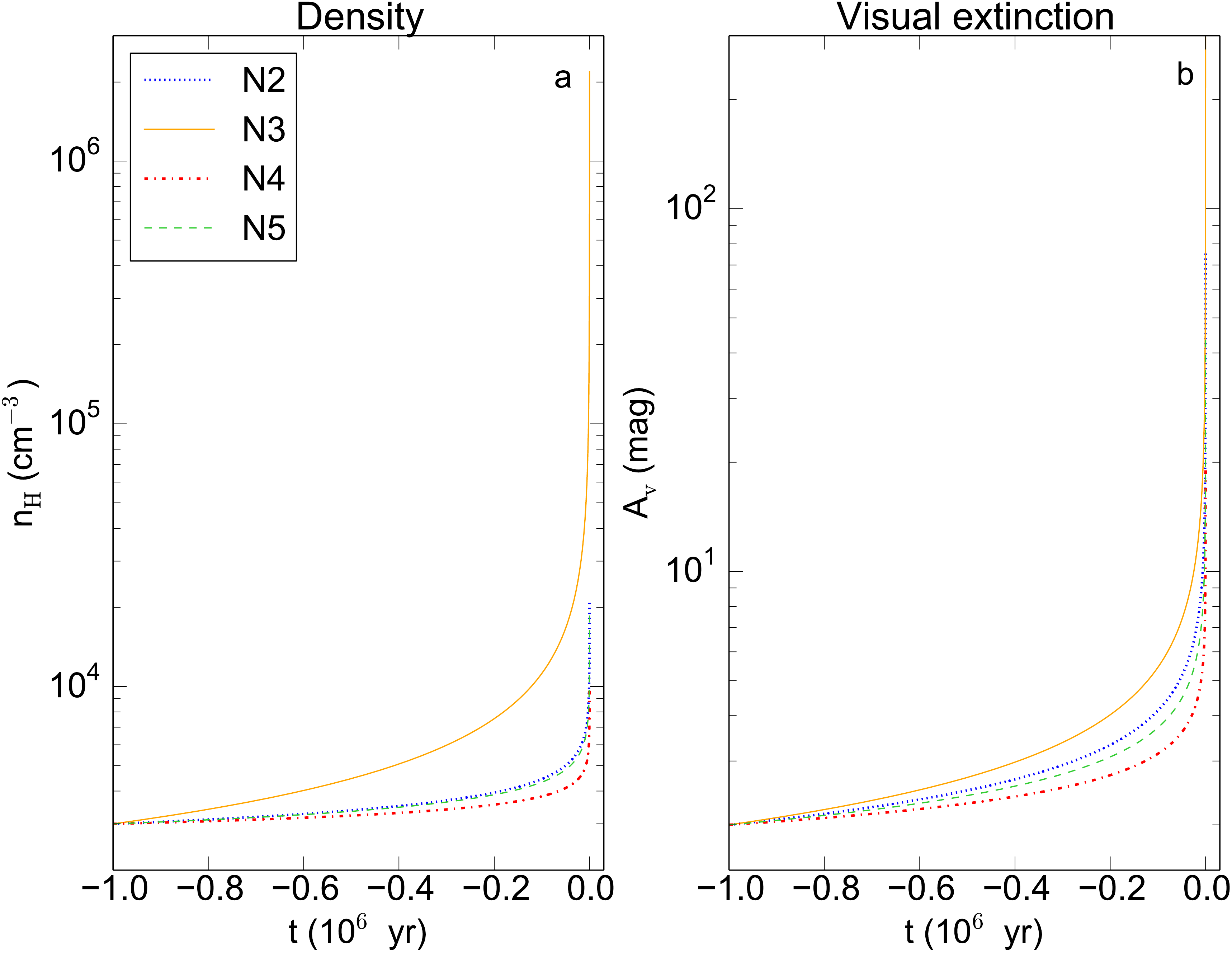} 
    \caption{\label{FIG-appendix-nH-Av-time} \textbf{a} Total hydrogen density as a function of time during the quasi-static contraction phase for Sgr~B2(N2-N5) for a parcel of gas at a fixed radius, $r = r_{\rm start}$ (see Sect.~\ref{section-collapse-phase} and Table~\ref{TAB-power-law-index}). \textbf{b} Same as (a) but for the visual extinction.}
   \end{center}
\end{figure}

\begin{figure*}[!h]
   \resizebox{\hsize}{!}
   {\begin{tabular}{cc}
       \includegraphics[width=\hsize]{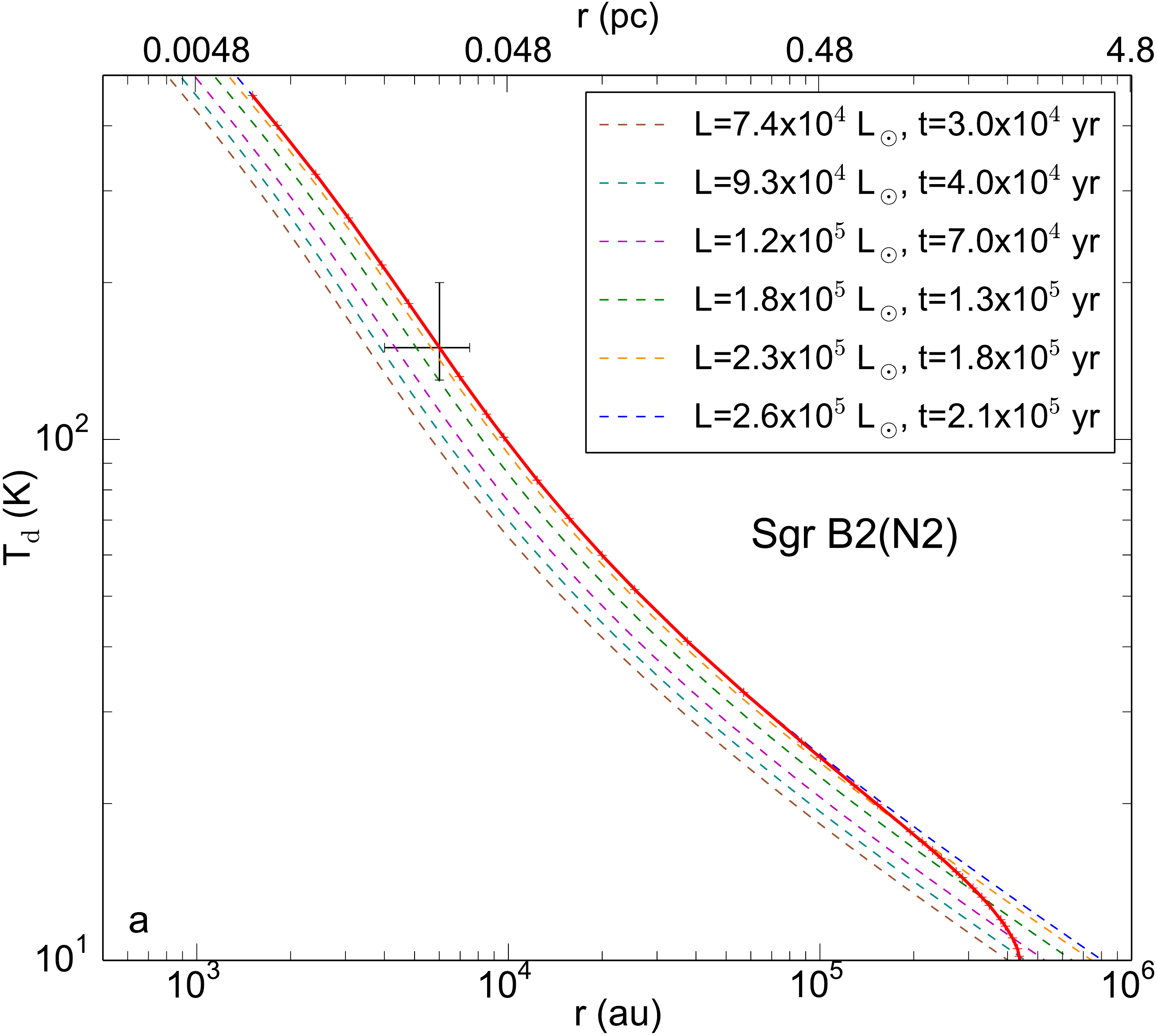} &
       \includegraphics[width=\hsize]{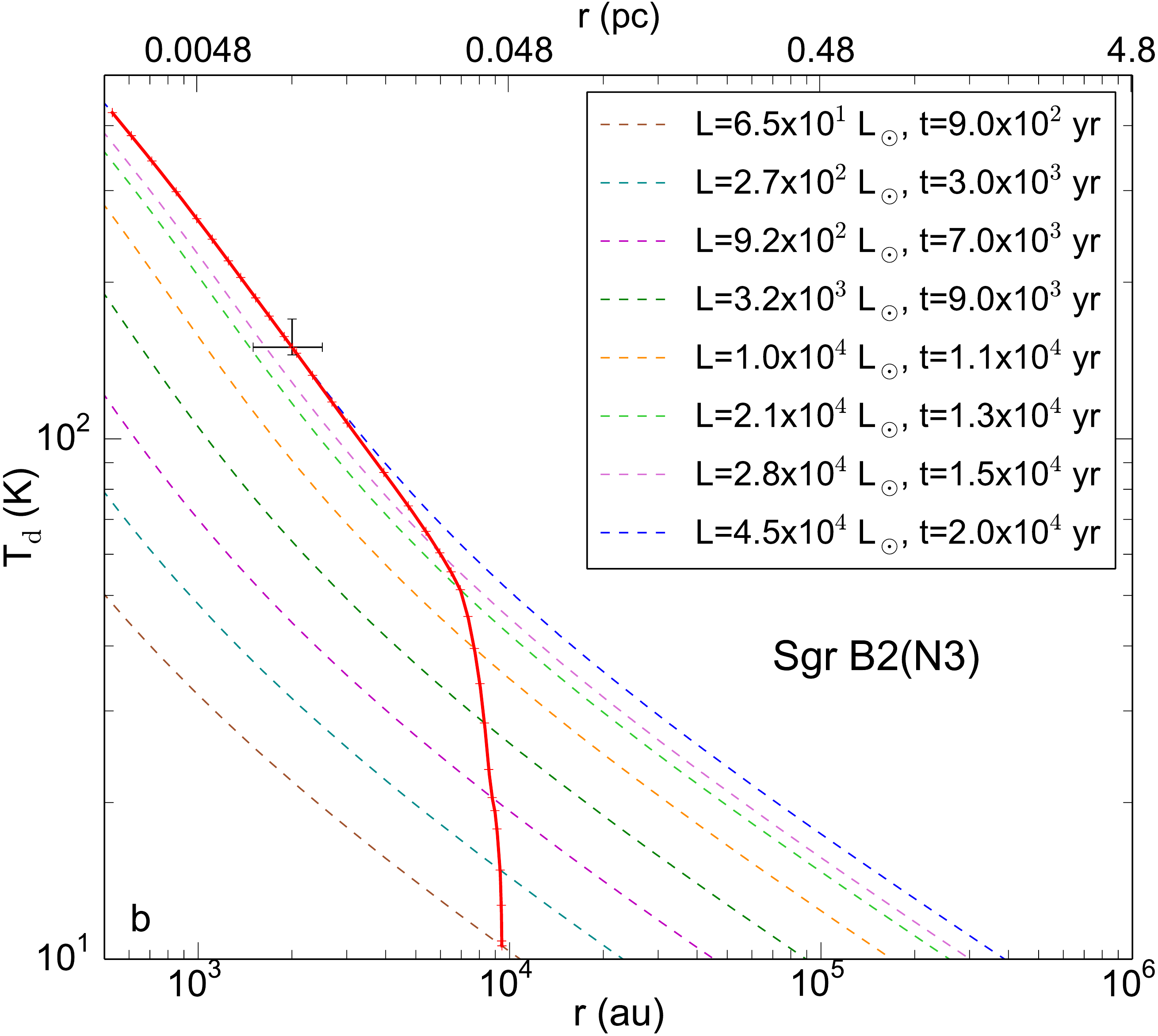}  \\
       \includegraphics[width=\hsize]{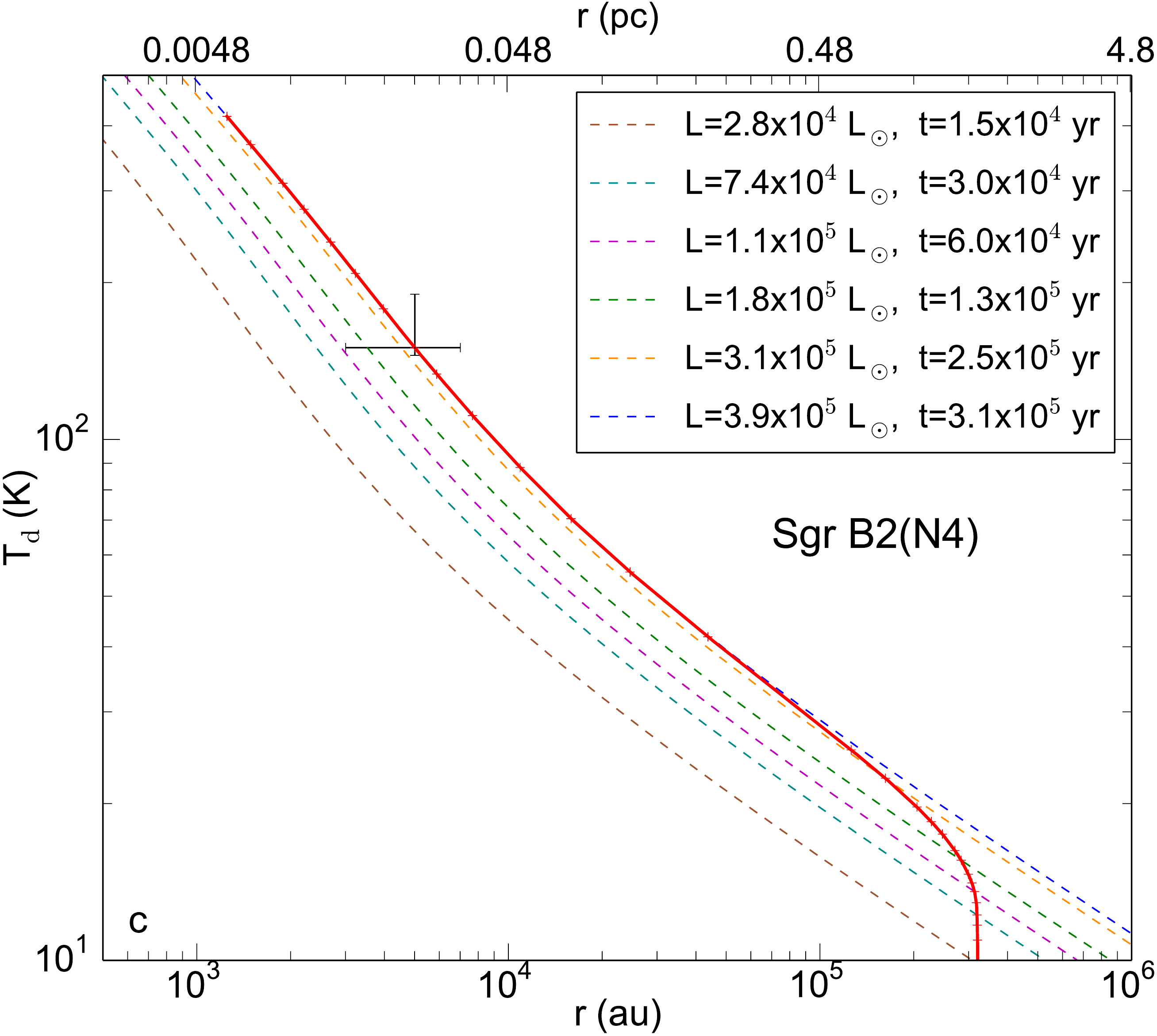} &
       \includegraphics[width=\hsize]{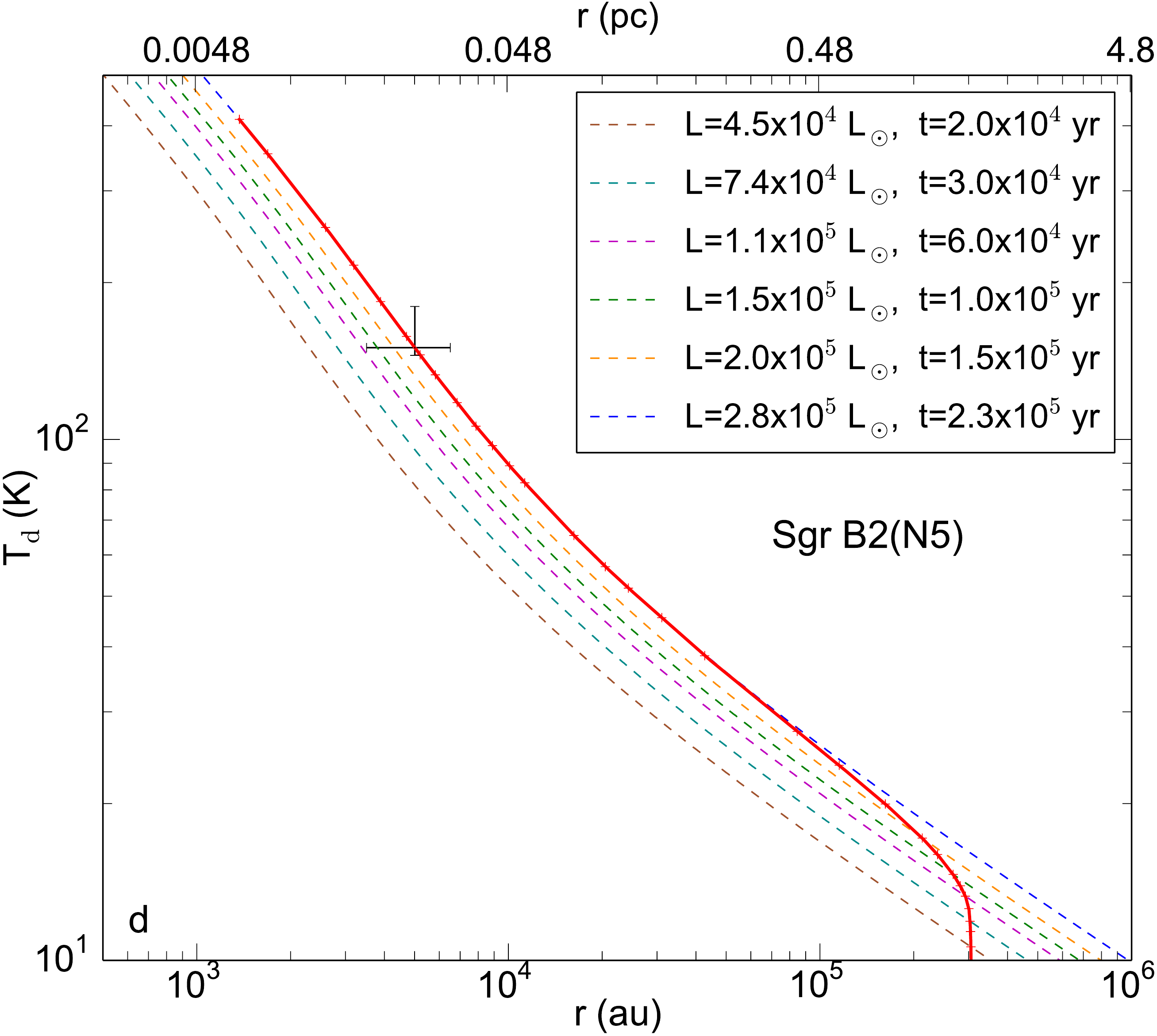}  \\
    \end{tabular}}
    \caption{\label{FIG-appendix-interpolated-T-profiles} Dust temperature evolution along the trajectory of a parcel of gas infalling toward the central protostar during the free-fall collapse phase for Sgr~B2(N2-N5) (red line). In each panel the dashed lines show the dust temperature profiles computed using RADMC-3D with different luminosities (that is at different evolutionary stages). They show how the dust temperature at a given radius in the envelope of each source increases as the total luminosity of the central protostar rises with time (as indicated in the upper right corner). The observational constraint, $\mathbf{T_{\rm d}}(r_0)$~=~$T_0$, used to derive the current luminosity of the source is plotted with errorbars (1$\sigma$).}
\end{figure*}

\begin{figure*}[!h]
\begin{center}
       \includegraphics[width=\hsize]{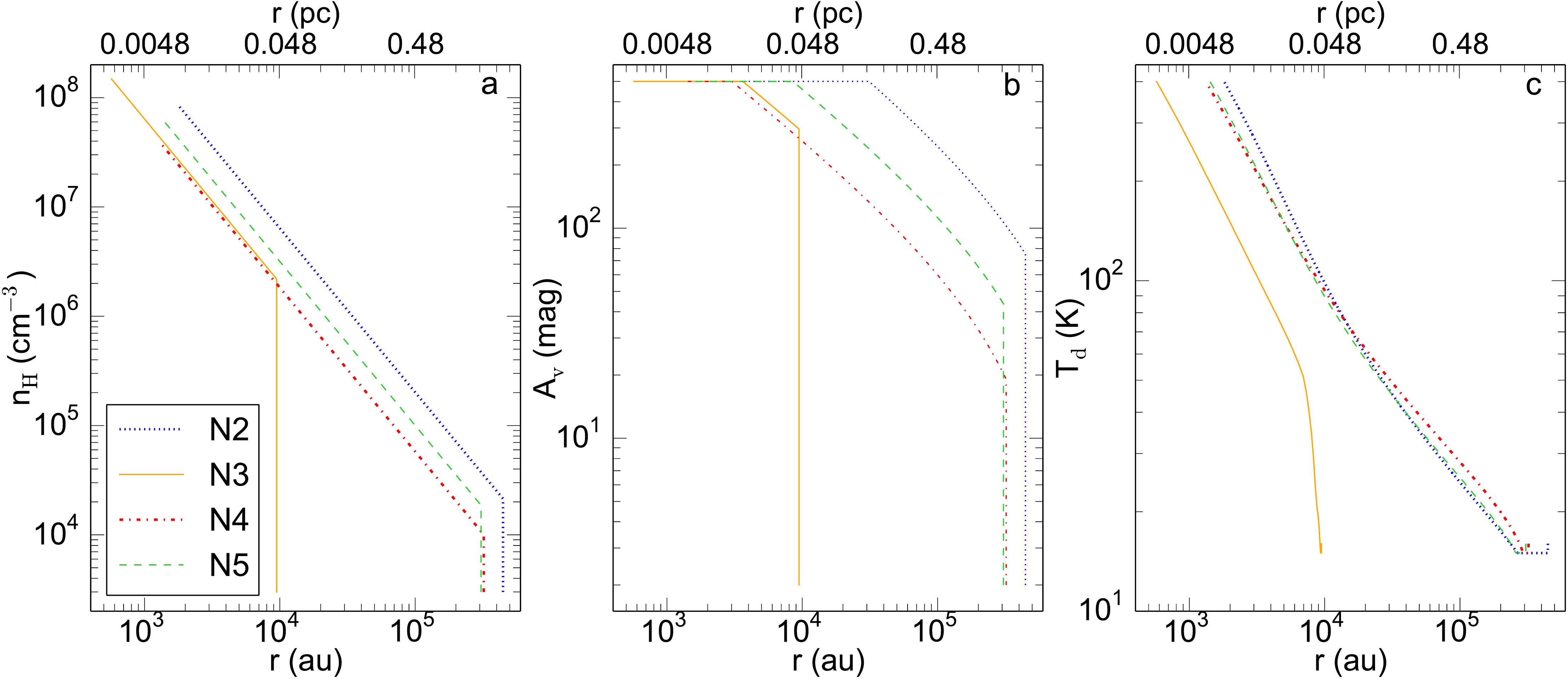} 
 \caption{\label{FIG-appendix-all-physical-profiles}\textbf{a} Gas densities along the trajectory of a parcel of gas gradually infalling through the envelopes of Sgr~B2(N2-N5). The quasi-static contraction phase is characterized by the density increasing over time at a constant radius (Fig.~\ref{FIG-appendix-nH-Av-time}a). The subsequent free-fall collapse phase assumes a density distribution propotional to $r^{-1.5}$. \textbf{b} Visual extinction calculated as a function of the density along the trajectory of a parcel of gas gradually infalling through the envelope. The quasi-static contraction phase is characterized by the extinction increasing over time at a constant radius (Fig.~\ref{FIG-appendix-nH-Av-time}b). During the free-fall collapse phase the extinction increases with the density (Eq.~\ref{physical-profiles-eq7}) until $A_{\rm v-max}$~=~500~mag. \textbf{c} Temperature evolution along the trajectory of a parcel of gas gradually infalling through the envelope (see also Fig.~\ref{FIG-appendix-interpolated-T-profiles}). The evolution of the temperature during the quasi-static contraction phase is shown for a minimum temperature $T_{\rm min}$~=~15~K (see Fig.~\ref{FIG-Td-time-pre-collapse}).}
\end{center}
\end{figure*}

\begin{figure*}[!h]
       \includegraphics[width=\hsize]{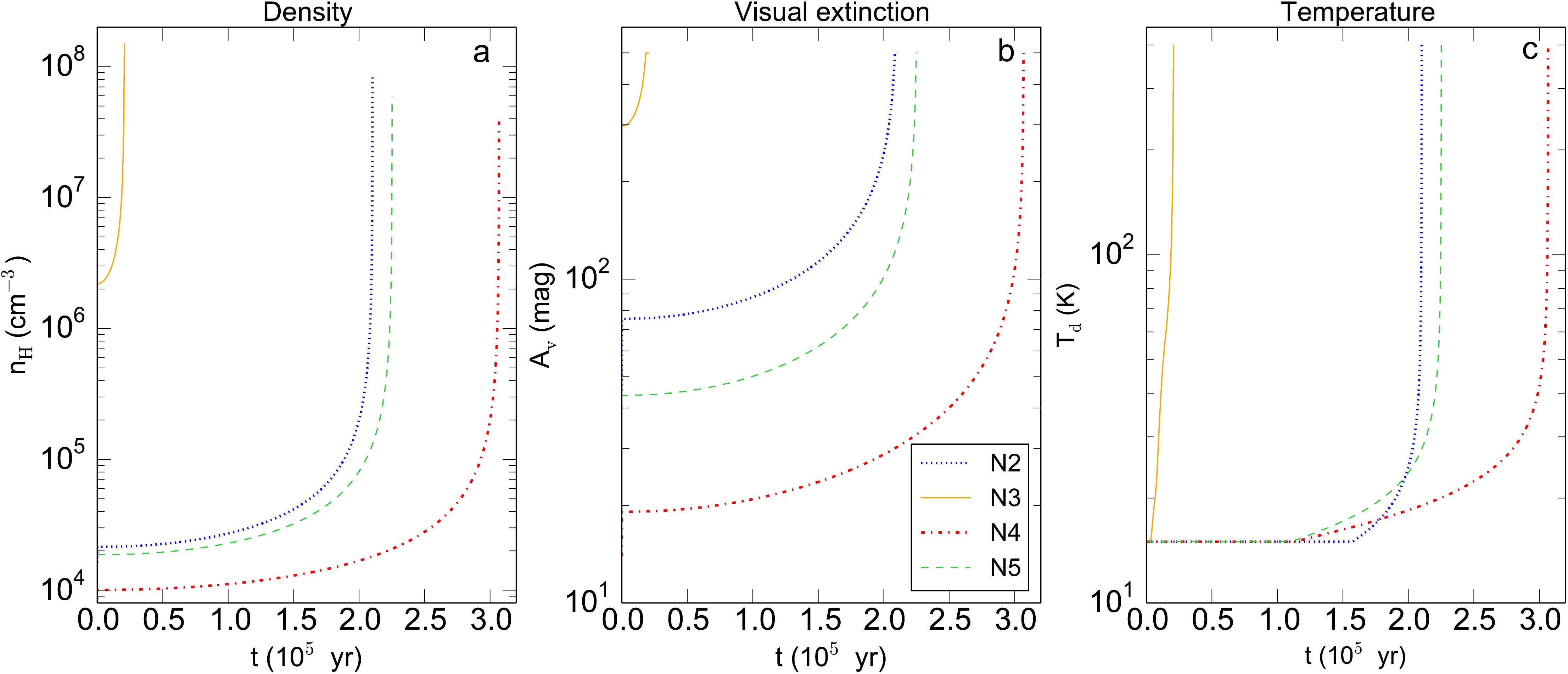} 
 \caption{\label{FIG-appendix-profiles-time}  \textbf{a} Gas densities as a function of time along the trajectory of a parcel of gas gradually infalling through the envelope of Sgr~B2(N2-N5) (see also Fig.~\ref{FIG-appendix-radius-time}). \textbf{b} Same as (a) for the visual extinction. \textbf{c} Same as (a) for the dust temperature for $T_{\rm min}$~=~15~K, up to $T_{\rm max}$~=~400~K.}
\end{figure*}

\begin{figure*}[!h]
   \resizebox{\hsize}{!}
   {\begin{tabular}{cc}
       \includegraphics[scale=0.5]{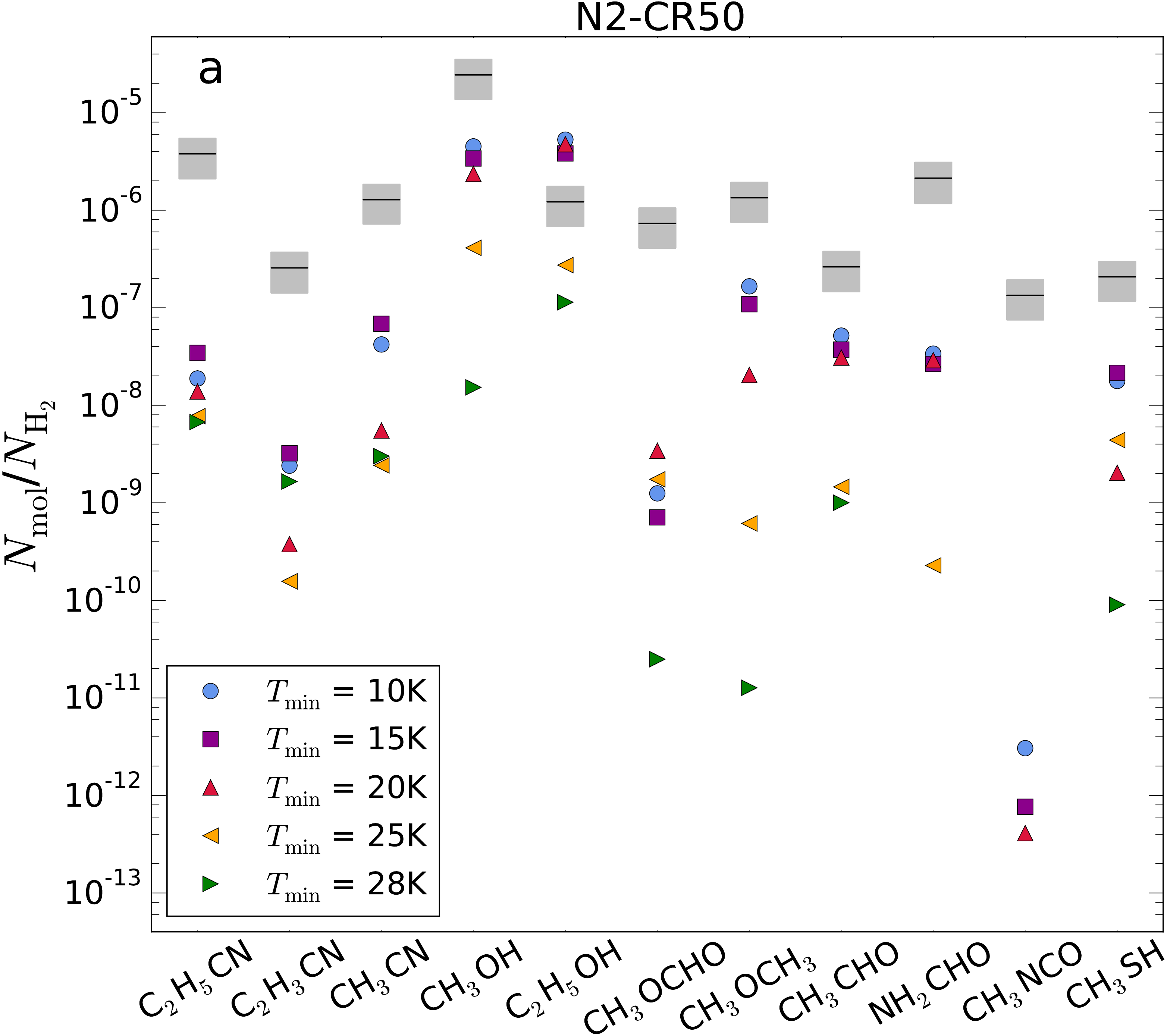} &
       \includegraphics[scale=0.5]{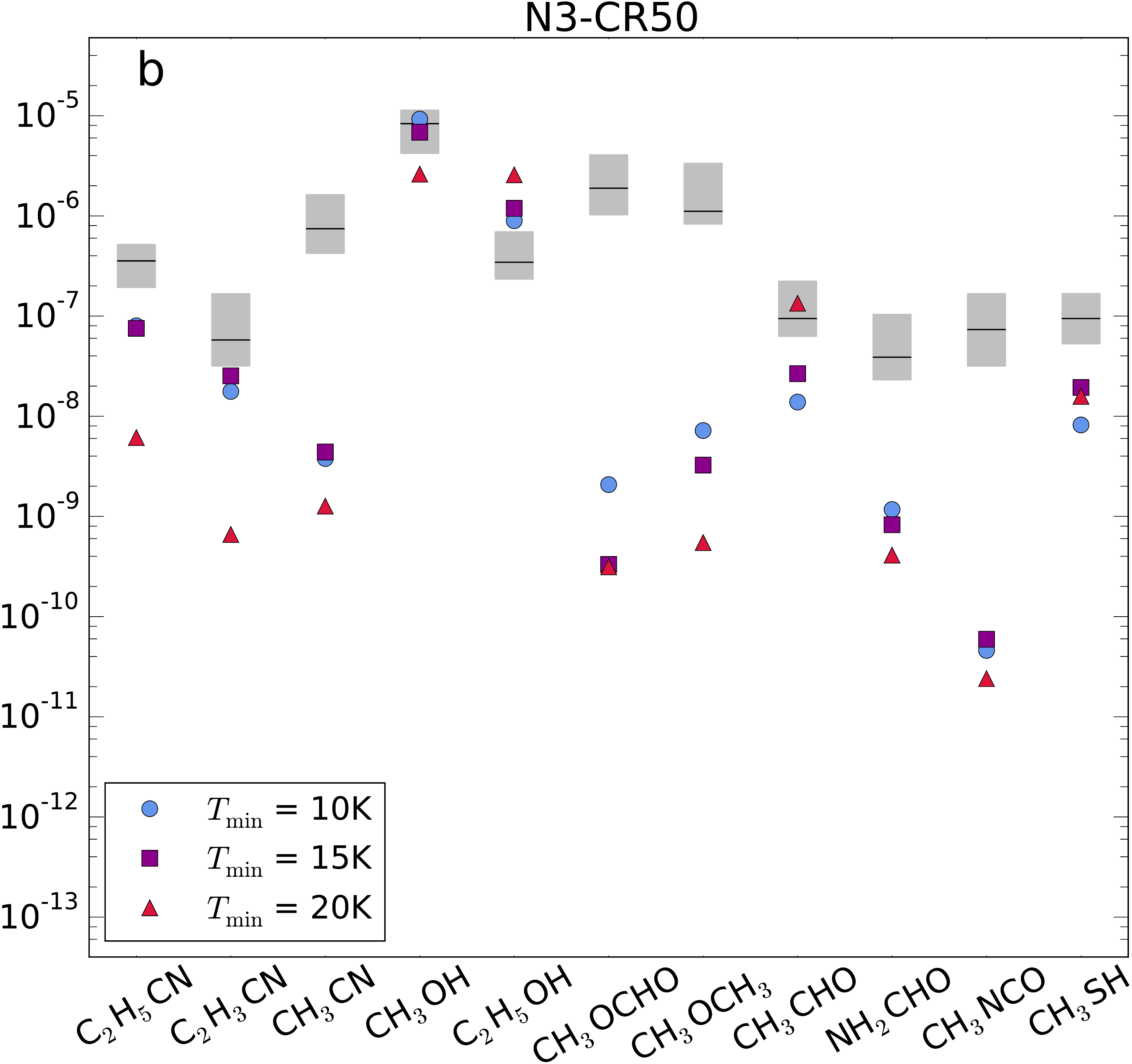}  \\
       \includegraphics[scale=0.5]{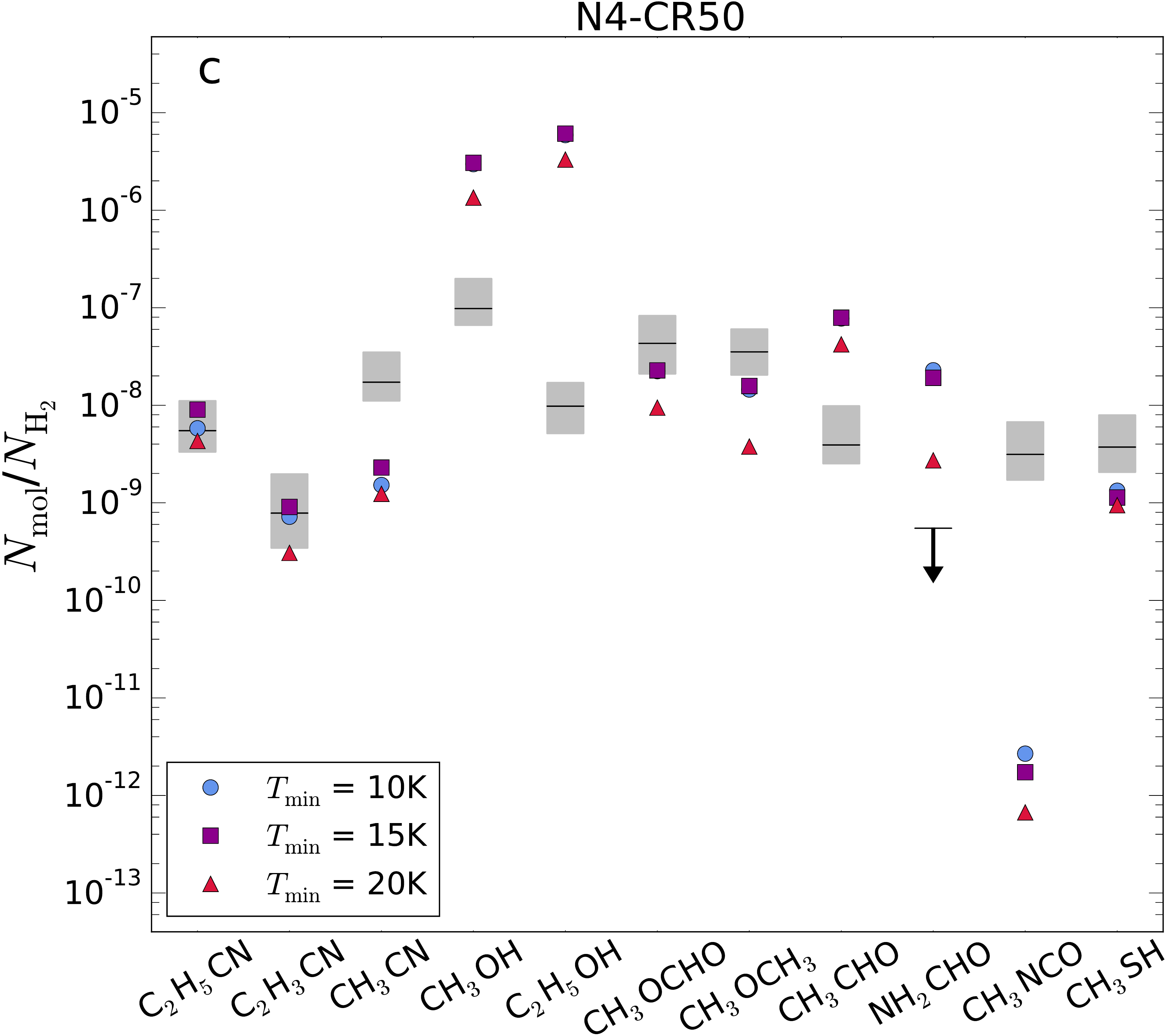} &
       \includegraphics[scale=0.5]{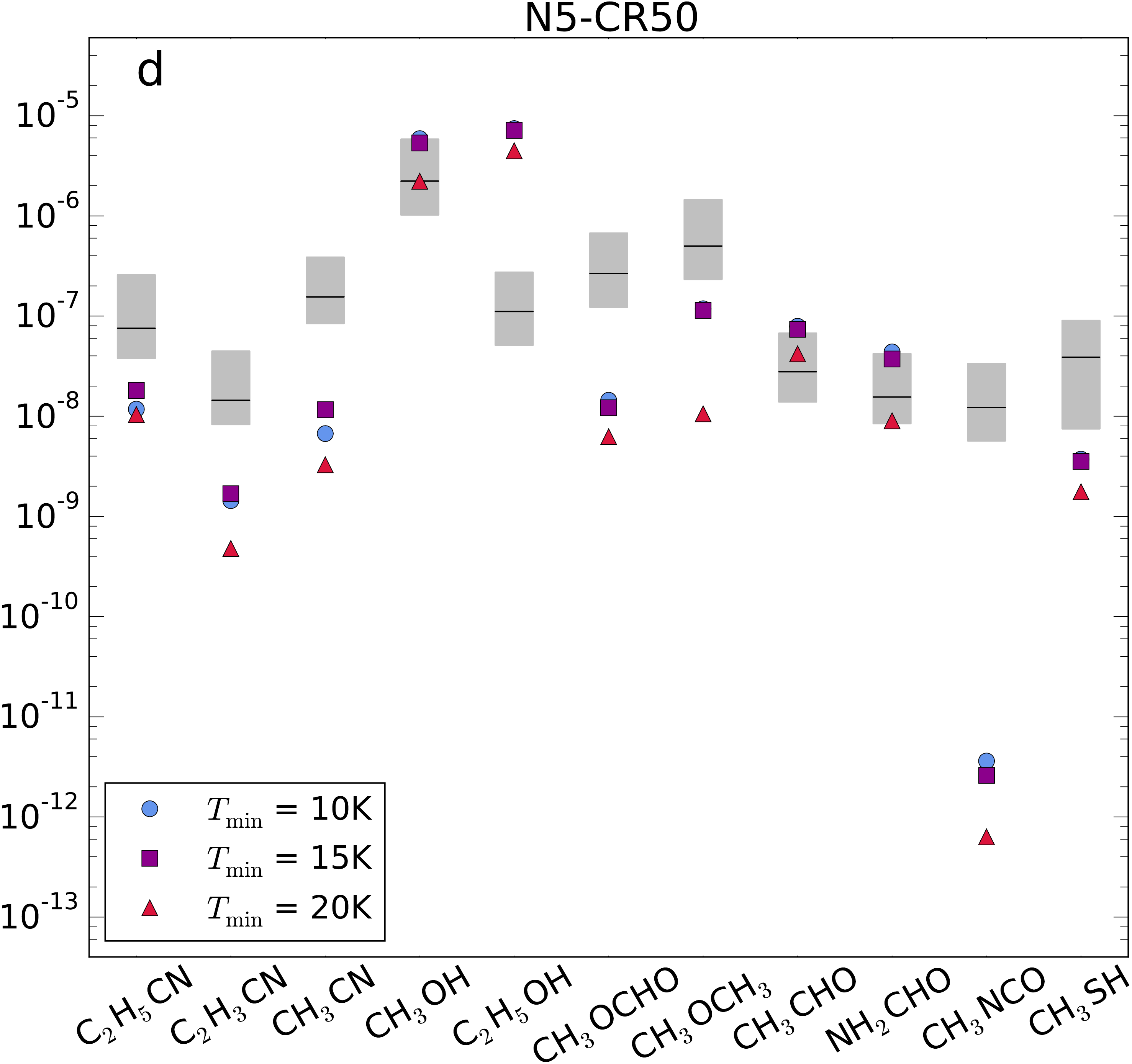} \\ 
    \end{tabular}}
    \caption{\label{FIG-comparison-H2} Abundances of 11 COMs with respect to H$_2$ for models N2-CR50 (a), N3-CR50 (b), N4-CR50 (c), and N5-CR50 (d). In each panel the different symbols indicate the abundances calculated at $T$~=~150~K for different minimum dust temperatures. In each panel the horizontal black lines show the observed abundances relative to H$_2$ (Table~\ref{TAB-appendix-H2}). The gray boxes show the 1$\sigma$ uncertainties. The arrow indicates an upper limit. In panel a the chemical abundances lower than 4$\times$10$^{-14}$ are not visible.}
\end{figure*}

\end{appendix}

\end{document}